\documentclass[12pt, twoside, letterpaper]{book}
\usepackage{graphicx,graphics}
\usepackage{amsmath}
\usepackage{amssymb}
\usepackage{epstopdf}
\usepackage{amstext}
\usepackage{amsthm}
\usepackage{amsfonts}
\usepackage{latexsym}
\usepackage{array}
\usepackage{xfrac}
\usepackage[use  names, dvipsnames]{color}
\usepackage{subfigure}
\usepackage{multirow}
\usepackage[T1]{fontenc}
\usepackage{bm}
\usepackage{adjustbox}
\usepackage{dcolumn}
\usepackage{soul}
\usepackage{fancyhdr}
\usepackage{hyperref}
\usepackage[Glenn]{fncychap}
\usepackage{emptypage}
\usepackage{cite}
\usepackage[toc,page]{appendix}
\usepackage{dcolumn}
\usepackage{afterpage}
\usepackage{eso-pic}
\usepackage{transparent}
\newcolumntype{d}{D{.}{.}{2.5}}
\newcolumntype{s}{D{.}{.}{2}}
\newcolumntype{p}{D{.}{.}{2}}

\usepackage[a4paper, width=160mm, top=25mm, bottom=25mm, bindingoffset=6mm]{geometry}

\def\blankpage{%
      \clearpage%
      \thispagestyle{empty}%
      \addtocounter{page}{0}%
      \null%
      \clearpage}

\font\myfont=cmr16 at 35pt

\title{
\vspace{-4cm}
{ \textbf{  \Huge NEUTRON-RICH MATTER IN ATOMIC NUCLEI AND NEUTRON STARS}}\\
\vspace{1cm}
{\bf \it{{\Large Mem\`oria presentada per optar al grau de doctor per la Universitat de Barcelona}}}\\
\vspace{2cm}
{ {\bf{\huge Author}} \\ \textsc{ \Large Claudia Gonz\'alez Boquera}}\\
\vspace{1cm}
{ {\bf{\huge Supervisors}} \\ \textsc{ \Large Mario Centelles Aixal\`a and Xavier Vi{\~n}as Gaus\'i}}\\
\vspace{1cm}
{ {\bf{\huge Tutor}} \\ \textsc{ \Large Dom\`enec Espriu Climent}}\\
\vspace{4cm}
{\huge September 2019}\\
\vspace{2cm}
{\includegraphics[clip=true,width=0.4\paperwidth]{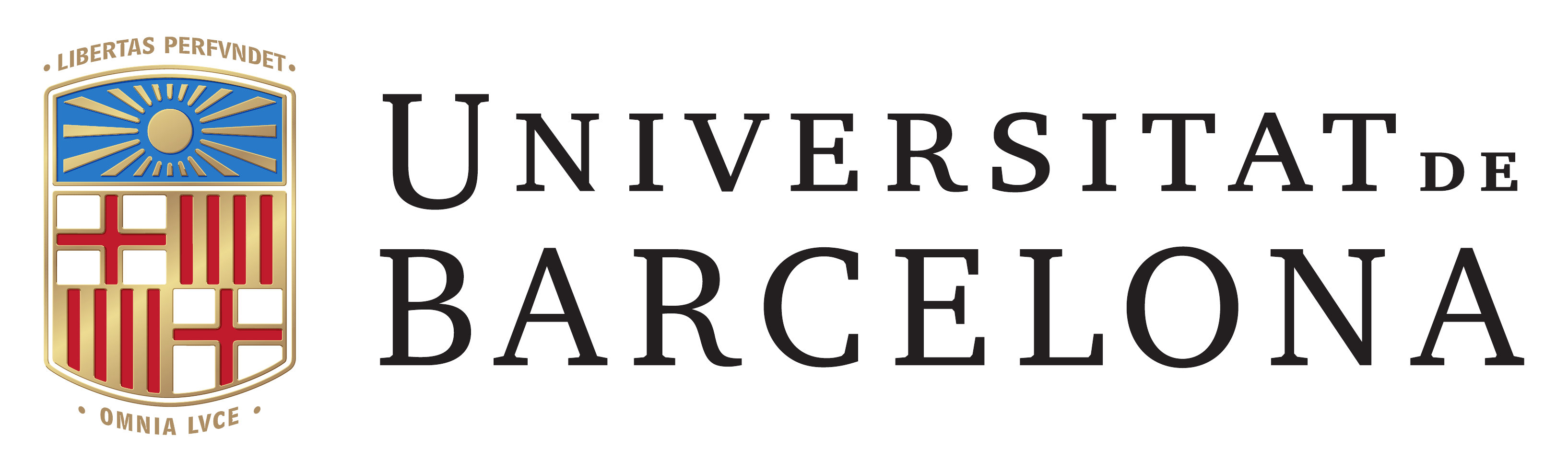}}
\hspace{1cm}
{\includegraphics[clip=true,width=0.25\paperwidth]{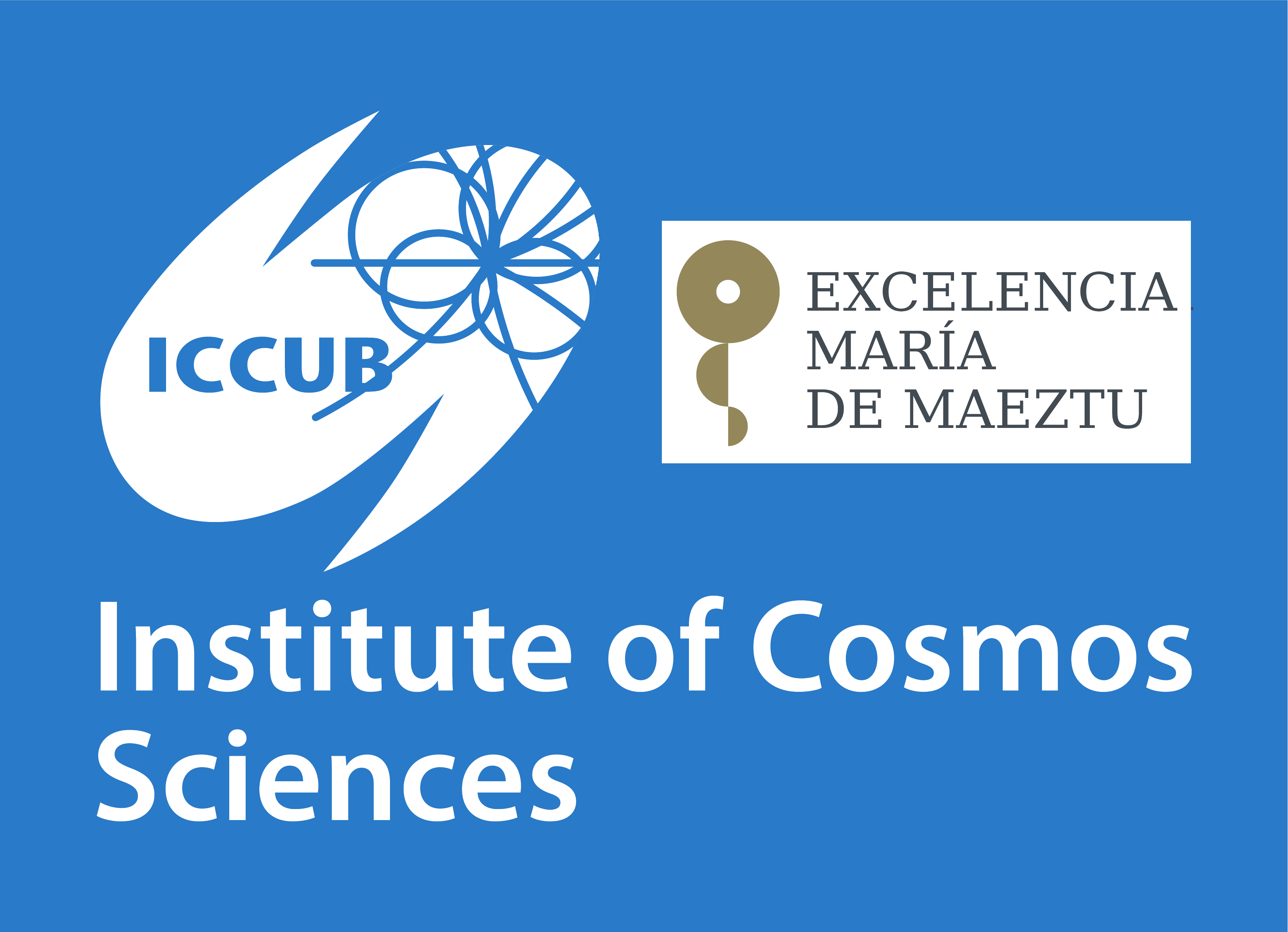}}
\vspace{-6cm}
}
\date{ }

\newcommand\BackgroundPic{%
\put(0,0){%
\parbox[b][1.05\paperheight]{1.4\paperwidth}{%
\vfill
\centering
{\transparent{0.3}\includegraphics[ scale=1.2 ]{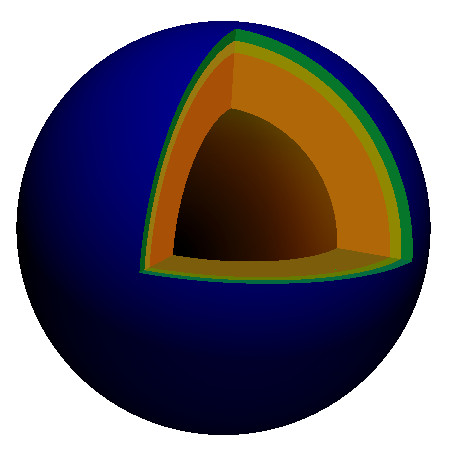}}
\vfill
}}}

\makeatletter
\renewcommand\tableofcontents{%
    \if@twocolumn   
      \@restonecoltrue\onecolumn
    \else
      \@restonecolfalse
    \fi
    \chapter*{\contentsname}%
    \@mkboth{\MakeUppercase\contentsname}{\MakeUppercase\contentsname}%
    \@starttoc{toc}%
    \if@restonecol\twocolumn\fi
    }
\makeatother

\begin{document}

\AddToShipoutPicture*{\BackgroundPic}
\maketitle

\fancyhead[RE, LO]{  }
\pagenumbering{roman}
\thispagestyle{plain}
\begin{center}
\bf{ {\myfont Neutron-rich matter in atomic nuclei and neutron stars}}\\
\vspace{2cm}
{\Large Mem\`oria presentada per optar al grau de doctor per la Universitat de Barcelona}\\
\vspace{2cm}
{\Large Programa de Doctorat en F\'isica}\\
\vspace{2cm}
{ {\Large Autora} \\ { \large Claudia Gonz\'alez Boquera}}\\
\vspace{2cm}
{ {\Large Directors} \\ { \large Mario Centelles Aixal\`a i Xavier Vi{\~n}as Gaus\'i}}\\
\vspace{2cm}
{ {\Large Tutor} \\ { \large Dom\`enec Espriu Climent}}\\
\vspace{1.5cm} 
{\Large Departament de F\'isica Qu\`antica i Astrof\'isica}\\
\vspace{1cm}
{\Large Setembre 2019}\\
\vspace{0.5cm}
{\includegraphics[clip=true,width=0.4\paperwidth]{./grafics/marca_pos_cmyk}
\hspace{1cm}
{\includegraphics[clip=true,width=0.25\paperwidth]{./grafics/LogoICCUB_MdM_prefered_negative_EN_cmyk}}}
\vspace{10cm}
\end{center}

\thispagestyle{empty}
\begin{center}
\newpage
\end{center}

\thispagestyle{plain}
\begin{center}
\vspace*{5cm}

\hspace{8cm}{\it Dedicat a vosaltres, els que sempre}\\
\hspace{8cm}{\it heu estat al meu costat}\\
\end{center}

\blankpage
\thispagestyle{plain}
\begin{center}
\small{
\hspace{7cm}You gave me the best of me\\
\hspace{7cm}So I'll give you the best of you\\
\hspace{7cm}You found me. You knew me\\
\hspace{7cm}You gave me the best of me\\
\hspace{7cm}So you'll give you the best of you\\
\hspace{7cm}You'll find it, the galaxy inside you\\
\hspace{9cm} {\it Magic Shop, BTS}}

\small{
\hspace{-7cm}Starlight that shines brighter in the darkest night\\
\hspace{-7cm}Starlight that shines brighter in the darkest night\\
\hspace{-7cm}The deeper the night, the brighter the starlight\\
\hspace{-7cm}One history in one person\\
\hspace{-7cm}One star in one person\\
\hspace{-7cm}7 billion different worlds\\
\hspace{-7cm}Shining with 7 billion lights\\
\hspace{-7cm}7 billion lives, the city's night view\\
\hspace{-7cm}Is possibly another city's night\\
\hspace{-7cm}Our own dreams, let us shine\\
\hspace{-7cm}You shine brighter than anyone else\\
\hspace{-7cm}One\\
\hspace{-5cm} {\it Mikrokosmos, BTS}}

\small{
\hspace{7cm}When you're standing on the edge\\
\hspace{7cm}So young and hopeless\\
\hspace{7cm}Got demons in your head\\
\hspace{7cm}We are, we are\\
\hspace{7cm}No ground beneath your feet\\
\hspace{7cm}Now here to hold you\\
\hspace{7cm}'Cause we are, we are\\
\hspace{7cm}The colors in the dark\\
\hspace{9cm} {\it We are, One OK Rock}}

\small{
\hspace{-7cm}We'll fight fight till there's nothing left to say\\
\hspace{-7cm}(Whatever it takes)\\
\hspace{-7cm}We'll fight fight till your fears, they go away\\
\hspace{-7cm}The light is gone and we know once more\\
\hspace{-7cm}We'll fight fight till we see another day\\
\hspace{-5cm} {\it Fight the night, One OK Rock}}

\small{
\hspace{7cm}So you can throw me to the wolves\\
\hspace{7cm}Tomorrow I will come back\\
\hspace{7cm}Leader of the whole pack\\
\hspace{7cm}Beat me black and blue\\
\hspace{7cm}Every wound will shape me\\
\hspace{7cm}Every scar will build my throne\\
\hspace{9cm} {\it Throne, Bring me the horizon}}

\small{
\hspace{-7cm}Please just once\\
\hspace{-7cm}If I can just see you\\
\hspace{-7cm}I'm OK with losing everything I have\\
\hspace{-7cm}I'll meet you, even if it's in a dream\\
\hspace{-7cm}And we can love again\\
\hspace{-7cm}Just as we are\\
\hspace{-5cm} {\it UNTITLED, 2014, G-Dragon}}
\end{center}

\blankpage

\chapter*{Abstract}
The proper understanding of the equation of state (EoS) of highly asymmetric nuclear matter is essential 
when studying systems such as neutron stars (NSs). Using zero-range Skyrme interactions and finite-range 
interactions such as Gogny forces, momentum-dependent 
interactions (MDI) and simple effective interactions (SEI), we analyze the properties of the EoS and the
influence they may have on the calculations for NSs. 

We start by studying the convergence properties of the Taylor series expansion of the EoS in powers of the isospin asymmetry.
Next, we analyze the accuracy of the results for $\beta$-stable nuclear matter, which is found in the interior of NSs, 
when it is computed using the Taylor expansion of the EoS. The agreement with the results obtained using the full expression of the EoS 
is better for interactions with small-to-moderate values of the slope of the symmetry energy, $L$.
We also obtain the results for the $\beta$-equilibrated matter when the Taylor expansion of the EoS is performed up to 
second order in the potential part of the interaction, while the kinetic part is used in its full form. In this case, 
one almost recovers the exact results. 

The mass and radius relation for an NS is obtained by integrating the so-called Tolman-Oppenheimer-Volkoff (TOV) equations, where the
input is the EoS of the system. We have studied the mass-radius relation for Skyrme and Gogny interactions, and we see  
that very soft forces are not able to give stable solutions of the TOV equations and only the stiff enough
parametrizations can provide $2$ solar mass ($M_\odot$) NSs.
We also notice that none of the existing parametrizations of the  
standard Gogny D1 interaction are able to provide an NS inside the observational constraints. 
Because of that, we propose a new parametrization, which we name D1M$^*$, that is able to 
provide NSs of $2 M_\odot$ while still providing the same good description of finite nuclei as D1M. A parametrization
D1M$^{**}$ is also presented, which is fitted in the same way as D1M$^*$ and provides NSs up to $1.91 M_\odot$.

An accurate determination of the core-crust transition point, which is intimately related to the isospin dependence 
of the nuclear force at low baryon densities, 
is necessary for the modeling of NSs for astrophysical 
purposes. We estimate the core-crust transition in NSs by finding 
where the nuclear matter in the core is unstable against fluctuations of the density. 
To do that, we employ two methods, the thermodynamical method and the dynamical method. 
The first one considers the mechanical and chemical stability conditions for the core, and neglects 
the surface and Coulomb effects in the stability conditions.
Moreover, we obtain the core-crust transition using the dynamical method, 
where one considers bulk, surface and Coulomb effects when studying 
the stability of the uniform matter.
In the case of finite-range interactions, such as the Gogny forces, we have had to derive the explicit expression of the energy curvature matrix in 
momentum space for this type of interactions. 
We observe a decreasing trend of the transition density with the slope $L$ of the symmetry energy, while 
the correlation between the transition pressure and $L$ is much lower.
The results of the core-crust transition properties
obtained with the Taylor expansion of the EoS are close to the exact results only in the case of soft EoSs. 
For interactions with large values of $L$ and stiff EoSs, the results 
computed using the Taylor expansion, even after adding terms beyond the second-order in the expansion, are far from the exact values. 

Finally, different NS properties are studied. The crustal properties, such as the crustal mass, crustal thickness and crustal 
fraction of the moment of inertia, have lower values if one computes them using the core-crust transition density 
obtained with the dynamical method instead of the one obtained with the thermodynamical method,
pointing out the importance of the accurate evaluation of the transition density when studying observational phenomena. 
We have also studied  
the moment of inertia of NSs, which is compared to some constraints proposed in the literature. 
Finally, the tidal deformability for NSs is also calculated and compared with the constraints coming from the 
GW170817 event detected by the LIGO and Virgo observatories and which accounts for the merger of two NSs in a binary system.

\chapter*{Resum}\label{resum}
Aquesta tesi doctoral pret\'en estendre els estudis de l'equaci\'o d'estat (EoS)
de mat\`eria nuclear altament assim\`etria, utilitzant models de camp mig no relativistes, com per 
exemple les interaccions de contacte tipus Skyrme~\cite{skyrme56, vautherin72,sly41}, o models d'abast finit com les 
interaccions de Gogny~\cite{decharge80, berger91}, 
les anomenades {\it
momentum dependent interactions}, (MDI)~\cite{das03,li08} i les {\it simple effective
interactions} (SEI)~\cite{behera98, Behera05}. 

El Cap\'itol~\ref{chapter1} recull un breu resum de l'aproximaci\'o
de camp mig, a on un assumeix el sistema nuclear com a un conjunt
de quasi-part\'icules no interactuants que es mouen independentment
dins d'un potencial de camp mig efectiu. 
El mateix cap\'itol recull els conceptes b\`asics del m\`etode de 
Hartree-Fock utilitzat per tal de trobar l'energia del sistema.
A m\'es a m\'es, s'introdueixen els diferents potencials 
fenomenol\`ogics que s'utilitzaran al llarg d'aquesta tesi, tals com
les interaccions de Skyrme, Gogny, MDI i SEI.
Tots aquests funcionals, especialment els de Skyrme i els de Gogny,
reprodueixen amb bona qualitat les propietats dels nuclis finits. 
En aquest treball tamb\'e estudiarem 
mat\`eria nuclear a elevades densitats i a elevades 
assimetries d'isosp\'i.
Les definicions de diverses propietats de 
l'EoS de mat\`eria nuclear sim\`etrica
i de mat\`eria nuclear assim\`etrica tamb\'e estan incloses en 
aquest cap\'itol.

Es dedica el Cap\'itol~\ref{chapter2} a l'estudi de les
propietats de la mat\`eria nuclear assim\`etrica utilitzant un conjunt d'~interaccions de Skyrme i de Gogny.
Primerament s'analitza el comportament dels coeficients de 
l'energia de simetria que apareixen en l'expansi\'o de Taylor
de l'energia per part\'icula en termes de l'assimetria 
d'isosp\'i. L'EoS s'expandeix fins al des\`e ordre en el cas de les 
interaccions de Skyrme i fins al sis\`e ordre en el cas de les
forces de Gogny~\cite{gonzalez17}. 
El comportament del coeficient de segon ordre de l'energia de simetria, 
el qual se'l coneix com a energia de simetria, divideix les interaccions
de Skyrme (i també les de Gogny) en dos grups. 
El primer grup recull les parametritzacions que tenen una energia
de simetria que desapareix a una certa densitat sobre la saturaci\'o,
implicant una inestabilitat d'isosp\'i. 
El segon grup est\`a format per aquelles interaccions, normalment
amb un pendent $L$ a la densitat de saturaci\'o major, les
energies de simetria de les quals tenen sempre un pendent creixent. 
Tamb\'e s'estudia l'energia de simetria si s'ent\'en com la 
difer\`encia entre l'energia per part\'icula en mat\`eria
neutr\`onica i en mat\`eria sim\`etrica, la qual s'anomenar\`a
energia de simetria parab\`olica. 
Aquesta definici\'o tamb\'e coincideix amb la suma infinita 
de tots els coeficients de l'expansi\'o de Taylor de 
l'energia per part\'icula en termes de l'assimetria si aquesta
pren valors de la unitat. Al voltant del punt de saturaci\'o, 
les difer\`encies entre el coeficient de segon ordre i l'energia
de simetria parab\`olica es redueixen si es
consideren m\'es termes de l'expansi\'o de l'EoS~\cite{gonzalez17}.
A m\'es a m\'es, tamb\'e  s'evaluen 
els seus respectius pendents de l'energia de simetria, 
i veiem que poden sorgir algunes discrep\`ancies entre ells. 

L'interior de les estrelles de neutrons (NSs) est\`a format
per mat\`eria en $\beta$-equilibri. 
En el Cap\'itol~\ref{chapter2} d'aquesta tesi s'analitza la converg\`encia de l'expansi\'o
en s\`erie de Taylor de l'EoS en pot\`encies de l'assimetria
d'isosp\'i quan estudiem mat\`eria nuclear en $\beta$-equilibri.
L'acord de l'assimetria d'isosp\'i i de la pressi\'o al llarg de totes les 
densitats calculades amb l'expansi\'o de l'EoS millora si es consideren m\'es ordres, 
sent la millora m\'es lenta per interaccions amb par\`ametre de pendent $L$ m\'es gran. 
Aquestes difer\`encies s\'on rellevants en l'estudi de NSs, en el qual s'utilitza 
l'EoS de mat\`eria nuclear infinita per descriure el nucli d'una NS.
Si es duu a terme el desenvolupament en s\`erie de Taylor nom\'es en la part 
potencial de la for\c{c}a i s'utilitza l'expressi\'o completa per a la part cin\`etica, 
pr\`acticament es recobren els mateixos valors per l'assimetria i per la pressi\'o 
que en el cas que s\'on calculades amb l'expressi\'o completa de l'EoS~\cite{gonzalez17}.

La relaci\'o entre la massa i el radi de les NSs tamb\'e ha estat estudiada 
al Cap\'itol~\ref{chapter2}
considerant models de Skyrme i de Gogny. Es troba que forces que s\'on 
molt {\it soft}, i.e., 
amb un baix valor del pendent de l'EoS, no s\'on capaces de donar 
solucions estables de les equacions 
TOV, i nom\'es les interaccions suficientment {\it stiff}, i.e., amb 
un valor alt del pendent
de l'EoS, poden proveir NSs de $2$ masses solars ($M_\odot$).
En particular, notem que cap de les interaccions de Gogny que s'engloben dins de 
la familia D1 d\'ona NSs dins dels l\'imits observacionals~\cite{Sellahewa14, gonzalez17}. 
La converg\`encia de l'EoS tamb\'e \'es testejada quan s'estudien propietats de les NSs. 
Un troba que, si l'expansi\'o de Taylor es talla al segon ordre, els resultats 
poden quedar lluny dels obtinguts utilitzant l'EoS completa. 
Aquesta converg\`encia \'es m\'es lenta contra m\'es elevat \'es
el pendent de l'energia de simetria de la 
interacci\'o .
Aquest comportament senyala la necessitat d'utilitzar l'expressi\'o completa de l'EoS sempre
que es pugui. 

Tal i com s'ha esmentat, la fam\'ilia D1 d'interaccions Gogny no inclou cap for\c{c}a que
sigui capa\c{c} de donar una NS que arribi a $2 M_\odot$, ja que totes les parametritzacions
tenen energies de simetria {\it soft}~\cite{Sellahewa14, gonzalez17}. 
Dins del Cap\'itol~\ref{chapter3} proposem dues interaccions de Gogny noves, 
les quals anomenem D1M$^*$ i D1M$^{**}$, que s\'on capaces de donar una NS dins dels
lligams observacionals a la vegada que proveeixen una bona descripci\'o dels nuclis 
finits semblant a la de la interacci\'o D1M~\cite{gonzalez18, gonzalez18a,Vinas19}. 
La interacci\'o D1M$^{*}$ \'es capa\c{c} de donar una NS de $2M_\odot$, 
mentre que la interacci\'o D1M$^{**}$ \'es capa\c{c} de descriure NSs de fins a $1.91 M_\odot$.
Altres propietats estudiades amb les interaccions D1M$^*$ i D1M$^{**}$ estan en acord amb els resultats obtinguts
utilitzant l'EoS de SLy4~\cite{douchin01}.
En aquest cap\'itol s'analitzen algunes propietats de l'estat fonamental de nuclis finits, com per 
exemple energies de lligam, els radis de neutrons i protons, la resposta al moment quadrupolar i barreres de fissi\'o.
Aquestes dues noves parametritzacions D1M$^{*}$ i D1M$^{**}$ duen a terme igual de b\'e que D1M aquests estudis relacionats 
amb nuclis finits~\cite{gonzalez18, gonzalez18a}. Es pot dir que les interaccions D1M$^{*}$ i D1M$^{**}$ s\'on bones alternatives per descriure 
simult\`aniament els nuclis finits i les NSs, donant resultats molt bons en harmonia amb dades experimentals i observacionals. 

La determinaci\'o correcte de la transici\'o entre el nucli i l'escor\c{c}a en les NSs \'es clau en la comprensi\'o
de fen\`omens en les NSs, com per exemple {\it glitches} en els p\'ulsars, els quals depenen de la mida de l'escor\c{c}a~\cite{Link1999,Fattoyev:2010tb,Chamel2013,PRC90Piekarewicz2014,Newton2015}. 
En el Cap\'itol~\ref{chapter4} s'estima sistem\`aticament el punt de transici\'o entre el nucli i l'escor\c{c}a
buscant la densitat en que la mat\`eria nuclear del nucli estel·lar \'es inestable contra fluctuacions de densitat. 
Les inestabilitats s\'on determinades utilitzant dos m\`etodes. Primer, utilitzem l'anomenat m\`etode termodin\`amic, 
a on s'estudien les estabilitats mec\`anica i qu\'imica del nucli, i ho fem per interaccions de Skyrme i de Gogny. 
Tal i com s'ha esmentat en literatura anterior, es troba una tend\`encia a disminuir la densitat de transici\'o 
quan el pendent $L$ augmenta. 
Per altra banda, no es troben correlacions fortes entre la pressi\'o de transici\'o i $L$~\cite{gonzalez17}.
Tamb\'e s'ha estudiat la converg\`encia de les propietats de transici\'o quan s'utilitza el desenvolupament de Taylor de l'EoS. 
En general, quan s'afegeixen m\'es termes al desenvolupament, la densitat de transici\'o s'aproxima als resultats trobats amb l'EoS exacte. 
No obstant, es continuen trobant difer\`encies significatives quan s'utilitzen fins i tot termes d'ordre majors que dos, 
especialment en casos a on el pendent de l'energia de simetria \'es elevat. 
La densitat de transici\'o tamb\'e s'ha obtingut amb el m\`etode din\`amic, a on un considera,
a l'hora d'estudiar l'estabilitat del sistema, efectes de volum, superf\'icie i de Coulomb.
Al Cap\'itol~\ref{chapter4} es duen a terme els c\`alculs per a interaccions de Skyrme, i per diferents 
forces d'abast finit que s\'on, en el nostre cas, les interaccions de Gogny, MDI i SEI. 
En general els resultats per la densitat de transici\'o utilitzant el m\`etode din\`amic s\'on inferiors als que es troben quan 
s'utilitza el m\`etode termodin\`amic. 
La converg\`encia \'es millor per a EoSs {\it soft}.
Primer obtenim els resultats per a interaccions de Skyrme, i analitzem la converg\`encia de les propietats de transici\'o 
si s'utilitza el desenvolupament de Taylor quan es calculen. 
La converg\`encia de les propietats de transici\'o entre el nucli i l'escor\c{c}a \'es la mateixa que es troba quan s'utilitza 
el m\`etode termodin\`amic, \'es a dir, els resultats s\'on m\'es propers als exactes si s'utilitzen m\'es termes a l'expansi\'o de Taylor de l'EoS.
Si la transici\'o \'es obtinguda aplicant el desenvolupament de Taylor de l'EoS nom\'es en la part potencial i utilitzant 
l'energia cin\`etica exacte, els resultats tornen a ser pr\`acticament els mateixos que els exactes. 

Finalment, al Cap\'itol~\ref{chapter4} s'obtenen els valors de les propietats de transici\'o utilitzant el m\`etode din\`amic
amb interaccions d'abast finit. 
Contr\`ariament al cas de les interaccions de Skyrme, s'ha de derivar expl\'icitament l'expressi\'o de la matriu de curvatura de l'energia
en espai de moments per aquest tipus de forces~\cite{gonzalez19}. 
Les contribucions al terme de superf\'icie s'han pres tant de la part d'interacci\'o com de la part cin\`etica, fent aquesta derivaci\'o
m\'es autoconsistent comparada a la d'estudis previs. Les contribucions provinents de la part directa s\'on obtingudes a partir de 
l'expansi\'o en termes de distribucions dels seus factors de forma, i les contribucions provinents dels termes d'intercanvi i cin\`etics 
es troben expressant les seves energies com una suma del terme de volum m\'es una correcci\'o $\hbar^2$ en el marc de l'aproximaci\'o
{\it Extended Thomas Fermi}. Es troba que els efectes de la part d'abast finit de la interacci\'o sobre la matriu de curvatura venen
majorit\`ariament del terme directe de l'energia. 
Per tant, en l'aplicaci\'o del m\`etode din\`amic amb interaccions d'abast finit, utilitzar nom\'es la contribuci\'o del terme directe \'es una
aproximaci\'o acurada, al menys per les interaccions utilitzades en aquesta tesi. 
Tamb\'e s'ha analitzat el comportament global de la densitat de transici\'o i de la pressi\'o de transici\'o en funci\'o del
pendent de l'energia de simetria a la saturaci\'o. Els resultats per les interaccions MDI estan en acord amb resultats previs~\cite{xu10b} i tamb\'e, per MDI i SEI, 
la densitat de transici\'o i la pressi\'o de transici\'o estan altament correlacionades amb $L$. Tot i aix\'i, si els models tenen 
diferents propietats de saturaci\'o, com per exemple en el cas del grup de forces de Gogny que hem utilitzat en aquest treball, les correlacions es deterioren. 

El Cap\'itol~\ref{chapter5} de la tesi inclou l'an\`alisi de diferents propietats de les NSs.
Primer s'estudia la influ\`encia de l'EoS a l'escor\c{c}a interna quan s'estudien propietats globals, com per exemple masses i radis~\cite{gonzalez17, gonzalez19}. 
S'analitzen algunes propietats de l'escor\c{c}a, com poden ser la massa de  l'escor\c{c}a o el seu gruix. 
Aqu\'i es veu una altra vegada la import\`ancia de la bona determinaci\'o de la localitzaci\'o de la transici\'o entre el nucli i 
l'escor\c{c}a, ja que els resultats de les propietats de l'escor\c{c}a s\'on menors si la transici\'o s'ha estimat a dins de l'aproximaci\'o
din\`amica en comptes de la termodin\`amica. 
Aquestes propietats de l'escor\c{c}a juguen un paper crucial a l'hora de predir molts fen\`omens observacionals, com per exemple {\it glitches}, 
oscil{·}lacions {\it r-mode}, etc. 
Per tant, una bona estimaci\'o de les propietats de l'escor\c{c}a \'es clau en la comprensi\'o de les NSs. 

La detecci\'o d'ones gravitacionals ha obert una nova finestra a l'Univers. La senyal GW170817 detectada per la 
col{·}laboraci\'o LIGO i Virgo provinent d'un {\it merger} (fusi\'o) de dues NSs ha donat peu a un seguit de nous lligams tant en 
astrof\'isica com en f\'isica nuclear~\cite{Abbott2017, Abbott2018, Abbott2019}.
Un lligam directament observat de la senyal \'es el relacionat amb el que s'anomena la {\it tidal deformability}, o la 
deformaci\'o deguda a les forces de marea, representada per $\tilde{\Lambda}$, a una certa {\it chirp mass} del sistema binari. 
Despr\'es de l'an\`alisi de les dades, es van donar lligams a altres propietats, com per exemple a la {\it tidal deformability} d'una 
NS can\`onica de $1.4 M_\odot$ ($\Lambda_{1.4}$), a les masses, als radis, etc.~\cite{Abbott2017, Abbott2018, Abbott2019}. 
Dins del Cap\'itol~\ref{chapter5} s'han analitzat els valors de $\tilde{\Lambda}$ i $\Lambda_{1.4}$, i es veu que EoSs molt {\it stiff}
no s\'on capaces de predir valors dins de les restriccions observacionals. Utilitzant un grup de diverses interaccions de 
camp mig, s'estima el radi d' una NS de $1.4M_\odot$, els quals estan en conson\`ancia amb els valors donats per la col{·}laboraci\'o LIGO i Virgo.
Finalment, el moment d'in\`ercia tamb\'e \'es analitzat, trobant, una altra vegada, que EoSs molt {\it stiff} no proveeixen moments d'in\`ercia
dins dels lligams predits per Landry i Kumar per el sistema doble p\'ulsar PSR J0737-3039~\cite{Landry18}.
Les noves interaccions D1M$^{*}$ i D1M$^{**}$ donen molt bons resultats tant per les estimacions de la {\it tidal deformability} com 
per les estimacions del moment d'in\`ercia, confirmant el seu bon rendiment en el domini astrof\'isic. 
Hem analitzat la fracci\'o del moment d'in\`eria encl\`os en l'escor\c{c}a, utilitzant les densitats de transici\'o obtingudes 
amb els m\`etodes termodin\`amic i din\`amic. 
Tal i com passa amb la massa i el gruix de l'escor\c{c}a, la fracci\'o del moment d'in\`ercia encl\`os en l'escor\c{c}a 
\'es menor si la transici\'o s'ha obtingut amb el m\`etode din\`amic~\cite{gonzalez19}. 

Les conclusions de la tesi estan incloses al Cap\'itol~\ref{conclusions}.
S'afegeixen tres Ap\`endixs al final de la tesi. L'Ap\`endix~\ref{appendix_thermal} inclou les expressions expl\'icites 
de les derivades que es necessiten per tal d'obtenir la transici\'o entre el nucli i l'escor\c{c}a. L'Ap\`endix~\ref{app_taules}
 cont\'e els resultats de les propietats de transici\'o obtinguts utilitzant tant el m\`etode termodin\`amic com el din\`amic.
Facilitem a l'Ap\`endix~\ref{app_vdyn} detalls t\`ecnics sobre l'aproximaci\'o {\it Extended Thomas Fermi}, la qual 
s'utilitza per derivar la teoria del m\`etode din\`amic per a interaccions d'abast finit. 

\chapter*{Acknowledgements}
M'agradaria comen\c{c}ar donant les gr\`acies als meus dos directors de tesi, en Mario Centelles i en Xavier Vi{\~n}as, els quals des d'un primer moment 
em van instar i animar a comen\c{c}ar aquesta aventura. Els vull donar les gr\`acies per tota la paci\`encia que 
han tingut amb mi, i per tot el coneixement que m'han transm\`es al llarg d'aquests cinc anys (quatre de doctorat i 
un de m\`aster). Vaig fer b\'e d'escollir-los a ells com a mentors, gr\`acies a ells he apr\`es, no nom\'es de f\'isica, 
sin\'o tamb\'e com a persona. 
Tamb\'e m'agradaria fer esment aqu\'i a l'Elo\"isa, la qual sempre m'ha obert les portes de casa seva. Moltes gr\`acies per les teves 
s\`avies paraules. 

 A tota la gent del Departament, en especial al grup de F\'isica Nuclear i At\`omica, moltes gr\`acies per brindar-me 
la m\`a cada vegada que ho he necessitat. M'agradaria fer esment als Professors  Artur Polls, \`Angels Ramos, Assumpta Parre{\~n}o, 
Volodymir Magas, Bruno Juli\`a, Manuel Barranco, Jos\'e Maria Fern\'andez i Francesc Salvat per aportar amenes converses i ajut en tot el possible. 
No puc deixar de fer esment a totes les persones que he anat coneixent durant aquests anys, cada una aportant el seu gra de sorra en el meu creixement. 
En especial, m'agradaria remarcar l'ajut que sempre m'ha brindat el Professor Isaac Vida{\~n}a.

I would also like to thank my external collaborators with whom I have worked. Thank you Professors Arnau Rios, Luis Robledo, T.R. Tusar, L.Tolos,
O. Louren\c{c}̧o, M. Bhuyan, C. H. Lenzi, M. Dutra, for your very valuable inputs in the expansion of my knowledge. 

Als meus companys de despatx, els que ja han partit i els que es quedaran quan jo marxi: els Antonios, els Alberts, l'Ivan, la Clara, en Rahul i la Maria. 
Gr\`acies per crear un bon ambient de treball en el que un pot estar com a casa dins del seu despatx. 

Al grup del {\~n}am {\~n}am: Albert, Adri\`a, Alejandro, Andreu, Chiranjib, Gl\`oria, Ivan, Javi, Marc Illa i Marc Oncins. Gr\`acies per fer les hores de 
dinar tant divertides i que passin com si fossin cinc minuts. Que l' amistat que ha creat compartir quatre metres quadrats entre vint persones no la 
trenqui la dist\`ancia. Per moltes excursions m\'es a Montserrat i cal\c{c}otades al febrer. Recordeu que tenim pendent escapades a Osca, Finl\`andia, 
Cant\`abria, Nova Zelanda, India i a l'Ant\`artida. 

Moltes gr\`acies al Departament de F\'isica Qu\`antica i Astrof\'isica per brindar-me un lloc en un acollidor despatx i tot el material i eines 
necess\`aries per dur a terme aquesta tesi. 
Tamb\'e agraeixo el suport de l'ajut 
FIS2017-87534-P provinent del MINECO i de FEDER,
el projecte MDM-2014-0369 de l'ICCUB
(Unidad de Excelencia Marı́a de Maeztu) del MINECO i la beca FPI BES-2015-074210 provinent del Ministerio de Ciencia, Innovaci\'on y Universidades.

Ja cap al final, per\`o no menys important, m'agradaria donar les gr\`acies al meu pilar fonamental durant aquest anys: la meva fam\'ilia. 
Moltes gr\`acies per estar sempre amb mi, tant en els bons moments com en els no tant bons. Moltes gr\`acies per evitar que m'enfons\'es en els 
moments que pensava que ja no podia m\'es. Gr\`acies per aconsellar-me, per posar-me tot el m\'es f\`acil possible, encara que us perjudiqu\'es a vosaltres. 
Per riure amb mi, per\`o tamb\'e per haver plorat amb mi al llarg d'aquests anys.
Un remarc especial ha d'anar cap a la meva mare, la qual sempre ha estat al meu costat, tant quan s'ha tingut el vent a favor, com quan anava en contra.
 Espero que hagi valgut la pena. Gr\`acies per tot.

Finally, I would like to thank you, the reader, to take the time to read the work of four years of my life.

A tots els que heu passat en algun moment per la meva vida durant aquests anys, per tot el que hi ha hagut i pel que vindr\`a, gr\`acies. 

\tableofcontents

\chapter{Introduction}\label{intro}
\fancyhead[RE, LO]{Chapter \thechapter}
\pagenumbering{arabic}
The presence of neutrons, neutral-charged particles, inside atomic nuclei 
was proposed by Ernest Rutherford in 1920 and experimentally proved in 1932 by James Chadwick, who 
received the Nobel prize for it in 1935. 
During that time, the existence of compact stars with a density comparable to the one of an atomic nucleus 
was first discussed by Landau, Bohr, and Rosenfeld. 
In 1934, Walter Baade and Fritz Zwicky wrote~\cite{Baade34}:
{\it ``With all 
reserve we advance the view that supernovae represent the transition from ordinary stars into 
neutron stars, which in their final stages consist of extremely closely packed neutrons.''}
With this sentence, they pointed out the origin of neutron stars to be supernova explosions. 
This led Richard Tolman~\cite{Tolman39} and, independently, Robert Oppenheimer and his student George Volkoff~\cite{Oppenheimer39} to perform 
the first neutron star calculations by proposing a set of equations describing static spherical stars in General Relativity. 
In order to find the relation between the pressure and the energy density, i.e., the equation of state (EoS) of the system,
they considered neutron stars as spheres of a degenerate gas of free neutrons. This led Oppenheimer and Volkoff to 
find that static neutron stars could not have masses larger than $\sim 0.7$ solar masses ($M_\odot$), a value that is
much lower than the Chandrasekhar
mass limit of white dwarfs $\sim 1.44 M_\odot$. This result pointed out the 
importance of considering nuclear forces in the description of the neutron star interior. 
Around 1960, John Weeler and collaborators presented~\cite{Wheeler58} the first results for neutron stars considering their interiors composed of 
neutron, proton, and electron Fermi gases. 
In 1959, Cameron used Skyrme interactions to study the effect of nuclear interactions on the structure of neutron stars, finding 
solutions of maximum masses around $2 M_\odot$~\cite{Cameron59}. More works related to the possible new ingredients to the neutron star EoS followed, where 
other particles like muons, mesons, hyperons, or even deconfined quark matter were considered~\cite{Vidana18}. 

Neutron stars were expected to be seen in X-rays, but the observations were inconclusive until the detection of pulsars. 
In 1967, Jocelyn Bell, a Ph.D. student under the supervision of Anthony Hewish, was observing quasars with a radio telescope at Cambridge University
when she detected an extremely regular pulsating signal of $81.5$ MHz and a period of $1.377$s~\cite{Hewish68}.
After eliminating possible man-made sources for those regularly-spaced bursts of radio source, she realized that this emission 
had come from outer space. 
One possible explanation she and their collaborators jokingly gave for the signal was that they perhaps had observed extraterrestrial life, 
and named the signal as LGM: Little Green Men. Later on, they realized that the source of the signal had to come from a rapidly spinning neutron star. 
For that, Anthony Hewish was awarded the Nobel Prize in 1974. 

\begin{figure}[!t]
\centering
\includegraphics[clip=true, width=0.7\linewidth]{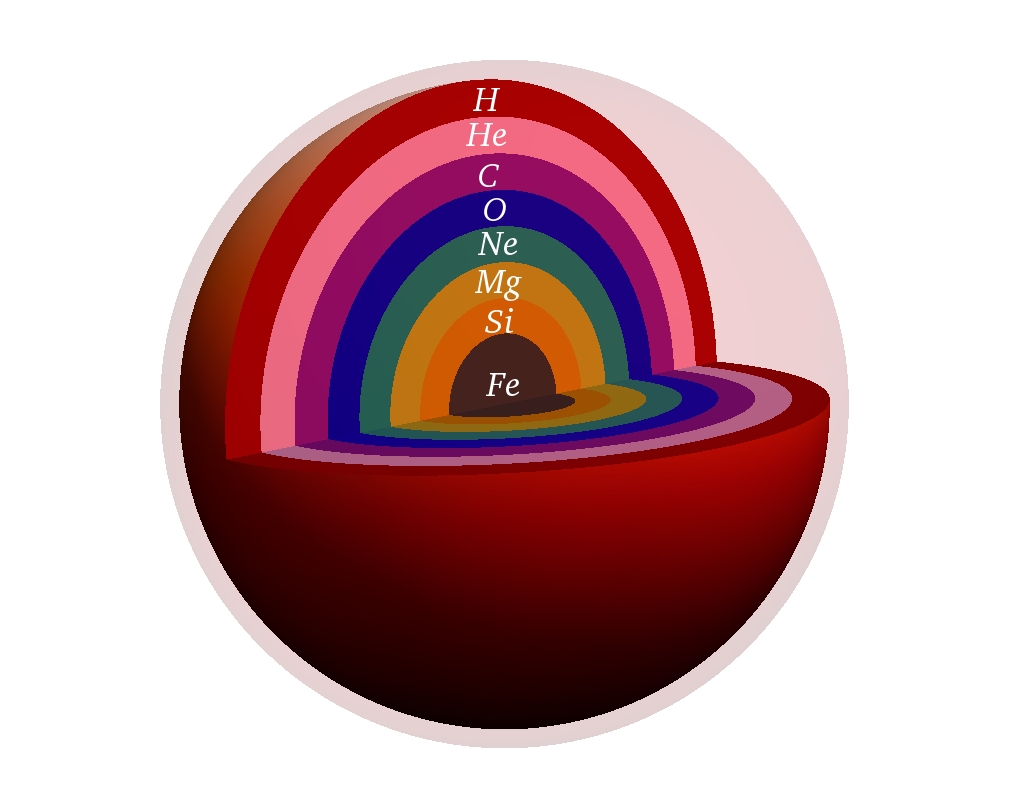}
\caption{Scheme of the onion-like structure of a burning star.}\label{fig:Slayers}
\end{figure}

During the majority of their lives, stars are in thermal and gravitational equilibrium, 
fusing hydrogen~\cite{shapiro83,Glendenning2000, haensel07}. When the hydrogen is all burned, the core of the star, which is now mostly composed of helium, contracts until it reaches such 
temperatures that the helium can be burned, leading the star to a new gravitational and thermal equilibrium. 
Surrounding the core, the star has a hydrogen shell, which is also burning, and the envelope is expanded to such dimensions that the star becomes a red giant. 
When the helium in the core is exhausted, the star has a carbon core, and the same process of finding a new equilibrium where this element can be used as fuel is started. 
For stars of mass greater than about $10 M_\odot$, this process is repeated several times, obtaining each time a core composed of a heavier element, 
obtaining an onion-like structure inside the star (see Fig.~\ref{fig:Slayers}). 
This process is stopped when the core is formed of iron, which is the most tightly bound element in the universe. 
Because of that, the star cannot produce energy through iron fusion. As the other shells are still burning lighter elements, more matter 
will be falling to the core, 
until there is a point that the electrons become ultrarelativistic. The mass of the core will continue growing until reaching 
the Chandrasekhar mass when the electrons cannot avoid the gravitational collapse. 
Inside the core, which has reached temperatures of $ \sim 10^{10}$ K, highly energetic photons are able to photodissociate 
the iron nuclei and the core starts to cool and further contract, increasing its density. 
Moreover, there are inverse $\beta$-decay processes, in which the electrons are captured by protons, releasing neutrinos and forming neutrons, which at such densities 
become degenerate. 
Contrary to the electrons, which cannot leave the core, the neutrinos can escape, giving an additional energy loss to the system and speeding the 
collapse. When reaching densities around \mbox{$\sim 4 \times 10^{11}-10^{12}$ g cm$^{-3}$}, the core becomes opaque to the neutrinos. The energy 
cannot be freed, reheating nuclei that start to burn again. The collapse continues until the core reaches densities of 
$\sim 2-3$ times the saturation density, which is of the order of $ \sim 10^{14}$ g cm$^{-3}$. At this stage, the radius of the core
is around $10$ km and the core consists of $A \sim 56$ nuclei, neutrons, protons, and electrons. The material falling in the core bounces 
releasing a shock-wave outwards from the interior of the proto-neutron star and the material produced in the previous stages is expelled at very high energies
in the form of a supernova explosion. 
For stars of a mass of about $10 M_\odot \lesssim M \lesssim 40 M_\odot$, the result of the supernova explosion will be a neutron star. These stellar objects will have 
radii of $\sim 10-16$ km, masses of the order of $\sim 1-2 M_\odot$ and average densities around $\sim 10^{14}-10^{15}$ g cm$^{-3}$. Some of them can present strong 
magnetic fields and highly precise rotational periods. 
Pulsars are magnetized neutron stars that emit focused beams of electromagnetic radiation through their magnetic axis. 
If the rotational axis is not aligned with its magnetic axis, a ``lighthouse effect'' will arise which from the Earth will be seen 
as radio pulses.
\begin{figure}[!t]
\centering
\includegraphics[clip=true, width=0.7\linewidth]{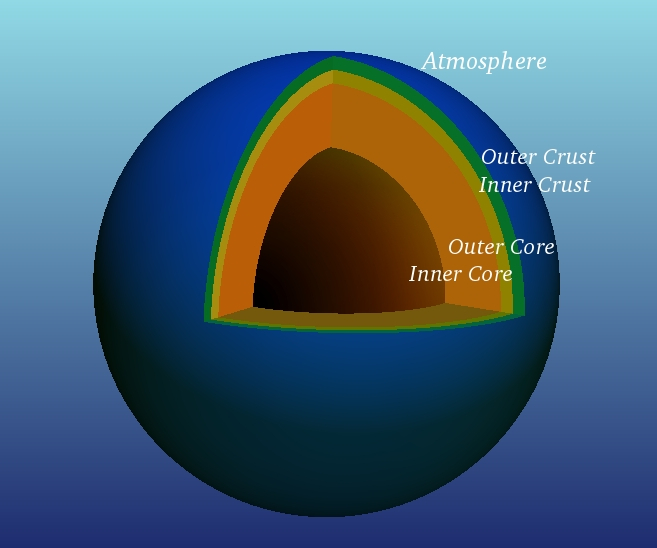}
\caption{Scheme of the structure of a neutron star.}\label{fig:NSlayers}
\end{figure}

A neutron star can be divided into different regions, namely the atmosphere, the crust, and the core. 
The atmosphere contains a negligible amount of mass compared to the crust and the core. 
It influences the photon spectrum and the thermal energy released from the surface of the star~\cite{Lattimer2004}. 
After the atmosphere, we find the crust, which can be separated into two different regions, the outer crust and the inner crust. 
The outer crust consists of nuclei distributed in a solid body-centered cubic (bcc) lattice of 
positive-charged clusters permeated by a 
free electron gas. It goes from a density around the one of the terrestrial iron $\sim 7.5$ g cm$^{-3}$, to densities 
around $\sim 4.3 \times 10^{11}$ g cm$^{-3}$, where the density and the pressure are so high that the nuclear force repels the neutrons inside the nuclei
 and they start to drip, i.e., the system has reached the neutron drip line. 
 The transition between the outer and inner crust parts is essentially determined by nuclear masses~\cite{baym71}, 
 which are experimentally known up to average densities of $\sim 4 \times 10^{11}$ g cm$^{-3}$ of the outer crust. 
 From this density on, masses are predicted theoretically by using finite-tuned mass formulas~\cite{Duflo95, Moller95} or successful mean-field models 
 of nuclear masses~\cite{ruster06, XaviRoca08, Pearson11, sharma15}. 
 After these densities, one enters the region of the inner crust, where the lattice structure of nuclear clusters remains, 
 but now is embedded in free neutron and electron gases. The fraction of free neutrons grows when the density increases up to 
 about one half of the nuclear matter saturation density. At this density, the transition to the core
 occurs because it is energetically favorable for the system to change from a solid 
 to a liquid phase. In the deepest layers of the inner crust, the nuclear clusters may adopt shapes different from the spherical one, i.e., 
 the so-called ``pasta phases'', in order to minimize the Coulomb energy. 
Since the inner crust is largely dominated by the neutron gas and shell effects are to a certain extent marginal, semiclassical approaches 
 are very useful to describe the inner crust of neutron stars including non-spherical shapes. 
 Finally, we have the core, which can be separated into the outer core and the inner core.
 It constitutes around the $99\%$ of the neutron star mass. The outer core is formed of uniform matter composed of neutrons, protons, 
 electrons and eventually muons in $\beta$-equilibrium. The composition of the core is yet to be fully determined. Due to energetic 
 reasons, more exotic particles, such as hyperons, which contain strange quarks, may appear. Also, at those densities and pressures, the transition
 to a phase of hadronic and deconfined quark matter could be feasible~\cite{Lattimer2004}. 
  
The study of the EoS is one of the central issues in nuclear physics as well as in astrophysics. The EoS of symmetric nuclear matter has been studied, 
through experiments based on nuclear masses and sizes, giant resonances of finite nuclei, heavy-ion collisions, etc., 
for more than half a century, becoming relatively well-determined. 
On the other hand, the EoS of asymmetric nuclear matter, which characterizes the 
isospin-dependent part of the EoS, is less known. 
Many facilities have been constructed, or are under construction, around the world with the purpose of constraining the asymmetric nuclear matter properties. 
Some of them are the Radioactive Ion Beam (RIB) Factory at RIKEN in Japan, the FAIR/GSI in Germany, the SPIRAL2/GANIL in France, the Facility 
for Rare Isotope Beams (FRIB), the FRIB/NSCL, the T-REX/TAMU and the Jefferson Lab in the USA, the CSR/Lanzou and BRIF-II/Beijing in China, the SPES/LNL in Italy, the RAON in Korea, etc.
These facilities aim to extract information of the isovector part of effective interactions, as well as of the EoS of 
asymmetric nuclear matter, studying nuclear matter at high densities through radioactive beam physics, heavy-ion collisions, giant  
resonances, isobar analog states, parity-violating phenomena, etc. 
\begin{figure}[!t]
\centering
\includegraphics[clip=true, width=0.5\linewidth]{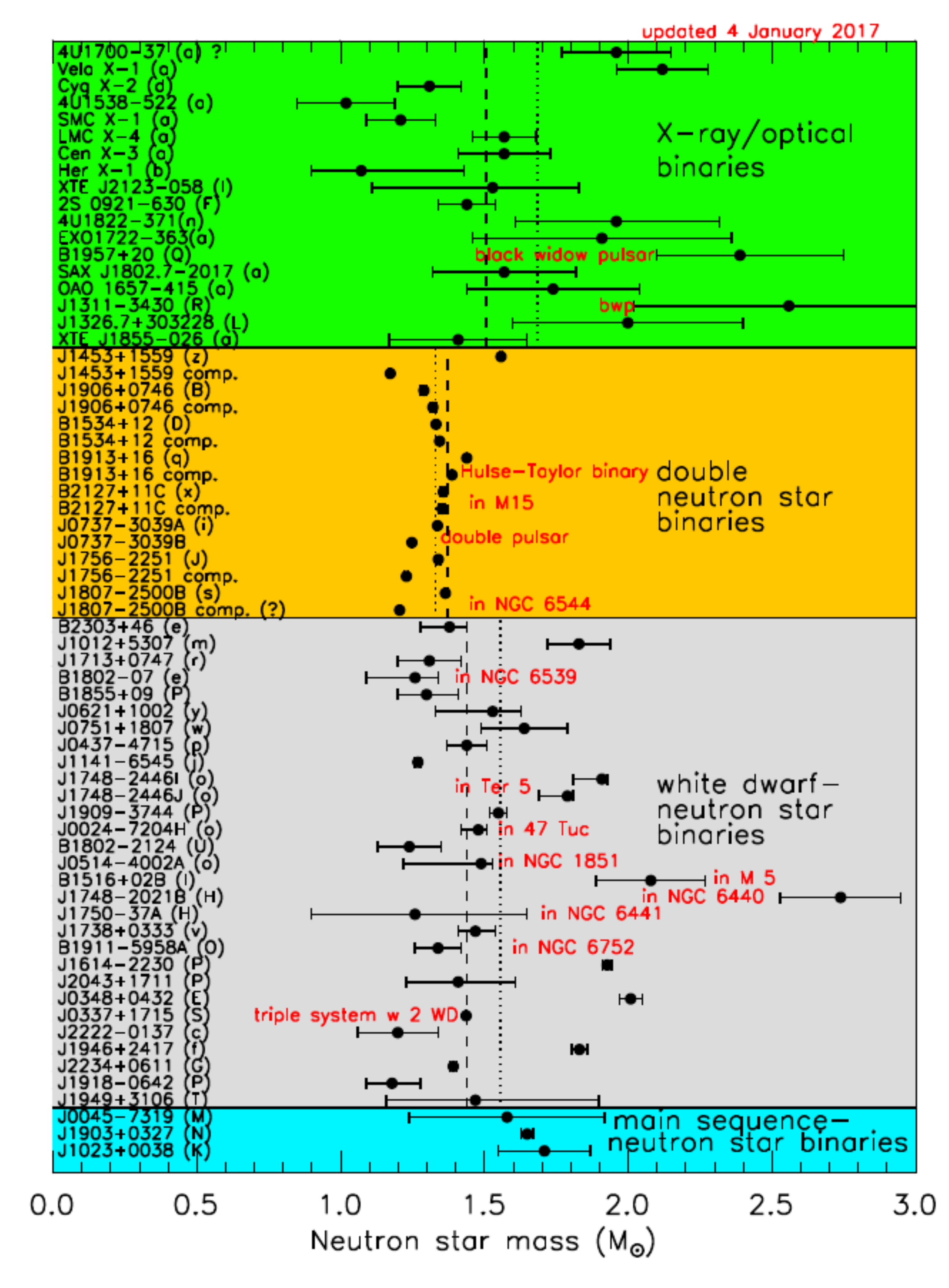}
\caption{Masses measured from pulsar timing. Vertical dashed (dotted) lines indicate category error-weighted
(unweighted) averages. Figure extracted from Ref.~\cite{Lattimer19}. }\label{fig:Latimerpulsars}
\end{figure}

In the astrophysical domain, it is known that various properties of neutron stars, such as the mass-radius relation, the moment of inertia or the tidal deformability, 
are very sensitive to the properties of nuclear matter at saturation and at supra-nuclear densities. 
As the present facilities available for laboratory experiments still cannot reach high densities such as the ones found in the interior of
neutron stars, it is important to study theoretically the EoS inside them, especially in their core, which in the center can attain densities several 
times the saturation density. 
A variety of different functionals (and many-body theories) have been used to 
determine the properties of neutron stars, including Brueckner--Hartree--Fock interactions~\cite{Wiringa95,Vidana2009,Ducoin11,Li2016},
Skyrme forces \cite{xu09a,ducoin07,Pearson12,Newton2014}, 
finite-range functionals \cite{routray16}, relativistic mean-field (RMF) models 
\cite{horowitz01a,carriere03,Klahn06,Moustakidis10,Fattoyev:2010tb,Cai2012,Newton2014} and
momentum-dependent interactions \cite{xu09a,Moustakidis12, routray16}.  
 
The total mass is one of the best well-established observables of neutron stars from 
many observational studies. Among them, there are the recent accurate observations of highly massive 
neutron stars, corresponding to $(1.97 \pm 0.04)M_{\odot}$ and $(2.01 \pm 0.04)M_{\odot}$ for the 
\mbox{PSR J1614-2230} and \mbox{PSR J0348+0432} pulsars, respectively~\cite{Demorest10,Antoniadis13}.
The mass of \mbox{PSR J1614-2230} has been revised in Ref.~\cite{Fonseca_2016} constraining it to $(1.928 \pm 0.017)M_{\odot}$.
A very recent observation~\cite{Cromartie19} of $(2.14^{+0.10}_{-0.09})M_{\odot}$ for the pulsar PSR J0740+6620 
 would correspond to the heaviest neutron star detected up to date. 
Still, these are preliminary results with high error bars, and the earlier Shapiro delay mass measurements 
are being revised~\cite{Zhang_2019}. Hence, in our work, we will restrict ourselves to the observational mass constraints of Refs.~\cite{Demorest10,Antoniadis13}.
These masses revoke many of the proposed theoretical EoSs for neutron stars if the calculated maximum neutron star mass does
not reach the observed values.
As a result, a great effort has been addressed to 
derive nuclear models able to generate EOSs that predict such massive objects (see~\cite{Oertel:2016bki,Li2014} 
and references therein). However, a precise mass measurement is not enough to completely constrain 
the underlying EoS. One would also need a precise measurement of the radius of the neutron star whose mass has 
been obtained. The uncertainties in the determination of the neutron star radius are still an open question 
for observational studies~\cite{Stein_2014,Nattila16,De2018}. The Neutron Star Interior Composition Explorer 
(NICER) mission is already set up with the aim to provide a measurement of the radius with an accuracy 
of order 5\%. 

The detection of gravitational waves (GWs) has opened a new window to explore the Universe and, specifically, neutron stars, with the 
help of the new generation of gravitational observatories like the Laser Interferometer Gravitational-Wave Observatory (LIGO), 
the Virgo laboratory from the European Gravitational Observatory or the future European Space Agency mission LISA (Laser Interferometer 
Space Antenna), planned to be launched around 2034. 
The LIGO and Virgo collaboration detected GWs from the GW170817 event~\cite{Abbott2017}, which accounted for the 
first time or a merger of two neutron stars. This detection led to a whole new set of constraints in both astrophysics and nuclear physics, as it has 
enhanced the present interest to examine the sensitivity of the EoS at large values of the density and of the isospin asymmetry.

In this thesis we further extend the studies of the EoS of highly asymmetric nuclear matter, using non-relativistic
mean-field models, such as zero-range density-dependent Skyrme interactions~\cite{skyrme56, vautherin72,sly41}, or 
finite-range forces like Gogny interactions~\cite{decharge80, berger91}, momentum-dependent interactions (MDI)~\cite{das03,li08} and simple effective interactions (SEI)~\cite{behera98, Behera05}.
The theoretical calculations obtained using these models are compared to experimental data from finite nuclei and from
astrophysical observations. 

Chapter~\ref{chapter1} collects a brief summary of the mean-field approximation, where one assumes the nuclear system as a set 
of non-interacting quasi-particles that move independently inside an effective mean-field potential. 
We also collect the basic concepts of the Hartree-Fock method used to find the energy of the system. 
Moreover, we introduce the different phenomenological potentials we are going to use through this thesis, namely the Skyrme, 
Gogny, MDI and SEI interactions. All these functionals, especially Skyrme and Gogny forces, reproduce with 
good quality the properties of finite nuclei. In this work, we will also study nuclear matter at large densities and isospin asymmetries
with them.
The definition of some properties of the EoS of symmetric nuclear matter and of asymmetric nuclear matter that we will use in the following
chapters are also given for the different models.

We devote Chapter~\ref{chapter2} to study the properties of asymmetric nuclear matter
using a set of Skyrme and Gogny interactions~\cite{gonzalez17}. Firstly, we analyze the behaviour of the different symmetry energy 
coefficients appearing in the Taylor expansion of the energy per particle in terms of even powers of the isospin asymmetry.
We expand the EoS up to tenth-order for Skyrme interactions and up to sixth-order for Gogny forces. 
The behaviour of the second-order coefficient, commonly known as the symmetry energy, divides the Skyrme (and also the Gogny)
interactions in two groups. 
The first group contains the parametrizations that have a symmetry energy that vanishes at some density above saturation, implying 
an isospin instability. The second group is formed by those interactions, usually with a larger slope parameter $L$, that have an increasing 
trend for the symmetry energy.
We also study the symmetry energy understood as the difference between the 
energy per particle in pure neutron matter and in symmetric nuclear matter, which we call parabolic symmetry energy. 
This definition also coincides with the infinite sum of all the coefficients of the Taylor expansion of the energy per particle in terms of the asymmetry
if the isospin asymmetry is equal to one.
Around saturation, the differences between the parabolic and the second-order symmetry energy  coefficients are reduced when more terms of the expansion 
are considered~\cite{gonzalez17}. 
Moreover, the corresponding slopes of the symmetry energy are also evaluated, and we see that some discrepancies can arise between them. 
The interior of neutron stars is composed of matter that is in $\beta$-equilibrium. We test the convergence of the Taylor expansion of the EoS in 
powers of the isospin asymmetry when studying $\beta$-stable nuclear matter~\cite{gonzalez17}. 
The agreement of the isospin asymmetry and pressure along all densities calculated with the EoS expansion improves if more orders are considered, 
being the improvement slower for interactions with larger slope parameter $L$.
These differences are relevant when studying neutron star properties, where one uses the EoS of infinite nuclear matter to describe the neutron star core. 
If the Taylor expansion is performed only in the potential part of the force and using the full expression for the kinetic part, 
one almost recovers the same values for the 
isospin asymmetry and for the pressure as if they are calculated using the full expression of the EoS. 

The mass and radius relation of neutron stars has been studied also in Chapter~\ref{chapter2} considering the EoSs of Skyrme and Gogny models. 
We find that very soft forces are not able to give stable solutions of the TOV equations and only the stiff enough
parametrizations can provide $2 M_\odot$ neutron stars.
In particular, remark that none of the Gogny interactions of the D1 family can provide a neutron star inside the observational bounds~\cite{Sellahewa14, gonzalez17}. 
The convergence of the EoS is also tested when studying neutron star properties. One finds that, if the Taylor expansion is cut at second order, the 
results may lay quite far from the ones obtained using the full EoS. This convergence is slower as larger is the slope of the symmetry 
energy of the interaction.
This behaviour points out the necessity of using the full expression of the EoS whenever possible. 

As said previously, the Gogny D1 family does not include any parametrization able to provide a neutron star that reaches $2 M_\odot$, as they 
have very soft symmetry energies~\cite{Sellahewa14, gonzalez17}. 
In Chapter~\ref{chapter3} we propose two new Gogny forces, which we name D1M$^{*}$ and D1M$^{**}$, that are able to 
provide neutron stars inside the observational constraints while still providing the same good description of finite nuclei as D1M~\cite{gonzalez18, gonzalez18a,Vinas19}.
The D1M$^{*}$ interaction is able to provide a neutron star of $2M_\odot$, while the D1M$^{**}$ is able to describe up to $1.91 M_\odot$ neutron stars.
Other stellar properties studied with the D1M$^{*}$ and D1M$^{**}$ are in agreement with the Douchin-Haensel SLy4 EoS~\cite{douchin01}. 
We analyze some ground state properties of finite nuclei, such as binding energies, neutron and proton radii, response to 
quadrupole deformation and fission barriers. 
The two new parametrizations D1M$^{*}$ and D1M$^{**}$ perform as well as D1M in all these studies of finite nuclei~\cite{gonzalez18, gonzalez18a}. 
We can say that the D1M$^{*}$ and D1M$^{**}$ are good alternatives to describe simultaneously finite nuclei and neutron stars
providing excellent results in harmony with the experimental and observational data.

The correct determination of the transition between the core and the crust in neutron stars is key in the understanding of 
neutron star phenomena, such as pulsar glitches, which depend on the size of the crust~\cite{Link1999,Fattoyev:2010tb,Chamel2013,PRC90Piekarewicz2014,Newton2015}.
In Chapter~\ref{chapter4} we estimate systematically the core-crust transition searching 
for the density where the nuclear matter in the core is unstable against fluctuations of the density. 
The instabilities are determined using two methods. First, we use the so-called thermodynamical method, 
where one studies the mechanical and chemical stabilities of the core. We find the corresponding results for Skyrme and Gogny
interactions. As stated in previous literature, we find a downward trend when the transition density is plotted against the slope $L$.
On the other hand, we do not find strong correlations between the transition pressure and $L$~\cite{gonzalez17}.
We have also studied the convergence of the transition properties when the Taylor expansion of the EoS is used. In general, adding 
more terms to the expansion brings the transition density closer to the values found with the exact EoS. 
However, we still find significant differences even using terms of order higher than two, especially in cases where the 
slope of the symmetry energy is large. 
The transition density is also obtained by the dynamical method, where one considers bulk, surface, and Coulomb effects when studying 
the stability of the uniform matter. 
We perform the calculations for Skyrme interactions, and for different finite-range forces, which, in our case, are the Gogny, MDI and SEI models. 
In general, the results for the transition density using the dynamical method are lower than the ones obtained with the thermodynamical method.
The convergence is better for softer EoSs. 
We first obtain the results for Skyrme interactions, and we analyze the convergence of the transition properties 
if the Taylor expansion of the EoS is used to calculate them. 
The convergence of the core-crust transition properties is the same as the one obtained using the thermodynamical method, that is, 
the results are closer to the exact ones as higher-order terms in the expansion are considered.
If the transition density is obtained using the Taylor expansion of the EoS only in the potential part of the interaction
and the exact kinetic energy, the results obtained 
are almost the exact ones. 

Finally, in Chapter~\ref{chapter4} we obtain the values of the transition properties using the dynamical method with finite-range interactions. 
Contrary to the case of Skyrme interactions, we have had to derive the explicit expression of the energy curvature matrix in 
momentum space for these types of forces~\cite{gonzalez19}. 
The contributions to the surface term have been taken from both the interaction part and the kinetic part, making this 
derivation more self-consistent compared to earlier studies. The contributions coming from the direct part are obtained through the 
expansion of their finite-range form factors in terms of distributions, and the contributions coming from the exchange and 
kinetic parts are found expressing their energies as a sum of a bulk term plus a $\hbar^2$ correction within the Extended Thomas Fermi approximation. 
We find that the effects of the finite-range part of the interaction on the curvature matrix arise mostly from the direct part of the energy. 
Therefore, in the application of the dynamical method with finite-range forces, it is an accurate approximation to use only the direct contribution to the energy, 
at least for the forces used in this thesis.
We have also analyzed the global behaviour of the core-crust transition density and pressure as a function of the slope of the symmetry energy at saturation
for these finite-range interactions. The results for MDI are in agreement with previous literature~\cite{xu10b}, and for MDI and SEI, the transition density and pressure
are highly correlated with $L$. However, if the models have different saturation properties, as the set of Gogny interactions we have used in this work, 
the correlations are deteriorated.

Chapter~\ref{chapter5} encloses the analysis of different neutron star properties. 
We first study the influence of the inner crust part of the EoS when analyzing global properties such as the mass and radius~\cite{gonzalez17, gonzalez19}. 
We analyze some crustal properties, such as the crustal mass and the crustal radius. We see again the importance of the 
good determination of the location of the core-crust transition, as the results for the crust are way lower 
if the core-crust transition is estimated within the dynamical approach instead of within the thermodynamical approach. 
These crustal properties play a crucial role when predicting several observed phenomena, like glitches, r-mode oscillations, etc. 
Hence, a good estimation of these properties is key in the understanding of neutron stars. 

As said previously in this Introduction, the detection of GWs opened a new window to look at the Universe. 
The GW170817 signal detected by the LIGO and Virgo collaboration coming from a merger of two neutron stars has established a new set of constraints in 
astrophysics and in nuclear physics~\cite{Abbott2017, Abbott2018, Abbott2019}.
One constraint directly measured form the signal is on the dimensionless mass-weighted tidal deformability, $\tilde{\Lambda}$, at a certain chirp mass of the binary system. 
After a data analysis, constraints on other properties, such as on the dimensionless tidal deformability of a canonical neutron star ($\Lambda_{1.4}$), on masses, on radii, etc. 
were provided~\cite{Abbott2017, Abbott2018, Abbott2019}. 
We have analyzed the values of $\tilde{\Lambda}$ and $\Lambda_{1.4}$, and we see that very stiff EoSs are not able to predict values 
inside the observational bounds. With a set of mean-field interactions, we are also able to roughly estimate a radius of a neutron star of $1.4M_\odot$
in consonance with the values given by the LIGO and Virgo collaboration. 
Finally, the moment of inertia is also analyzed, finding that, again, very stiff EoSs are not able to provide moments of inertia
inside the constraints predicted by Landry and Kumar for the binary double pulsar PSR J0737-3039~\cite{Landry18}. 
The newly D1M$^{*}$ and D1M$^{**}$ interactions provide very good results for both the tidal deformability and the moment of inertia, confirming their
good performance in the astrophysical domain. 
We have analyzed the fraction of the moment of inertia enclosed in the crust, using the core-crust transition density obtained either 
with the thermodynamical and the dynamical methods. As happened with the crustal thickness and crustal mass, the crustal fraction of the 
moment of inertia is lower if the transition is obtained with the dynamical approach~\cite{gonzalez19}. 

Finally, the conclusions of this work are given in Chapter~\ref{conclusions}. 
Three Appendices are added at the end of the thesis. Appendix~\ref{appendix_thermal} collects the explicit 
expressions of the derivatives needed to obtain the core-crust transition using either the thermodynamical or the 
dynamical methods. Appendix~\ref{app_taules} contains the numerical results of the core-crust transition properties 
for Skyrme and finite-range interactions computed using both the thermodynamical and the dynamical methods. 
We provide in Appendix~\ref{app_vdyn} technical details about the Extended Thomas Fermi 
approximation, which is used to derive the theory of the dynamical method for finite-range interactions. 

\chapter{Non-Relativistic Mean Field Models}\label{chapter1}
\section{Mean field approximation within the Hartree-Fock framework}
The system of a nucleus composed of $A$ nucleons (neutrons and protons) can be described by a many-body Hamiltonian $H$, which
consists of a kinetic part plus a potential part. 
There are different approaches to compute the nuclear structure~\cite{Bender03}. 
The ab-initio calculation of the properties of a nuclear system starts with a nucleon-nucleon potential, 
which describes the nucleon-nucleon scattering data~\cite{Machleidt_2001}. It is characterized for having a highly
repulsive core, and for reproducing the basic features of nuclear saturation. However, if only the nucleon-nucleon
potential is employed to reproduce the properties of the system, the ab-initio approach fails to determine 
quantitatively the saturation point, and additional three-body forces have to be considered. 
The study of a nuclear many-body system with ab-initio methods requires highly-developed many-body theories
like, e.g., the Brueckner-Hartree-Fock~\cite{Serot:1984ey, Brockmann90, Dickhoff_1992}, 
correlated basis functions~\cite{pandh81, Heiselberg00} or self-consistent Green's functions~\cite{MarioArtur2, Artur2,Artur1, 
MarioArtur4, MarioArtur3, MarioArtur1}. Thus, it is a highly complicated and challenging
endeavour. 

In the other extreme of the existing approaches for describing nuclear systems, one finds 
the macroscopic nuclear liquid-drop model~\cite{myers82}. 
In this case, the energy of the system is phenomenologically parametrized in terms of global properties of nuclei, 
such as volume energy, asymmetry energy, surface energy, etc. 
Usually, shell correction energies that approximate quantal effects are added to these macroscopic models, 
giving rise to the so-called microscopic-macroscopic (mic-mac) models. 

In between these two approaches one finds, on the one hand, the shell model, in which one considers a phenomenological
single-particle potential and performs a configuration mixing calculation involving all many-body
states that can be constructed using a band of the possible single-nucleon states around the Fermi energy~\cite{brown88}.
On the other hand, between the ab-initio and macroscopic models, there is the mean-field approximation,
which we will use in this thesis.
A way to circumvent the determination of the full potential describing the whole system is to assume the nucleus as a set of 
quasiparticles that do no interact between them and where each nucleon moves independently within an 
effective mean-field (MF) created by the nucleons themselves. 
We will restrict ourselves to an effective two-body Hamiltonian, whose potential part $V$ can be decomposed 
as the sum of a single-particle potential for each nucleon $i$ plus a 
residual potential,
\begin{equation}\label{eq:pot}
 V = \sum_i v ({\bf r}_i) + V^\mathrm{res}=V^\mathrm{MF} + V^\mathrm{res}.
\end{equation}
The sum of all $ v ({\bf r}_i)$ will be denoted as the mean-field potential $V^\mathrm{MF}$ of the whole system.
The contribution of the residual potential $V^\mathrm{res}$ is supposed to be much weaker than the  
contribution of the original potential $V$, 
and usually is treated in perturbation theory.
Taking Eq.~(\ref{eq:pot}) into account, we can rewrite the Hamiltonian as
\begin{equation}
 H= T+V^\mathrm{MF} + V^\mathrm{res}= H^\mathrm{MF}+ V^\mathrm{res},
\end{equation}
where the sum of the kinetic term $T$ and the mean-field potential $V^\mathrm{MF}$ is denoted as the mean-field Hamiltonian $H^\mathrm{MF}$:
\begin{equation}\label{HH}
 H^\mathrm{MF} = T+ V^\mathrm{MF} = \sum_{i} t({\bf r}_i) + \sum_i  v({\bf r}_i) = \sum_i h({\bf r}_i),
\end{equation}
where ${\bf r}_i$ is the coordinate of the $i$-th nucleon, with $i,j=1\cdots A$ and $h({\bf r}_i)$ are the single-particle Hamiltonians. 

If we write the Hamiltonian $H$ in second quantization, the Hamiltonian operator becomes \cite{rin80}
\begin{equation}\label{eq:Ham}
H = \sum_{ij} t_{ij} \hat{c}^\dagger_i \hat{c}_j + \frac{1}{4} \sum_{ijkl} \bar{v}_{ijkl} \hat{c}^\dagger_i \hat{c}^\dagger_j \hat{c}_l \hat{c}_k,
\end{equation}
where $\hat{c}^\dagger_i$ and $\hat{c}_i$ are the single-particle creation and annihilation operators in a single-particle state $i$, 
and 
\begin{equation}
 \bar{v}_{ijkl} = v_{ijkl}- v_{ijlk}
\end{equation}
is the antisymmetrized two-body interaction matrix elements.
The indexes ${i,j,k,l}$ run over a complete set of states.

The Schr\"{o}dinger equation associated to the Hamiltonian can be written as
\begin{equation}\label{eq:sch}
 H |\phi \rangle = E  |\phi \rangle,
\end{equation}
where $ |\phi \rangle$ is the total wave-function of the system and $E$ the corresponding energy. 
The solution of Eq.~(\ref{eq:sch}) will be found using the variational principle, which states that the 
exact  Schr\"{o}dinger equation is equivalent to the variational equation~\cite{rin80}
\begin{equation}\label{eq:var1}
 \delta E[\phi] = 0,
\end{equation}
where 
\begin{equation}\label{eq:var2}
 E[\phi] = \frac{ \langle \phi | H | \phi \rangle}{ \langle \phi| \phi \rangle}
\end{equation}
is the expectation value of the energy.
The variation equation~(\ref{eq:var1}) can be expanded as
\begin{equation}\label{eq:var3}
 \delta E[\phi] = \langle \delta \phi | H - E |\phi \rangle + \langle \phi | H - E | \delta \phi \rangle = 0,
\end{equation}
where $E$ is a Lagrange multiplier which can be understood as the energy corresponding to $|\phi \rangle$.
The wave-function $|\phi \rangle$ can be a complex function, and therefore the variation has to be performed in the 
real and imaginary parts independently, which is equivalent to carry out the variation over $|\delta \phi \rangle$
and $ \langle \delta \phi|$ independently. 
This yields Eq.~(\ref{eq:var3}) to be reduced to
\begin{equation}
 \langle \delta \phi | H- E |\phi \rangle =0
\end{equation}
and its respective complex conjugate equation. 
The determination of the ground state, with corresponding energy $E_0$, and for a trial $|\phi \rangle$ satisfies
\begin{equation}
 E [\phi] \geq E_0.
\end{equation}
Therefore, the goal of the variational principle will be to find a wave-function $|\phi \rangle$ that minimizes the value of $E[\phi]$.

Up to now, it has been assumed that the Hamiltonian $H$ does not depend on the wave function $|\phi \rangle$.
However, many effective interactions do depend on the density and therefore on $|\phi \rangle$. Hence, the problem has to be 
solved in a self-consistent way until the solution converges. 

In the Hartree-Fock (HF) approach, one solves self-consistently the variational problem considering the many-body wave-functions 
of the type of Slater determinants
\begin{equation}\label{Slater}
 |\phi^\mathrm{HF} \rangle = | \phi (1 \cdots A) \rangle= \prod_{i=1}^A \hat{c}_i^\dagger |0 \rangle,
\end{equation}
where $c_k^\dagger$ and $c_k$ correspond to the single-particle wave-functions $\phi_k$ which, at the
same time are eigenfunctions of the single-particle Hamiltonian $h$, i.e., 
\begin{equation}\label{hMF}
 h(\mathbf{r}, \sigma, \tau) \phi_k (\mathbf{r}, \sigma, \tau) = \varepsilon_k \phi_k(\mathbf{r}, \sigma, \tau).
\end{equation}
Eq.~(\ref{hMF}) is determined with the variational condition (\ref{eq:var1}), where $\phi_k(\mathbf{r}, \sigma, \tau)$ are the
respective eigenstates.
The $|\phi^\mathrm{HF} \rangle$ wave-function will describe the fermions in the nuclear system and therefore it has to be antisymmetrized.

The Hartree-Fock energy will be given by \cite{rin80}
\begin{equation}
 E^\mathrm{HF} =  \langle \phi^\mathrm{HF} | H | \phi^\mathrm{HF} \rangle , 
\end{equation}
where the Hamiltonian $H$ is given in Eq.~(\ref{eq:Ham}). Using Wick's theorem, 
the energy can be calculated as a functional of the single-particle density 
$\rho_{ij} =  \langle \phi^\mathrm{HF} |  c_j^\dagger c_i | \phi^\mathrm{HF} \rangle$: 
\begin{eqnarray}\label{EHF}
 E^\mathrm{HF} &=& \sum_{ij} t_{ij}  \langle \phi^\mathrm{HF} | c_i^\dagger c_j | \phi^\mathrm{HF} \rangle + 
 \frac{1}{4} \sum_{ijkl} \bar{v}_{ijkl} \langle \phi^\mathrm{HF} | c_i^\dagger c_j^\dagger c_l c_k |\phi^\mathrm{HF} \rangle \nonumber\\
 &=& \sum_{ij} t_{ij} \rho_{ji} + \frac{1}{2} \sum_{ijkl} \rho_{ki} \bar{v}_{ijkl} \rho_{lj}.
 \end{eqnarray}
If the HF energy in Eq.~(\ref{EHF}) is solved using the Slater wave-functions (\ref{Slater}), it can be written 
in coordinate space as 
\begin{eqnarray}
 E^\mathrm{HF} &=& - \sum_{i} \frac{\hbar^2}{2m} \int \phi^*_ i (\mathbf{r}) \nabla^2 \phi_i (\mathbf{r}) d \mathbf{r}\nonumber\\
 &+& \frac{1}{2} \sum_{ij} \int \phi_i^* (\mathbf{r}) \phi_j^* (\mathbf{r}') V(\mathbf{r}, \mathbf{r}') \phi_i (\mathbf{r}) \phi_j (\mathbf{r}')
 d \mathbf{r} d \mathbf{r}'\nonumber\\
 &-& \frac{1}{2} \sum_{ij} \int \phi_i^* (\mathbf{r}) \phi_j^* (\mathbf{r}') V(\mathbf{r}, \mathbf{r}') \phi_j (\mathbf{r}) \phi_i (\mathbf{r}')
 d \mathbf{r} d \mathbf{r}',
\end{eqnarray}
where $m$ is the nucleon mass and $V(\mathbf{r}, \mathbf{r}') $ is the two-body interaction between two nucleons at $\mathbf{r}$ and $\mathbf{r}'$.

From Eqs.~(\ref{HH}) and (\ref{hMF}), the Hartree-Fock equations in coordinate space for the single-particle wave functions
will be given by
\begin{equation}
 \frac{-\hbar^2}{2m} \nabla^2 \phi_k({\bf r}) + V_H ({\bf r}) \phi_k ({\bf r})+
 \int d {\bf r}' V_{F} ({\bf r}, {\bf r}')\phi_k ({\bf r}') = \varepsilon_k \phi_{k}({\bf r}).
\end{equation}
The Hartree term is the local part of the potential, depends on the one-body density
\begin{equation}
 \rho ({\bf r}) = \sum_k \phi_k^* ({\bf r}) \phi_k ({\bf r}) 
\end{equation}
and is defined as
\begin{equation}
 V_H({\bf r})=\int d {\bf r}' v({\bf r},{\bf r}') \rho({\bf r}').
\end{equation}
On the other hand, the Fock potential gives the non-locality of this type of systems. It depends 
on the non-local one-body density matrix
\begin{equation}
 \rho ({\bf r},{\bf r}') = \sum_k \phi_k^* ({\bf r}') \phi_k ({\bf r}) 
\end{equation} 
and is given by
 \begin{equation}
  V_{F}  ({\bf r},{\bf r}')= -v({\bf r},{\bf r}') \rho({\bf r},{\bf r}') .
 \end{equation}

 The potential term $ V({\bf r}, {\bf r}')$ includes all possible nucleon-nucleon forces, as well Coulomb interactions.
 In our case, the Coulomb interaction in the system will be represented by
\begin{equation}
 V_\mathrm{Coul} ({\bf r})= \frac{e^2}{2} \int \frac{\rho_p ({\bf r}') d^3 r'}{|{\bf r}-{\bf r}'| } - \frac{e^2}{2}
 \left( \frac{3}{\pi}\right)^{1/3} \rho_p^{1/3} ({\bf r}'),
\end{equation}
where the Slater approximation has been used in the exchange part \cite{sly42}.

\section{Infinite matter properties with phenomenological potentials}
In the MF approach, the interaction between nucleons is characterized by a phenomenological 
potential $V(\mathbf{r}, \mathbf{r}') $ which depends on several free parameters which will be fitted
to reproduce the experimental data of some nuclear properties. 
Typically, these properties are the observables related to nuclear masses, radii, binding energies, 
shell structure properties, etc., or to infinite nuclear matter properties, such as the saturation energy, the 
nuclear matter incompressibility, etc.
These phenomenological potentials can be of zero-range type, such as 
the Skyrme interactions \cite{skyrme56, vautherin72,sly41}, or may include finite-range terms, such as the 
Gogny interactions~\cite{decharge80, berger91}, the MDI forces~\cite{das03,li08} and the SEI functionals~\cite{behera98, Behera05},
which we will use in the present thesis.

 In the fitting procedure of phenomenological potentials, it is very usual to consider the saturation density $\rho_0$ in symmetric nuclear matter (SNM)
 and the energy per particle $E_b$ in SNM at $\rho_0$,  which have values of $\rho_0 \simeq 0.16$ fm$^{-3}$ and $E_b (\rho_0)\simeq-16$ MeV, respectively.
The saturation density $\rho_0$ is given by the minimum of the energy per particle in SNM, 
\begin{equation}
 \left.\frac{\partial E_b (\rho)}{\partial \rho}\right|_{\rho_0}=0 .
\end{equation}
Moreover, the pressure and the incompressibility are given, respectively, by 
\begin{equation}\label{eq:press0}
  P (\rho)= 
\rho^2 \frac{\partial E_b (\rho)}{\partial \rho}
\end{equation}
and 
\begin{equation}\label{eq:K00}
 K (\rho)= 9\rho^2 \frac{\partial^2 E_b (\rho)}{\partial \rho^2}.
\end{equation}
At the saturation density $\rho_0$ there is a cancellation of the pressure, i.e., 
\begin{equation}\label{eq:press}
  \hspace{0.5cm}P_0 (\rho_0)= 
 \left.\rho_0^2 \frac{\partial E_b (\rho)}{\partial \rho}\right|_{\rho_0}=0,
\end{equation}
and the nuclear matter incompressibility at the saturation point, 
\begin{equation}\label{eq:K0}
 K_0 (\rho_0)= \left.9\rho_0^2 \frac{\partial^2 E_b (\rho)}{\partial \rho^2}\right|_{\rho_0},
\end{equation}
is usually considered as a constraint over the equation of state, as well as the effective mass of the system,
\begin{equation}\label{eq:effmass}
 \frac{m^*}{m}= \left[1+ \frac{m}{\hbar k} \frac{\partial V_{F} (k)}{\partial k} \right]^{-1}_{k_F},
\end{equation}
where $V_F (k)$ is the Fock potential in momentum space and $k_F = (3 \pi^2 \rho/2)^{1/3}$ is the Fermi momentum of the system.

On the other hand, in asymmetric nuclear matter, where the neutron and proton densities take different values
$\rho_n \neq \rho_p$, the energy per particle will be a function of them, $E_b (\rho_n, \rho_p)$.
Also, it can be rewritten as a function of the total density 
\begin{equation}
 \rho= \rho_n+\rho_p
\end{equation}
and of the isospin asymmetry
\begin{equation}
 \delta= (\rho_n-\rho_p)/\rho,
\end{equation}
 i.e., $E_b (\rho, \delta)$.
This way, the energy density is given by 
\begin{equation}
 \mathcal{H}_b ( \rho, \delta) = \rho E_b( \rho, \delta), 
\end{equation}
and the neutron and proton chemical potentials are defined 
as the derivative of the baryon
energy density 
 with respect to the neutron and proton densities:
 \begin{equation}\label{chempot}
  \mu_n = \frac{\partial \mathcal{H}_b}{\partial \rho_n} \hspace{2cm} 
  \mu_p = \frac{\partial \mathcal{H}_b}{\partial \rho_p}.
 \end{equation}
 Finally, the pressure of the system can be defined either as a function of the derivative of the energy per particle or as a function of the chemical potentials, i.e.,  
\begin{equation}\label{eq:pre}
 P (\rho, \delta) = \rho^2 \frac{\partial E_b (\rho, \delta)}{\partial \rho} = \mu_n \rho_n + \mu_p \rho_p - \mathcal{H}_b (\rho, \delta).
\end{equation} 

If one expands the EoS around isospin asymmetry $\delta=0$, the energy per particle of a nuclear system can be rewritten as 
\begin{equation}\label{eq:eosexp}
 E_b(\rho, \delta) = E_b(\rho, \delta=0) + E_{\mathrm{sym}}(\rho) \delta^2 +\mathcal{O}(\delta^{4}),
\end{equation}
where the lowest term $E_b(\rho, \delta=0)$ is the energy of the system in SNM and 
$E_{\mathrm{sym}}(\rho)$ is the symmetry energy of the system, which reads

\begin{equation}\label{eq:esym}
 E_\mathrm{sym} (\rho) = \left.\frac{1}{2} \frac{\partial^2 E_b (\rho, \delta)}{\partial \delta^2} \right|_{\delta=0}.
\end{equation}
Notice that, due to the charge symmetry, assumed in the nuclear interactions, 
only even powers of $\delta$ can appear in the expansion of the symmetry energy~(\ref{eq:eosexp}).
If we expand the symmetry energy around the 
saturation density $\rho_0$ one obtains the expression
\begin{equation}\label{esymexp}
E_{\mathrm{sym}} (\rho)= E_{\mathrm{sym}} (\rho_0) + L \epsilon + K_\mathrm{sym} \epsilon^2+\mathcal{O}(\epsilon^3),
\end{equation}
where the density displacement from the saturation density $\rho_0$ is given by
\begin{equation}
 \epsilon = (\rho - \rho_ 0)/3\rho_0
\end{equation}
and $L$ is the slope of the symmetry energy at saturation, defined as 
\begin{eqnarray}\label{eq:L}
L\equiv  L (\rho_0)&=& 3\rho_0 \left.\frac{\partial E_{\mathrm{sym}} (\rho)}{\partial \rho} \right|_{\rho_0} 
\end{eqnarray}
and which gives information about the stiffness of the equation of state.
In Eq.~(\ref{esymexp}), the coefficient $K_\mathrm{sym}$ is the symmetry energy curvature, defined as
\begin{eqnarray}
K_\mathrm{sym}\equiv K_\mathrm{sym} (\rho_0) &=& 9 \rho_0^2 \left.\frac{\partial^2 E_{\mathrm{sym}} (\rho) }{\partial \rho^2} \right|_{\rho_0}.
\end{eqnarray}

In the following sections, we will introduce the phenomenological interactions we have used through all this work.
First, in Section \ref{Skyrme}, we will introduce the Skyrme zero-range forces, in Section \ref{Gogny} we introduce the Gogny finite-range 
interactions and finally, in Section \ref{MDISEI} we will introduce the finite-range momentum-dependent interactions
(MDI) and simple effective interactions (SEI).

 \section{Skyrme interactions}\label{Skyrme}
Skyrme interactions were first proposed considering that the functional of the energy could be expressed 
as a minimal expansion in momentum space compatible with the underlying symmetries
in terms of a zero-range expansion~\cite{skyrme56, Skyrme58}.
The standard Skyrme two-body effective nuclear interaction in coordinate space reads as \cite{Skyrme58, skyrme56, vautherin72,sly41,sly42}
 \begin{eqnarray}\label{VSkyrme}
  V (\mathbf{r}_1 , \mathbf{r}_2) &=& t_0 (1+x_0 P_\sigma) \delta (\mathbf{r})\nonumber
   \\
 &+& \frac{1}{2} t_1 (1 + x_1 P_\sigma) \left[ \mathbf{k'}^2 \delta (\mathbf{r}) + 
\delta (\mathbf{r}) \mathbf{k}^2 \right]\nonumber
 \\ 
  &+& t_2 (1+ x_2 P_\sigma) \mathbf{k'} \cdot \delta (\mathbf{r}) \mathbf{k} \nonumber
 \\
  &+&\frac{1}{6} t_3 (1+ x_3 P_\sigma) \rho^\alpha (\mathbf{R}) \delta (\mathbf{r}) \nonumber
 \\
  &+& i W_0 (\bm{\sigma}_1 + \bm{\sigma}_2) [\mathbf{k'} \times \delta (\mathbf{r}) \mathbf{k}], 
\end{eqnarray}
where $\mathbf{r} =\mathbf{r}_1 - \mathbf{r}_2 $ is the relative distance between two nucleons and
$\mathbf{R}= (\mathbf{r}_1 + \mathbf{r}_2)/2$ is their  center of mass coordinate. 
The two-body spin-exchange operator is defined as 
$P_\sigma = (1+ \bm{\sigma}_1 \cdot\bm{\sigma}_2)/2$, 
$\mathbf{k}= (\overrightarrow{\nabla}_1-\overrightarrow{\nabla}_2)/2i$ is the relative momentum between 
two nucleons, and $\mathbf{k}'$ is its complex conjugate,
$\mathbf{k}'= -(\overleftarrow{\nabla}_1-\overleftarrow{\nabla}_2)/2i$. 
 The first term in Eq.~(\ref{VSkyrme}) is the central term and the second and third ones are 
the non-local contributions, which simulate the finite range. 
The three-body force is also assumed as a zero-range force, which provides a simple 
phenomenological representation of many-body effects, and describes how
the interaction between two nucleons is influenced by the presence of others~\cite{vautherin72}.
Finally, the last term in Eq.~(\ref{VSkyrme}) is the spin-orbit 
contribution, which depends on the gradients of the density, and does not contribute in the case of homogeneous systems. 

In an infinite symmetric nuclear system the energy per baryon is given by~\cite{sly41}
\begin{equation}\label{eq:ebsnm}
 E_b (\rho) = \frac{3 \hbar^2}{10m} \left( \frac{3 \pi^2}{2}\right)^{2/3} \rho^{2/3} + 
 \frac{3}{8} t_0 \rho + \frac{3}{80} \left[3 t_1 + (5+4 x_2)t_2 \right] \left( \frac{3 \pi^2}{2}\right)^{2/3}
 \rho^{5/3} + \frac{1}{16} t_3 \rho^{\alpha+1}.
\end{equation}
 Using the definition in Eq.~(\ref{eq:pre}) the pressure in SNM reads
 \begin{eqnarray}
  P(\rho)&=& 
  \frac{\hbar^2}{5m} \left(\frac{3 \pi^2}{2} \right)^{2/3} \rho^{5/3} +\frac{3}{8} t_0 \rho^2 +  \frac{1}{16} t_3 (\alpha + 1) \rho^{\alpha+2}\nonumber\\
  &+& \frac{1}{16}  
  \left[3 t_1 + (5+4 x_2)t_2 \right] \left(\frac{3 \pi^2}{2} \right)^{2/3} \rho^{8/3} ,
 \end{eqnarray}
and the incompressibility for a Skyrme interaction is given by 
\begin{eqnarray}
 K(\rho)&=& 
 -\frac{3\hbar^2}{5m} \left(\frac{3 \pi^2}{2} \right)^{2/3} \rho^{2/3}+ \frac{9}{16}\alpha (\alpha+1) t_3 \rho^{\alpha+1} 
 \nonumber\\
 &+& \frac{3}{8} \left[3 t_1 + (5+4 x_2)t_2 \right]
 \left( \frac{3 \pi^2}{2}\right)^{2/3} \rho^{5/3} .
\end{eqnarray}

In asymmetric nuclear matter of density $\rho$ and isospin asymmetry $\delta$, the energy per particle for
a Skyrme interaction becomes
\begin{eqnarray}\label{eq:Ebanm}
E_b( \rho, \delta) &=& \frac{ 3 \hbar^2}{10m} \left(\frac{3 \pi^2}{2}\right)^{2/3} \rho^{2/3} 
F_{5/3} + \frac{1}{8} t_0 \rho \left[ 2 (x_0 +2) - (2 x_0 +1) F_2 \right] \nonumber\\
&+& \frac{1}{48} t_3 \rho^{\alpha + 1} \left[ 2 (x_3 +2) 
- (2 x_3 +1) F_2  \right] + \frac{3}{40} \left(\frac{3 \pi^2}{2} \right)^{2/3} 
\rho^{5/3} \\
&\times& \left[ \vphantom{\frac{1}{2}} \left[ t_1 (x_1 + 2) + t_2 (x_2 + 2) \right] F_{5/3} 
+ \frac{1}{2} \left[ t_2 (2 x_2 + 1 ) - t_1 (2 x_1 + 1) \right] F_{8/3}  \right], \nonumber
\end{eqnarray}
where the function $F_m$ is defined as
\begin{equation}
F_m = \frac{1}{2} \left[ (1 + \delta)^m + (1- \delta)^m \right].
\end{equation}
For Skyrme interactions of the type (\ref{VSkyrme}), the neutron ($n$) and proton ($p$) chemical
potentials, defined as the derivative of the energy density $\mathcal{H}_b$ with respect to 
the density of each kind of nucleons take form of
\begin{eqnarray}\label{eq:chempotskyrme}
 \mu_\tau (\rho_\tau, \rho_{\tau'})&=& \frac{\hbar^2}{2m}  
 (3 \pi^2)^{2/3} \rho_\tau^{2/3} + \frac{1}{2} t_0 \rho_\tau \left[ 2(x_0+2) - (2x_0+1) \right]
 \nonumber\\
 &+& \frac{1}{24} t_3 \left[ 2 (x_3+2) - (2 x_3+1) \right] \left[ \alpha \left( \rho_\tau^2+ \rho_{\tau'}^2 \right)
 \rho^{\alpha-1} + 2 \rho^\alpha \rho_\tau\right]\nonumber\\
 &+& \frac{3}{40} (3 \pi^2)^{2/3} \left\{  \left[t_1 (x_1+2)+t_2 (x_2+1) \right] \left[
  \rho_\tau^{5/3}+ \rho_{\tau'}^{5/3} + \frac{5}{3} \rho \rho_\tau^{2/3}\right] \right. \nonumber\\
 &+& \left. \frac{8}{3} \left[  t_2 (2 x_2 + 1) - t_1 (2 x_1+1)\right] \rho_\tau^{5/3}  \right\},
\end{eqnarray}
being $\tau, \tau' = n,p$, $\tau \neq \tau'$ the isospin indexes.
Moreover, the pressure in asymmetric nuclear matter, obtained through the derivative of $E_b (\rho, \delta)$ in 
Eq.~(\ref{eq:Ebanm}) with respect to the density is defined as 
\begin{eqnarray}\label{eq:press_skyrme}
 P(\rho, \delta)&=& \frac{\hbar^2}{5m} 
 \left(\frac{3 \pi^2}{2}\right)^{2/3} \rho^{5/3} 
F_{5/3} +  \frac{1}{8} t_0 \rho^2 \left[ 2 (x_0 +2) - (2 x_0 +1) F_2 \right]\nonumber\\
&+&\frac{1}{48} (\alpha+1) t_3 \rho^{\alpha + 2} \left[ 2 (x_3 +2) 
- (2 x_3 +1) F_2  \right] + \frac{1}{8} \left(\frac{3 \pi^2}{2} \right)^{2/3} 
\rho^{8/3} \\
&\times& \left[ \vphantom{\frac{1}{2}} \left[ t_1 (x_1 + 2) + t_2 (x_2 + 2) \right] F_{5/3} 
+ \frac{1}{2} \left[ t_2 (2 x_2 + 1 ) - t_1 (2 x_1 + 1) \right] F_{8/3}  \right]. \nonumber
\end{eqnarray}

If one expands the EoS in Eq.~(\ref{eq:Ebanm}) around isospin asymmetry $\delta=0$ [see Eq.~(\ref{eq:eosexp})],  the 
symmetry energy for the case of Skyrme forces is found to be of the form~\cite{sly41}
\begin{eqnarray}\label{eq:esym2skyrme}
 E_{\mathrm{sym}} (\rho)&=&
  \frac{\hbar^2}{6m} \left(  \frac{3 \pi^2}{2}\right)^{2/3} \rho^{2/3} - \frac{1}{8} t_0 \rho (2x_0 +1) 
  \\
&-& \frac{1}{48} t_3 \rho^{\alpha+1} (2x_3 +1) 
+ \frac{1}{24} \left(  \frac{3 \pi^2}{2}\right)^{2/3}   
 \times \rho^{5/3} \left[ -3 x_1 t_1 + t_2 (5 x_2 + 4) \right],  \nonumber
\end{eqnarray}
with a slope parameter $L$
\begin{eqnarray}
 L &=& \frac{\hbar^2}{3m}
 \left(  \frac{3 \pi^2}{2}\right)^{2/3} \rho_0^{2/3} -\frac{3}{8} t_0 \rho_0 (2 x_0+1) 
-\frac{1}{16}(\alpha+1) t_3 \rho_0^{\alpha+1} (2 x_3+1) 
\\
&+& \frac{5}{24} 
\left(  \frac{3 \pi^2}{2}\right)^{2/3} \rho_0^{5/3} \left[ -3 x_1 t_1 + t_2 (5 x_2+4)\right]\nonumber
\end{eqnarray}
and symmetry energy curvature 
\begin{eqnarray}
 K_\mathrm{sym} &=& 
 -\frac{\hbar^2}{3m}
 \left(  \frac{3 \pi^2}{2}\right)^{2/3} \rho_0^{2/3} -\frac{3}{16}\alpha(\alpha+1) t_3 \rho_0^{\alpha+1} (2 x_3+1)
\\
 &+& \frac{5}{12} 
\left(  \frac{3 \pi^2}{2}\right)^{2/3} \rho_0^{5/3} \left[ -3 x_1 t_1 + t_2 (5 x_2+4)\right].\nonumber
\end{eqnarray}

Finally, if we consider pure neutron matter with isospin asymmetry $\delta=1$, the energy per particle for 
Skyrme interactions becomes
\begin{eqnarray}
  E_b(\rho, \delta=1)&=&  \frac{3\hbar^2}{10m} \left( 3 \pi^2\right)^{2/3} \rho^{2/3} + \frac{1}{4}\rho t_0 (1-x_0)
  +\frac{1}{24} \rho^{\alpha+1} t_3 (1-x_3) \nonumber\\
  &+& \frac{3}{40} \left(3 \pi^2 \right)^{2/3} \rho^{5/3} \left[ t_1 (1-x_1) + 3 t_2 (1+x_2) \right].
\end{eqnarray}

In the following chapters, we will use several Skyrme parametrizations, that have been fitted to different
infinite nuclear matter properties and to properties of finite nuclei.
They are collected in Table~\ref{table:Skyrmeprops}, along with some of their properties of symmetric and asymmetric matter, 
and the references where 
their fittings are explained. We can see that all of the considered parametrizations have saturation densities around $\rho_0\simeq 0.16$ fm$^{-3}$, 
with energies of SNM at saturation around $E_b (\rho_0) \simeq -16$ MeV. Their incompressibilities are mostly
between $220 \lesssim K (\rho_0) \lesssim 270$ MeV.
The symmetry energy of these forces are within $ 26 \lesssim E_\mathrm{sym}  (\rho_0)\lesssim 37$ MeV, 
and their slopes within $9 \lesssim L \lesssim 130$ MeV,
covering the range given by some estimates coming from experimental data~\cite{BaoAnLi13,Vinas14}.
Finally, we see that the symmetry energy incompressibility ($K_\mathrm{sym}$) is the less constrained parameter, as also
there are not many constraints on it coming from experiments. The range of $K_\mathrm{sym} $
when using the interactions in Table \ref{table:Skyrmeprops} is $ -275 \lesssim K_\mathrm{sym}  \lesssim 71$ MeV.
All interactions in this Table~\ref{table:Skyrmeprops} are of the type described in Eq.~(\ref{VSkyrme}), 
excepting the Skyrme interactions Sk$\chi$414, Sk$\chi$450 and Sk$\chi$500, which
have an additional density-dependent term
\begin{equation}
V (\mathbf{r}_1 , \mathbf{r}_2) \rightarrow V (\mathbf{r}_1 , \mathbf{r}_2)
+\frac{1}{6} t_4 (1+ x_4 P_\sigma) \rho^{\alpha'} (\mathbf{R}) \delta (\mathbf{r})
\end{equation}
that takes into account the chiral N3LO asymmetric matter equation of state~\cite{skyrmechiral}. In these cases, the 
equations of the energy per particle, pressure and chemical potentials are changed accordingly, 
adding the new zero-range density-dependent term.
\begin{table}[t!]
\centering
\begin{tabular}{c|cdddddd}
\hline
\multirow{2}{*}{Skyrme force} &\multirow{2}{*}{Ref.}& \multicolumn{1}{c}{$\rho_0$}    & 
\multicolumn{1}{c}{$E_b (\rho_0)$} & \multicolumn{1}{c}{$K (\rho_0)$}& 
\multicolumn{1}{c}{$E_\mathrm{sym} (\rho_0)$}& \multicolumn{1}{c}{$L$ }   &  
\multicolumn{1}{c}{$K_\mathrm{sym}$} \\
             & & \multicolumn{1}{c}{(fm$^{-3}$)} & \multicolumn{1}{c}{(MeV)}  
             & \multicolumn{1}{c}{(MeV)}                     &\multicolumn{1}{c}{(MeV)} 
             & \multicolumn{1}{c}{(MeV)}  & \multicolumn{1}{c}{(MeV)}              \\\hline\hline
MSk7                      &\cite{Msk7}    & 0.158       & -15.80         & 231.21         & 27.95              & 9.41   & -274.62                   \\
SIII                      &\cite{SIII}     & 0.145       & -15.85         & 355.35         & 28.16              & 9.91   & -393.72                   \\
SkP                      &\cite{SkP}      & 0.163       & -15.95         & 200.96         & 30.00              & 19.68  & -266.59                   \\
HFB-27                  &\cite{HFB27}      & 0.159       & -16.05         & 241.63         & 30.00              & 28.50  & -221.41                   \\
SKX                      &\cite{SkX}     & 0.156       & -16.05         & 271.05         & 31.10              & 33.19  & -252.11                   \\
HFB-17                   &\cite{HFB17}     & 0.159       & -16.06         & 241.68         & 30.00              & 36.29  & -181.83                   \\
SGII                     & \cite{sgii}     & 0.158       & -15.59         & 214.64         & 26.83              & 37.63  & -145.90                   \\
UNEDF1                    &\cite{unedf1}    & 0.159       & -15.80         & 220.00         & 28.99              & 40.01  & -179.46                   \\
Sk$\chi$500                    &\cite{skyrmechiral}    & 0.168       & -15.99         & 238.14         & 29.12              & 40.74  & -77.40                    \\
Sk$\chi$450                    &\cite{skyrmechiral}    & 0.156       & -15.93         & 239.51         & 30.64              & 42.06  & -142.68                   \\
UNEDF0                    &\cite{unedf0}     & 0.161       & -16.06         & 230.00         & 30.54              & 45.08  & -189.67                   \\
SkM*                     &\cite{skms}     & 0.161       & -15.77         & 216.60         & 30.03              & 45.78  & -155.93                   \\
SLy4                     &\cite{sly41, sly42}     & 0.160       & -15.97         & 229.90         & 32.00              & 45.96  & -119.70                   \\
SLy7                     &\cite{sly41, sly42}     & 0.158       & -15.90         & 229.69         & 31.99              & 47.22  & -113.32                   \\
SLy5                     &\cite{sly41, sly42}     & 0.160       & -15.98         & 229.92         & 32.03              & 48.27  & -112.34                   \\
Sk$\chi$414                   &\cite{skyrmechiral}     & 0.170       & -16.20         & 243.17         & 32.34              & 51.92  & -95.71                    \\
MSka                     &\cite{mska}     & 0.154       & -15.99         & 313.32         & 30.35              & 57.17  & -135.34                   \\
MSL0                     &\cite{msl0}     & 0.160       & -16.00         & 229.99         & 30.00              & 60.00  & -99.33                    \\
SIV                      &\cite{SIII}     & 0.151       & -15.96         & 324.54         & 31.22              & 63.50  & -136.71                   \\
SkMP                      &\cite{skmp}     & 0.157       & -15.56         & 230.86         & 29.89              & 70.31  & -49.82                    \\
SKa                      &\cite{ska}     & 0.155       & -15.99         & 263.14         & 32.91              & 74.62  & -78.45                    \\
R$_\sigma$               &\cite{rsgs}     & 0.158       & -15.59         & 237.35         & 30.58              & 85.69  & -9.14                     \\
G$_\sigma$               &\cite{rsgs}     & 0.158       & -15.59         & 237.22         & 31.37              & 94.01  & 13.98                     \\
SV                       &\cite{SIII}      & 0.155       & -16.08         & 305.68         & 32.83              & 96.09  & 24.18                     \\
SkI2                    &\cite{ski2}      & 0.158       & -15.78         & 240.92         & 33.38              & 104.33 & 70.68                     \\
SkI5                    &\cite{ski2}       & 0.156       & -15.78         & 255.78         & 36.64              & 129.33 & 70.68         \\\hline            
\end{tabular}
\caption{Compilation of the Skyrme interactions used through this work, some of their SNM properties, such as the 
saturation density $\rho_0$, energy per particle $E_b (\rho_0)$ and incompressibility $K (\rho_0)$ at saturation density, and some ANM properties, 
such as the symmetry energy at the saturation point, $E_\mathrm{sym} (\rho_0)$, and its slope $L$ and curvature $K_\mathrm{sym}$ at saturation.
\label{table:Skyrmeprops}}
\end{table}
\newpage

\section{Gogny interactions}\label{Gogny}
Gogny interactions were proposed by D. Gogny with the aim of describing the mean-field and the pairing field in the same interaction.
The standard Gogny two-body effective nuclear interaction reads as\cite{decharge80, berger91, chappert08, goriely09,Sellahewa14}
\begin{eqnarray}\label{VGogny}
  V (\mathbf{r}_1 , \mathbf{r}_2) &=&  \sum_{i=1,2} \left( W_i + B_i P_\sigma - H_i P_\tau - M_i P_\sigma P_\tau \right)e^{-r^2 /\mu_i^2} \nonumber
   \\
 &+& t_3 \left( 1+x_3 P_\sigma \right) \rho^\alpha(\mathbf{R})\delta(\mathbf{r}) \nonumber
 \\
  &+ & i W_0 \left( \bm{\sigma}_1 + \bm{\sigma}_2 \right) \left[ \mathbf{k}' \times \delta(\mathbf{r}) \mathbf{k} \right].
\end{eqnarray}
The first term in Eq.~(\ref{VGogny}) is the finite-range part of the interaction, and it is modulated by two
Gaussian form-factors of long- and short-ranges. 
The following term is the zero-range density-dependent contribution to the interaction. The last term 
of the interaction corresponds to the spin-orbit force, which is also zero-range as in the 
case of Skyrme interactions and does not contribute in infinite nuclear matter. 
Gogny forces describe nicely ground-state systematics of finite nuclei, nuclear excitation properties, and fission phenomena.

The energy per baryon $E_b(\rho, \delta)$ in the Hartree--Fock approximation 
in asymmetric infinite nuclear matter for Gogny forces
as a function of the total baryon number density $\rho $ and 
of the isospin asymmetry $\delta$ can be decomposed as a 
sum of four different contributions, namely, a kinetic and a zero-range contributions, 
and the direct and exchange finite-range terms:
\begin{eqnarray}
 E_b (\rho, \delta)=  E_b^{\mathrm{kin}} (\rho, \delta)+ E_b^{\mathrm{zr}} (\rho, \delta)  
 \mbox{} + E_b^{\mathrm{dir}} (\rho, \delta) + E_b^{\mathrm{exch}} (\rho, \delta) \label{eq:eb.terms} \, ,
\end{eqnarray}
which read as
\begin{eqnarray}
 E_b^{\mathrm{kin}} (\rho, \delta)&=& \frac{ 3 \hbar^2}{20m} \left(\frac{3 \pi^2}{2}\right)^{2/3} \rho^{2/3}
 \left[ (1+\delta)^{5/3} + (1-\delta)^{5/3} \right] \label{eq:eb.kin}
\\
E_b^{\mathrm{zr}} (\rho, \delta)&=&  \frac{1}{8} t_3 \rho^{\alpha+1} \left[ 3-(2x_3+1)\delta^2 \right] \label{eq:eb.zr}
\\
E_b^{\mathrm{dir}} (\rho, \delta)&=&  \frac{1}{2} \sum_{i=1,2} \mu_i^3 \pi^{3/2} \rho  \left[ {\cal A}_i 
+{\cal B}_i \delta^2 \right] \label{eq:eb.dir}
\\
E_b^{\mathrm{exch}} (\rho, \delta)&= & -\sum_{\mathrm{i}=1,2}\frac{1}{2  k_F^3 \mu_i^3} 
\Big\{ {\cal C}_i \left[ {\mathsf  e} (k_{Fn} \mu_i ) + {\mathsf  e} (k_{Fp} \mu_i ) \right]
-  {\cal D}_i  \bar {\mathsf  e}( k_{Fn} \mu_i,k_{Fp} \mu_i )  \Big\}, \label{eq:eb.exch}
\end{eqnarray}
 with 
 \begin{equation}
{\mathsf e}(\eta) = \frac{\sqrt{\pi}}{2} \eta^3 \mathrm{erf}(\eta) 
+ \left(\frac{\eta^2}{2} - 1 \right) e^{-\eta^2} - \frac{3 \eta^2}{2} +  1 \, ,
\end{equation}
and 
 \begin{eqnarray}
\bar {\mathsf  e}(\eta_1,\eta_2)&=& \sum_{s=\pm 1} s 
\left[ 
\frac{ \sqrt{\pi}}{2} (\eta_1 + s \eta_2 ) \left( \eta_1^2 + \eta_2^2 - s \eta_1 \eta_2 \right) 
\mathrm{erf} \left( \frac{\eta_1 + s \eta_2 }{2}  \right) \right. \nonumber \\
&+& \left. \left( \eta_1^2 + \eta_2^2  - s \eta_1 \eta_2  -2 \right) e^{ - \frac{1}{4} (\eta_1 + s \eta_2)^2 } 
\right] ,
\end{eqnarray}
where 
\begin{equation}
 \displaystyle \mathrm{erf}(x) =  \frac{2}{\sqrt{\pi}} \int_0^x  e^{-t^2} dt
\end{equation}
 is the error function.
The function $\bar {\mathsf  e}(\eta_1,\eta_2)$ is a symmetric function of its arguments, satisfying 
$\mathsf{\bar e}(\eta,\eta)= 2 \mathsf{e}(\eta)$ and $\mathsf{\bar e}(\eta,0) = 0$.
The value of the parameter $x_3$ is considered equal to one, $x_3=1$, for all Gogny parametrizations of the D1 family, in order to 
avoid the zero-range contributions to the pairing field~\cite{decharge80}.

The term $E_b^{\mathrm{kin}} (\rho, \delta)$ is the sum of the contributions of the neutron and proton kinetic energies, and 
$E_b^{\mathrm{zr}} (\rho, \delta)$ comes from the zero-range interaction. The term in Eq.~(\ref{eq:eb.dir}) defines
the direct contribution of the finite-range part of the force, whereas the term in Eq.~(\ref{eq:eb.exch}) defines its exchange contribution. 
The kinetic, zero-range and finite-range direct terms can be expressed as functions of the density $\rho$ and the asymmetry $\delta$, 
whereas the finite-term exchange contribution is defined as a function of the neutron, $k_{Fn} = k_F (1+\delta)^{1/3}$ ,
and proton, $k_{Fp} = k_F (1-\delta)^{1/3}$, Fermi momenta. The Fermi momentum 
of symmetric nuclear matter is given by $k_F = (3 \pi^2 \rho/2 )^{1/3}$.
Moreover, the combinations of parameters appearing in the different terms of Eq.~(\ref{eq:eb.terms}) are the following:
\begin{eqnarray}
{\cal A}_i &=& \frac{1}{4} \left( 4 W_i + 2 B_i - 2H_i -M_i \right) \label{Ai}
\\
{\cal B}_i&=&  -\frac{1}{4}\left( 2 H_i + M_i \right)\label{Bi}
\\
{\cal C}_i&=& \frac{1}{\sqrt{\pi}} \left( W_i + 2 B_i - H_i -2 M_i\right)\label{Ci}
\\
{\cal D}_i&=& \frac{1}{\sqrt{\pi}} \left( H_i + 2 M_i \right).\label{Di}
\end{eqnarray}
The constants $ {\cal A}_i $ and ${\cal B}_i$ define, respectively, the isoscalar and isovector 
part of the direct term. For the exchange terms, the matrix elements ${\cal C}_i $ relate to 
neutron-neutron and proton-proton interactions, whereas 
the matrix elements ${\cal D}_i$ take care of neutron-proton interactions. 

If the energy per particle in asymmetric nuclear matter is expanded 
in terms of the isospin asymmetry $\delta$ [see Eq.~(\ref{eq:eosexp})] the symmetry energy coefficient 
for the Gogny parametrization is defined as~\cite{Sellahewa14}
\begin{eqnarray}
E_{\mathrm{sym}} (\rho) &=& 
\frac{\hbar^2}{6m} \left(  \frac{3 \pi^2}{2}\right)^{2/3} \rho^{2/3}   -
\frac{1}{8} t_3 \rho^{\alpha+1} (2x_3 +1) \nonumber
\\
&+& \frac{1}{2} \sum_{i=1,2} \mu_i^3 \pi^{3/2}  {\cal B}_i  \rho  
+ \mbox{} \frac{1}{6}\sum_{i=1,2}  \left[-{\cal C}_i  G_1 ( k_F \mu_i)+ {\cal D}_i G_2 ( k_F \mu_i)  \right] \label{eq:esym2gogny}
\end{eqnarray}
with
\begin{eqnarray}
 G_1 (\eta)&=& \frac{1}{\eta} -\left( \eta + \frac{1}{\eta} \right) e^{-\eta^2} \label{G1}
\\
 G_2 (\eta)&=& \frac{1}{\eta} -\bigg( \eta + \frac{e^{-\eta^2}}{\eta} \bigg), \label{G2}
\end{eqnarray}
and its symmetry energy slope $L$ at saturation density $\rho_0$ is given by 
 \begin{eqnarray}
 L &=& 3 \frac{\hbar^2}{3m} \left(\frac{3 \pi^2}{2}\right)^{2/3} \rho_0^{2/3} -
  \frac{3(\alpha + 1)}{8} t_3 \rho_0^{\alpha+1} (2 x_3 +1)\nonumber \\
  &+& \frac{3}{2} \sum_{i=1,2} \mu_i^3 \pi^{3/2} \mathcal{B}_i \rho_0 
  + \mbox{}  \frac{1}{6} \sum_{i=1,2} \left[ - \mathcal{C}_i L_1(\mu_i k_{F0}) + \mathcal{D}_i L_2(\mu_i k_{F0}) \right] ,
 \end{eqnarray}
where $k_{F0} = (3 \pi^2 \rho_0/2)^{1/3}$ is the Fermi momentum at saturation and the $L_n (\eta)$ functions are
\begin{eqnarray}
 L_1 (\eta) &=& -\frac{1}{\eta} +e^{-\eta^2} \left( \frac{1}{\eta} + \eta + 2 \eta^3\right) \label{G1'}
 \\
 L_2 (\eta) &=&  -\frac{1}{\eta} + e^{-\eta^2} \left( \frac{1}{\eta} +2\eta \right) - \eta.
\end{eqnarray}
Also, for Gogny interactions, the symmetry energy curvature is defined as
\begin{eqnarray}
 K_\mathrm{sym}  &=&  
  -\frac{\hbar^2}{3m} \left(\frac{3 \pi^2}{2}\right)^{2/3} \rho_0^{2/3} -
  \frac{9(\alpha + 1)}{8} t_3 \rho_0^{\alpha+1} (2 x_3 +1)\nonumber \\
  &-& \mbox{}  \frac{2}{3} \sum_{i=1,2} \left[ - \mathcal{C}_i K_1(\mu_i k_{F0}) + \mathcal{D}_i K_2(\mu_i k_{F0}) \right] ,
\end{eqnarray}
with the functions $K_n (\eta)$ reading as
\begin{eqnarray}
 K_1 (\eta) &=& -\frac{1}{\eta} + \left( \frac{1}{\eta} + \eta + \frac{\eta^3}{2} + \eta^5\right)e^{-\eta^2}
 \\
 K_2 (\eta) &=&  -\frac{1}{\eta} - \frac{\eta}{2} + \left( \frac{1}{\eta} + \frac{3 \eta}{2} + \eta^3 \right)e^{-\eta^2}.
\end{eqnarray}

As stated previously, the proton and chemical potentials in asymmetric matter will be given by the 
derivatives of the nuclear energy density $\mathcal{H}_b = \rho E_b$ with respect to the neutron ($\rho_n$) and
proton ($\rho_p$) densities. For Gogny interactions, the neutron ($\tau = +1$) and proton ($\tau = -1$) chemical potentials are 
\begin{eqnarray}
 \mu_\tau (\rho_\tau, \rho_{\tau'})&=&  \frac{\hbar^2}{2m} \left( 3 \pi^2 \right)^{2/3} \rho_\tau^{2/3} + \frac{t_3}{8} \rho^{\alpha + 1} 
 \left[ 3 \left( \alpha + 2\right) - 2 \tau \left( 2 x_3 +1\right) \delta - \alpha \left( 2 x_3 + 1\right) 
 \delta^2\right] \nonumber
 \\
 &+&  \sum_{i=1,2} \mu_i^3 \pi^{3/2} \rho \left( {\cal A}_i + \tau {\cal B}_i \delta \right) 
 - \sum_{i=1,2} \left[ {\cal C}_i\,  \bar {\mathsf w}( k_F^\tau \mu_i , k_F^\tau \mu_i  ) - 
 {\cal D}_i\, \bar {\mathsf  w} (k_F^\tau \mu_i , k_F^{-\tau} \mu_i )\right],\nonumber\\ 
\label{mu-tau-gog}
\end{eqnarray}
where $\bar {\mathsf w} \left( \eta_1 , \eta_2\right)$ is the dimensionless function 
\begin{equation}
\bar {\mathsf w} \left( \eta_1 , \eta_2\right) =
\sum_{s=\pm 1} s \left[ \frac{\sqrt{\pi}}{2} \mathrm{erf}\left( \frac{\eta_1 + s \eta_2}{2}\right) 
+ \frac{1}{\eta_1} e^{- \frac{1}{4} ( \eta_1 + s \eta_2 )^2} \right] \, .
\end{equation}

From the derivative of the energy per particle with respect to the density, one can obtain the pressure of the baryons 
in the system, which for Gogny forces reads
\begin{eqnarray}\label{eq:pressure_bars}
P_b(\rho,\delta) &=& 
\frac{\hbar^2}{10m} \left(\frac{3 \pi^2}{2}\right)^{2/3} \rho^{5/3} \left[ (1+\delta)^{5/3} + (1-\delta)^{5/3} 
\right] \nonumber \\
&+& \frac{(\alpha+1)}{8} t_3 \rho^{\alpha+2} \left[ 3 - (2x_3+1) \delta^2 \right] 
 + \frac{\rho^2}{2} \sum_{i=1,2} \pi^{3/2} \mu_i^3 \left( {\cal A}_i + {\cal B}_i \delta^2 \right)  \\
& -&  \frac{\rho}{2}  \sum_{i=1,2} \left\{ 
{\cal C}_i
\left[ (1+\delta) \mathsf{p} ( k_{Fn} \mu_i ) 
     + (1-\delta) \mathsf{p} ( k_{Fp} \mu_i ) \right]   
- {\cal D}_i \mathsf{ \bar p} ( k_{Fn} \mu_i,k_{Fp}\mu_i )  
     \right\} \, .\nonumber
\end{eqnarray}
The function $\mathsf{p}(\eta)$ contains the density dependence of the pressure in both symmetric and neutron 
matter \cite{Sellahewa14}:
\begin{align}
\mathsf{p}\left(\eta \right) &= -\frac{1}{\eta^3} + \frac{1}{2 \eta} + \left( \frac{1}{\eta^3} + 
\frac{1}{2 \eta} \right) e^{- \eta^2} \, .
\end{align}
In asymmetric matter, the double integral on the exchange terms leads to the appearance of a term that 
depends on the two Fermi momenta: 
\begin{align}
 \mathsf{\bar p} \left( \eta_1, \eta_2 \right)  = &
\frac{2}{\eta_1^3 + \eta_2^3}
\sum_{s=\pm 1}  (\eta_1 \eta_2 +2s ) e^{- \frac{1}{4} \left(\eta_1 + s\eta_2\right)^2 } .
 \label{eq:p_function}
 \end{align}

In SNM, where the neutron and proton densities are equal ($\delta=0$),
the energy per baryon (\ref{eq:eb.terms}) can be 
rewritten as~\cite{Sellahewa14}
\begin{equation}\label{Ebgognysnm}
 E_{b} (\rho) = \frac{3}{5} \frac{\hbar^2}{2 m} k_F^2 + \frac{3}{8} t_3 \rho^{\alpha+1} + \frac{1}{2} \sum_{i=1,2}\left[ \mu_i^3
 \pi^{3/2} \rho \mathcal{A}_i + \mathcal{B}_{0i}  g (\mu_i k_F)\right], 
\end{equation}
where the coefficient $\mathcal{A}_i$ is defined in Eq.~(\ref{Ai}) and $\mathcal{B}_{0i}$ is
\begin{equation}\label{Boigogny}
 \mathcal{B}_{0i} = -\frac{1}{\sqrt{\pi}} \left( W_i + 2 B_i - 2 H_i -4 M_i \right).
\end{equation}
Similarly to asymmetric nuclear matter, one can find the pressure in SNM as the derivative of the energy per particle
with respect to the density, obtaining
\begin{equation}\label{PbsnmGogny}
 P (\rho) = 
 \frac{2}{5} \frac{\hbar^2}{2 m} k_F^2 \rho + \frac{3}{8} t_3 (\alpha + 1)\rho^{\alpha+2} + \frac{1}{2} \sum_{i=1,2}\left[ \mu_i^3
 \pi^{3/2} \rho^2 \mathcal{A}_i + \rho \mathcal{B}_{0i}   p (\mu_i k_F)\right].
\end{equation}
The nuclear matter incompressibility is given by the curvature of the energy per particle in SNM. 
In the case of Gogny interactions, the expression for the nuclear matter incompressibility is
\begin{equation}\label{K0Gogny}
 K (\rho) = 
 -\frac{6}{5} \frac{\hbar^2}{2 m} k_F^2  + \frac{9}{8} t_3 (\alpha + 1)\alpha\rho^{\alpha+1} - 3 \sum_{i=1,2}
  \mathcal{B}_{0i}   k (\mu_i k_F).
\end{equation}
The functions $g(\eta)$, $p(\eta)$ and $k(\eta)$ appearing in the expressions of the energy per particle, pressure and 
incompressibility in SNM are given, respectively, by 
\begin{eqnarray}
 g(\eta) &=& \frac{2}{\eta^3} - \frac{3}{\eta} - \left( \frac{2}{\eta^3}-\frac{1}{\eta}\right)e^{-\eta^2} +
 \sqrt{\pi} \mathrm{erf} (\eta)
\\
 p(\eta) &=& - \frac{1}{\eta^3} + \frac{1}{2\eta} + \left( \frac{1}{\eta^3} + \frac{1}{2\eta} \right) e^{-\eta^2}
\\
 k(\eta) &=& - \frac{6}{\eta^3} + \frac{2}{\eta} + \left( \frac{6}{\eta^3} + \frac{4}{\eta}+\eta \right) e^{-\eta^2}.
\end{eqnarray}
If one considers pure neutron matter, with $\delta=1$, the expression of the energy per particle can be
written as~\cite{Sellahewa14}
\begin{equation}
 E_{b} (\rho, \delta=1) = \frac{3}{5} \frac{\hbar^2}{2 m} k_{Fn}^2 + \frac{1}{4} t_3 \rho^{\alpha+1} 
 (1-x_3)+ \frac{1}{2} \sum_{i=1,2}\left[ \mu_i^3
 \pi^{3/2} \rho \mathcal{A}_i - \mathcal{C}_{i}  g (\mu_i k_{Fn})\right].
\end{equation}
\begin{table}[!t]
\centering
\begin{tabular}{c|ccccccc}
\hline
\multirow{2}{*}{Gogny Force} & \multirow{2}{*}{Ref.} & $\rho_0$    & $E_b$ ($\rho_0$) & $K(\rho_0)$  & $E_\mathrm{sym} (\rho_0)$ & $L$ & $K_\mathrm{sym} $ \\
                             &                       & (fm$^{-3}$) & (MeV)                            & (MeV)  & (MeV)                     & (MeV)        & (MeV)                     \\\hline\hline
D260                         &   \cite{NPA591Blaizot1995}                    & 0.1601      & -16.26                           & 259.49 & 30.11                     & 17.57        & 259.49                    \\
D1                           &    \cite{decharge80}                   & 0.1665      & -16.31                           & 229.37 & 30.70                     & 18.36        & 229.37                    \\
D1S                          &     \cite{berger91}                  & 0.1633      & -16.01                           & 202.88 & 31.13                     & 22.43        & 202.88                    \\
D1M                          &    \cite{goriely09}                  & 0.1647      & -16.03                           & 224.98 & 28.55                     & 24.83        & 224.98                    \\
D250                         &    \cite{NPA591Blaizot1995}                   & 0.1577      & -15.80                           & 249.41 & 31.54                     & 24.90        & 249.41                    \\
D300                         &    \cite{NPA591Blaizot1995}                   & 0.1562      & -16.22                           & 299.14 & 31.22                     & 25.84        & 299.14                    \\
D1N                          &     \cite{chappert08}                  & 0.1612      & -15.96                           & 225.65 & 29.60                     & 33.58        & 225.65                    \\
D1M$^{**}$                        &     \cite{gonzalez18a}                  & 0.1647      & -16.02                           & 225.38 & 29.37                     & 33.91        & 224.98                    \\
D1M$^*$                         &     \cite{gonzalez18}                  & 0.1650      & -16.06                           & 224.98 & 30.25                     & 43.18        & 225.38                    \\
D2                           &   \cite{chappert15}                    & 0.1628      & -16.00                           & 209.26 & 31.11                     & 44.83        & 209.26                    \\
D280                         &    \cite{NPA591Blaizot1995}                   & 0.1525      & -16.33                           & 285.20 & 33.14                     & 46.53        & 285.20                   \\\hline
\end{tabular}
\caption{Compilation of the Gogny interactions used through this work, some of their SNM properties, such as the 
saturation density $\rho_0$, energy per particle $E_b (\rho_0)$ and incompressibility $ K  (\rho_0)$ at saturation density, and some ANM properties, 
such as the symmetry energy at the saturation point $E_\mathrm{sym} (\rho_0)$, and its slope $L$ and curvature $K_\mathrm{sym}$ at saturation.\label{table:Gognyprops}}
\end{table}
We collect in Table~\ref{table:Gognyprops} some isoscalar and isovector properties of the different Gogny interactions we have used in this work.
The first proposed Gogny interaction was D1 \cite{decharge80}, fitted to some properties of nuclear matter properties and of few closed-shell nuclei.
After D1, D1S \cite{chappert08} was proposed with the aim of getting a better description of nuclear fission \cite{berger91}. 
The interactions  D250, D260, D280, and D300 \cite{NPA591Blaizot1995} were devised to have different 
nuclear matter incompressibility for calculations of the breathing mode in nuclei. 
In order to improve the isovector part of the Gogny interactions, the D1N \cite{chappert08} and D1M 
\cite{goriely09} interactions were fitted to reproduce the microscopic neutron matter EoS of  
Friedman and Pandharipande~\cite{Friedman81}. 
The interactions D1M$^*$~\cite{gonzalez18} and D1M$^{**}$~\cite{gonzalez18a} will be introduced later in 
Chapter~\ref{chapter3}. They are fitted in such a way that, 
while preserving the description of finite nuclei similar to the one obtained with D1M, they are able to provide NSs inside 
the observational constraints for the neutron star mass~\cite{Demorest10, Antoniadis13}. 
Finally, the D2 \cite{chappert15} interaction is a recent Gogny interaction where the zero-range density-dependent  
term in Eq.~(\ref{VGogny}) has 
been replaced by a finite-range density-dependent term of Gaussian type, i.e., 
\begin{eqnarray}  
 V_\mathrm{dens}^{D1} &=& t_3 \left( 1+x_3 P_\sigma \right) \rho^\alpha(\mathbf{R})\delta(\mathbf{r}) \nonumber\\
 &&\downarrow\nonumber\\
 V_\mathrm{dens}^{D2} &=& \left( W_3 + B_3 P_\sigma - H_3 P_\tau - M_3 P_\sigma P_\tau \right)\times \frac{e^{-r^2 /\mu_3^2}}{(\mu_3 \sqrt{\pi})^3}
 \frac{\rho^\alpha ({\bf r}_1) + \rho^\alpha ({\bf r}_2)}{2},
 \end{eqnarray}
and therefore the equations listed above are switched accordingly. 
From Table~\ref{table:Gognyprops} we see that Gogny interactions have similar saturation densities around $\rho_0 \simeq 0.16$ fm$^{-3}$, 
and SNM energy per particle  at saturation around \mbox{$E_b (\rho_0) \simeq-16$ MeV}. Their incompressibilities vary in the range 
$202$ MeV $ \lesssim K_0  (\rho_0)\lesssim 286$ MeV and their symmetry energies lay within 
$ 28$ MeV $   \lesssim E_\mathrm{sym}(\rho_0) \lesssim 34$ MeV. Finally, Gogny interactions have 
slope parameters $L$ in the low-moderate regime between $17$ and $47$ MeV, and asymmetric incompressibilities 
$K_\mathrm{sym}$ that go from $202$ to $300$ MeV.

\section{Momentum-dependent interactions and Simple effective interactions}\label{MDISEI}

The momentum-dependent interactions (MDI)~\cite{das03} have been extensively used to study  transport calculations in heavy-ion 
collisions~\cite{das03,li08}, and have also been applied to other different scenarios,
in particular to neutron stars \cite{xu09a,xu09b,xu10a,xu10b,li08,Krastev19}.
The  potential energy density that one uses for MDI interactions in an asymmetric nuclear matter system 
is \cite{das03}
\begin{eqnarray}
 V (\rho, \delta) &=& \frac{A_1}{2 \rho_0} \rho^2 + \frac{A_2}{2 \rho_0} \rho^2 \delta^2 + 
 \frac{B}{\sigma+1} \frac{\rho^{\sigma+1}}{\rho_0^\sigma} (1-x \delta^2) \nonumber\\
 &+& \frac{1}{\rho_0} \sum_{\tau, \tau'} C_{\tau, \tau'} \int \int d^3p d^3p' \frac{f_\tau ({\bf r}, {\bf p})
 f_{\tau'} ({\bf r}, {\bf p}')}{1+({\bf p} - {\bf p}')^2/\Lambda^2},
\end{eqnarray}
where $\tau$ and $\tau'$ refer to the nucleon isospin and $f_\tau ({\bf r}, {\bf p}) $ is the nucleon phase-space
distribution function at the position ${\bf r}$. The parameters $A_1= (A_l+A_u)/2$, $A_2= (A_l-A_u)/2$, $B$, 
$\sigma$, $\Lambda$, $C_l=C_{\tau, \tau}$ and $C_u=C_{\tau, \tau'}$ are fitted as explained in Refs. \cite{das03, chen14},
subject to the constraint that the 
momentum-dependence of the single-particle potential reproduces the behaviour predicted by the Gogny
interaction, which gives a reasonable parametrization of the real part of the optical potential in nuclear
matter as a function of the incident energy. 
In the fitting protocol of the MDI parameters, it is  
required that in SNM the saturation density has values of $\rho_0=0.16$ fm$^{-3}$, the binding energy of $E_b(\rho_0)= 16$ MeV 
and the incompressibility
of $K (\rho_0)=211$ MeV. Moreover, it is imposed that the symmetry energy at saturation takes a value of 
$E_\mathrm{sym}(\rho_0)=30.5$ MeV \cite{das03, chen14}. 
In the original MDI interaction~\cite{das03}, the parameter $x$ could only take values of $x=0$ and $x=1$.
In order to be able to have a larger range of the dependence of the symmetry energy with the density
while keeping its fitted value $E_\mathrm{sym} (\rho_0)=30.5$ MeV at saturation, the parameters $A_l$ and $A_u$ \cite{Chen05} are
redefined as
\begin{equation}
 A_l=-120.57 + x \frac{2B}{\sigma+1} \hspace{2cm} A_u=-95.98-x \frac{2B}{\sigma+1}.
\end{equation}
Hence, a change in the parameter $x$ of the MDI interaction will modify the density 
dependence of the symmetry energy and of the neutron matter EoS, without changing the EoS of 
SNM and the symmetry energy at the saturation density.

The MDI interaction can be rewritten as a zero-range contribution plus a finite range term
with a single Yukawa form factor \cite{das03}, 
\begin{eqnarray}\label{VMDI}
V (\mathbf{r}_1 , \mathbf{r}_2) &=& \frac{1}{6} t_3 \left( 1+x_3 P_\sigma \right) \rho^\sigma \left({\bf R} \right) \delta({\bf r})
+ \left( W+B P_\sigma -H P_\tau- M P_\sigma P_\tau \right)\frac{e^{-\mu r}}{ r}.
\end{eqnarray}

The simple effective interaction (SEI) is an effective nuclear force constructed with a minimum number of parameters to 
study the momentum and density dependence of the nuclear mean-field,
defined as \cite{behera98, Behera05}
\begin{eqnarray}\label{VSEI}
V (\mathbf{r}_1 , \mathbf{r}_2) &=& t_0(1+x_0P_\sigma) \delta({\bf r})
+\frac{1}{6} t_3 \left( 1+x_3 P_\sigma \right) \left(
\frac{\rho({\bf R})}{1+b \rho ({\bf R}) }\right)^\gamma  \delta({\bf r})\nonumber\\
&+& \left( W+B P_\sigma -H P_\tau- M P_\sigma P_\tau \right)v (r) .
\end{eqnarray}
This interaction has a finite-range part, which can be of any conventional form factor $v(r)$ of Gaussian, Yukawa or exponential types, and 
two zero-range terms, one of them density-dependent, containing altogether eleven parameters plus the range $\mu$ of the form factor
and the spin-orbit strength parameter $W_0$ which enters in the description of finite nuclei. 
We see that the density-dependent
term of SEI contains the factor (1+b$\rho$)$^\gamma$ in the denominator, 
where the parameter $b$ is fixed to prevent the supra-luminous behavior in nuclear matter
at high densities~\cite{Behera_1997}.
The study of asymmetric 
nuclear matter involves altogether nine parameters, 
namely, $\gamma$, $b$, $\varepsilon_{0}^{l}$, $\varepsilon_{0}^{ul}$,
$\varepsilon_{\gamma}^{l}$,$\varepsilon_{\gamma}^{ul}$, $\varepsilon_{ex}^{l}$,
$\varepsilon_{ex}^{ul}$ and $\alpha$, whose connection to the interaction 
parameters is given in the following pages.
However, the description of SNM requires only the following combinations
of the strength parameters in the like and unlike channels
\begin{eqnarray}
\left(\frac{\varepsilon_{0}^{l}+\varepsilon_{0}^{ul}}{2}\right)=\varepsilon_0,
\left(\frac{\varepsilon_{\gamma}^{l}+\varepsilon_{\gamma}^{ul}}{2}\right)=\varepsilon_{\gamma},
\left(\frac{\varepsilon_{ex}^{l}+\varepsilon_{ex}^{ul}}{2}\right)=\varepsilon_{ex},
\label{eq.18}
\end{eqnarray}
together with $\gamma$, $b$ and $\alpha$, 
i.e., altogether six parameters.
The coefficients are fitted, considering $\gamma$ as a free parameter, to properties of finite nuclei, such as 
the nucleon mass, saturation density or 
binding energy per particle at saturation.
Moreover, it is required that 
 the nuclear mean-field in SNM at saturation density 
vanishes for a kinetic energy of the nucleon of $300$ MeV, a result extracted from optical 
model analysis
of nucleon-nucleus data~\cite{behera98}. 
The splitting of $\varepsilon_{ex}$ into
$\varepsilon_{ex}^{l}$ and $\varepsilon_{ex}^{ul}$ is decided from the 
condition that the entropy density in pure neutron matter should not exceed 
that of the symmetric nuclear matter, prescribing a critical value of
$\varepsilon_{ex}^{l}=2\varepsilon_{ex}/3$~\cite{Behera_2011}. 
The splitting of the remaining two strength parameters 
$\varepsilon_{\gamma}$ and $\varepsilon_{0}$,
is obtained from the values of the symmetry energy and 
its derivative with respect to the density
at saturation density $\rho_0$.
In our study, we will also consider the slope parameter $L$ as a free parameter.

In this work we will mostly use the Yukawian form factor $v(r)=\frac{e^{-\mu r}}{ \mu r}$ for the SEI force.
In this case, if we compare the MDI interaction defined in Eq.~(\ref{VMDI}) and the SEI interaction given in Eq.~(\ref{VSEI}),
we can write the expressions of the 
energy in a general way to useful for both MDI and SEI model with the following caveats in mind:
\begin{itemize}
 \item Firstly, MDI interactions only have a zero-range density-dependent term, contrary to SEI interactions that have two 
zero-range terms. Therefore, in the following expressions, we will consider, for MDI the parameters $t_0=0$ and $x_0=0$.
\item The expression for the SEI the zero-range density-dependent term is defined differently with 
respect to the MDI one:
\begin{equation}
 V_\mathrm{SEI}^{zr} (\mathbf{r}_1 , \mathbf{r}_2) = \frac{V_\mathrm{MDI}^{zr} (\mathbf{r}_1 , \mathbf{r}_2)}{(1+b\rho({\bf R}))^\gamma}.
\end{equation}
For this reason, from here on, we will consider the value of $b=0$ when using the MDI force.
\item The Yukawian form factor for SEI interaction is defined in dimensionless form, 
\begin{equation}
 v(r) = \frac{e^{-\mu r}}{ \mu r}.
\end{equation}
On the other hand, the MDI form factor has units of $\mathrm{fm}^{-1}$, i.e., 
\begin{equation}
 v(r) = \frac{e^{-\mu r}}{r}.
\end{equation}
Thus, the contribution of the potential part of the SEI interactions will be divided with an additional parameter $\mu$ 
with respect to the contribution given by the MDI interaction. We will further emphasize this point 
when we define the energy, pressure and chemical potentials in the following lines.  
\end{itemize} 

Hence, from here on, we present the expressions of different properties of both symmetric and asymmetric nuclear matter for the case of SEI interactions. 
Taking into account the stated differences between the SEI and MDI functionals, one can use the same 
equations to describe the corresponding properties for the case of MDI interactions
\footnote{Notice that a similar procedure could be applied for Gogny interactions and for 
SEI interactions with a Gaussian form factor instead of the Yukawa form factor considered here.}.

In the limit of an infinite symmetric nuclear system $\rho_n=\rho_p$, the energy per particle of SEI (MDI) interactions can be written as~\cite{routray16}
\begin{eqnarray}
 E_b (\rho) &=& \frac{3 \hbar^2 k_F^2}{10 m } + \frac{\left(\varepsilon_0^l + \varepsilon_0^{ul}\right)}{4 \rho_0} \rho +
 \frac{\left(\varepsilon_\gamma^l + \varepsilon_\gamma^{ul} \right)}{4 \rho^{\gamma+1}} \rho \left(\frac{\rho}{1+ b \rho} \right)^\gamma
 + \frac{(\varepsilon_{ex}^l+\varepsilon_{ex}^{ul})}{4 \rho_0} \rho\mathcal{A} (x_F), \nonumber
\end{eqnarray}
where the exchange term $\mathcal{A}(x_F)$ is
\begin{equation}
 \mathcal{A} (x_F)=  \left(\frac{3}{32 x_F^6} + \frac{9}{8 x_F^4} \right)\ln (1+4 x_F^2)
 -\frac{3}{8x_F^4} + \frac{9}{4 x_F^2} - \frac{3}{x_F ^3} \tan^{-1} (2 x_F).
\end{equation}
 The dimensionless quantity $x_F$ is defined as $x_F = k_F/\mu$, being $k_F=(3 \pi^2\rho/2 )^{1/3}$ the Fermi momentum of the system.
 
On the other hand, in an infinite pure neutron matter system, that is, with $\rho= \rho_n$ and $\rho_p=0$,
the energy per particle will be defined as
\begin{eqnarray}\label{SEIe0}
 E_b (\rho, \delta=1) &=& \frac{3 \hbar^2 k_{Fn}^2}{10 m} + \frac{\varepsilon_0^l}{2 \rho_0} \rho + \frac{\varepsilon_\gamma^l}{2 \rho_0^{\gamma+1}}
 \rho \left( \frac{\rho}{1+ b \rho} \right)^\gamma
 + \frac{\varepsilon_{ex}^l}{2 \rho_0} \rho  \mathcal{B} (x_F),
\end{eqnarray}
with
\begin{equation}\label{SEIen}
 \mathcal{B} (x_n) =  \left( \frac{3}{32 x_n^6}  + \frac{9}{8 x_n^4}
 \right) \ln (1+4 x_n^2) - \frac{3}{8 x_n^4} + \frac{9}{4 x_n^2 } - \frac{3}{x_n}^3 \mathrm{tan}^{-1} (2x_n).
\end{equation}
In this case, $x_n$ is defined as $x_n= k_{Fn}/\mu$, with $k_{Fi}= (3 \pi^2 \rho_i)^{1/3}$ ($i=n,p$) the Fermi momentum of each type of nucleon.

The coefficients $\varepsilon_i^{l (ul)}$ include the parameters of the interaction in the following way
\begin{eqnarray}
 \varepsilon_{0}^l &=& \frac{t_0}{2} \rho_0 (1-x_0) +\left( W+ \frac{B}{2} -H-\frac{M}{2} \right)\rho_0 \int v(r) d^3 r
 \\
 \varepsilon_{0}^{ul} &=& \frac{t_0}{2} \rho_0 (2+x_0)+ \left( W+ \frac{B}{2} \right)\rho_0 \int v(r) d^3 r
 \\
 \varepsilon_{\gamma}^l &=&  \frac{t_3}{12} \rho_0^{\gamma+1} (1-x_3)
 \\
 \varepsilon_{\gamma}^{ul} &=& \frac{t_3}{12} \rho_0^{\gamma+1} (2+x_3)
 \\
 \varepsilon_{ex}^l &=&   \left( M + \frac{H}{2} -B - \frac{W}{2} \right)\rho_0 \int v(r) d^3 r
 \\
 \varepsilon_{ex}^{ul} &=&  \left(M + \frac{H}{2} \right) \rho_0 \int v(r) d^3 r.
\end{eqnarray}
Notice that the integration of the form factor of MDI interactions is
\begin{equation}\label{ffMDI}
 \int v(r) d^3 r = \int \frac{e^{-\mu r}}{r} d^3r= \frac{4\pi^2}{\mu^2} ,
\end{equation}
whereas for SEI interactions is
\begin{equation}\label{ffS}
 \int v(r) d^3 r =\int \frac{e^{-\mu r}}{\mu r} d^3r= \frac{4\pi^2}{\mu^3}.
\end{equation}
Hence, defining the expressions of the energy using the  parameters $\varepsilon_i^{l (ul)}$
allows one to write the properties of nuclear matter in a general form for both MDI and SEI  without the worry
of the definition of the form factor, which is 
taken into account only when obtaining the $\varepsilon_i^{l (ul)}$ parameters.

The pressure in SNM will be given by 
\begin{eqnarray}
 P (\rho) &=& \frac{\hbar^2 k_F^2\rho}{5m } + \rho^2\frac{\varepsilon_0^l+\varepsilon_0^{ul}}{4 \rho_0} + 
 \frac{\varepsilon_\gamma^l+\varepsilon_\gamma^{ul}}{4 \rho_0^{\gamma+1}}\rho^2 \left[  \frac{(\gamma+1) \rho^{\gamma}}{(1+b\rho)^{\gamma}} 
-\frac{\gamma b \rho^{\gamma+1}}{(1+b\rho)^{\gamma+1}}\right] 
 \nonumber\\
 &+&\frac{\varepsilon_{ex}^l+\varepsilon_{ex}^{ul}}{4 \rho_0} \rho^2 \left[ \mathcal{A}(x_F) + \frac{k_F}{3 \mu} \frac{\partial A(x_F)}{\partial x_F}  \right]
\end{eqnarray}
and its nuclear matter incompressibility is defined as
\begin{eqnarray}
 K (\rho) &=&  -\frac{3\hbar^2 k_F^2}{5 m} +  \frac{\varepsilon_\gamma^l +\varepsilon_\gamma^{ul}}{4 \rho_0^{\gamma+1}}  \rho^2
 \left[ \frac{\gamma(\gamma+1) \rho^{\gamma-1}}{(1+b \rho)^{\gamma}} - \frac{2 b \gamma (\gamma+1) \rho^\gamma}{(1+b \rho)^{\gamma+1}} + 
 \frac{\gamma b^2 (\gamma+1) \rho^{\gamma+1}}{(1+b \rho)^{\gamma+2}}\right] \nonumber\\
 &+& 
 \frac{\varepsilon_{ex}^l + \varepsilon_{ex}^{ul}}{4 \rho_0} \left[ \frac{4 k_F \rho}{\mu}\frac{\partial \mathcal{A}(x_F)}{\partial x_F}
 + \frac{k_F^2 \rho}{\mu^2}  \frac{\partial^2 \mathcal{A}(x_F)}{\partial x_F^2}\right].
\end{eqnarray}
The derivatives of $\mathcal{A} (x_F)$  appearing in the expressions for the pressure $ P (\rho) $ and incompressibility 
$ K (\rho)$ are given by
\begin{eqnarray}
\frac{\partial \mathcal{A}}{\partial x_F} (x_F)&=& -9 \left[ \left( \frac{1}{16 x^7} + \frac{1}{2 x_F^5} \right) 
\ln (1+4x_F^2) - \frac{1}{4 x_F^5} + \frac{1}{2 x_F^3} - \frac{1}{x_F^4} \mathrm{tan}^{-1} (2 x_F)
\right]\nonumber
\\
\\
 \frac{\partial^2 \mathcal{A}}{\partial x_F^2} (x_F) &=& 9 \left[ 
 \left( \frac{7}{16x_F^8} + \frac{5}{2 x_F^6}\right) \ln (1+4 x_F^2) - \frac{7}{16 x_F^6} + \frac{3}{2 x_F^4} - 
 \frac{4}{x_F^5} \mathrm{tan}^{-1} (2 x_F)
 \right].\nonumber\\
\end{eqnarray}

On the other hand, in an asymmetric nuclear system with $\rho_n \neq \rho_p$, 
the energy per particle is given by~\cite{routray16}
\begin{eqnarray}
 E_b (\rho_n, \rho_p) &=& \frac{3 \hbar^2}{10 m } \left( k_n^2 \frac{\rho_n}{\rho} + k_p^2 \frac{\rho_p}{\rho} \right) + 
 \frac{\varepsilon_0^l}{2\rho_0 } \left( \frac{\rho_n^2}{\rho} + \frac{\rho_p^2}{\rho} \right) + \frac{\varepsilon_0^{ul}}{\rho_0} \frac{\rho_n \rho_p}{\rho}
 \nonumber \\
 &+& \left[ \frac{\varepsilon_\gamma^l}{2\rho_0^{\gamma +1} } \left( \frac{\rho_n^2}{\rho} + \frac{\rho_p^2}{\rho} \right) + 
 \frac{\varepsilon_\gamma^{ul}}{\rho_0} \frac{\rho_n \rho_p}{\rho} \right] \left( \frac{\rho}{1+ b \rho} \right)^\gamma
 + \frac{\varepsilon_{ex}^l}{2 \rho_0} \left( \frac{\rho_n^2}{\rho} J(k_{Fn}) +  \frac{\rho_p^2}{\rho} J(k_{Fp}) \right)
 \nonumber\\
 &+& \frac{\varepsilon_{ex}^{ul}}{2 \rho_0} \frac{1}{\pi^2} \left[\frac{\rho_n}{\rho} \int_0^{k_{Fp}} I (k, k_{Fn}) k^2 dk +
 \frac{\rho_p}{\rho} \int_0^{k_{Fn}} I (k, k_{Fp}) k^2 dk\right] ,
\end{eqnarray}
with
\begin{equation}
 J(k_{Fi}) = \left( \frac{3}{32 x_i^6} + \frac{9}{8 x_i^4}\right)\ln (1+4 x_i^2) - \frac{3}{8 x_i^4} + \frac{9}{4 x_i^2}
 - \frac{3}{x_i^3} \mathrm{tan}^{-1} (2 x_i)
\end{equation}
and
\begin{equation}
 I (k, k_{Fi}) = \frac{3 (1+ x_i^2 - x^2)}{8 x_i^3 x} \ln \left[ \frac{1 + (x+x_i)^2}{1+ (x-x_i)^2} \right]
 + \frac{3}{2 x_i^2} - \frac{3}{2 x_i^3} \left[ \mathrm{tan}^{-1} (x+x_i) - \mathrm{tan}^{-1} (x-x_i) \right].
\end{equation}

The neutron and proton chemical potentials are given by~\cite{routray16}
\begin{eqnarray}
 \mu_\tau (\rho_\tau, \rho_{\tau'})&=& \frac{\hbar^2 k_{F \tau}}{2m} \left[  \frac{\varepsilon_0^l}{\rho_0} + \frac{\varepsilon_\gamma^l}{\rho_0^{\gamma+1}} \left( 
 \frac{\rho}{1+b \rho}\right)^\gamma\right] \rho_\tau + \left[  \frac{\varepsilon_0^{ul}}{\rho_0} + \frac{\varepsilon_\gamma^{ul}}{\rho_0^{\gamma+1}} \left( 
 \frac{\rho}{1+b \rho}\right)^\gamma\right] \rho_{\tau'} \nonumber \\
 &+& \varepsilon_{ex}^l \frac{\rho_\tau}{\rho_0} \left[ \frac{3 \ln (1+4 x_{\tau}^2)}{8 x_\tau^4} + \frac{3}{2 x_\tau^2} -
 \frac{3 \mathrm{tan}^{-1} (2 x_\tau)}{2 x_\tau^3}\right] \nonumber\\
 &+& \varepsilon_{ex}^{ul} \frac{\rho_{\tau'}}{\rho_0} \left[ \frac{3 (1+x_{\tau'}^2-x_\tau^2)}{8 x_\tau x_{\tau'}^3}
 \ln \left[ \frac{1+ (x_{\tau} + x_{\tau'})^2}{1+ (x_{\tau} - x_{\tau'})^2}\right] + \frac{3}{2 x_{\tau'}^2} \right.
 \nonumber\\
 &-& \left.\frac{3}{2 x_{\tau'}^3} \left( \mathrm{tan}^{-1} (x_\tau+x_{\tau'}) -\mathrm{tan}^{-1} (x_\tau-x_{\tau'}) \right)\right] \nonumber\\
 &+& \left[ \frac{\varepsilon_\gamma^l (\rho_n^2+\rho_p^2)}{2 \rho_0^{\gamma+1}} + \frac{\varepsilon_\gamma^{ul}(\rho_n \rho_p)}{\rho_0^{\gamma+1}} \right]
 \frac{\gamma \rho^{\gamma-1}}{(1+b \rho)^(\gamma+1)},
\end{eqnarray}
where $\tau$ and $\tau'$ refer to either protons or neutrons.

If the energy per particle of asymmetric matter, rewritten in terms of the total density $\rho$ and the isospin 
asymmetry $\delta$, is expanded around $\delta=0$ [see Eq.~(\ref{eq:eosexp})], the symmetry energy coefficient of the system
for SEI (MDI) interactions is defined as 
\begin{eqnarray}
 E_\mathrm{sym} (\rho) &=& 
 \frac{\hbar^2 k_F^2}{6 m} +
  \frac{\rho}{4} \left( \frac{\varepsilon_0^l - \varepsilon_0^{ul}}{\rho_0} \right)
  \nonumber\\
&+& \frac{\rho}{4} \left( \frac{\varepsilon_\gamma^l - \varepsilon_\gamma^{ul}}{\rho_0^{\gamma+1}} \right)
 \left( \frac{\rho}{1+b\rho} \right)^\gamma 
 +\frac{\rho}{4} \left(\frac{\varepsilon_{ex}^l - \varepsilon_{ex}^{ul}}{\rho_0} \right) \mathcal{C} (x_F)
 \nonumber\\
 &-& \frac{\rho}{4} \left(  \frac{\varepsilon_{ex}^l + \varepsilon_{ex}^{ul}}{\rho_0}\right) \mathcal{D} (x_F),
\end{eqnarray}
where 
\begin{equation}
 \mathcal{C} (x_F) =   \frac{\ln(1+4x_F^2)}{4 x_F^2}
\end{equation}
and
\begin{equation}
 \mathcal{D} (x_F)=
  \left( 
 \frac{1}{4 x_F^2} + \frac{1}{8 x_F^4}\right) \ln(1+4 x_F^2) - \frac{1}{2x_F^2} .
\end{equation}
The slope of the symmetry energy $L$ for SEI (MDI) interactions reads
\begin{eqnarray}
 L  &=&  
 \frac{\hbar^2 k_F^2}{3 m} + \frac{3}{4} \rho_0 \left( \frac{\varepsilon_0^l- \varepsilon_0^{ul}}{\rho_0} \right) \nonumber\\
  &+& \frac{3}{4} \rho_0 \left( \frac{\varepsilon_\gamma^l- \varepsilon_\gamma^{ul}}{\rho_0^{\gamma+1}} \right)  
  \left[ \frac{(\gamma+1) \rho_0^\gamma}{(1+b\rho_0)^\gamma}-
  \frac{b \gamma \rho_0^{\gamma+1}}{(1+b\rho_0)^{\gamma+1}}\right] \nonumber\\
  &+& \frac{3}{4} \rho_0\left( \frac{\varepsilon_{ex}^l- \varepsilon_{ex}^{ul}}{\rho_0} \right) \left[ \mathcal{C}(x_F) 
  + \left.\frac{k_F}{3 \mu}\frac{\partial \mathcal{C}(x_F)}{\partial x_F}\right|_{\rho_0}
 \right] \nonumber\\
 &-&  
  \frac{3}{4} \rho_0\left( \frac{\varepsilon_{ex}^l+ \varepsilon_{ex}^{ul}}{\rho_0} \right) \left[ \mathcal{D}(x_F) +  
  \left.\frac{k_F}{3 \mu} \frac{\partial \mathcal{D}(x_F)}{\partial x_F}\right|_{\rho_0}
  \right]\nonumber\\
\end{eqnarray}
and the isospin curvature of the system is 
\begin{eqnarray}
 K_\mathrm{sym}  &=& -\frac{\hbar^2 k_F^2}{3m } + \frac{9}{4} \rho_0^2 
 \left( \frac{\varepsilon_\gamma^l- \varepsilon_\gamma^{ul}}{\rho_0} \right)\nonumber\\
 &\times&\left[  \frac{\gamma (\gamma+1) \rho_0^{\gamma-1}}{(1+b \rho_0)^{\gamma}} - 
 \frac{2 \gamma b (\gamma+1) \rho_0^\gamma}{(1+b \rho_0)^{\gamma+1}} + \frac{b^2 \gamma (\gamma+1) 
 \rho_0^{\gamma+1}}{(1+b\rho_0)^{\gamma+2}}\right]\nonumber\\
 &+& \frac{1}{4} \left( \frac{\varepsilon_{ex}^l- \varepsilon_{ex}^{ul}}{\rho_0} \right) \left[ \frac{4 k_{F0} \rho_0 }{\mu}
 \left.\frac{\partial \mathcal{C}(x_F)}{\partial x_F}\right|_{\rho_0}
 + \frac{k_{F0}^2 \rho_0}{\mu^2} \left.\frac{\partial^2 \mathcal{C}(x_F)}{\partial x_F^2}\right|_{\rho_0} \right]\nonumber\\
  &+& \frac{1}{4} \left( \frac{\varepsilon_{ex}^l- \varepsilon_{ex}^{ul}}{\rho_0} \right)
  \left[ \frac{4 k_{F0} \rho_0 }{\mu} \left.\frac{\partial \mathcal{D}(x_F)}{\partial x_F}\right|_{\rho_0}
 + \frac{k_{F0}^2 \rho_0}{\mu^2} \left.\frac{\partial^2 \mathcal{D}(x_F)}{\partial x_F^2}\right|_{\rho_0} \right],
\end{eqnarray}
where $k_{F0}$ is the Fermi momentum of the system calculated at the saturation density $\rho_0$.
The derivatives of the expressions $\mathcal{C} (x_F)$ and $\mathcal{D} (x_F)$ needed to implement $L$ and $K_\mathrm{sym}$
are, respectively, 
\begin{eqnarray}
 \frac{\partial \mathcal{C}(x_F)}{\partial x_F}  &=& \frac{2}{4 x_F^3 + x_F} - \frac{1}{2 x_F^3} \ln (1+4 x_F^2)
\\
 \frac{\partial^2 \mathcal{C}(x_F)}{\partial x_F^2}  &=& \frac{3}{2 x_F^4} \ln (1+4x_F^2) - \frac{40 x_F^2 +6}{(4 x_F^3+x_F)^2}
\\
 \frac{\partial \mathcal{D}(x_F)}{\partial x_F}  &=& \frac{6 x_F^2+2}{4 x_F^5+x_F^3} - \frac{x_F^2+1}{2x_F^5} \ln (1+4x_F^2)
\\
 \frac{\partial^2\mathcal{D}(x_F)}{\partial x_F^2}  &=& -\frac{2 (5+33x_F^2+44 x_F^4)}{x_F^4 (1+4x_F^2)^2} + \frac{5+3x_F^2}{2 x_F^6} \ln (1+4x_F^2).
\end{eqnarray}

\chapter{Asymmetric nuclear matter studied with  Skyrme and Gogny interactions}\label{chapter2}
The EoS of asymmetric nuclear matter (ANM) is not completely established
and, sometimes, it is not trivial to compute as, for example, in the case
of finite-range interactions, such as Gogny, MDI or SEI forces.
In order to prove the main features of the isospin dependence of the EoS, it can useful to expand the energy per particle
given by such interactions, which may be written in terms of 
the total density $\rho$ and the isospin asymmetry $\delta$
of the system,
around asymmetry $\delta=0$ \cite{gonzalez17}:
\begin{eqnarray}\label{eq:EOSexpgeneral}
 E_b(\rho, \delta) &=& E_b(\rho, \delta=0) + E_{\mathrm{sym}, 2}(\rho)\delta^{2} +... 
 + E_{\mathrm{sym}, 2k}(\rho)\delta^{2k} + \mathcal{O}(\delta^{2k+2}), \nonumber\\
\end{eqnarray}
with $k \geq  1 $, and where each symmetry energy coefficient is defined as  
\begin{equation}\label{eq:esymgen}
 \left. E_{\mathrm{sym}, 2k} (\rho) = \frac{1}{(2k)!} \frac{\partial^{2k} E_b(\rho, \delta)}
{\partial \delta^{2k}}\right|_{\delta=0}.
\end{equation} 
The coefficients in Eq.~(\ref{eq:EOSexpgeneral}) are directly connected to the 
properties of the single-nucleon potential 
in ANM~\cite{Xu11,Chen12}. At low densities, such as the ones found in 
terrestrial nuclei, the system has   
small isospin asymmetry values, 
and the expansion~(\ref{eq:EOSexpgeneral}) can be cut at second order,  
neglecting higher-order terms [see Eq.~(\ref{eq:eosexp}) of Chapter~\ref{chapter1}].
The coefficient $E_\mathrm{sym,2} (\rho)$ is the quantity we have previously 
defined in Eq.~(\ref{eq:esym}) as
the symmetry energy of the system. 
Sometimes in the literature, the second-order
symmetry energy is also denoted as $S(\rho)$~\cite{Sellahewa14,Piekarewicz08}.
The second-order symmetry energy coefficient, which we will also refer to as symmetry energy, has been relatively well constrained both 
experimentally and theoretically at low values of the density below and around the saturation density~\cite{Danielewicz13,Tsang08,Zhang15, Chen_15}. 

On the other hand, when studying systems with high isospin asymmetry, such as the 
interior of neutron stars (NSs), where $\beta$-stable 
nuclear matter shows large differences in its neutron and proton contributions,  
higher-order terms in the expansion asymmetry $\delta$ may play a significant role when obtaining the equation of state 
by adding corrections to the parabolic law \cite{xu09a,Xu11, Cai2012, Moustakidis12,Seif14, gonzalez17, Liu18}. 

To analyze the density dependence of the symmetry energy, 
one can expand the coefficients $E_{\mathrm{sym},2k} (\rho)$
around the saturation density $\rho_0$ as follows:
\begin{equation}\label{esymexpgen}
E_{\mathrm{sym},2k} (\rho)= E_{\mathrm{sym}, 2k} (\rho_0) + L_{2k} \epsilon + \mathcal{O}(\epsilon^2),
\end{equation}
where $ \epsilon = (\rho - \rho_ 0)/3\rho_0$ is the density displacement from the saturation density $\rho_0$ and 
the coefficients $L_{2k}$ are the slope parameters of the symmetry energy coefficients, which are computed at the
saturation density as
\begin{eqnarray}\label{eq:L2k} 
L_{2k}\equiv  L_{2k} (\rho_0)&=& 3\rho_0 \left.\frac{\partial E_{\mathrm{sym}, 2k} (\rho)}{\partial \rho} \right|_{\rho_0}.
\end{eqnarray}
Recalling Eq.~(\ref{eq:EOSexpgeneral}) and the saturation condition of nuclear forces, we see that the density slope at 
saturation of the energy per particle $E_b (\rho, \delta)$ of asymmetric matter can be parametrized as~\cite{gonzalez17}
\begin{equation}
\left. \frac{\partial E_b (\rho, \delta)}{\partial \rho} \right|_{\rho_0} =
\frac{1}{3\rho_0} \left( L_2 \delta^2 + L_4 \delta^4 + L_6 \delta^6 + \cdots \right) .
 \label{eq:slope2k}
\end{equation}
In this notation, $L_2$ is the commonly known slope of the symmetry energy coefficient and we will refer to it as $L$ [cf. Eq.~(\ref{eq:L})].

On the other hand, if the $\delta$-expansion is truncated at second order and one considers an asymmetry $\delta=1$,  we can 
define the symmetry energy as the difference between the energy per particle in neutron matter and in symmetric matter:
\begin{equation}\label{eq:esympa}
 E_{\mathrm{sym}}^{PA}(\rho) = E_b(\rho, \delta=1)- E_b(\rho, \delta=0).
\end{equation}
This definition of the parabolic symmetry energy can be understood as the energy 
cost for converting all protons into neutrons in SNM, which can be 
useful in some cases as for instance in microscopic calculations of 
Brueckner-Hartree-Fock type.
If one considers the Taylor expansion of the EoS~(\ref{eq:EOSexpgeneral}), one gets that the parabolic symmetry energy 
corresponds to the sum of the whole series of the symmetry energy coefficients when the isospin asymmetry is $\delta=1$, 
assuming that this series is convergent:
\begin{equation}  
E_{\mathrm{sym}}^{PA}(\rho)= \sum_{k=1}^{\infty} E_{\mathrm{sym}, 2k} (\rho).
\label{eq:PAgen}
\end{equation}
Analogously to the definition in Eq.~(\ref{eq:L2k}), the slope parameter to the symmetry energy using the PA can be computed as 
\begin{eqnarray}
   L_{PA} \equiv L_{PA}(\rho_0)&=& \left.3 \rho_0 \frac{\partial E_{\mathrm{sym}}^{PA} (\rho)}{\partial \rho} \right|_{\rho_0} .
\end{eqnarray}

Recent calculations in many-body perturbation theory have shown  
that the isospin asymmetry expansion~(\ref{eq:EOSexpgeneral}) may not be convergent at zero temperature 
when the many-body corrections beyond the Hartree-Fock mean-field level are incorporated~\cite{Wellenhofer2016}. 
We do not deal with this complication here since we will be working at the Hartree-Fock level, 
where no non-analyticities are found in the equation of state.

In this Chapter, we will analyze the behaviour of higher-order terms in the Taylor expansion  
of different Skyrme and Gogny EoSs, and their influence when studying $\beta$-equilibrated matter. 
Also, we will check how the isovector characteristics of the EoS affect the 
relation between the mass and the radius of NSs. 
\section{Convergence of the isospin Taylor expansion of the EoS for Skyrme interactions}
In this Section, we compute the contributions to the symmetry energy up to 
$10^\mathrm{th}$ order in the expansion of the energy per particle~(\ref{eq:EOSexpgeneral}) for Skyrme interactions, and we 
study their influence on the ANM EoS. 
Applying Eq.~(\ref{eq:esymgen}) to Skyrme forces, the coefficients up to second-, fourth-, sixth-, eighth- and tenth- order in the expansion of the EoS
for Skyrme interactions are obtained as
\begin{eqnarray}
E_{\mathrm{sym}, 2} (\rho)&=&\frac{\hbar^2}{6m} \left(  \frac{3 \pi^2}{2}\right)^{2/3} \rho^{2/3} - \frac{1}{8}
t_0 \rho (2x_0 +1) 
- \frac{1}{48} t_3 \rho^{\sigma+1} (2x_3 +1) \nonumber 
\\
&+& \frac{1}{24} \left(  \frac{3 \pi^2}{2}\right)^{2/3}   
\times \rho^{5/3} \left[ -3 x_1 t_1 + t_2 (5 x_2 + 4) \right], \label{eq:esym2skyrme2}
\\
E_{\mathrm{sym}, 4}(\rho) &=& \frac{\hbar^2}{162m} \left(  \frac{3 \pi^2}{2}\right)^{2/3} \rho^{2/3} +
\frac{1}{648} \left(  \frac{3 \pi^2}{2}\right)^{2/3}  
 \nonumber\\
 &\times& \rho^{5/3} \left[ 3 t_1 (x_1 +1) - t_2 ( x_2 -1) \right], 
\\
E_{\mathrm{sym}, 6} (\rho)&=&  \frac{7\hbar^2}{4374m} \left(  \frac{3 \pi^2}{2}\right)^{2/3} \rho^{2/3} +
\frac{7}{87840} \left(  \frac{3 \pi^2}{2}\right)^{2/3}  
\nonumber\\
&\times& \rho^{5/3} \left[  t_1 (9x_1 +12) + t_2 ( x_2 +8) \right], 
\\
 E_{\mathrm{sym}, 8}(\rho) &=&  \frac{13\hbar^2}{19683m} \left(  \frac{3 \pi^2}{2}\right)^{2/3} \rho^{2/3}+ 
 \frac{13}{314928} \left(  \frac{3 \pi^2}{2}\right)^{2/3} 
\nonumber \\
 &\times&  \rho^{5/3} \left[ 3 t_1 (2x_1 +3)  + t_2 (2  x_2 +7) \right],
\\
E_{\mathrm{sym},10} (\rho)&=&  \frac{2717\hbar^2}{7971615m} \left(  \frac{3 \pi^2}{2}\right)^{2/3} \rho^{2/3}+
\frac{247}{31886460} \left(  \frac{3 \pi^2}{2}\right)^{2/3} \nonumber
\\
& \times& \rho^{5/3} \left[t_1 (15x_1 +24) + t_2 (7  x_2 +20) \right]. \label{eq:esym10skyrme}
\end{eqnarray}
Note that, in contrast to $E_{\mathrm{sym},2} (\rho)$, the higher-order symmetry energy coefficients, i.e, 
$E_{\mathrm{sym}, 4,6,8,10} (\rho)$, arise exclusively from the kinetic term and from the momentum-dependent term 
of the interaction, which in Skyrme forces is the term with the usual $t_1$ and $t_2$ parameters \cite{sly41,sly42}.

\begin{figure}[t]
 \centering
 \includegraphics[width=0.8\linewidth, clip=true]{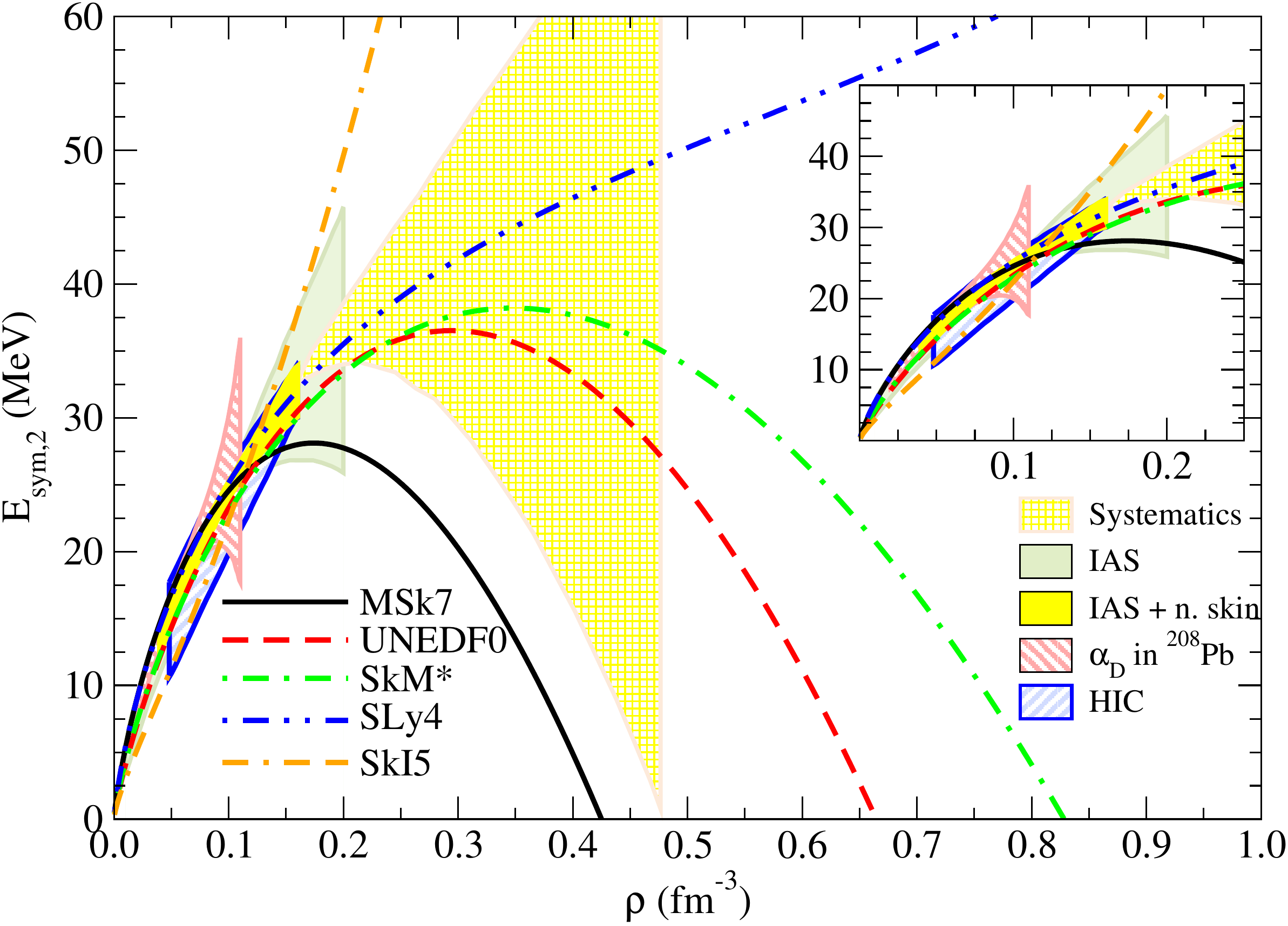}
  \caption{Density dependence of the second-order symmetry energy coefficient 
  $E_{\mathrm{sym}, 2}(\rho)$ for the Skyrme 
forces MSk7 ($L=9.41$ MeV), UNEDF0 ($L=45.08$ MeV), SkM$^*$ ($L=45.78$ MeV), SLy4 ($L=45.96$ MeV) and SkI5 ($L=129.33$ MeV).
Also represented are the symmetry energy constraints extracted from the analysis of data on isobaric analog states (IAS)~\cite{Danielewicz13}
and of IAS data combined with neutron skins (IAS+n.skin)~\cite{Danielewicz13}, 
the constraints from the electric dipole polarizability in lead ($\alpha_D$ in $^{208}$Pb)~\cite{Zhang15},
 from transport simulations of heavy-ion collisions of tin isotopes (HIC)~\cite{Tsang08} and from systematics of the symmetry energy at high densities~\cite{Chen_15}.}
 \label{fig:esym2}
\end{figure}

We plot in Fig.~\ref{fig:esym2} the second-order symmetry energy coefficient against the density of the system 
for the MSk7, UNEDF0, SkM$^*$, SLy4 and SkI5 Skyrme interactions. To carry out our study, we have 
chosen these parametrizations as representative ones
because they cover a wide range of values of the slope of the symmetry energies $L$, from $L=10$ MeV to $L=130$ MeV. 
In the same figure, we plot various existing empirical constraints for the symmetry energy at
subsaturation density \cite{Danielewicz13,Zhang15,Tsang08, Chen_15}.
At subsaturation densities, the considered Skyrme forces fit inside the majority of the 
constraints. On the other hand, we see that at higher densities, neither the MSk7 interaction, with 
a very small value of $L$ and with a soft EoS, nor the SkI5 interaction, with a very large $L$ value and a stiff EoS,
fit inside the bands coming from systematics at high density~\cite{Chen_15}.  

At low densities below $\rho \sim0.1$ fm$^{-3}$ the coefficients $E_{\mathrm{sym},2} (\rho)$ of all interactions behave in a
similar way. The reason behind it is the fact that this is the density regime where experimental 
data of finite nuclei exist, and which has been used to fit the majority of the interactions. 
The second-order symmetry energy of all interactions intersect at a density around $\rho \sim0.1$ fm$^{-3}$
and from this density onwards they show a more model-dependent behaviour. 
Some of them show an increasing trend as a function of the density, such as the ones 
calculated with the SkI5 and SLy4 interactions. However, 
there are other parametrizations, such as the MSk7, UNEDF0 and SkM$^*$ forces,
whose symmetry energy coefficients $E_{\mathrm{sym},2} (\rho)$ 
reach a maximum and then decrease until vanishing. This implies an isospin instability, as 
the energy per particle in neutron matter becomes more bound than in SNM. 

\begin{figure}[t]
 \centering
 \includegraphics[width=1\linewidth, clip=true]{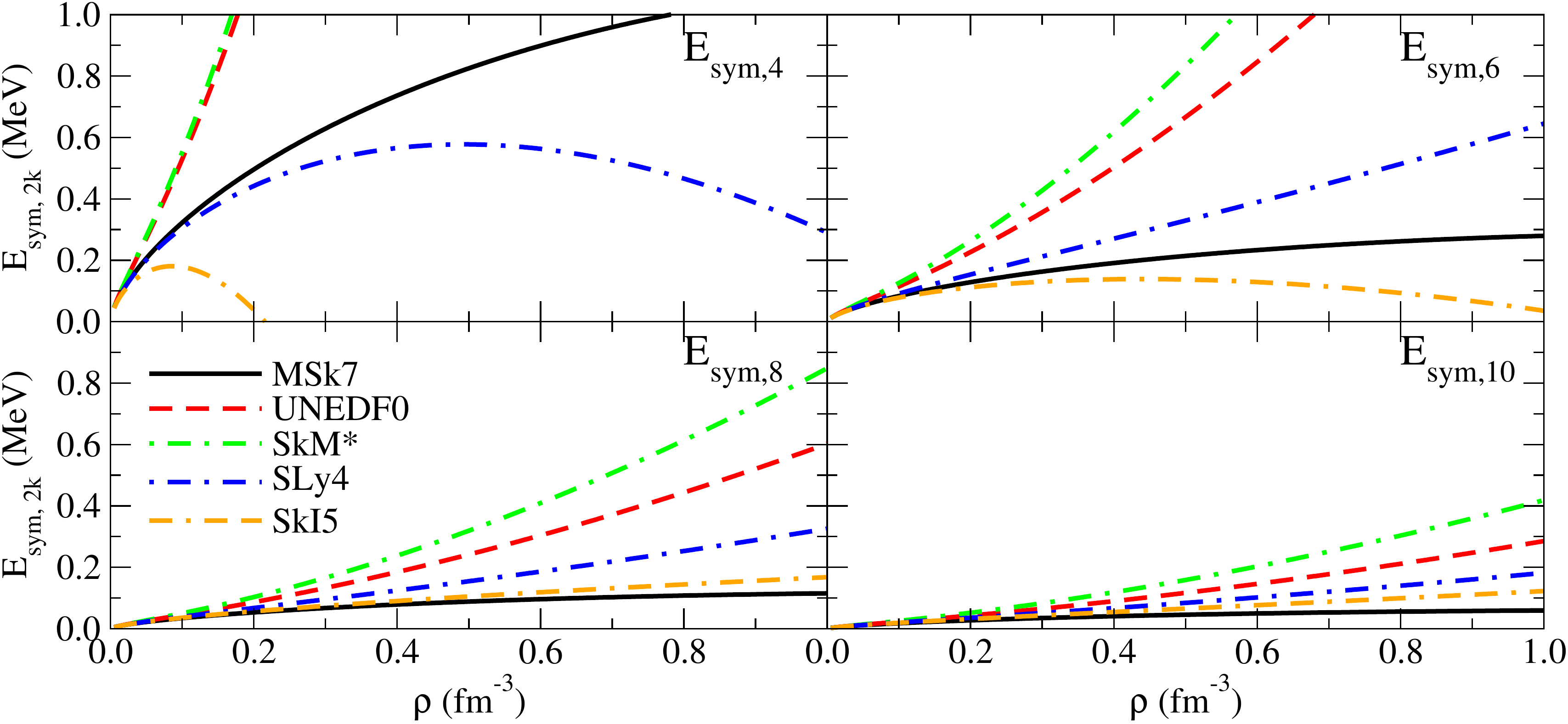}
  \caption{Density dependence of the fourth-, sixth-, eight- and tenth-order symmetry energy coefficients, 
  $E_{\mathrm{sym}, 4}(\rho)$, $E_{\mathrm{sym}, 6}(\rho)$, $E_{\mathrm{sym}, 8}(\rho)$ and $E_{\mathrm{sym}, 10}(\rho)$ respectively, for the Skyrme 
forces MSk7, UNEDF0, SkM$^*$, SLy4 and SkI5.}
 \label{fig:esym2k}
\end{figure}

In Fig.~\ref{fig:esym2k} we plot the fourth-, sixth-, eighth- and tenth-order symmetry energy coefficients 
for the same interactions as in Fig.~\ref{fig:esym2}. As happens with the second-order
symmetry energy coefficient, the behaviour of the higher-order coefficients is model-dependent.
The interaction SkI5 is, of the ones plotted, the one that has the stiffest second-order symmetry energy at saturation, with 
a slope parameter of $L=129.33$ MeV. 
However, its fourth-order symmetry energy coefficient bends and vanishes at a density
$\rho \sim0.2$ fm$^{-3}$, giving a small or even negative contribution to the expansion of the symmetry energy.
The $E_{\mathrm{sym},6} (\rho)$ coefficient for SkI5 has a similar behaviour as its $E_{\mathrm{sym},4} (\rho)$.
It bends at $\rho \sim0.4$ fm$^{-3}$
and becomes negative at densities larger than the ones considered in the plot. 
In this case, the sixth-order contribution may have a larger impact on the expansion of the EoS, at least for 
values of isospin asymmetry $\delta$ close to the unity. 
The interactions SLy4, SkM$^*$ and UNEDF0 have similar values of the slope of the symmetry energy $L$, which are,
respectively, $L=45.96$ MeV, $L=45.78$ MeV and $L= 45.08$ MeV. However, 
the $E_{\mathrm{sym},4} (\rho)$ of the UNEDF0 and SkM$^*$ interactions are very stiff having similar trends,
while the $E_{\mathrm{sym},4} (\rho)$ coefficient of the 
SLy4 interaction bends at a certain density $\rho \sim 0.45$ fm$^{-3}$ and then 
decreases, becoming negative at a density larger than the ones shown in the figure. This scenario is 
different from the one found in Fig.~\ref{fig:esym2}, 
where the $E_{\mathrm{sym},2} (\rho)$ of the SLy4 interaction does not bend, while the ones of the UNEDF0 and SkM$^*$
models bend at relatively small densities, presenting isospin instabilities. 
The sixth-order symmetry energy coefficients of these three interactions increase at all values of the density 
that are considered in the plot. 
For the MSk7 interaction, which has a small value of the slope parameter of $L=9.41$ MeV, we find a fourth-order 
symmetry energy coefficient that is rather stiff inside the range 
of densities considered, and a sixth-order coefficient that is also positive in this same density regime. 
The $E_{\mathrm{sym},8} (\rho)$ and $E_{\mathrm{sym},10} (\rho)$ coefficients of all the above interactions are 
positive and do not bend inside the range of densities up to $ \rho =1$ fm$^{-3}$.
At subsaturation densities the $E_{\mathrm{sym},8} (\rho)$ coefficient has values below $\sim 0.08$ MeV and the $E_{\mathrm{sym},10} (\rho)$ 
do not exceed values of $\sim 0.05$ MeV.

\begin{figure}[t]
 \centering
 \includegraphics[width=1\linewidth, clip=true]{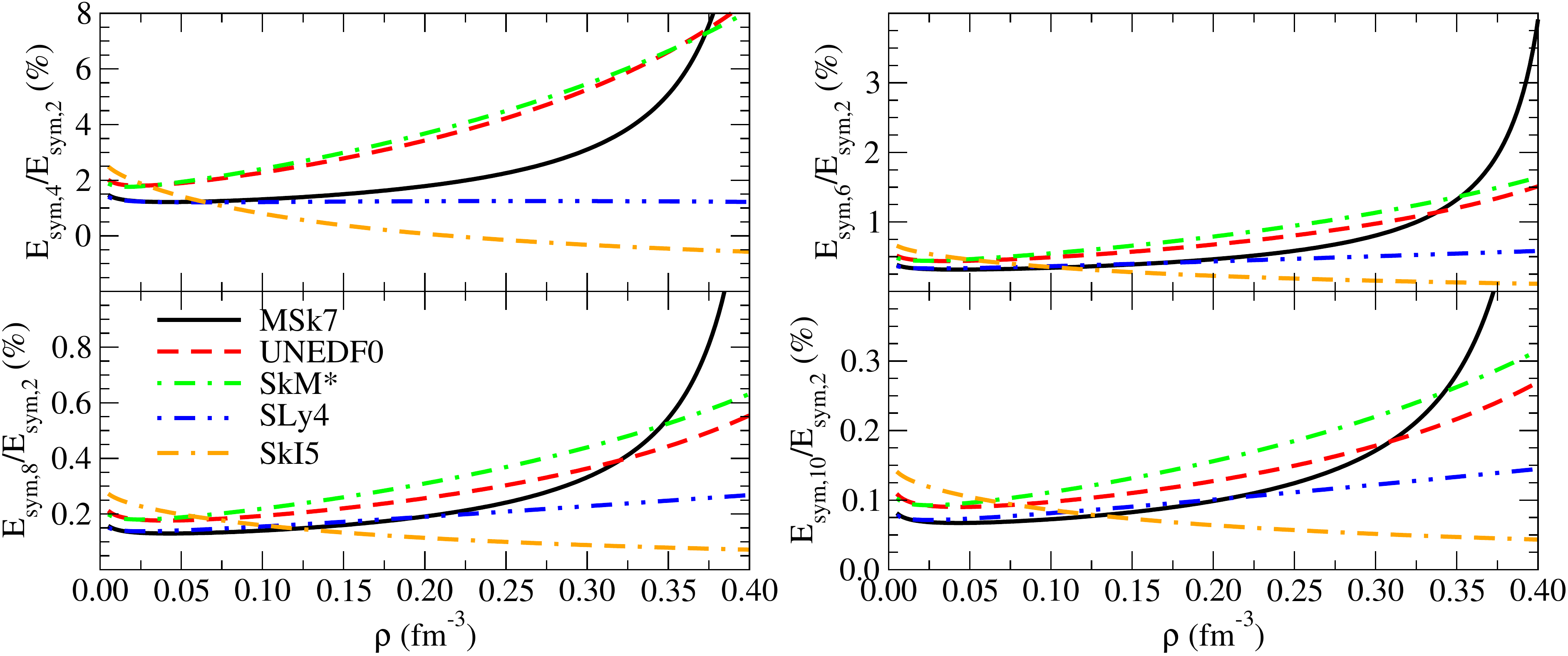}
  \caption{Ratios $E_{\mathrm{sym},2k}(\rho)/E_{\mathrm{sym},2}(\rho)$ ($k=2,3,4,5$) as a function of the 
  density for the Skyrme forces MSk7, UNEDF0, SkM$^*$, SLy4 and SkI5.}
 \label{fig:esym2ratio}
\end{figure}
The density dependence of the 
ratios of E$_{\mathrm{sym}, 4} (\rho)$,  E$_{\mathrm{sym}, 6}(\rho)$, E$_{\mathrm{sym},8}(\rho)$ and 
E$_{\mathrm{sym},10}(\rho)$ with respect to E$_{\mathrm{sym}, 2}(\rho)$ is plotted in Fig.~\ref{fig:esym2ratio} 
for the Skyrme forces MSk7, UNEDF0,
SkM$^*$, SLy4 and SkI5
up to a density $\rho=0.4$ fm$^{-3}$.
At low densities $\rho \sim 0.1$ fm$^{-3}$, the fourth-order symmetry energy 
is not bigger than 3$\%$ of the symmetry energy at second order, 
and the sixth-, eighth- and tenth-order terms are, respectively, less than $0.8\%$, $0.3\%$ and $0.15\%$.
However, as we go to higher densities, the contributions of these coefficients increase, 
up to the point that for some interactions they may not be negligible. If they are not considered, 
the calculations of the equation of state may lead to non-realistic results far from the ones 
obtained if one uses the exact expression of the EoS. 
Notice that the sudden increase of the ratios $E_{\mathrm{sym},2k}(\rho)/E_{\mathrm{sym},2}(\rho)$ of the MSk7 interaction 
at densities close to $\rho=0.4$ fm$^{-3}$ is caused by the low values of the second-order symmetry energy coefficient 
at that density regime.

\subsection{Comparison between the parabolic approximation $E_{\mathrm{sym}}^{PA}(\rho)$ and the $E_{\mathrm{sym},2k}(\rho)$
coefficients}
\begin{figure}[!t]
 \centering
 \includegraphics[width=0.8\linewidth, clip=true]{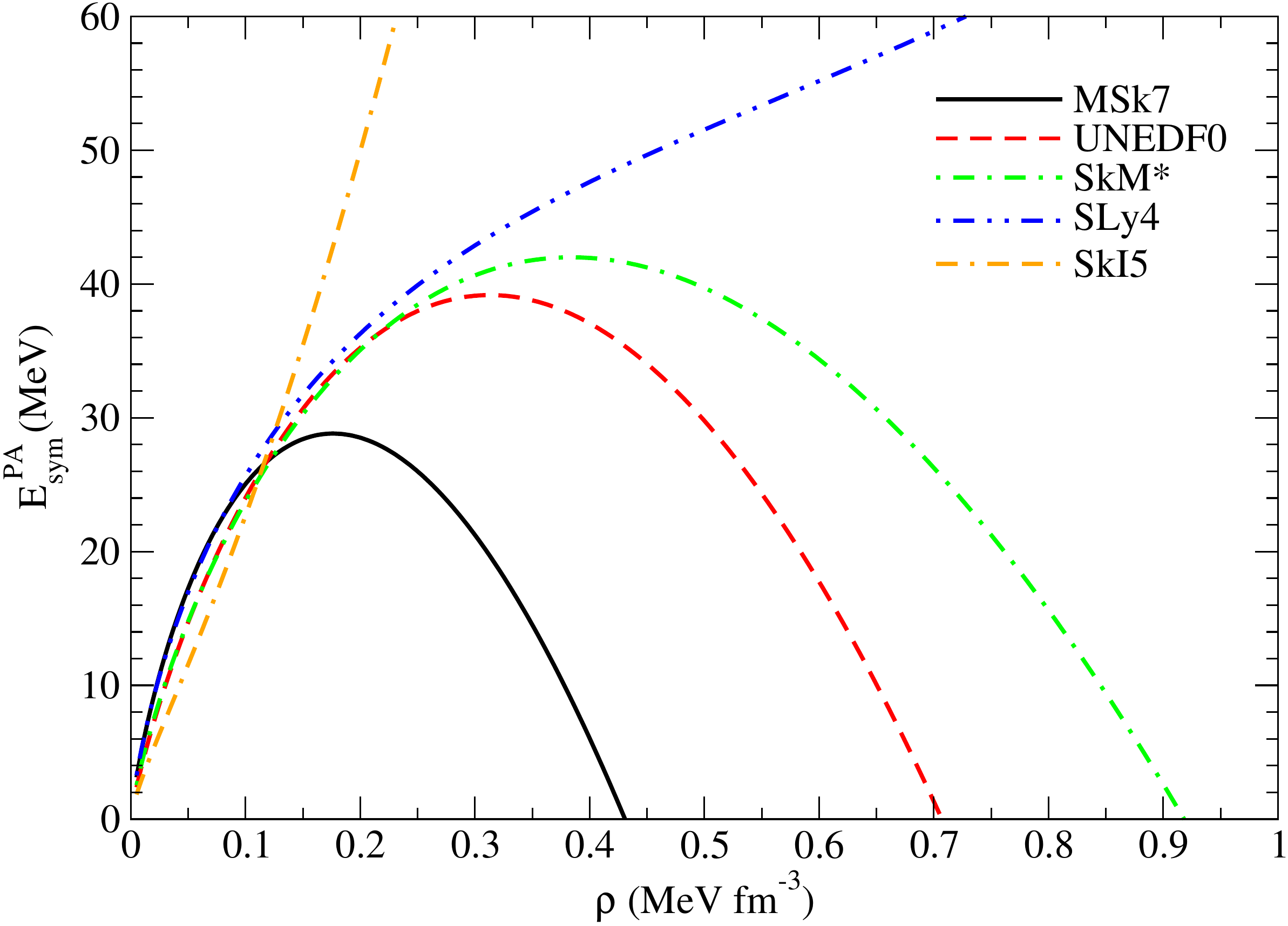}
  \caption{Density dependence of the symmetry energy coefficient calculated within the parabolic approximation
  $E_{\mathrm{sym}}^{PA}(\rho)$ for the Skyrme 
forces MSk7, UNEDF0, SkM$^*$, SLy4 and SkI5.}
 \label{fig:esympaskyrme}
\end{figure}

\begin{figure}[!b]
\centering     
\subfigure{\label{fig:esymPAesym2skyrme}\includegraphics[width=0.4\linewidth]{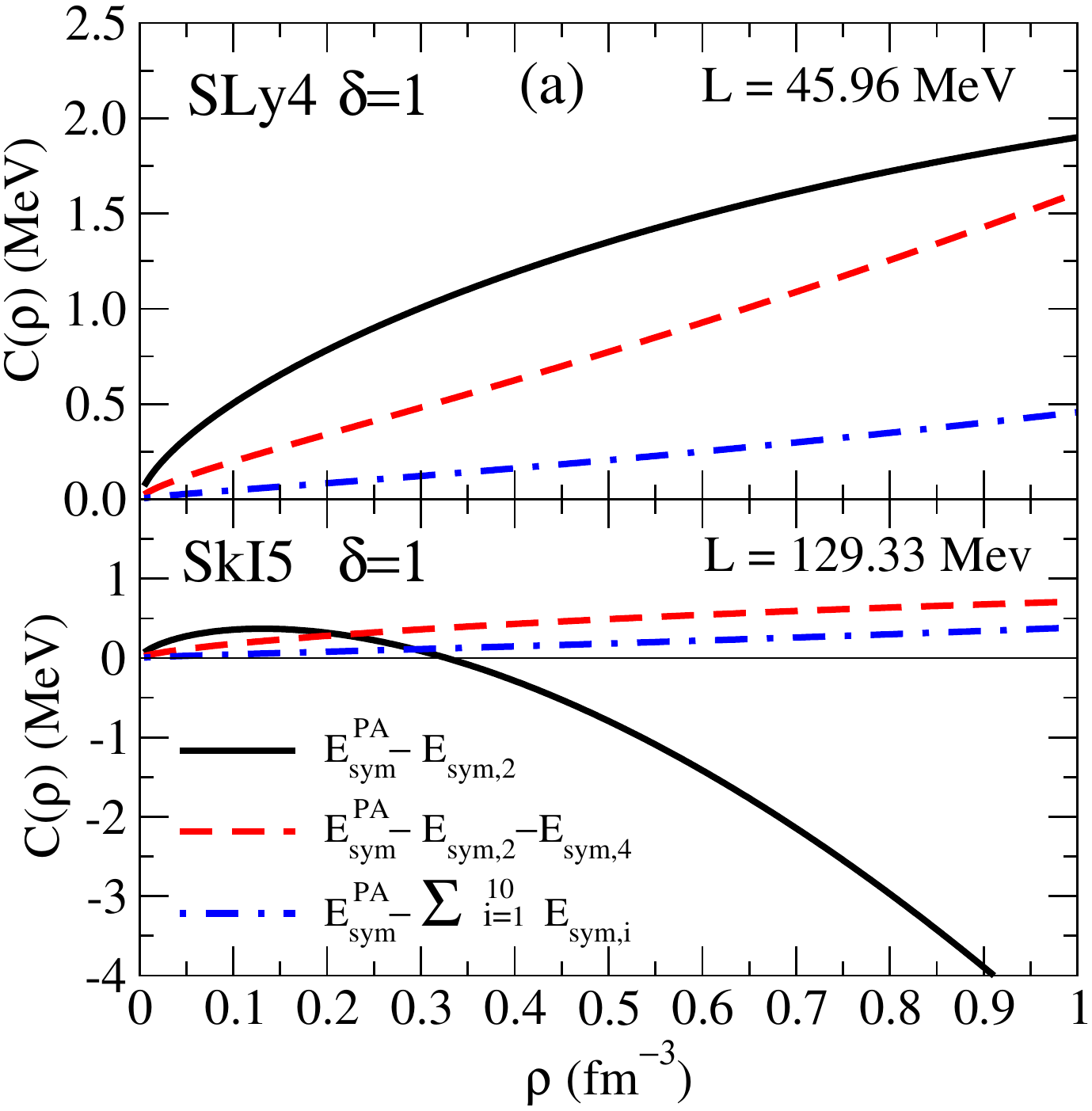}}
\subfigure{\label{fig:esymPaesym205skyrme}\includegraphics[width=0.422\linewidth]{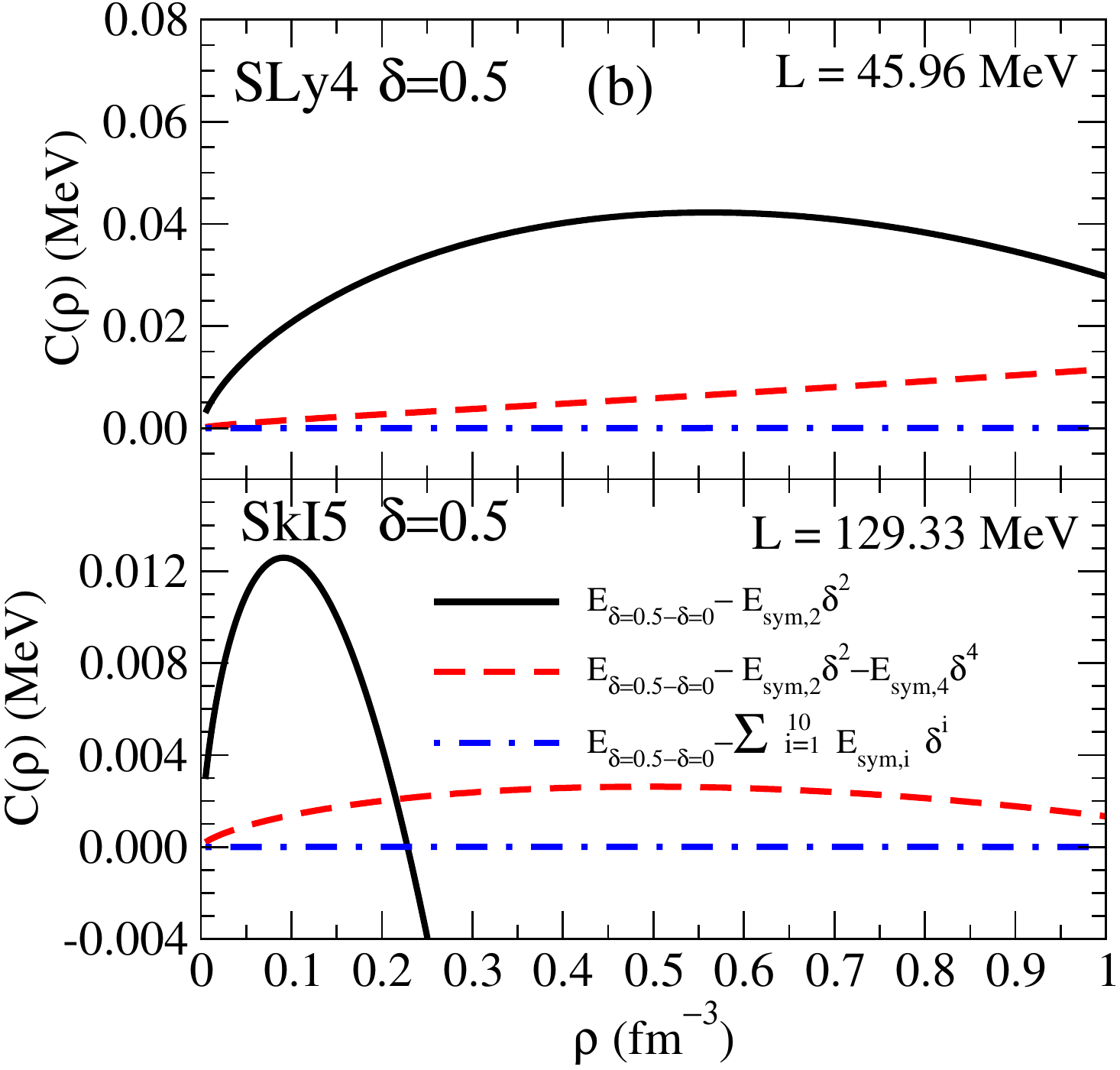}}
\caption{Panel a: Density dependence of the parabolic approximation $E_{\mathrm{sym}}^{PA}=E(\delta=1)-E(\delta=0)$ 
minus the sum of the symmetry energy contributions at 
  second, fourth and tenth-order in neutron matter ($\delta=1$).
  Panel b: Same as in Panel (a) but for $\rho_n=3\rho_p$ matter ($\delta=0.5$).}
\end{figure}

In order to analyze up to which extent the parabolic approximation compares to the EoS expansion in asymmetry,
we plot in Fig.~\ref{fig:esympaskyrme} the 
symmetry energy coefficient $E_{\mathrm{sym}}^{PA}(\rho)$ 
calculated within the parabolic approximation 
 with the same Skyrme interactions MSk7, UNEDF0, SkM$^*$, SLy4 and SkI5 against the baryon density.
We observe that the behaviour of $E_{\mathrm{sym}}^{PA}(\rho)$ is considerably similar in general trends 
to the one we find in Fig.~\ref{fig:esym2} for $E_{\mathrm{sym},2}(\rho)$. 
In the case of $E_{\mathrm{sym}}^{PA}(\rho)$ we see again that the SkI5 and SLy4 interactions show stiff trends, while the 
softer SkM$^*$, UNEDF0 and MSk7 interactions bend at a certain point, presenting isospin instabilities.  
To study the convergence of the series in Eq.~(\ref{eq:PAgen}), 
we plot in Fig.~\ref{fig:esymPAesym2skyrme}
the differences for the SLy4 and SkI5 interactions between the symmetry energy calculated with the parabolic approximation, 
and the sum of the symmetry energy coefficients up to a given order, i.e., 
\begin{equation}\label{eq:ccoeff}
 C \left(  \rho \right) =  E_\mathrm{sym}^{PA} (\rho) - \sum_{k} E_{\mathrm{sym}, 2k} (\rho).
\end{equation}
The different symmetry energy coefficients entering in the sum of the 
right hand side of Eq.~(\ref{eq:PAgen}) are calculated using the definition~(\ref{eq:esymgen}), and the difference between the
energies of neutron and symmetric nuclear matter is obtained using the exact energy per particle, which in the case of 
Skyrme interactions is defined in Eq.~(\ref{eq:Ebanm}) of Chapter~\ref{chapter1}.
The differences $C(\rho)$ should go to zero when the number of symmetry energy coefficients considered in the sum increases. 

In the case where one only considers the second-order symmetry energy coefficient, the differences between $E_\mathrm{sym}^{PA}(\rho)$
and $E_{\mathrm{sym}, 2}(\rho)$ can reach up to $2-4$ MeV in the range of the considered densities. 
These differences are reduced to values of $0.5$ MeV when higher-order contributions are added to the sum of the symmetry energy coefficients.
From Fig.~\ref{fig:esymPAesym2skyrme} we can extract some conclusions. Firstly, we see that the 
convergence of the series as a function of the density shows a strong model dependence. Moreover, we observe that 
the full convergence of the series is not completely achieved, even taking up to tenth-order 
coefficients in the expansion~(\ref{eq:EOSexpgeneral}), at least using the SLy4 and SkI5 Skyrme interactions. This result points out that 
the expansion~(\ref{eq:EOSexpgeneral}) is slowly convergent in the case of neutron matter computed with these two forces.   
To further explore the convergence of the asymmetry expansion of the EoS, we now 
consider the energy per particle of a system with an asymmetry $\delta$ intermediate between symmetric matter and neutron matter. From Eq.~(\ref{eq:EOSexpgeneral}) we 
obtain 
   \begin{equation} \label{Ed0g2}
 E_\mathrm{\delta - \delta=0} (\rho,\delta) \simeq \sum_{k} E_{\mathrm{sym}, 2k}  (\rho) \delta^{2k},
 \end{equation}
where $E_{\delta - \delta=0} (\rho,\delta) \equiv E_b(\rho, \delta) - E_b(\rho, \delta=0)$ is the difference between the exact energy per particle calculated 
in asymmetric nuclear matter and in symmetric matter. 
We plot in Fig.~\ref{fig:esymPaesym205skyrme} the differences 
 \begin{equation}\label{eq:ccoeff05}
 C \left(  \rho \right) =  E_\mathrm{\delta - \delta=0} (\rho) - \sum_{k} E_{\mathrm{sym}, 2k} (\rho) \delta^{2k}
\end{equation}
for the same forces as in Fig.~\ref{fig:esymPAesym2skyrme} considering a system where $\delta=0.5$, which corresponds to $\rho_n=3 \rho_p$.
In this case, the differences between the two sides of 
Eq.~(\ref{Ed0g2}) are much smaller than the ones obtained in pure neutron matter, and 
become almost zero using the expansion of the energy per particle up to tenth order. This points out that the convergence of the 
expansion (\ref{eq:EOSexpgeneral}) becomes slower as the isospin asymmetry $\delta$ of the system increases.

\subsection{Convergence of the expansion of the slope of the symmetry energy for Skyrme interactions}
In previous Sections, we have discussed two possible definitions for the slope of the symmetry energy. 
The first one, which we call $L$, results from considering the symmetry energy as the second-order coefficient of the Taylor expansion~(\ref{eq:EOSexpgeneral})
and defined in Eq.~(\ref{eq:esymgen}).
The second definition (see Eq.~(\ref{eq:esympa})) comes from assuming a parabolic expansion of the EoS and defining the symmetry energy as the 
difference between the energy per particle in pure neutron matter and in symmetric nuclear matter. 
This last definition can be also understood as the infinite sum of the $L_{2k}$ (\ref{eq:L2k}).
The slopes of the fourth-, sixth-, eighth- and tenth- order symmetry energy coefficients 
are reported in Table~\ref{table_L}, together with the slope $L_{\mathrm{PA}}$ of the parabolic symmetry energy. 
To test the convergence of this sum, we also include in the same table the results of the sum of the $L_{2k}$ series up to the
tenth order. We observe that, indeed, there is a good convergence of the $L_{2k}$ sum to $L_\mathrm{PA}$ if the slopes 
of the coefficients with higher order than two are considered.

\begin{table}[t]
\centering
\begin{tabular}{c|ppppppp}
\hline
\multirow{2}{*}{Force}  & \multicolumn{1}{c}{$L$}      & \multicolumn{1}{c}{$L_4$}   & \multicolumn{1}{c}{$L_6$}   & \multicolumn{1}{c}{$L_8$}  & \multicolumn{1}{c}{$L_{10}$} &
\multicolumn{1}{c}{$\sum_{k=1}^5 L_{2k}$} & \multicolumn{1}{c}{$L_\mathrm{PA}$} \\ 
  & \multicolumn{1}{c}{(MeV)}      & \multicolumn{1}{c}{(MeV)}   & \multicolumn{1}{c}{(MeV)}   & \multicolumn{1}{c}{(MeV)}  & \multicolumn{1}{c}{(MeV)} &
\multicolumn{1}{c}{(MeV)} & \multicolumn{1}{c}{(MeV)} \\ \hline\hline
MSk7   & 9.41  & 0.79  & 0.21 & 0.08 & 0.04  & 10.53  & 10.63\\
SIII   & 9.91   & 2.89  & 0.60 & 0.24 & 0.12  & 13.76 & 14.02\\
SkP    & 19.68  & 3.33  & 0.61 & 0.23 & 0.11  & 23.96 & 24.20\\
HFB-27 & 28.50 & 2.44  & 0.53 & 0.21 & 0.11  & 31.78 & 32.02\\
SkX    & 33.19 & 3.10  & 0.57 & 0.21 & 0.11  & 37.18  & 37.40\\
HFB-17 & 36.29 & 1.66  & 0.41 & 0.17 & 0.09  & 38.61 & 38.81\\
SGII   & 37.63   & 3.01  & 0.62 & 0.24 & 0.12 & 41.63 & 41.90\\
UNEDF1 & 40.00  & 2.63  & 0.50 & 0.19 & 0.09  & 43.42 & 43.62\\
Sk$\chi$500 & 40.74   & -0.58 & -0.01 & -0.01& 0.01 &40.17  &40.20 \\
Sk$\chi$450 & 42.06   & 1.30 & 0.29 &0.12 & 0.06 &43.83  &43.96 \\
UNEDF0 & 45.08  & 3.08  & 0.55 & 0.20 & 0.10  & 49.00 & 49.21\\
SkM*   & 45.78  & 3.32  & 0.67 & 0.26 & 0.13  & 50.16 & 50.44\\
SLy4   & 45.96  & 0.61  & 0.29 & 0.13 & 0.07  & 47.08 & 47.25\\
SLy7   & 47.22  & 0.54  & 0.28 & 0.13 & 0.07  & 48.25 & 48.42\\
SLy5   & 48.27  & 0.64  & 0.30 & 0.14 & 0.07  & 49.41 & 49.59\\
Sk$\chi$414 &  51.92 & 0.84 & 0.21 & 0.09& 0.04 &53.11  &53.21 \\
MSka   & 57.17  & 2.98  & 0.61 & 0.24 & 0.12  & 61.12 & 61.38\\
MSL0   & 60.00  & 2.70  & 0.57 & 0.22 & 0.11  & 63.60 & 63.85\\
SIV    & 63.50   & 5.51  & 1.20 & 0.47 & 0.24 & 70.92 & 71.45\\
SkMP   & 70.31   & 3.30  & 0.73 & 0.29 & 0.15 & 74.77 & 75.10\\
SKa    & 74.62  & 4.33  & 0.91 & 0.36 & 0.18  & 80.40 & 80.79\\
R$_\sigma$     & 85.69 & 2.88  & 0.60 & 0.24 & 0.12 & 89.53  & 89.79 \\
G$_\sigma$     & 94.01  & 2.87  & 0.60 & 0.24 & 0.12 & 97.84 & 98.10 \\
SV     & 96.09  & 7.18  & 1.58 & 0.62 & 0.32 & 105.78 & 106.49\\
SkI2   & 104.33 & 0.48  & 0.28 & 0.13 & 0.07 & 105.29 & 105.46\\
SkI5   & 129.33 & -0.72 & 0.15 & 0.10 & 0.06 & 128.91 & 129.06\\\hline
\end{tabular}
\caption{Values of the slope of the symmetry energy coefficients appearing in the Taylor expansion of the 
energy per particle up to tenth order in the isospin asymmetry $\delta$, and the parabolic approximation.}
\label{table_L}
\end{table}

\newpage
\section{Convergence of the isospin Taylor expansion of the EoS for Gogny interactions}
We proceed to study the contributions to the symmetry energy up to 6$^{\mathrm{th}}$ order for Gogny interactions
in the energy per particle Taylor expansion~(\ref{eq:EOSexpgeneral})~\cite{gonzalez17}. 
The calculation of the second-, fourth- and sixth-order symmetry energy coefficients for Gogny interactions 
is much more involved than in the case of the zero-range Skyrme interactions. We have obtained the 
following expressions for $ E_{\mathrm{sym}, 2} (\rho)$, $E_{\mathrm{sym}, 4} (\rho)$ and $E_{\mathrm{sym}, 6} (\rho)$
coefficients for Gogny forces~\cite{gonzalez17}:
\begin{eqnarray}
E_{\mathrm{sym}, 2} (\rho) &=& 
\left. \frac{1}{2!} \frac{\partial^{2} E_b(\rho, \delta)}{\partial \delta^{2}}\right|_{\delta=0} =
\frac{\hbar^2}{6m} \left(  \frac{3 \pi^2}{2}\right)^{2/3} \rho^{2/3}  - 
\frac{1}{8} t_3 \rho^{\alpha+1} (2x_3 +1)
\nonumber
\\
&+& \frac{1}{2} \sum_{i=1,2} \mu_i^3 \pi^{3/2}  {\cal B}_i  \rho  
 \mbox{} +\frac{1}{6}\sum_{i=1,2}  \left[-{\cal C}_i  G_1 ( k_F \mu_i)+ {\cal D}_i G_2 ( k_F \mu_i)  \right] , \label{eq:esym2gog}
\\
E_{\mathrm{sym}, 4} (\rho) &=&
\left. \frac{1}{4!} \frac{\partial^{4} E_b(\rho, \delta)}{\partial \delta^{4}}\right|_{\delta=0} =
\frac{\hbar^2}{162m} \left(  \frac{3 \pi^2}{2}\right)^{2/3} \rho^{2/3} 
\nonumber
\\
&+&
\frac{1}{324} \sum_{i=1,2}
\left[ {\cal C}_i  G_3 ( k_F \mu_i)  + {\cal D}_i G_4 ( k_F \mu_i) \right] \, ,\label{eq:esym4gog}
\\
E_{\mathrm{sym}, 6} (\rho) &=&
\left. \frac{1}{6!} \frac{\partial^{6} E_b(\rho, \delta)}{\partial \delta^{6}}\right|_{\delta=0} =
\frac{7\hbar^2}{4374m} \left(  \frac{3 \pi^2}{2}\right)^{2/3} \rho^{2/3} 
\nonumber
\\
&+& 
\frac{1}{43740} \sum_{i=1,2}
 \left[ {\cal C}_i  G_5(k_F \mu_i )  - {\cal D}_i  G_6( k_F \mu_i) \right] , \label{eq:esym6gog}
\end{eqnarray}
with $G_1 (\eta)$ and $G_2(\eta)$ already given, respectively, in Eqs.~(\ref{G1}) and (\ref{G2}) and with 
\begin{eqnarray}
 G_3 (\eta)&=&  -\frac{14}{\eta}  + e^{-\eta^2} \left( \frac{14}{\eta} + 14 \eta + 7 \eta^3 + 2\eta^5 \right) \label{G3}
\\
 G_4 (\eta)&=&   \frac{14}{\eta} - 8 \eta + \eta^3 - 2 e^{-\eta^2} \left( \frac{7}{\eta}+3\eta\right)\label{G4}
\\
 G_5 (\eta)&=&  -\frac{910}{\eta} + e^{-\eta^2}\left( \frac{910}{\eta} +   910\eta  + 455 \eta^3 + 147 \eta^5 
 + 32\eta^7 + 4 \eta^9 \right) \label{G5}
\\
 G_6 (\eta)&=&  -\frac{910}{\eta} + 460 \eta - 65 \eta^3+ 3 \eta^5 +e^{-\eta^2} \left( \frac{910}{\eta} 
 + 450 \eta + 60\eta^3    \right). \label{G6}
\end{eqnarray}
The expressions for the constants ${\cal B}_i$, ${\cal C}_i$, and ${\cal D}_i$ have been given in Section~\ref{Gogny} of Chapter~\ref{chapter1}.
\begin{figure}[t!]
 \centering
 \includegraphics[width=0.8\linewidth, clip=true]{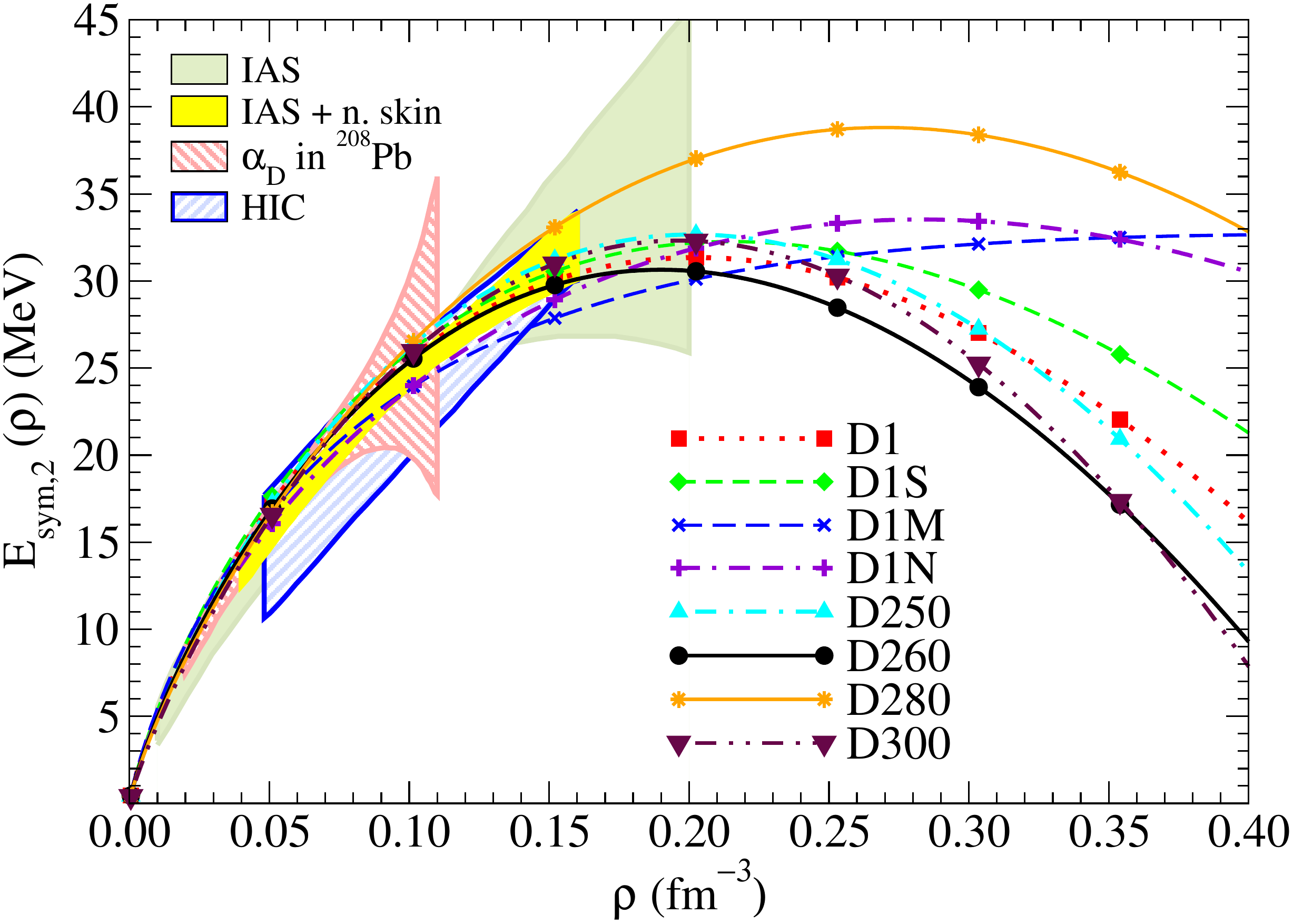}
  \caption{Density dependence of the second-order symmetry energy coefficient $E_{\mathrm{sym}, 2}(\rho)$ for different Gogny interactions. 
Also represented are the constraints on the symmetry energy extracted from the analysis of data on isobaric analog states (IAS)~\cite{Danielewicz13}
and of IAS data combined with neutron skins (IAS+n.skin)~\cite{Danielewicz13}, 
the constraints from the electric dipole polarizability in lead ($\alpha_D$ in $^{208}$Pb) \cite{Zhang15},
and from transport simulations of heavy-ion collisions of tin isotopes (HIC) \cite{Tsang08}.}
 \label{fig:esym2gog}
\end{figure}

\begin{figure}[t]
\centering     
\subfigure{\label{fig:esym4gog}\includegraphics[width=0.6\linewidth]{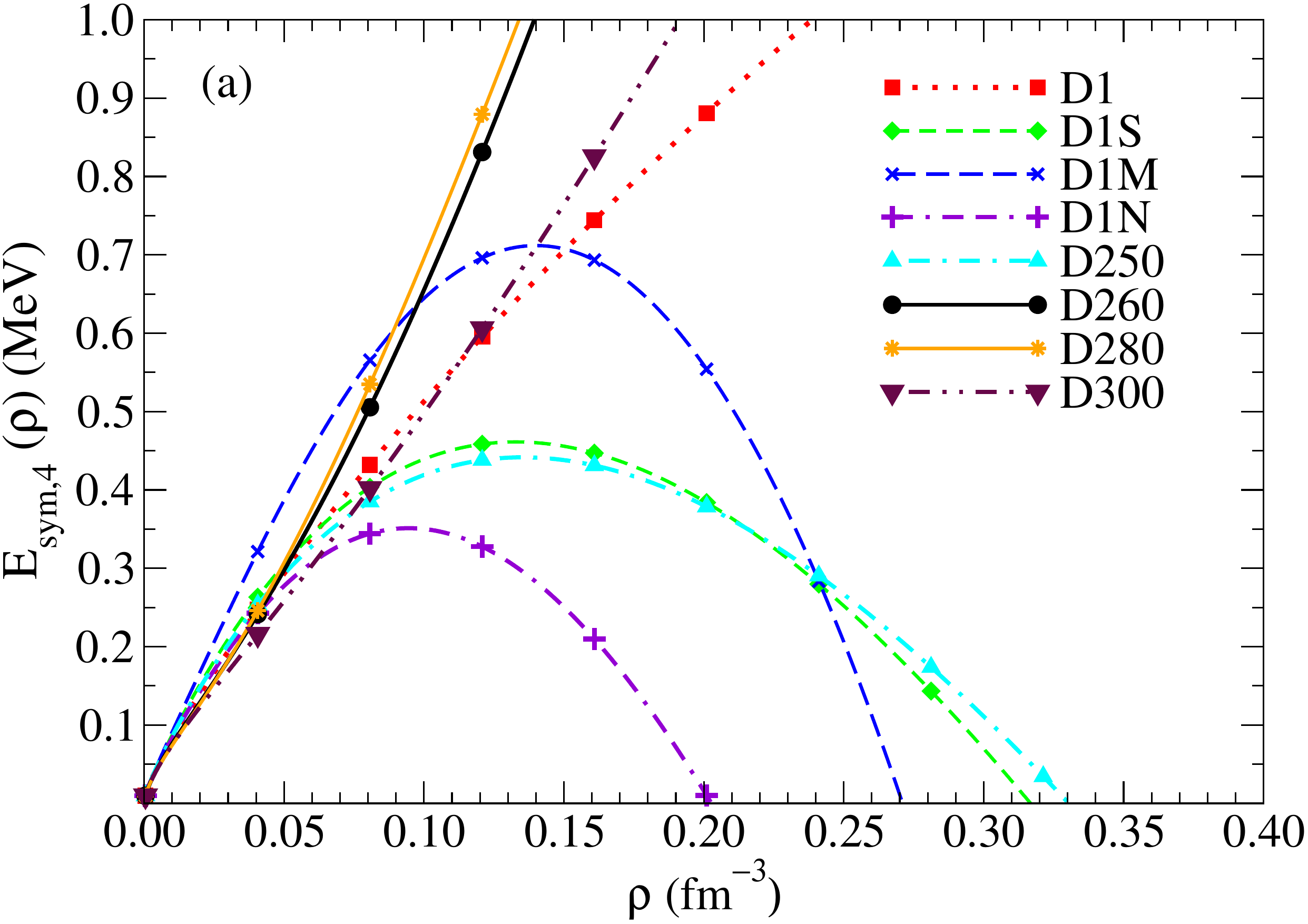}}
\subfigure{\label{fig:esym6gog}\includegraphics[width=0.6 \linewidth]{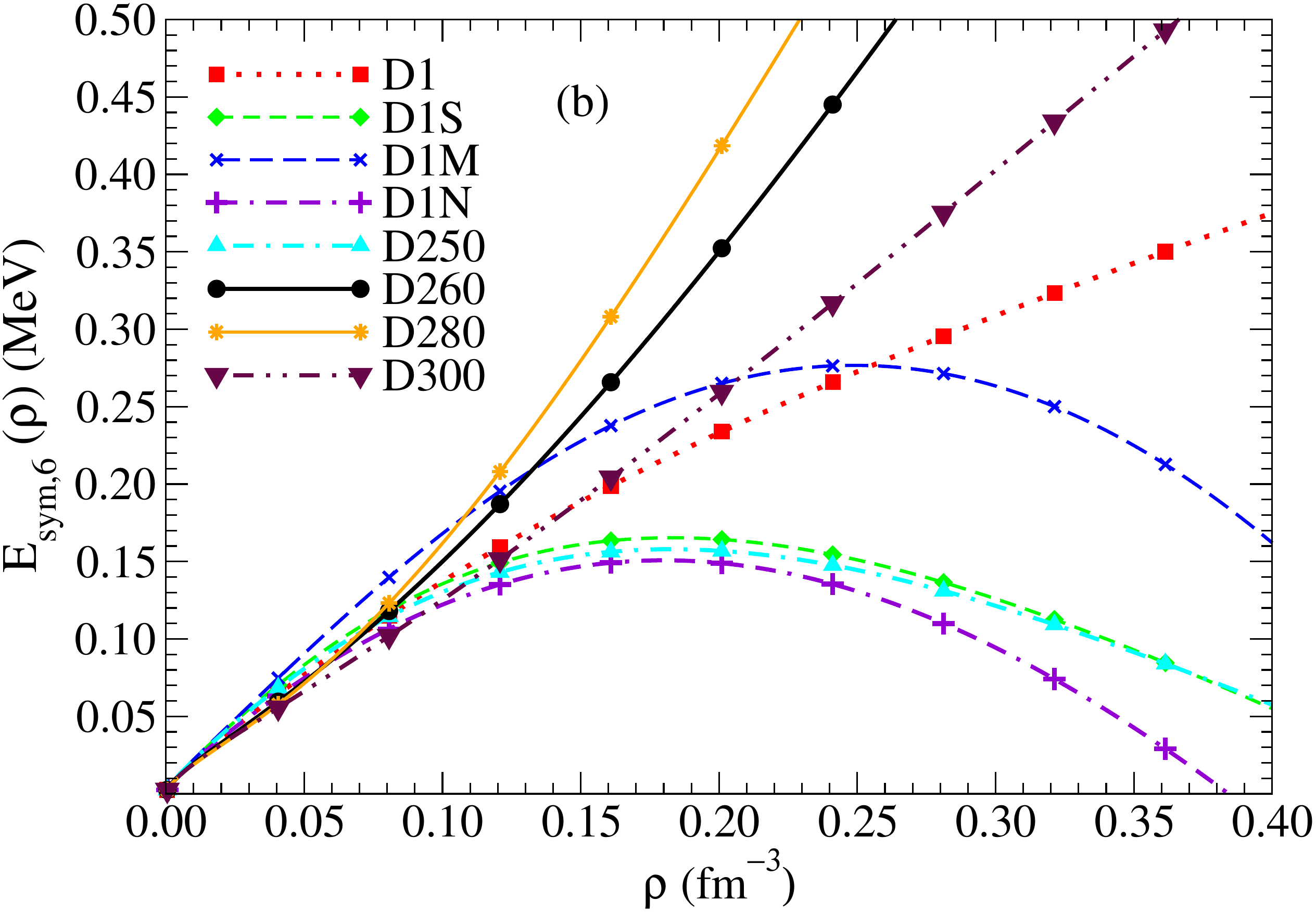}}
\caption{Density dependence of the fourth-order symmetry energy coefficient $E_{\mathrm{sym}, 4}(\rho)$ 
(panel (a)) and of the sixth-order symmetry energy coefficient $E_{\mathrm{sym}, 6}(\rho)$ (panel (b))
 for different Gogny interactions.}
\end{figure}

In Fig.~\ref{fig:esym2gog} we show the density dependence of the second-order symmetry energy coefficient $E_{\mathrm{sym}, 2} (\rho)$ for a set of
different Gogny interactions. As happened with Skyrme forces, at low densities $\rho \lesssim 0.1$ fm$^{-3}$, the
$E_{\mathrm{sym}, 2} (\rho)$ coefficient has similar trends for all interactions and 
increases with density. 
This is because most of the Gogny interactions are fitted to properties of finite nuclei, which are found in this subsaturation density regime.
On the other hand, from $\rho \gtrsim 0.1$ fm$^{-3}$ onwards, there are substantial differences between the trends of the different 
parametrizations. In comparison with existing empirical constraints for the symmetry energy 
at subsaturation density \cite{Danielewicz13, Zhang15, Tsang08}, 
one finds that the Gogny functionals in general respect them.
The symmetry energy coefficients of all considered interactions bend at values $E_{\mathrm{sym}, 2} (\rho) \sim 30-40$ MeV right above saturation density, 
and from this maximum value all parametrizations decrease with density.
In all interactions, their $E_{\mathrm{sym}, 2} (\rho)$ eventually becomes 
negative beyond $0.4$ fm$^{-3}$ (in D1M this happens only at a very large density of $1.9$ fm$^{-3}$), signaling the onset of an isospin instability. 
We show the symmetry energy coefficients of fourth-order, $E_{\mathrm{sym}, 4}(\rho)$, and sixth-order, 
$E_{\mathrm{sym}, 6}(\rho)$, in Figs.~\ref{fig:esym4gog} and \ref{fig:esym6gog}, respectively.
On the one hand, at subsaturation densities both terms are relatively small: below saturation density, $E_{\mathrm{sym}, 4}(\rho)$ is below $\approx$\,1 MeV and 
$E_{\mathrm{sym}, 6}(\rho)$ does not go above $\approx$\,0.3 MeV. These values can be compared with the larger values of 
$E_{\mathrm{sym}, 2}(\rho) > 10$ MeV in the same density regime. 
One should also consider that in the expansion of Eq.~(\ref{eq:EOSexpgeneral}) 
the terms $E_{\mathrm{sym}, 4}(\rho)$ and $E_{\mathrm{sym}, 6}(\rho)$ carry additional factors $\delta^2$ and $\delta^4$ with respect to 
$E_{\mathrm{sym}, 2}(\rho)$, and their overall magnitude will therefore be smaller.
On the other hand, above saturation density, we observe two markedly different behaviours for the density 
dependence of $E_{\mathrm{sym}, 4}(\rho)$ and $E_{\mathrm{sym}, 6} (\rho)$. 
For both $E_{\mathrm{sym}, 4}(\rho)$ and $E_{\mathrm{sym}, 6}(\rho)$, we find a group of parametrizations 
(D1S, D1M, D1N, and D250) that reach a maximum and then decrease with density. 
We call this set of forces as  ``group~1" from now on. A second set of forces, ``group~2'', 
is formed of D1, D260, D280, and D300,
which yield $E_{\mathrm{sym}, 4}(\rho)$ and $E_{\mathrm{sym}, 6}(\rho)$ terms
that do not reach a maximum and increase steeply in the range of the studied densities.
\begin{figure}[t!]
 \centering
 \includegraphics[width=0.7\linewidth, clip=true]{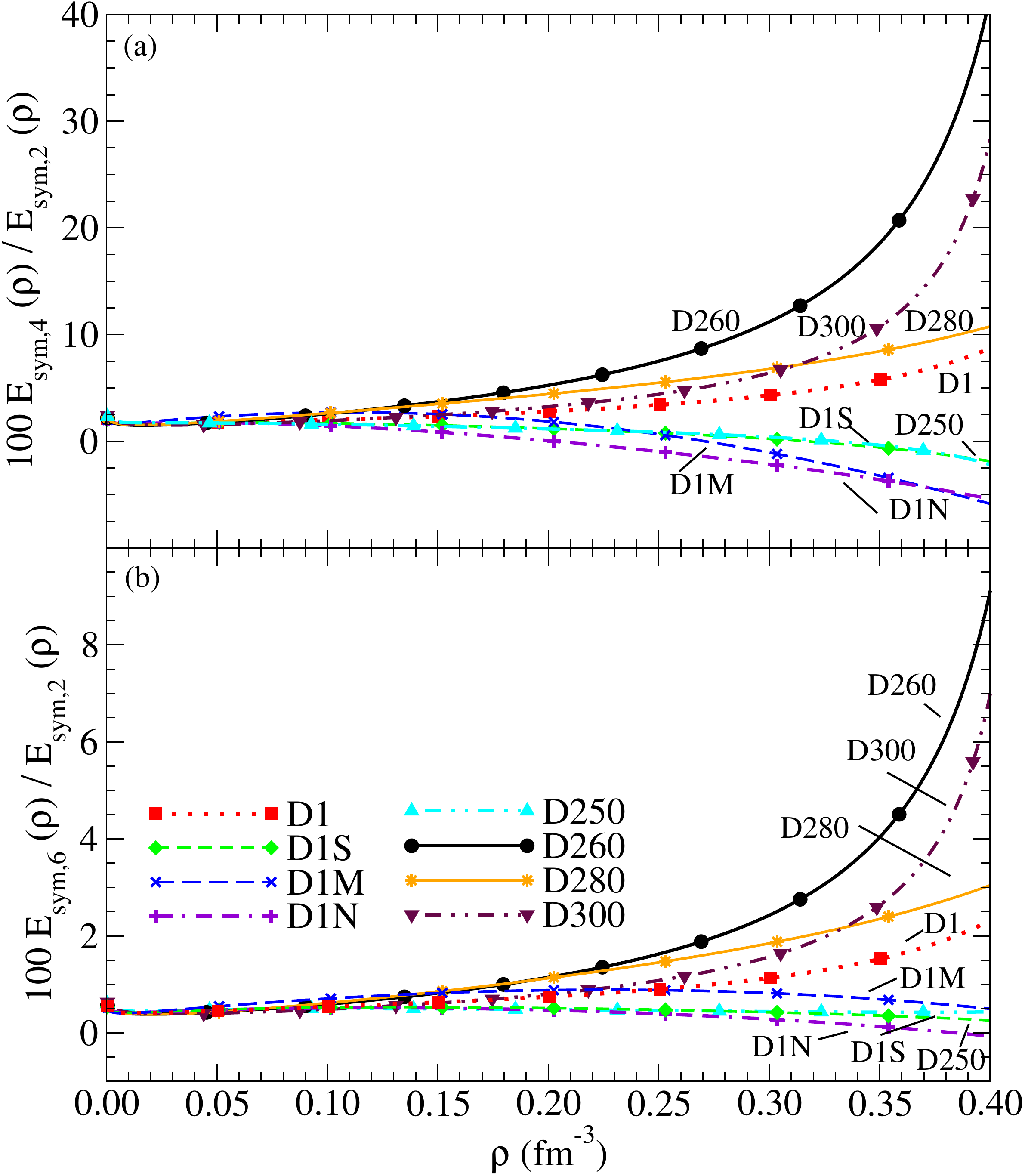}
  \caption{Density dependence of the ratios $E_{\mathrm{sym}, 4} (\rho)/E_{\mathrm{sym}, 2} (\rho)$ (panel (a)) 
  and $E_{\mathrm{sym}, 6} (\rho)/E_{\mathrm{sym}, 2} (\rho)$ (panel (b)) for different Gogny interactions.}
\label{fig:E46E2_Gogny}
\end{figure}

The difference in density dependence between the second-order symmetry energy and its higher-order corrections can be understood by 
decomposing them into terms associated to the different contributions from the nuclear Hamiltonian.
All three coefficients $E_{\mathrm{sym}, 2}(\rho)$, $E_{\mathrm{sym}, 4}(\rho)$ and $E_{\mathrm{sym}, 6}(\rho)$ include a 
kinetic component, which decreases substantially as the order increases. The
$E_{\mathrm{sym}, 2}(\rho)$ coefficient also receives contributions 
from the zero-range term of the force [Eq.~(\ref{eq:eb.zr})] as well as from the finite-range direct and exchange terms
[Eqs.~(\ref{eq:eb.dir})--(\ref{eq:eb.exch})].
We note that the direct terms of the finite-range contribution to $E_{\mathrm{sym}, 2}(\rho)$ are directly proportional to the constants ${\cal B}_i$ and to the density $\rho$. 
The functions $G_n (\mu_i k_F)$ are due solely to the exchange contribution in the matrix elements of the Gogny force. 
One can equally say that they reflect the contribution of the momentum dependence of the interaction to the symmetry energy. 
As discussed in Ref.~\cite{Sellahewa14}, the zero-range term, the direct term, and the exchange (momentum-dependent) term contribute 
with similar magnitudes to the determination of $E_{\mathrm{sym}, 2}(\rho)$ with Gogny forces. However, they contribute with different signs, which leads to 
cancellations in $E_{\mathrm{sym}, 2}(\rho)$ between the power-law zero-range term, the linear density-dependent direct term, and the exchange term. 
Depending on the parametrization, the sum of the zero-range and direct terms is positive and the exchange term is negative, 
or the other way around. In any case, there is a balance between terms, 
which gives rise to a somewhat similar density dependence of the symmetry energy coefficient $E_{\mathrm{sym}, 2}(\rho)$ for all parameter sets. 

In contrast to the case of the $E_{\mathrm{sym}, 2}(\rho)$ coefficient, neither the zero-range nor the direct term contribute to 
the $E_{\mathrm{sym}, 4}(\rho)$ and $E_{\mathrm{sym}, 6}(\rho)$ coefficients, see Eqs.~(\ref{eq:esym4gog}) and (\ref{eq:esym6gog}).
The cause is that both the zero-range and the direct components of the energy per particle [cf.\ Eqs.~(\ref{eq:eb.zr})--(\ref{eq:eb.dir})] 
depend on the square of the isospin asymmetry, $\delta^2$. 
In other words, the higher-order corrections to the symmetry energy are only sensitive to the kinetic term and to 
the momentum-dependent term, i.e., the exchange term of the Gogny force. 
We note that the same pattern is found in zero-range Skyrme forces, but in that case, 
the functional dependence of the momentum-dependent contribution to the symmetry energy coefficients is   
proportional to $\rho^{5/3}$, whereas in the Gogny case it has a more intricate density dependence due to the finite range of the interaction,
which is reflected in the $G_n (\mu_i k_F)$ functions~\cite{gonzalez17}.

In both $E_{\mathrm{sym}, 4}(\rho)$ and $E_{\mathrm{sym}, 6}(\rho)$ of Gogny forces, cf.\ Eqs.~(\ref{eq:esym4gog}) and (\ref{eq:esym6gog}), 
the exchange term is given by the product of two parametrization-dependent constants, ${\cal C}_i$ and ${\cal D}_i$, 
and two density-dependent functions, 
$G_3$ and $G_4$, or $G_5$ and $G_6$. Because the density dependence of these functions is similar, one does expect 
that comparable density dependences arise for the fourth- and the sixth-order symmetry energy coefficients, as observed in Figs.~\ref{fig:esym4gog} and \ref{fig:esym6gog}.
This simple structure also provides an explanation for the appearance of two distinct groups of forces in terms of the density dependence 
 of $E_{\mathrm{sym}, 4}(\rho)$ and $E_{\mathrm{sym}, 6}(\rho)$. In group~1 forces, the fourth- and sixth-order contributions to the symmetry 
energy change signs as a function of density, whereas group~2 forces produce monotonically increasing functions of density. The change of sign
 is necessarily due to the exchange contribution, which in the case of group~1 forces must also be attractive 
 enough to overcome the kinetic term. 

For further insight into the relevance of $E_{\mathrm{sym}, 4} (\rho)$ and $E_{\mathrm{sym}, 6} (\rho)$ for 
the Taylor expansion of the EoS at each density, we plot in Fig.~\ref{fig:E46E2_Gogny} their ratios with respect 
to $E_{\mathrm{sym}, 2} (\rho)$. In the zero density limit, we see that both ratios tend to a constant value. 
This is expected in the non-interacting case, although the actual values of these ratios are modified by the 
exchange contributions. In this limit, we find $E_{\mathrm{sym}, 4}(\rho)/E_{\mathrm{sym}, 2}(\rho) \approx 1.5 \%$ and 
$E_{\mathrm{sym}, 6}(\rho)/E_{\mathrm{sym}, 2}(\rho) \approx 0.4 \%$. 
At low, but finite densities, $\rho \lesssim 0.1$ fm$^{-3}$, the ratio $E_{\mathrm{sym}, 4}(\rho)/E_{\mathrm{sym}, 2}(\rho)$ 
is relatively flat and not larger than $3\%$. The ratio for the sixth-order term is also mildly density-dependent 
and less than $0.6 \%$. Beyond saturation, both ratios increase in absolute value, to the point that for some 
parametrizations the ratio of the fourth- (sixth-) order term to the second-order term is not negligible and 
of about $10-30 \%$ ($2-8 \%$) or even more. In particular, this is due to the decreasing trend of 
$E_{\mathrm{sym}, 2} (\rho)$ with increasing density for several interactions when $\rho$ is above saturation.
We may compare these results for the ratios with previous literature. For example, the calculations of 
Ref.~\cite{Moustakidis12} with the momentum-dependent interaction (MDI) and with the Skyrme forces 
SLy4, SkI5 and Ska find values of $ \left| E_{\mathrm{sym}, 4} (\rho)/E_{\mathrm{sym}, 2} (\rho) \right| < 8 \%$ 
at $\rho \sim 0.4 $ fm$^{-3}$. 
Our results for Skyrme interactions give similar ratios as the ones presented in Ref.~\cite{Moustakidis12}.
In the same reference~\cite{Moustakidis12}, the Thomas-Fermi model of Myers and Swiatecki 
yields a ratio $ \left| E_{\mathrm{sym}, 4} (\rho)/E_{\mathrm{sym}, 2} (\rho) \right|$ reaching 60\% already at a
moderate density $\rho= 1.6\rho_0$.
With RMF models such as FSUGold or IU-FSU, at densities $\rho \sim 0.4 $ fm$^{-3}$ one has ratios of $ \left| 
E_{\mathrm{sym}, 4} (\rho)/E_{\mathrm{sym}, 2} (\rho) \right| < 4 \%$ \cite{Cai2012}. All in all, 
it appears that Gogny parametrizations provide ratios that are commensurate with previous literature.

\subsection{Parabolic approximation for Gogny interactions}
\begin{figure}[!b]
 \centering
 \includegraphics[width=0.8\linewidth, clip=true]{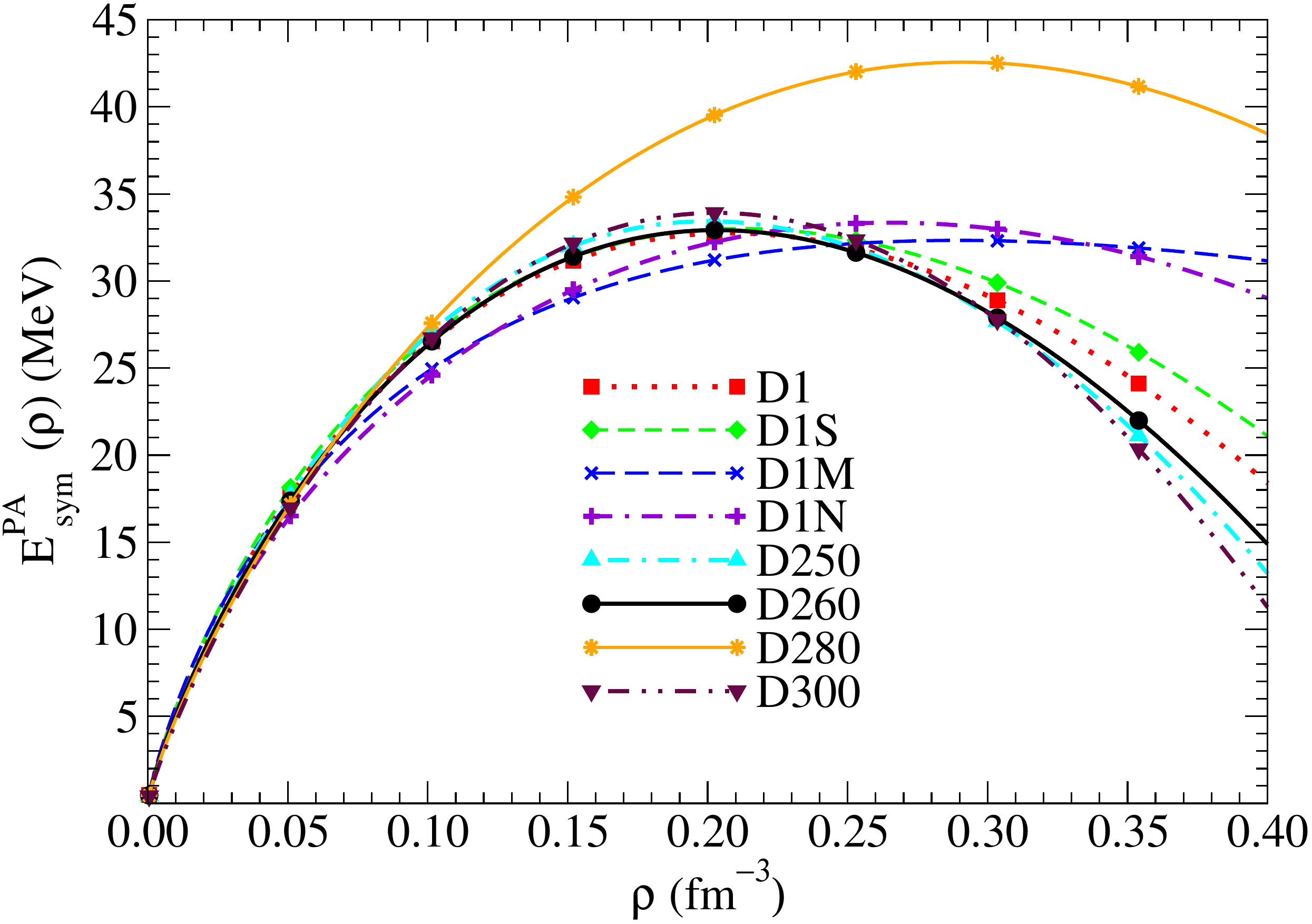}
  \caption{Density dependence of the symmetry energy coefficient in the parabolic approximation [Eq.~(\ref{eq:PAgen})] for different Gogny interactions.}
\label{fig:esymPA}
\end{figure}

In Fig.~\ref{fig:esymPA} we show the results for $E_\mathrm{sym}^{PA}(\rho)$ from the different Gogny
functionals. In general we find a similar picture to that of Fig.~\ref{fig:esym2gog} for the second-order symmetry energy $E_{\mathrm{sym},2} (\rho)$.  
At subsaturation densities, the symmetry energies $E_{\mathrm{sym}}^{PA}(\rho)$ of all the forces are 
quite close to each other presenting similar trends. On the other hand, at densities 
above $\rho \gtrsim 0.1$ fm$^{-3}$, there are markedly different behaviours between the interactions. 
Usually, $E_{\mathrm{sym}}^{PA}(\rho)$ reaches a maximum and then starts to decrease up to a given density 
where it becomes negative, presenting an isospin instability.

\begin{figure}[t!]
 \centering
 \includegraphics[width=0.8\linewidth, clip=true]{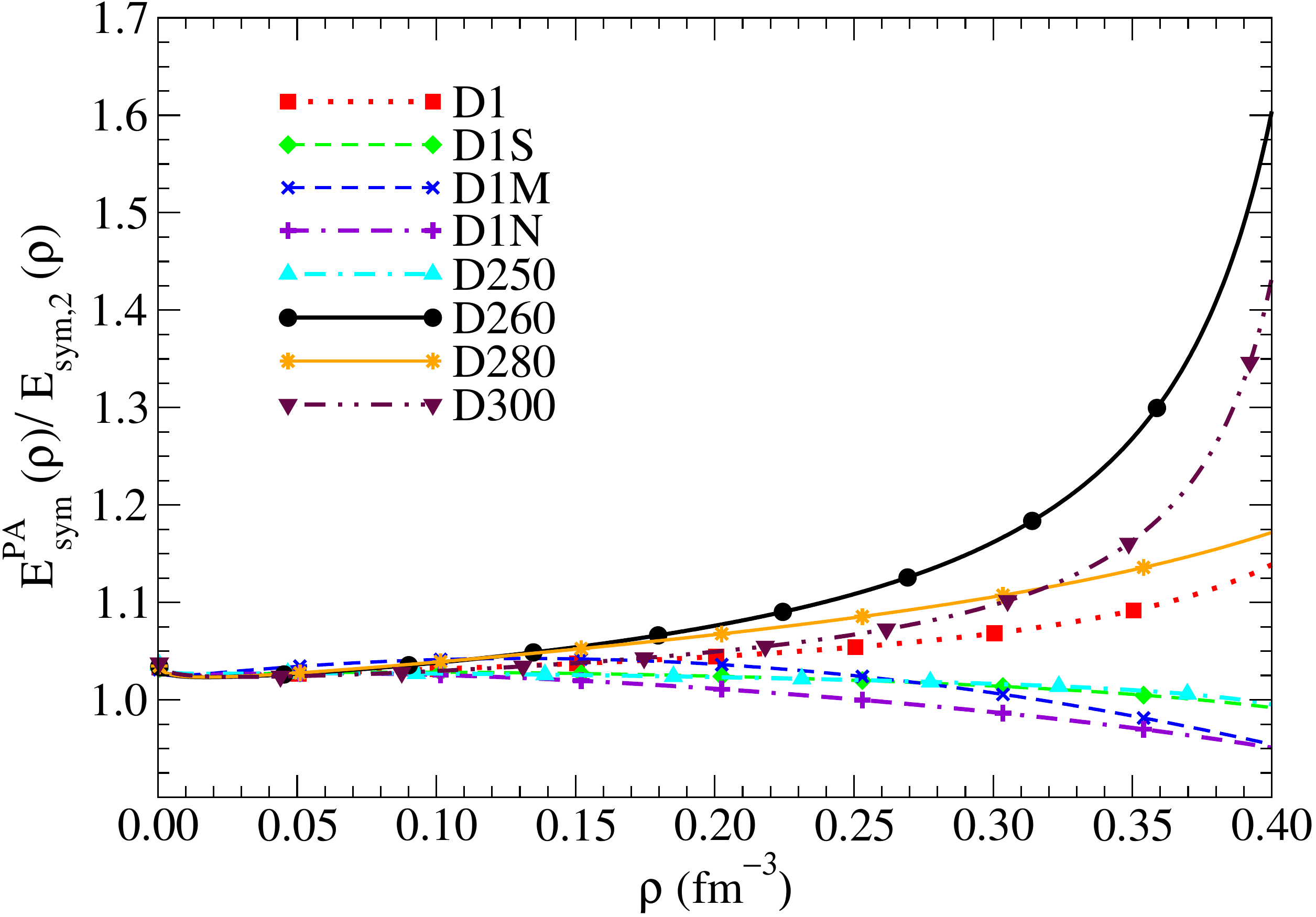}
  \caption{Density dependence of the ratio $E_{\mathrm{sym}}^{PA} (\rho)/E_{\mathrm{sym}, 2} (\rho)$ for different Gogny interactions.}
\label{fig:esymPAesym2}
\end{figure}

The correspondence between the second-order symmetry energy coefficient and the symmetry 
energy calculated within the parabolic approach may be spoiled by the influence of the
higher-order terms in the expansion of the energy per particle~(\ref{eq:EOSexpgeneral}).
In order to analyze better the differences between $E_\mathrm{sym}^{PA}(\rho)$ and $E_{\mathrm{sym}, 2}(\rho)$,
we plot in Fig.~\ref{fig:esymPAesym2} the ratio $E_{\mathrm{sym}}^{PA}(\rho)/E_{\mathrm{sym}, 2}(\rho)$.  
At low densities $\rho \lesssim 0.1$ fm$^{-3}$, the symmetry energy calculated with the parabolic law is always a little
larger than calculated with Eq.~(\ref{eq:esym4gog}) and (\ref{eq:esym6gog}) for $k=1$. The ratio is approximately 1.025 irrespective of the functional. 
This is relatively consistent with the zero-density limit of a free Fermi gas,  which has a ratio 
$E_{\mathrm{sym}}^{PA}(\rho)/E_{\mathrm{sym}, 2}(\rho) = \frac{9}{5}(2^{2/3}-1) \approx 1.06$. 
At densities $\rho \gtrsim 0.1$ fm$^{-3}$, the ratios change depending on the Gogny force. 
Here, group~1 and group~2 parametrizations again show two distinct behaviours. In group~1 (D1S, D1M, D1N, D250), 
the ratio becomes smaller than $1$ at large densities, whereas in group~2 (D1, D260, D280, D300), it increases with density.
There is a clear resemblance between Fig.~\ref{fig:esymPAesym2} and Fig.~\ref{fig:E46E2_Gogny}(a). 
Indeed, Eq.~(\ref{eq:PAgen}) suggests that the two ratios are connected~\cite{gonzalez17},
\begin{equation}
\frac{ E_{\mathrm{sym}}^{PA} (\rho) }{ E_{\mathrm{sym},2} (\rho) } = 1 + 
\frac{E_{\mathrm{sym}, 4} (\rho) }{E_{\mathrm{sym}, 2} (\rho) } + \cdots \, ,
\end{equation} 
as long as the next-order term $\frac{E_{\mathrm{sym}, 6} (\rho) }{E_{\mathrm{sym}, 2} (\rho) }$ 
is small. The behaviour of the ratio $\frac{ E_{\mathrm{sym}}^{PA} (\rho) }{ E_{\mathrm{sym},2} (\rho) }$ 
can therefore be discussed in similar terms as the ratios shown in Fig.~\ref{fig:E46E2_Gogny}. 
As discussed earlier in the context of Eqs.~Eq.~(\ref{eq:esym4gog}) and (\ref{eq:esym6gog}), $E_{\mathrm{sym}, 4}(\rho)$ and $E_{\mathrm{sym}, 6}(\rho)$
are entirely determined by the exchange contributions that are proportional to the 
constants ${\cal C}_i$ and ${\cal D}_i$ and the functions $G_n (\mu_i k_F)$.

\subsection{Convergence of the expansion of the slope of the symmetry energy for Gogny interactions}
\begin{table}[!b]
\centering
\resizebox{\columnwidth}{!}{%
\begin{tabular}{ddddddddd}	
\hline
 \multicolumn{1}{c}{Force}  &  \multicolumn{1}{c}{D1}  &  \multicolumn{1}{c}{D1S}  &  \multicolumn{1}{c}{D1M}  &  \multicolumn{1}{c}{D1N}  &  \multicolumn{1}{c}{D250}  &  \multicolumn{1}{c}{D260}  &  \multicolumn{1}{c}{D280}  &  \multicolumn{1}{c}{D300}  \\
 \hline\hline
 \multicolumn{1}{c}{$E_{\mathrm{sym}, 2}$($\rho_0$)}  &   30.70    &   31.13    &   28.55    &   29.60    &   31.54    &   30.11    &   33.14    &   31.23    \\
 \multicolumn{1}{c}{$E_{\mathrm{sym}, 4}$($\rho_0$)}  &   0.76     &   0.45     &   0.69     &   0.21     &   0.43     &   1.20     &   1.18     &   0.80     \\
 \multicolumn{1}{c}{$E_{\mathrm{sym}, 6}$($\rho_0$)}  &   0.20     &   0.16     &   0.24     &   0.15     &   0.16     &   0.27     &   0.29     &   0.20     \\
 \multicolumn{1}{c}{$L$}  &   18.36    &   22.43    &   24.83    &   33.58    &   24.90    &   17.57    &   46.53    &   25.84    \\
 \multicolumn{1}{c}{$L_4$}  &   1.75     &   -0.52    &   -1.04    &   -1.96    &   -0.33    &   4.73     &   4.36     &   2.62     \\
 \multicolumn{1}{c}{$L_6$}  &   0.46     &   0.08     &   0.42     &   0.08     &   0.09    &   0.99     &   1.19     &   0.63     \\
 \multicolumn{1}{c}{$\sum_{k=1}^3 L_{2k}$} &   20.57    &   21.99     &   24.21    &   31.7     &   24.66   &   23.26     &   52.08     &   29.09    \\
 \hline
 \multicolumn{1}{c}{$E_\mathrm{sym}^{PA}$($\rho_0$)}  &   31.91    &   31.95    &   29.73    &   30.14    &   32.34    &   31.85    &   35.89    &   32.44    \\
 \multicolumn{1}{c}{$L_{PA}$}  &   21.16    &   22.28    &   24.67    &   31.95    &   24.94    &   24.33    &   53.25    &   29.80    \\\hline
\end{tabular}}
\caption{Values of the $E_{\mathrm{sym}, 2k}$($\rho_0$) symmetry energy coefficients and their corresponding 
slopes $L_{2k}$ parameters at the saturation density $\rho_0$ for Gogny interactions. The values for the parabolic approximation, $E_\mathrm{sym}^{PA}$($\rho_0$) and $L_{PA}$,
are also included.}
\label{Table-saturation}
\end{table}
The values of $L_{2k}$ 
provide a good handle on the density dependence of the corresponding $E_{\mathrm{sym}, 2k} (\rho)$ contributions~\cite{gonzalez17}.
At second order, the slope 
parameter $L$ is positive in all the considered Gogny interactions. It goes from $L=17.57$ MeV in D260 to $46.53$ MeV in D280.
This large variation of the $L$ value indicates that the density dependence of the symmetry energy is poorly 
constrained with these forces \cite{Sellahewa14}. 
We also emphasize that all forces in Table \ref{Table-saturation} have a low slope parameter, under $50$ MeV, 
and thus correspond to soft symmetry energies \cite{Tsang2012,Lattimer2013,BaoAnLi13,Vinas14,Roca-Maza15,Lattimer2016}.
Indeed, we see that the $L$ values in Table \ref{Table-saturation}  are below or on the 
low side of recent results proposed from microscopic calculations, such as the ones coming from the study of the
electric dipole polarizability of $^{48}$Ca,
$L=43.8$--48.6 MeV \cite{Birkhan16}, or the ones coming from chiral effective field theory, $L=20$--65 MeV \cite{Holt2017} and 
$L=45$--70 MeV \cite{Drischler1710.08220}.
The higher-order slope parameters $L_4$ and $L_6$ are in keeping with the density dependence of 
$E_{\mathrm{sym}, 4}(\rho)$ and $E_{\mathrm{sym}, 6}(\rho)$, respectively.
$L_4$ goes from about $-2$ MeV (D1N) to 
$4.7$ MeV (D260) and $L_{6}$ is in the range of $0.1-1.2$ MeV for the different forces. 
Interestingly, we find a one-to-one correspondence between group 1 and group 2 forces and the sign of $L_4$. 
For group 1 forces, $E_{\mathrm{sym}, 4} (\rho)$ has already reached a maximum at saturation density and tends 
to decrease with density (cf.~Fig.~\ref{fig:esym4gog});
consequently, $L_4$ is negative. On the contrary, group 2 forces have positive $L_4$, reflecting the increasing nature 
of $E_{\mathrm{sym}, 4} (\rho)$ with density. In contrast to $L_4$, 
the values of $L_{6}$ are always positive. This is a reflection 
of the fact that the maximum of $E_{\mathrm{sym}, 6} (\rho)$ occurs somewhat above saturation density, as shown 
in Fig.~\ref{fig:esym6gog}. 
It is worth noting that in absolute terms the value of the $L_{2k}$ parameters decreases with increasing order 
of the expansion, i.e., we have $|L_6| < |L_4| < |L|$. This indicates that the dominant density dependence of the isovector part of the 
functional is accounted for by the second-order parameter $L$.
The $L_{PA}$ values are displayed in the last row of Table~\ref{Table-saturation}.
There are again differences between the two groups of functionals. 
In group 1 forces, such as D1S, D1M, D1N, or D250, the $L_{PA}$ values are fairly close to the slope parameter $L$. 
In contrast, group 2 forces have $L_{PA}$ values that are substantially larger than $L$. For example, the relative 
differences between $L_{PA}$ and $L$ are of the order of $40 \%$ for D260 and $15 \%$ for D280. This again may be 
explained in terms of the higher-order $L_{2k}$ 
contributions, which add up to give $L_{PA}$ analogously to Eq.~(\ref{eq:PAgen}).
For group 1 interactions, the addition of the higher-order terms to the $L$ parameter tends to disrupt a little the similarities
with $L_{PA}$. Nevertheless, the relative differences between $\sum_{k=1}^3 L_{2k}$ and $L_{PA}$ do not exceed values of $2\%$.
For group 2 interactions, on the other hand, the addition of higher-order terms to $L$ tend to reduce the differences 
between $\sum_{k=1}^3 L_{2k}$ and $L_{PA}$, achieving relative differences up to a maximum of $5\%$.

On the whole, for Gogny interactions, the parabolic approximation seems 
to work relatively well at the level of the symmetry energy. For the slope parameter, however, the contribution of 
$L_4$ can be large and spoil the agreement between the approximated $L_{PA}$ and $L$. 
$L_4$ is a density derivative of $E_{\mathrm{sym},4}(\rho)$, which, as shown in Eq.~(\ref{eq:esym4gog}), is entirely determined 
by the exchange finite-range terms in the Gogny force.
The large values of 
$L_4$ are given by the isovector finite-range exchange contributions. We therefore conclude that exchange contributions play 
a very important role in the slope parameter. These terms can provide substantial (in some cases of order $30 \%$) 
corrections and should be explicitly considered when it is possible to do so \cite{Vidana2009}.

\section{Beta-stable nuclear matter}
We now proceed to study the impact of the higher-order symmetry energy terms on the equation of state
of $\beta$-equilibrated matter. This condition is found in the interior 
of NSs, where the URCA reactions
\begin{align}
 n \rightarrow p + e^- + \bar{\nu}_e \qquad
 p+ e^- \rightarrow n + \nu_e 
\end{align}
take place simultaneously.
If one assumes that the neutrinos leave the system, the $\beta$-equilibrium leads to the condition 
\begin{equation}\label{betaeq}
 \mu_n - \mu_p = \mu_e ,
\end{equation}
where $\mu_n$, $\mu_p$, and $\mu_e$ are the chemical potentials of neutrons, protons, and electrons, respectively.
The expressions for the neutron, proton chemical potentials are given in Chapter~\ref{chapter1}.
The contribution given by the leptons, i.e., the electrons ($e$) and muons ($\mu$) in the system, to the 
energy density is
\begin{eqnarray} \label{He}
 \mathcal{H}_l (\rho, \delta) &=& \frac{m_l^4}{8\pi^2 \rho} \left[ x_F \sqrt{1+x_F^2} \left(1+ 2 x_F^2 
\right) - \mathrm{ln}\left(x_F + \sqrt{1+x_F^2}\right) \right],
\end{eqnarray}
where $m_l$ is the mass of each type of leptons $l=e, \mu$ and the dimensionless Fermi momentum is 
$x_F \equiv k_{Fl}/m_l = (3 \pi^2 \rho_l)^{1/3}/m_l$. 
The density $\rho_l$ defines the density
 of electrons ($\rho_e$) or muons ($\rho_m$).
 Due to charge neutrality, the density of the leptons is the same as the density of 
protons, i.e., $\rho_l=\rho_p$.
If the density regime allows the electron chemical potential to be larger than the 
mass of muons, $\mu_e \geq m_\mu$, the  appearance of muons in the system is energetically favorable.
In this case, the $\beta$-stability condition is given by 
\begin{equation}
 \mu_n - \mu_p = \mu_e =\mu_\mu
\end{equation}
and the charge neutrality establishes that $\rho_p=\rho_e+\rho_\mu$, 
where $\mu_\mu$ and $\rho_\mu$ are the muon chemical potential and muon density, respectively. 
 
The chemical potential of each type of leptons is defined as 
\begin{equation}\label{mu_lepton}
 \mu_l = \sqrt{k_{Fl}^2 + m_l^2}
\end{equation}
and their the pressure is given either by 
\begin{equation}\label{eq:P_lepton}
 P_l (\rho, \delta) = \rho_l^2 \frac{\partial E_{l} (\rho, \delta)}{\partial \rho_l} \hspace{1cm}
\mathrm{or~ by} \hspace{1cm}
 P_l (\rho, \delta)= \mu_l \rho_l-\mathcal{H}_l(\rho, \delta),
\end{equation}
where $E_{l} (\rho, \delta)$ is the energy per particle of each type of leptons $E_{l} (\rho, \delta)= \mathcal{H}_l (\rho, \delta)/\rho_l$.

Using the definitions of the baryon chemical potentials placed in Eqs.~(\ref{chempot}) and expressed in 
terms of the density and asymmetry of the system, the 
$\beta$-equilibrium condition can be rewritten as
\begin{eqnarray}\label{betamatter-full}
\mu_n - \mu_p=2 \frac{\partial E_b(\rho,\delta)}{\partial \delta} = \mu_e=\mu_\mu,
\end{eqnarray}
where $E_b(\rho,\delta)$ is the baryon energy per particle.
If one uses the Taylor expansion of the EoS [cf. Eq.~(\ref{eq:EOSexpgeneral})] instead of its exact expression, 
the $\beta$-equilibrium condition is also expressed in terms of the symmetry energy coefficients, and reads 
 \begin{eqnarray}\label{betamatter}
 \mu_n - \mu_p &=& 2 \frac{\partial E_b (\rho, \delta)}{\partial \delta} = 4 \delta E_{\mathrm{sym}, 2}(\rho) + 8 \delta^3 E_{\mathrm{sym}, 4} (\rho)
 + 12 \delta^5 E_{\mathrm{sym}, 6}(\rho) + 16 \delta^7 E_{\mathrm{sym}, 8} (\rho)\nonumber\\
 &+& 20\delta^9 E_{\mathrm{sym},10}(\rho) +\mathcal{O}(\delta^{11}) =\mu_e=\mu_\mu.  
 \end{eqnarray}
Moreover, if one uses the parabolic expression to calculate the energy per particle, the 
$\beta$-equilibrium condition takes the form
\begin{eqnarray}\label{betamatter-PA}
  \mu_{n} -\mu_p &=& 4 \delta E_{\mathrm{sym}}^{PA}(\rho) = \mu_e = \mu_\mu  .
\end{eqnarray}

\begin{figure}[t]
\centering     
\subfigure{\label{fig:delta_skyrme}\includegraphics[width=0.49\linewidth]{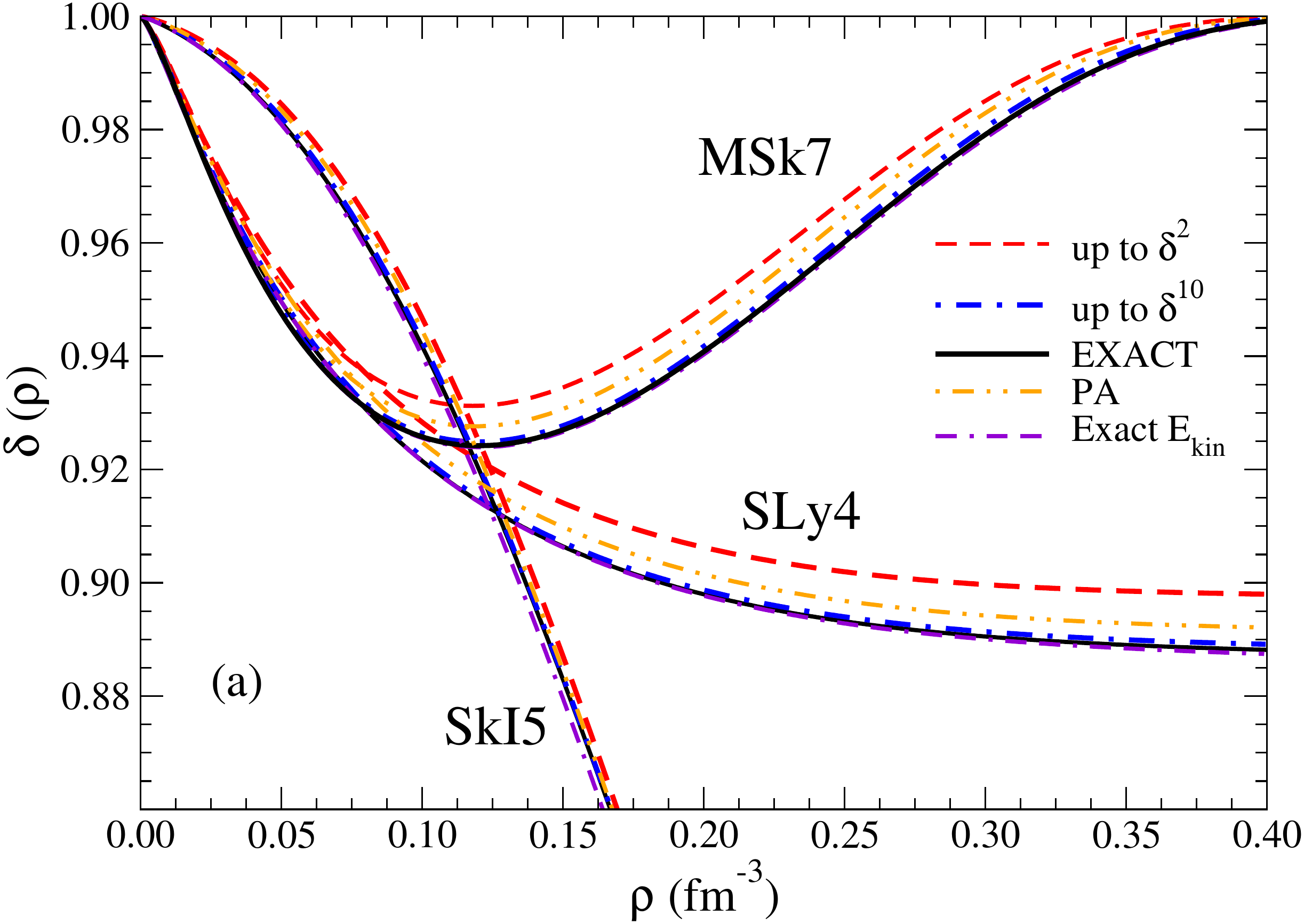}}
\subfigure{\label{fig:delta_gogny}\includegraphics[width=0.49 \linewidth]{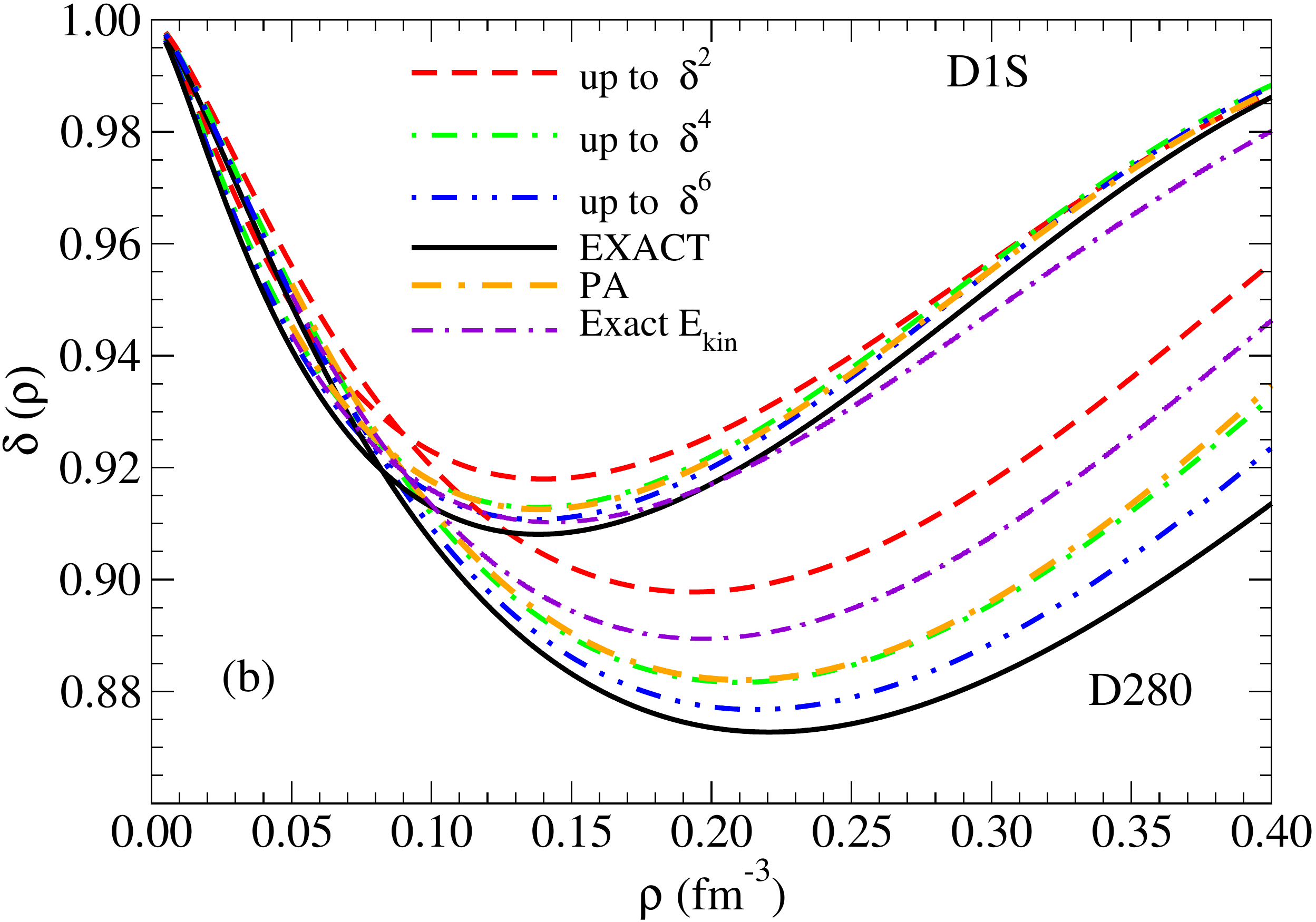}}
\caption{Panel a: Density dependence of the asymmetry in $\beta$-stable matter calculated using the
 exact expression of the energy per particle or the expression in Eq.~(\ref{eq:EOSexpgeneral}) up to second, fourth
  and tenth order for the Skyrme forces MSk7, SLy4 and SkI5.
  Panel b: Density dependence of the isospin asymmetry in $\beta$-stable matter calculated using the exact 
  expression of the EoS or the expression in Eq.~(\ref{eq:EOSexpgeneral})
  up to second, fourth, and sixth order for the D1S and D280 interactions. The results of the parabolic 
  approximation are also included in both panels, as well as the results using the second-order expansion of 
  for the potential part of the EoS up to second-order and 
  the full expression for the kinetic energy (label ``Exact E$_\mathrm{kin}$'').\label{fig:delta}}
\end{figure}

We display in Fig.~\ref{fig:delta_skyrme} the isospin asymmetry $\delta$ corresponding to the 
$\beta$-equilibrium of {\it npe} matter as a function of the density calculated
with the expansion of the EoS up to second- and tenth-order
for three representative Skyrme interactions, MSk7, with a very soft symmetry energy 
of slope $L=9.41$ MeV, SLy4, with with an intermediate slope of $L=45.96$ MeV and SkI5, with a stiff EoS of $L=129.33$ MeV.
Moreover, we plot in Fig.~\ref{fig:delta_gogny}~\cite{gonzalez17} the density dependence of the isospin asymmetry $\delta$
for two illustrative Gogny interactions, namely D1S ($L=18.36$ MeV) and D280 ($L=46.53$ MeV), if the density per particle is expanded up 
to second-, fourth- and sixth-order. 
A general trend for both types of functionals is that if one takes into account more terms in the expansion of the energy per particle,
the results are closer to the ones obtained using the exact 
EoS.
When one calculates the $\beta$-stability condition at second order in the case of Skyrme interactions,
one finds that the results for the isospin asymmetry $\delta$ calculated with the MSk7, SLy4, and SkI5 interactions do not exceed a
$5\%$ of relative differences with respect to the exact results. On the other hand, for interactions that find larger
relative differences, the second-order Taylor expansion is not enough to reproduce the exact values, and terms of 
a higher order than two are needed. 
In the case of Gogny interactions, we present the results for the D1S force
that has a low slope parameter $L=22.4$~MeV and the results for D280
that has $L=46.5$~MeV, the largest $L$ value of the analyzed Gogny forces.
For Gogny interactions, there is a trend of having an overall larger isospin 
asymmetry  at densities above $\sim 0.1$ fm$^{-3}$ for models with softer symmetry energies, 
that is, the system is more neutron-rich for these interactions. We can see this in Fig.~\ref{fig:delta_gogny},
where the isospin asymmetry of D1S is larger than the one of D280. This is 
in 
consonance with the fact that for the same density range the symmetry energy of D1S is smaller than in D280, 
as can be seen in Fig.~\ref{fig:esym2gog}.
For both D1S and D280 interactions, the results obtained when using the EoS up to second order are quite far 
from the exact results, hence, higher-order coefficients are needed to lessen these differences. Still, 
even using (\ref{eq:EOSexpgeneral}) up to sixth-order, the isospin asymmetries found with these interactions are not in line with 
the values obtained with the exact EoS.
The convergence of the symmetry parameter corresponding to $\beta$-equilibrium obtained starting from the 
expansion~(\ref{eq:EOSexpgeneral}) is also model dependent. 
 
In both panels of Fig.~\ref{fig:delta} we have also added the results obtained using the parabolic 
approximation~(\ref{eq:PAgen}) for the EoS.
It is interesting to note that they are significantly different from those obtained in the second-order 
approximation. 
In fact, for the functionals under consideration, the PA asymmetries are overall closer to the exact asymmetries 
than the second-order values. 

\begin{figure}[t]
\centering     
\subfigure{\label{fig:press_skyrme}\includegraphics[width=0.49\linewidth]{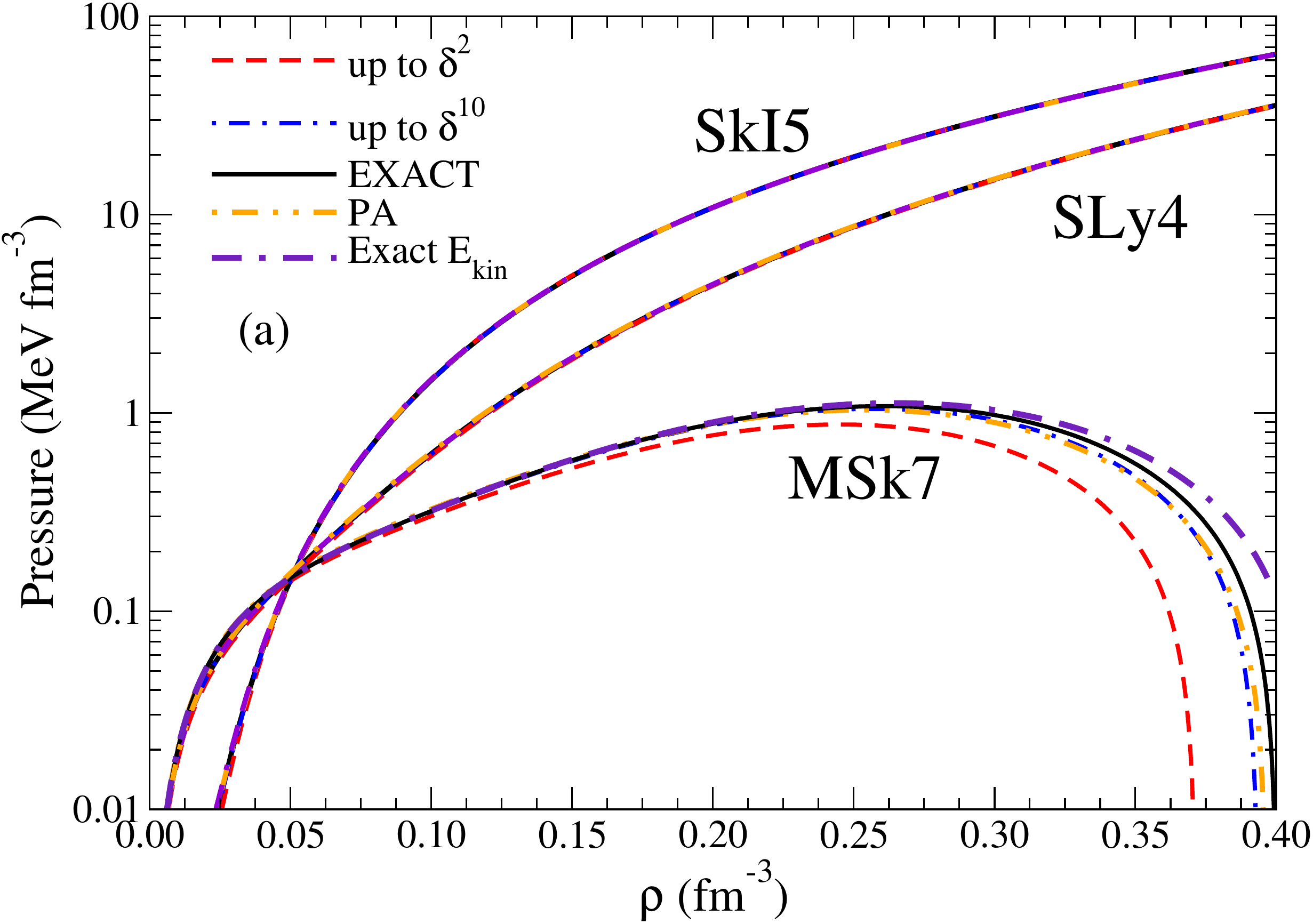}}
\subfigure{\label{fig:press_gogny}\includegraphics[width=0.49 \linewidth]{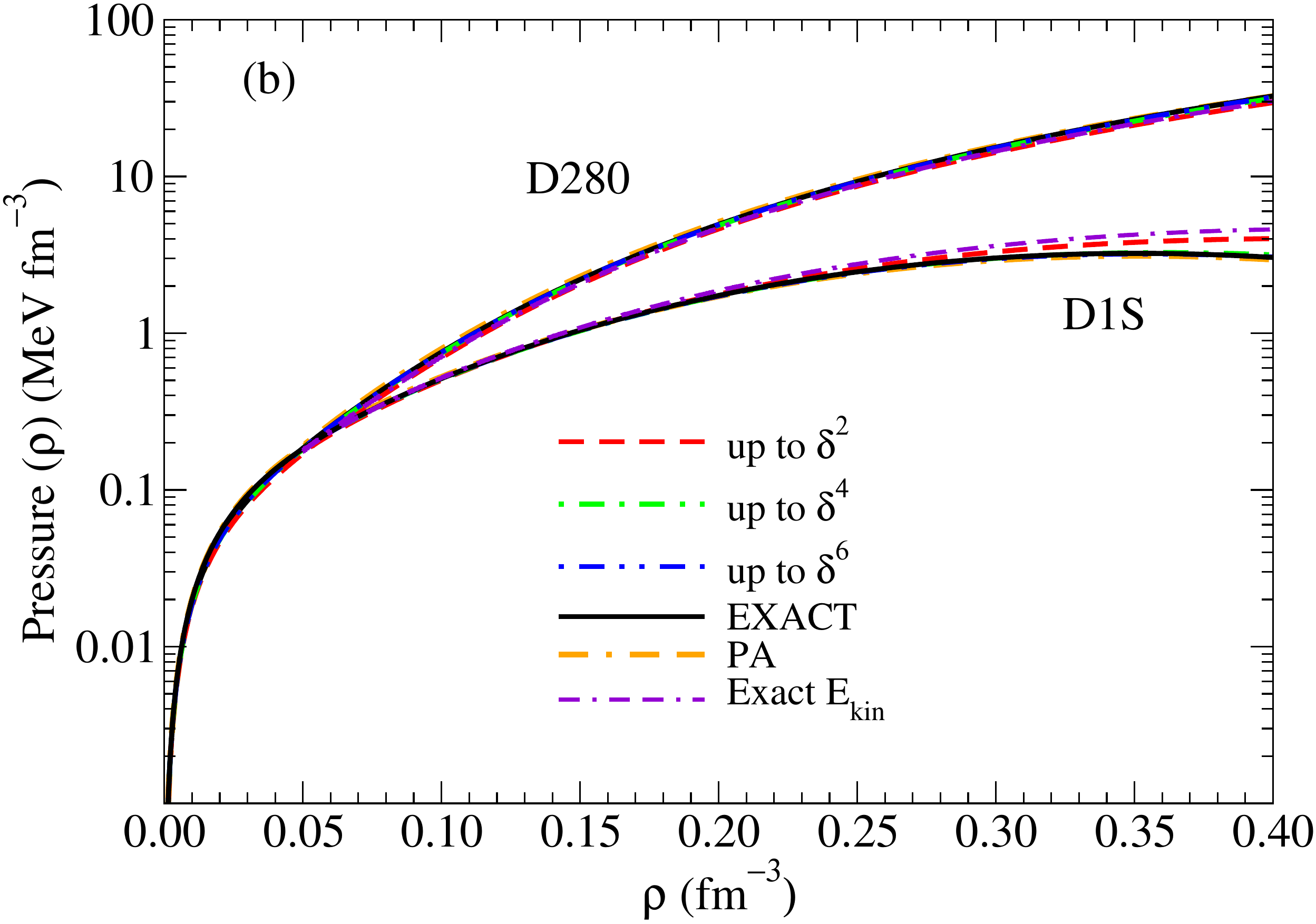}}
\caption{Panel a: Density dependence of the pressure in $\beta$-stable matter calculated using the exact expression 
  of the energy per particle 
or the expression in Eq.~(\ref{eq:EOSexpgeneral}) up to second, fourth, 
 and tenth order for three Skyrme forces, MSk7, SLy4, and SkI5.
  Panel b: Density dependence of the pressure in $\beta$-stable matter calculated using the 
  exact expression of the EoS or the expression in Eq.~(\ref{eq:EOSexpgeneral})
  up to second, fourth, and sixth order for the D1S and D280 interactions. The results of the parabolic 
  approximation are also included in both panels, as well 
  as the results obtained if applying the Taylor expansion only in the potential part of the EoS and using the full expression for its
  kinetic part (label ``Exact E$_\mathrm{kin}$''). In both panels the vertical axis is in logarithmic scale.\label{fig:press}}
\end{figure}

We display in Fig.~\ref{fig:press_skyrme} the total pressure of the system [cf. Eq.~(\ref{eq:press_skyrme})] in logarithmic scale  
at $\beta$-equilibrium as a function of the density calculated for the same 
three Skyrme forces for which we have analyzed the density dependence of the isospin asymmetry, namely the MSk7, SLy4, and SkI5 forces. 
Moreover, the same results for the D1S and D280 Gogny interactions are shown in Fig.~\ref{fig:press_gogny}~\cite{gonzalez17}, 
whose pressure is given in Eq.~(\ref{eq:pressure_bars}).
We plot the results calculated using the exact EoS and
the pressure obtained starting from  
the expansion~(\ref{eq:EOSexpgeneral}) up to tenth order for Skyrme models and up to sixth order for the Gogny ones. We observe that,
as happens for the isospin asymmetry $\delta$, all 
results obtained with the expanded EoS are closer to the exact values when more terms are included in
the calculation.
At a density $\rho \sim 0.1$ fm$^{-3}$, the relative difference between the pressure obtained
with the second-order approximation and the one calculated with the full EoS using 
the SkI5 interaction is of $1\%$, while if the expansion is pushed up to the tenth-order in 
$\delta$, the relative difference is only of $0.05\%$. The other forces displayed in
Fig.~\ref{fig:press_skyrme} show a similar behavior. The values of the relative differences 
between the exact and approximated pressures computed with the SLy4 interaction are, 
 of $4\%$ and $0.07\%$ when the expansions are pushed up to second- and tenth-order, respectively,
and the same differences are $6\%$ and $0.6\%$ when the pressure is computed using the MSk7 force.
As for the Gogny interactions, the relative differences for the D1S force between the pressure calculated at second 
order and the pressure of the exact EoS are of $30 \%$ at the largest density ($0.4$ fm$^{-3}$) of the figure.
With the corrections up to sixth order included, the differences reduce to $1 \%$. For D280, these 
differences are of $10 \%$ and $1.5 \%$, respectively.
In all cases, adding more terms to the expansion brings the results closer to the pressure of the exact EoS.
The results for the pressure are in keeping with the pure neutron matter predictions of 
Ref.~\cite{Sellahewa14} 
and the $\beta$-stable calculations of Ref.~\cite{Loan2011}.
Moreover, the $\beta$-stable nuclear matter has also been studied using the PA, and the results are
plotted in the same Fig.~\ref{fig:press_skyrme} and Fig.~\ref{fig:press_gogny} for Skyrme and Gogny forces, respectively. 
In all cases, the PA clearly improves the results calculated up to quadratic 
terms in the energy per baryon expansion, 
providing values close to the ones estimated when adding up to the tenth- order in $\delta$ in the 
Taylor series~(\ref{eq:EOSexpgeneral}). 

Some time ago, it was proposed in Ref.~\cite{Ducoin11} an improvement of 
the $E_b(\rho, \delta)$ expansion, 
consisting of using the expansion in powers of the asymmetry $\delta$ only in the potential energy 
part up to $\delta^2$, 
while using the exact kinetic energy. 
The underlying reason for this approach is the following. Although the quadratic expansion of the
energy per particle in asymmetric nuclear matter is quite accurate to describe the EoS even at high 
isospin asymmetry, this 
expansion fails in reproducing the spinodal contour in neutron-rich matter~\cite{Ducoin11} 
because the energy 
density curvature in the proton density direction diverges at small values of the proton density owing 
the kinetic energy term 
\cite{Baldo09}. In addition, we have also seen that a simple quadratic expansion of the 
energy per particle is not enough to predict accurately the exact isospin asymmetry corresponding to 
$\beta$-stable NS 
matter in all the range of considered densities, at least for some of the considered interactions. 
We plot in Fig.~\ref{fig:delta} and in Fig.~\ref{fig:press}, respectively, the isospin asymmetry $\delta$ and the 
total pressure for Skyrme and Gogny interactions if we only expand up to second-order the potential part of the 
interaction and we keep the full expression for the kinetic part. We have labeled these results as ``Exact E$_\mathrm{kin}$''.
This approximation works very well for most of the interactions, pointing out that the majority of the differences 
between the results obtained with the Taylor expansion~(\ref{eq:EOSexpgeneral}) and the full expression of the EoS come 
from the kinetic contribution to the energy.

\section{Influence of the symmetry energy on neutron star bulk properties}\label{MRbeta}

With access to the analytical expressions for the pressure and the energy density 
in asymmetric matter, one can compute the mass-radius relation 
of NSs by integrating the Tolman-Oppenheimer-Volkoff~(TOV) equations~\cite{shapiro83,Glendenning2000,haensel07}, given by
\begin{eqnarray}
 \frac{dP (r)}{dr} &=& \frac{G}{r^2 c^2} \left[\epsilon (r) + P(r) \right] \left[ m(r) + 4 \pi r^3 P (r)\right] 
 \left[ 1-\frac{2Gm(r)}{rc^2}\right]^{-1}\label{eq:TOV}\\
 \frac{dm(r)}{dr} &=& 4 \pi r^2 \epsilon(r),\label{eq:TOV2}
\end{eqnarray}
where $\epsilon(r)$, $P(r)$ and $m(r)$ are, respectively, the energy density, pressure and mass at each radius $r$ inside the NS. 
One starts from the center of the star at a given central energy density $\epsilon (0)$, central pressure $P(0)$ and mass $m(0)=0$
and integrates outwards until reaching the surface of the star, where the pressure is $P(R)=0$. The location of the 
surface of the NS will determine the total radius of the star $R$ and its total mass $M=m(R)$.

We have solved the TOV equations for a set of Skyrme and Gogny~\cite{SellahewaPhD, gonzalez17} forces using the $\beta$-equilibrium 
EoS with the exact isospin asymmetry dependence in the NS core.
Note that for interactions with very soft symmetry energies, at high densities these conditions may yield a pure NS with $\delta=1$, and we ignore the effects of an 
isospin instability at and beyond that point. At very low densities, we use the Haensel-Pichon EoS 
for the outer crust \cite{douchin01}. 

In the absence of microscopic calculations of the EoS of the inner crust for many Skyrme interactions
and for Gogny forces, 
we adopt the prescription of previous works \cite{Link1999, carriere03,xu09a,Zhang15, gonzalez17} by taking 
the EoS of the inner crust to be of the polytropic form 
\begin{equation}
 P=a+b\epsilon^{4/3},
\end{equation}
where $\epsilon$ 
denotes the mass-energy density.
The constants $a$ and $b$ are adjusted by demanding continuity at the inner-outer crust interface 
and at the core-crust transition point~\cite{carriere03,xu09a,Zhang15, gonzalez17}:
\begin{eqnarray}
 a&=& \frac{P_\mathrm{out} \epsilon_t^{4/3}-P_t \epsilon_\mathrm{out}^{4/3}}{\epsilon_t^{4/3}-\epsilon_\mathrm{out}^{4/3}}
\label{eq:coefa}\\
b&=& \frac{P_t -P_\mathrm{out}}{\epsilon_t^{4/3}-\epsilon_\mathrm{out}^{4/3}},\label{eq:coefb}
 \end{eqnarray}
 where $P_\mathrm{out}$ ($P_t$) and $\epsilon_\mathrm{out}$ ($\epsilon_t$) are the corresponding
 pressure and energy density at the outer crust-inner crust (core-crust) transition.
 We have calculated the transition density using the thermodynamical method,
 which will be explained later on in Chapter~\ref{chapter4}~\cite{gonzalez17}. 
At the subsaturation densities of the inner crust, the pressure of matter is dominated by the 
relativistic degenerate electrons. Hence, a polytropic form with an index of average value of about $4/3$ is found
to be a good approximation to the EoS in this region \cite{Link1999,Lattimer01,Lattimer2016}.
For more accurate predictions of the crustal properties, it would be of great interest to determine the 
microscopic EoS of the crust with the same interaction used for describing the core \cite{Than2011}.

\begin{figure}[!b]
\centering     
\subfigure{\label{fig:MR_skyrme}\includegraphics[width=0.49\linewidth]{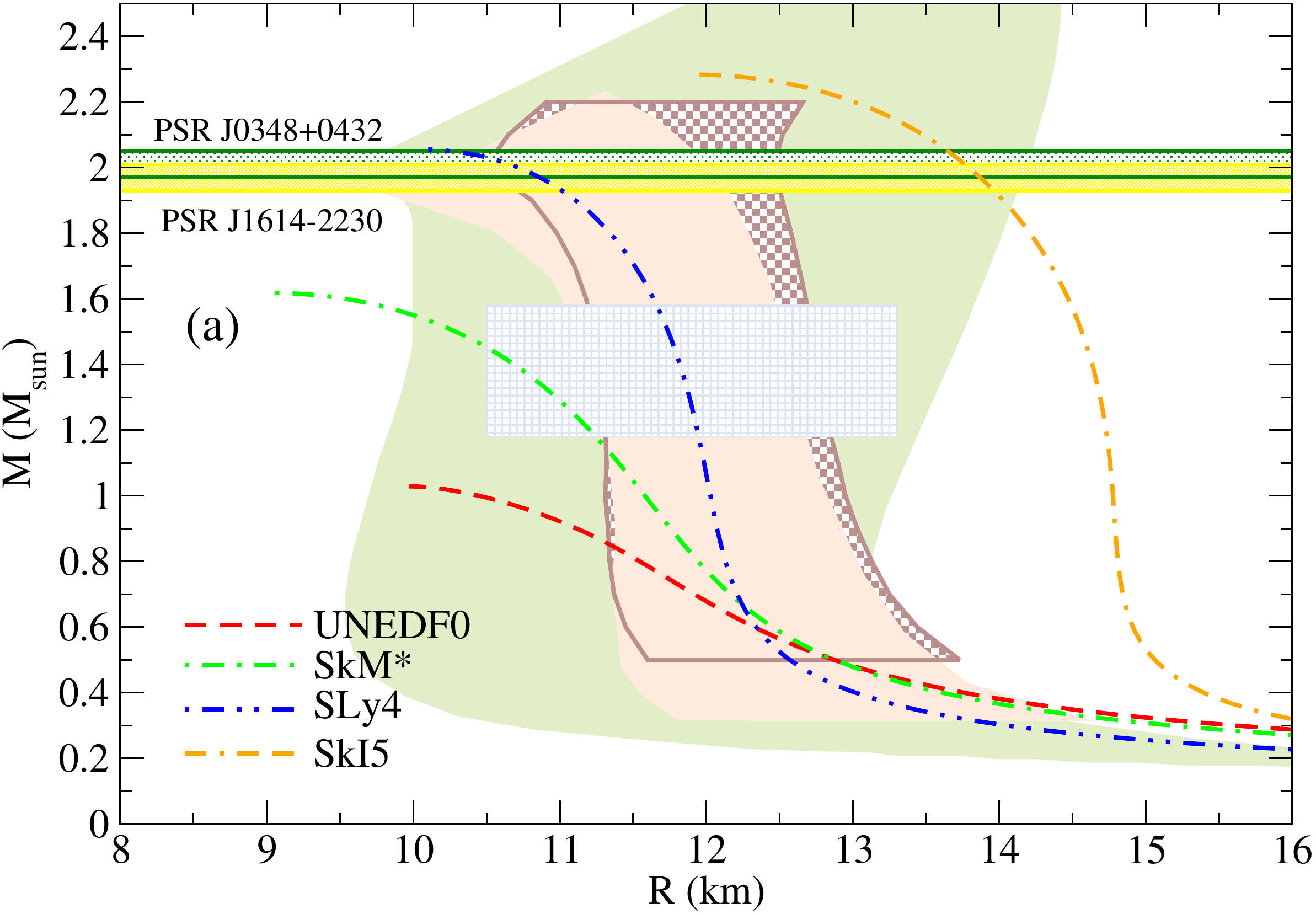}}
\subfigure{\label{fig:MR_gogny}\includegraphics[width=0.49 \linewidth]{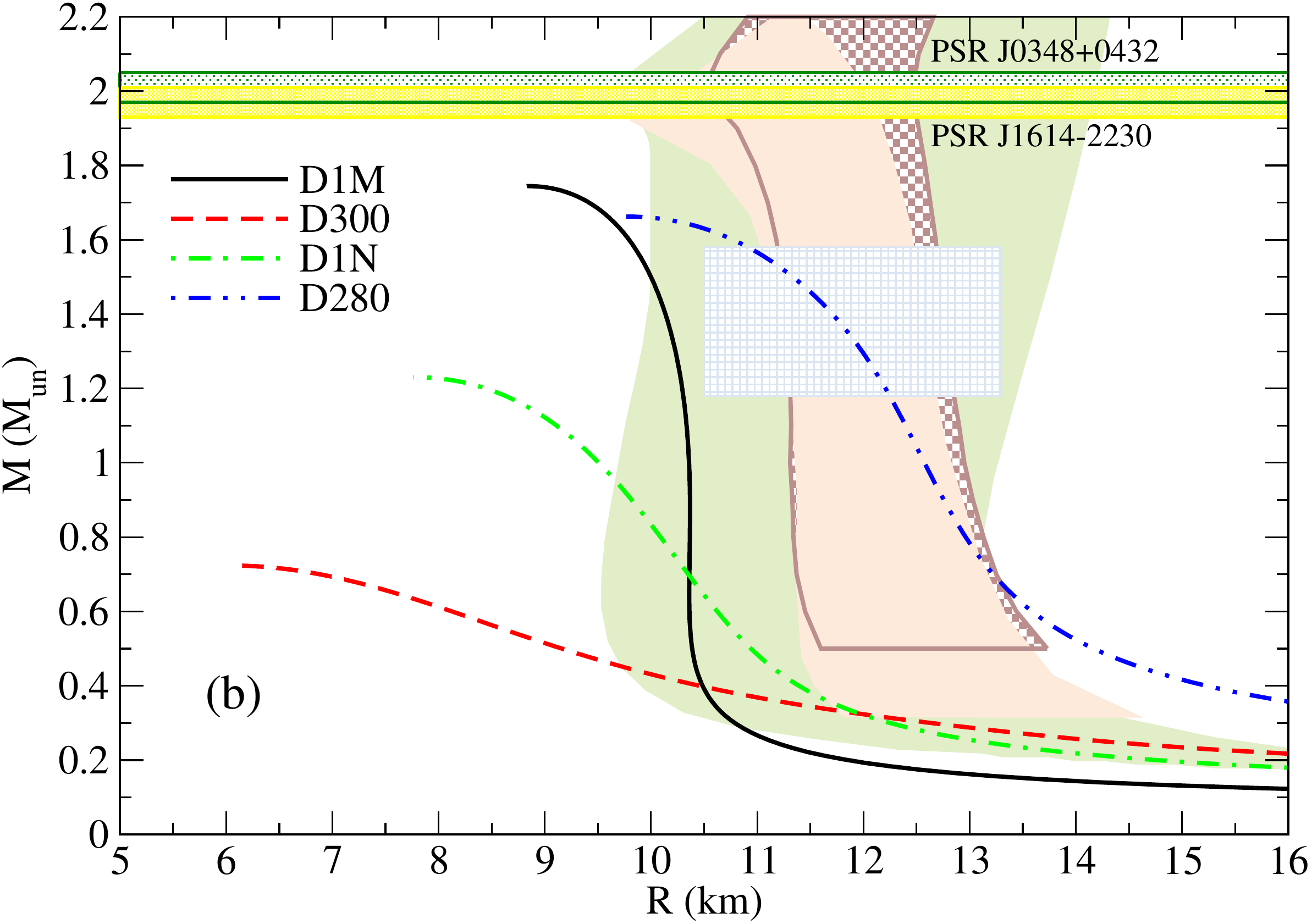}}
\caption{Mass-radius relation for the NSs for 4 stable Skyrme (panel a) and 4 Gogny (panel b) functionals. 
We show, with horizontal bands, the accurate 
  $M \approx 2 M_\odot$ mass measurements of highly massive NS~\cite{Demorest10,Antoniadis13}.
The vertical green band shows the \mbox{M-R} region deduced from chiral nuclear interactions up to normal density 
plus astrophysically constrained high-density EoS extrapolations \cite{Hebeler13}. 
The brown dotted band is the zone constrained by the cooling tails of type-I X-ray bursts 
in three low-mass X-ray binaries and a Bayesian analysis \cite{Nattila16}, and the beige 
constraint at the front is from five quiescent low-mass X-ray binaries and five photospheric
radius expansion X-ray bursters after a Bayesian analysis~\cite{Lattimer14}.
Finally, the squared blue band accounts for 
a Bayesian analysis of the data coming from the GW170817 detection of gravitational waves from a binary NS merger~\cite{Abbott2018}.
  \label{fig:MR}
  }
\end{figure}

The results for the mass-radius relation computed with 
Skyrme and Gogny interactions are presented in Fig.~\ref{fig:MR}. 
First, we compute the mass-radius relation for the same representative Skyrme models, and 
the results obtained are plotted in Fig.~\ref{fig:MR_skyrme}. 
We observe that only four of the five selected Skyrme interactions, namely UNEDF0, SkM$^*$,
SLy4 and SkI5 provide a stable solution 
of the TOV equations. Moreover, of these ones, only SLy4 and SkI5 can produce an NS
with a mass higher than the lower bound $M = 2 M_\odot$, given by observations of 
highly massive NSs~\cite{Demorest10,Antoniadis13}, and the UNEDF0 interaction cannot
converge to an NS of a canonical mass of $M = 1.4 M_\odot$. In order to study the radii 
obtained with Skyrme interactions, we add in Fig.~\ref{fig:MR_skyrme} boundary conditions for the NS radius coming from different analyses
\cite{Hebeler13, Nattila16, Lattimer14, Abbott2018}. 
From the plot, we see that the SLy4 Skyrme interaction, 
which gives  a radius of $\sim 10$ km for an NS with maximum mass and $\sim 11.8$ km for a 
canonical NS of mass $1.4 M_\odot$, is the only force that 
fits inside all constraints for the radii coming from low-mass X-ray binaries, X-ray burst sources, and gravitational waves,
which provide radii below 13 km for canonical 
mass stars~\cite{Nattila16, Lattimer14}. 
The good behaviour of SLy4 can be understood knowing that it was fitted to both properties of 
finite nuclei and neutron matter~\cite{douchin01}.
Other recent extractions of stellar 
radii from quiescent low-mass x-ray binaries and x-ray burst sources have suggested values in 
the range of $9-13$ km \cite{Guillot14,Heinke14,Ozel15,Ozel16}, which would include also the 
results for SkM$^*$.

\begin{table}[!t]
\resizebox{\columnwidth}{!}{%
\begin{tabular}{ccccccccc}
\hline 
\multirow{2}{*}{Force} & $L$    & $M_\mathrm{max}$ & $R(M_\mathrm{max})$ & $\rho_c(M_\mathrm{max})$ & $\epsilon_c(M_\mathrm{max})$ & $R(1.4M_\odot)$ & $\rho_c(1.4M_\odot)$ & $\epsilon_c(1.4M_\odot)$ \\
                       & (MeV)  & ($M_\odot$)      & (km)                & (fm$^{-3}$)              & ($10^{15}$ g cm$^{-3}$)         & (km)            & (fm$^{-3}$)          & ($10^{15}$ g cm$^{-3}$)                                                                       \\\hline \hline
MSk7                   & 9.41   & ---              & ---                 & ---                      & ---                             & ---             & ---                  & ---                                                                                           \\
SIII                   & 9.91   & 1.185            & 5.68                & 3.68                     & 8.81                            & ---             & ---                  & ---                                                                                           \\
SkP                    & 19.68  & ---              & ---                 & ---                      & ---                             & ---             & ----                 & ---                                                                                           \\
HFB-27                 & 28.50  & 1.530            & 7.91                & 2.04                     & 4.81                            & 9.00            & 1.29                 & 2.57                                                                                          \\
SKX                    & 33.19  & 1.396            & 7.99                & 2.16                     & 4.97                            & ---             & ----                 & ---                                                                                           \\
HFB-17                 & 36.29  & 1.767            & 8.96               & 1.56                     & 3.68                            & 10.65           & 0.77                 & 1.46                                                                                          \\
SGII                   & 37.63  & 1.663            & 8.88                & 1.65                     & 3.82                            & 10.49           & 0.84                 & 1.61                                                                                          \\
UNEDF1                 & 40.01  & 1.157            & 8.90                & 1.89                     & 3.94                            & ---             & ---                  & ---                                                                                           \\
Sk$\chi$500             & 40.74  & 2.142            & 10.47               & 1.09                     & 2.52                            & 11.73           & 0.49                 & 0.88                                                                                          \\
Sk$\chi$450             & 42.01  & 2.098            & 10.05               & 1.18                     & 2.80                            & 11.82           & 0.53                 & 0.97                                                                                          \\
UNEDF0                 & 45.08  & 1.029            & 9.97                & 1.47                     & 2.89                            & ---             & ---                  & ---                                                                                           \\
SkM*                   & 45.78  & 1.618            & 9.01                & 1.66                     & 3.81                            & 10.69           & 0.86                 & 1.65                                                                                          \\
SLy4                   & 45.96  & 2.056             & 10.03               & 1.20                     & 2.84                            & 11.82           & 0.53                 & 0.98                                                                                          \\
SLy7                   & 47.22  & 2.080            & 10.16               & 1.17                     & 2.77                            & 11.97           & 0.52                 & 0.94                                                                                          \\
SLy5                   & 48.27  & 2.060            & 10.09               & 1.19                     & 2.81                            & 11.91           & 0.53                 & 0.96                                                                                          \\
Sk$\chi$414             & 51.92  & 2.110            & 10.37               & 1.13                    & 2.64                            & 12.01           & 0.50                 & 0.91                                                                                          \\
MSka                   & 57.17  & 2.320            & 11.23               & 0.95                     & 2.24                            & 13.14           & 0.39                 & 0.70                                                                                          \\
MSL0                   & 60.00  & 1.955            & 10.15               & 1.24                     & 2.87                            & 12.21           & 0.53                 & 0.97                                                                                          \\
SIV                    & 63.50  & 2.380            & 11.60               & 0.90                     & 2.11                            & 13.61           & 0.36                 & 0.65                                                                                          \\
SkMP                   & 70.31  & 2.120            & 10.72               & 1.09                     & 2.54                            & 12.94           & 0.44                 & 0.79                                                                                          \\
SKa                    & 74.62  & 2.217            & 11.07               & 1.01                     & 2.38                            & 13.44           & 0.40                 & 0.73                                                                                          \\
R$_\sigma$             & 85.69  & 2.140            & 11.01               & 1.04                     & 2.42                            & 13.40           & 0.40                 & 0.72                                                                                          \\
G$\sigma$              & 94.01  & 2.151            & 11.14               & 1.02                     & 2.38                            & 13.62           & 0.38                 & 0.69                                                                                          \\
SV                     & 96.09  & 2.452            & 11.97               & 0.85                     & 1.99                            & 14.24           & 0.32                 & 0.57                                                                                          \\
SkI2                   & 104.33 & 2.201            & 11.46               & 0.97                     & 2.24                            & 13.96           & 0.35                 & 0.63                                                                                          \\
SkI5                   & 129.33 & 2.283            & 11.89               & 0.90                     & 2.09                            & 14.63           & 0.31                 & 0.55                                                                                          \\
\hline 
D260                   & 17.57  & ---              & ---                 & ---                      & ---                            & ---             & ---                  & ---                                                                                           \\
D1                     & 18.36  & ---              & ---                 & ---                      & ---                            & ---             & ---                  & ---                                                                                           \\
D1S                    & 22.43  & ---              & ---                 & ---                      & ---                           & ---             & ---                  & ---                                                                                           \\
D1M                    & 24.83  & 1.745            & 8.84                & 1.58                     & 3.65                            & 10.14           & 0.80                 & 1.51                                                                                          \\
D250                   & 24.90  & ---              & ---                 & ---                      & ---                            & ---             & ---                  & ---                                                                                           \\
D300                   & 25.84  & 0.724            & 6.04                & 4.71                     & 10.07                           & ---             & ---                  & ---                                                                                           \\
D1N                    & 33.58  & 1.230            & 7.77                & 2.38                     & 5.27                            & ---             & ---                  & ---                                                                                           \\
D280                   & 46.53  & 1.662            & 9.78                & 1.45                     & 3.27                            & 11.71           & 0.69                 & 1.30											\\ \hline                                                                                         
\end{tabular}
}
\caption{Properties (mass $M$, radius $R$, central density $\rho_c$ and central mass-energy density $\epsilon_c$) 
for the maximum mass and the $1.4 M_\odot$ configurations of NSs predicted by different Skyrme and Gogny functionals.
\label{Table-NSs}}
\end{table}

Fig.~\ref{fig:MR_gogny} contains the results of the mass versus radius only for 
the four Gogny functionals (of the ones we have studied) that provide 
solutions for NSs. 
Also, we see that all Gogny EoSs predict maximum NS masses that are well below 
the observational limit of $M \approx 2 M_\odot$ from Refs.~\cite{Demorest10,Antoniadis13}. 
As a matter of fact, only D1M and D280 are able to generate masses above the canonical $1.4 M_\odot$ value. 
The NS radii from these two EoSs are considerably different, however, with D1M producing 
stars with radii $R\approx 9-10.5$ km, and D280 stars with radii $R \approx 10-12$ km. 
These small radii for a canonical NS would be in line with recent extractions of stellar 
radii from quiescent low-mass X-ray binaries, X-ray burst sources and gravitational waves, that have suggested values in 
the range of $9-13$ km \cite{Hebeler13,Lattimer14, Guillot14,Heinke14,Ozel15,Lattimer2016,Ozel16,Nattila16}. 
It appears that a certain degree of softness of the nuclear symmetry energy is necessary in order 
to reproduce small radii for a canonical mass NS~\cite{Chen15,Jiang15,Tolos16}. 
The parameterizations D1N and D300, in contrast to D1M and D280, generate NSs which are
unrealistically small in terms of both mass and radius. 
One should of course be cautious in 
interpreting these results. Most Skyrme and Gogny forces have not been fitted to reproduce high-density, neutron-rich 
systems and it is not surprising that some parametrizations do not yield realistic NSs. 
One could presumably improve these results by guaranteeing that, 
at least around the saturation region, 
the pressure of neutron-rich matter is compatible with NS observations \cite{Lattimer2013}. 

We provide in Table~\ref{Table-NSs} data on the 
maximum mass and $1.4 M_\odot$ configurations 
of NSs produced by the different sets of Skyrme and Gogny models. The majority of
maximum mass 
configurations are reached at central baryon number densities close to $\sim 7-8 \rho_0$ for 
Skyrme interactions and $\sim 10 \rho_0$ for Gogny forces, 
whereas $1.4 M_\odot$ NSs have central baryon densities close to around $2-3 \rho_0$
and $4-5 \rho_0$ for Skyrme and Gogny forces, respectively. 
These large central density values  for Gogny are in keeping with the fact that the neutron matter Gogny 
EoSs are relatively soft, which requires larger central densities to produce realistic NSs.

\begin{figure}[!t]
\centering     
\subfigure{\label{fig:MR_skyrme_orders}\includegraphics[width=0.49\linewidth]{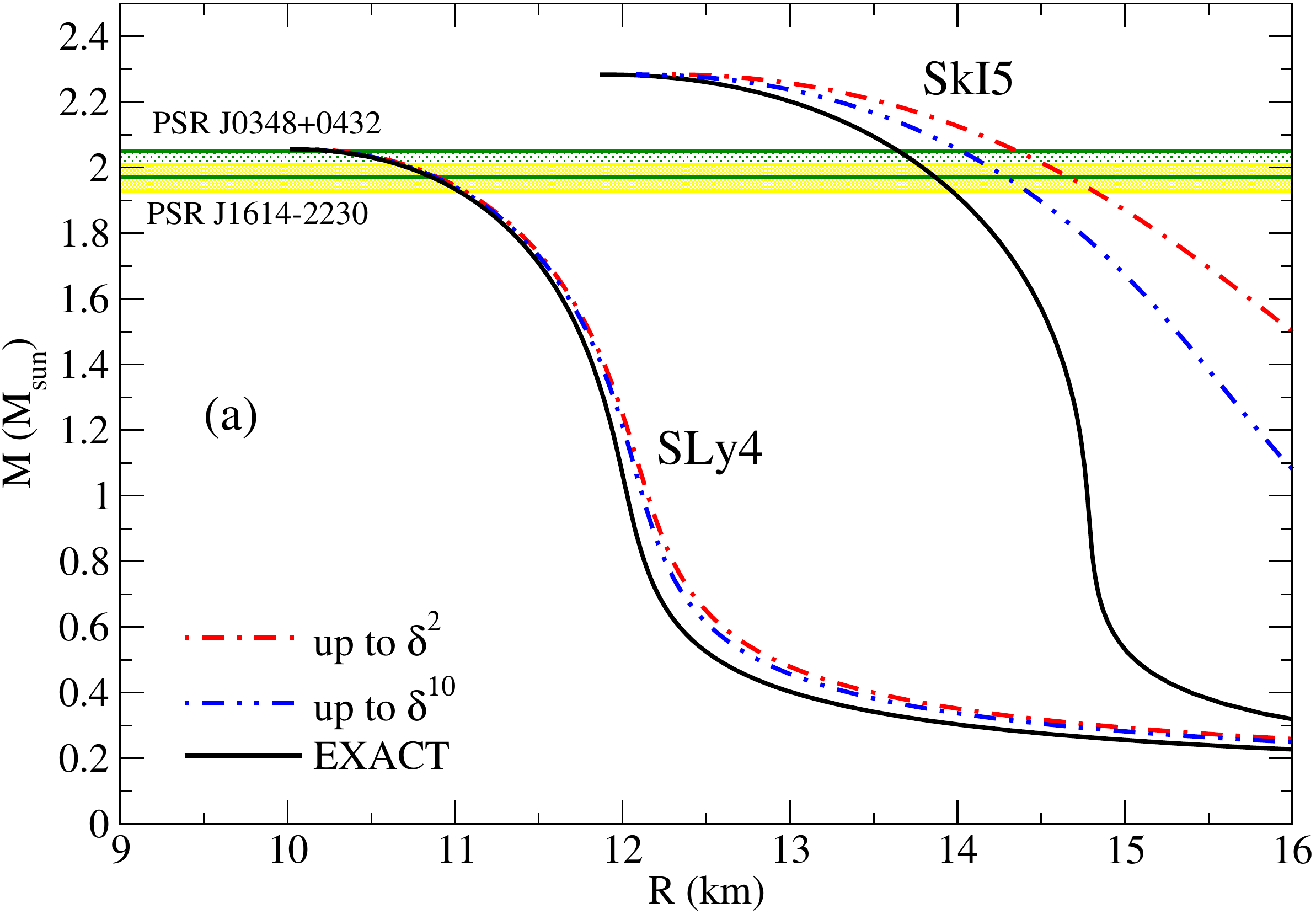}}
\subfigure{\label{fig:MR_gogny_orders}\includegraphics[width=0.49 \linewidth]{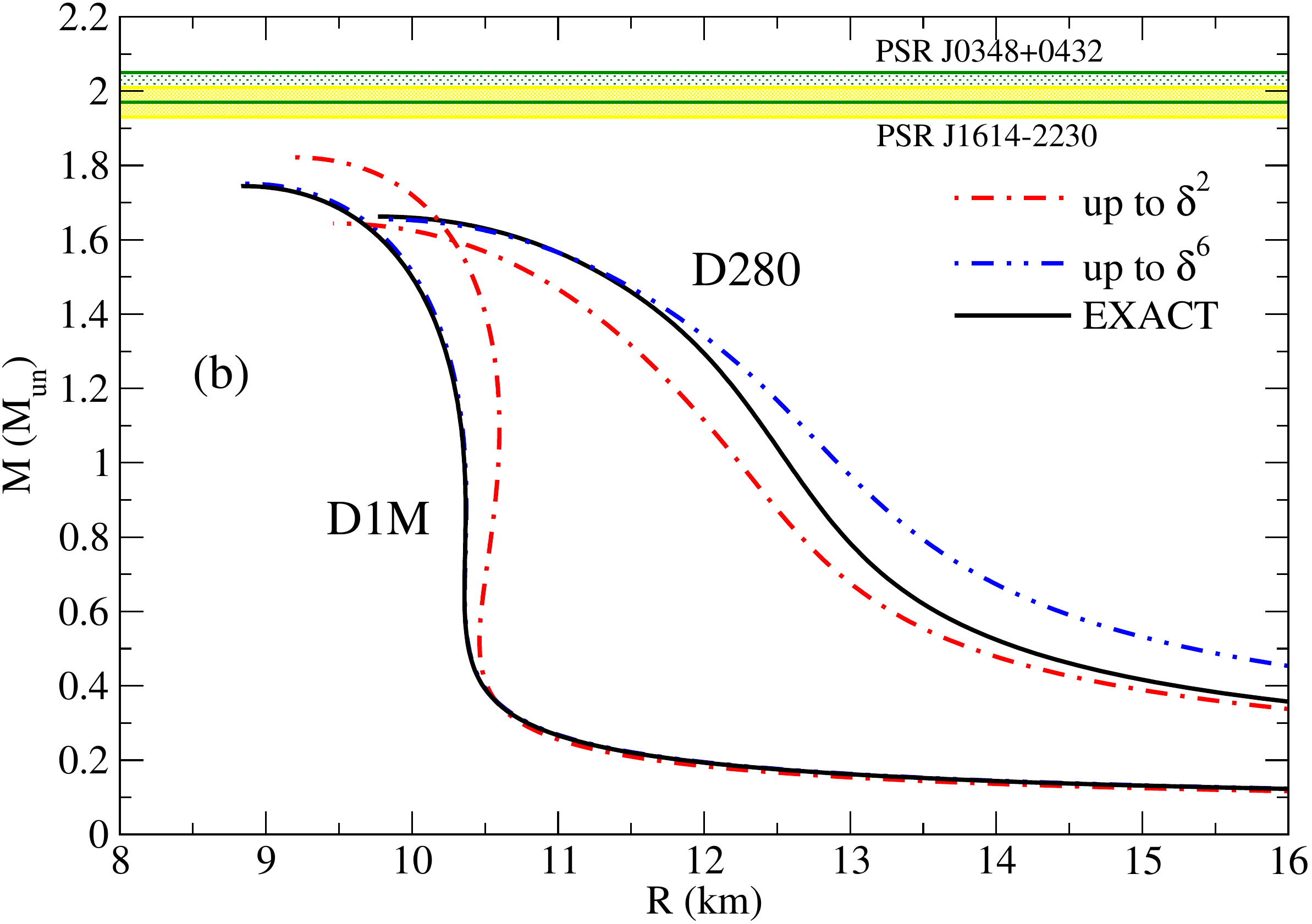}}
\caption{Mass-radius relation computed with two Skyrme (panel a) and two Gogny (panel b) forces. The results have been 
obtained using the EoS computed through its Taylor expansion up to second- and up to tenth- (Skyrme) or 
up to sixth- (Gogny) order. Moreover, the results calculated with the full EoS are also included. 
We show the constraints for the maximum mass of $2 M_\odot$ from the mass measurements of Refs.~\cite{Demorest10, Antoniadis13}.
  \label{fig:MRorders}}
\end{figure}

To analyze the impact on the expansion of the EoS of higher-order terms in its Taylor expansion, 
we plot in Fig.~\ref{fig:MRorders} the mass-radius relation for the SLy4 and SkI5 Skyrme interactions [see Fig.~\ref{fig:MR_skyrme_orders}]
and for the D1M and D280 Gogny forces [see Fig.~\ref{fig:MR_gogny_orders}].
In black solid lines, we have plotted the results when the full EoS of each type of interaction is used. These values
are the same as the ones in Fig.~\ref{fig:MR}. To test the accuracy of the Taylor expansion of the EoS, 
we first compute the TOV equations when the EoSs are expanded up to second order, and the mass-radius 
relation is plotted in the same Fig.~\ref{fig:MRorders} with red dashed-dotted lines. We also include 
the results if they are obtained using the expansion up to the highest order we have considered 
for each type of model. For Skyrme interactions, this is up to the tenth order and, for Gogny interactions,
it is up to the sixth order, and it is represented with blue dashed-double dotted lines.
In general, if we use the second-order EoS expansion, we get results that are relatively far from the ones
obtained using the full EoS, which has been labeled them as ``Exact'' in Fig.~\ref{fig:MRorders}.
The results using the approximated SLy4 EoS up to second-order are close to their corresponding exact mass-radius relation, 
only differing in the low-mass regime, giving higher values of the radii, and if the EoS up to tenth-order
is used, the results  are even closer. For the SLy4 interaction, 
we obtain relative differences in the radius of a canonical NS mass of $1.4 M_\odot$ of 
$0.6\%$ if we compare the second-order approach and the exact results and of $0.4\%$ if we compare the values
obtained with a tenth-order expansion of the EoS and the exact ones.
On the other hand, for the SkI5 interaction, the scenario is very different. 
If we plot the results using the EoS expanded even up to the tenth order, the results of the mass-radius relation
are quite different from the ones obtained using the full EoS, especially at the low-mass regime. In this case, the relative differences 
obtained between the approximated results for the mass and the exact ones are of $11\%$ if we consider a second-order
expansion and of $6\%$ if we consider a tenth-order approximation.
Such differences for low-mass NSs could be explained given the fact that the transition density 
given by the different EoS expansions may be quite different from the one obtained with the full EoS. 
This is more prominent for interactions with large values of the slope of the symmetry energy (see Chapter~\ref{chapter4}). 
Therefore, the polytropic equation of state used to reproduce the inner crust may bring some uncertainties given these
differences in the transition point. 
In the case of Gogny interactions, we see that the second-order EoS expansion gives a 
higher maximum mass for NSs at a higher radius for the case of the D1M interaction. Then, if instead of the second-order we consider
the sixth-order EoS expansion, the results are almost the same as the exact ones. For D1M we get 
relative differences for a canonical NS of $4\%$ if we use the EoS cut at second-order in asymmetry
and of $0.2\%$ if we cut the expansion at sixth-order. 
In the case of the D280 Gogny interaction, if we use the second-order 
approach, we get results for the mass-radius relation that are below the exact ones. On the other hand, 
using the sixth-order expansion, the mass-radius relation becomes very close to the exact one at high densities, 
while at low densities it gives higher values for the radii. 
The relative differences we get for the radius of a $1.4 M_\odot$ NS are of $4\%$ while 
comparing the EoS of second-order in $\delta$ and the exact results and of $0.8\%$ if comparing 
the values computed with the full EoS and with its sixth-order expansion. 

\chapter{D1M$^*$ and D1M$^{**}$ Gogny parametrizations}\label{chapter3}
In spite of their accurate description of ground-state properties of finite nuclei,
the extrapolation of Gogny interactions to the neutron star (NS) domain, as we have seen in Chapter~\ref{chapter2}, is not completely satisfactory. 
The successful Gogny D1S, D1N and D1M forces of the D1 family, which nicely reproduce the ground-state
properties of finite nuclei are unable to reach a maximal 
NS mass of 2$M_{\odot}$ as required by recent astronomical observations 
\cite{Demorest10,Antoniadis13}. Actually, only the D1M interaction predicts an NS mass 
above the canonical value of 1.4$M_{\odot}$ \cite{Sellahewa14, gonzalez17}.
As we have pointed out in the previous chapter, the basic underlying
problem lies in the fact that the symmetry energy, 
which determines the NS EoS and hence the maximal mass predicted by the model, 
is too soft in the high-density regime.

We have tried to reconcile this problem introducing two new Gogny 
interactions, which we call D1M$^*$\cite{gonzalez18} and D1M$^{**}$ \cite{gonzalez18a}, 
by reparametrizing one of 
the most recognized Gogny interactions, the  D1M force, to be able to explain the properties of 
NSs and, at the same time, preserving its 
successful predictions in the domain of finite nuclei.

\section{Fitting procedure of the new D1M$^*$ and D1M$^{**}$ Gogny interactions and properties of their EoSs}

To determine the new Gogny interactions D1M$^*$ and D1M$^{**}$, we have modified the values of the parameters that control  
the stiffness of the symmetry energy while retaining as much as possible the quality of D1M 
for the binding energies and charge radii of nuclei~\cite{gonzalez18, gonzalez18a}. 
The fitting procedure is similar to 
previous literature where families of Skyrme and RMF parametrizations were generated starting from accurate models, 
as for example the \mbox{SAMi-J}~\cite{Roca13}, KDE0-J~\cite{Agrawal05} or FSU-TAMU~\cite{Piekarewicz11,Fattoyev13} families.
The basic idea to obtain these  families is the following. Starting from a well calibrated and successful mean-field
model, one modifies the values of some parameters, which determine the symmetry energy, around their optimal values
retaining as much as possible values of the binding energies and radii of finite nuclei of the original model.

In our case, we readjust the eight parameters $W_i$, $B_i$, $H_i$, $M_i$ ($i=1,2$) of the 
finite-range part of the Gogny interaction (\ref{VGogny}),
while the other parameters, namely the ranges of the Gaussians, the zero-range part of the
interaction and the spin-orbit force, are kept fixed to the values of D1M.
The open parameters are constrained by requiring the same saturation density, energy per particle, incompressibility and effective mass 
in symmetric nuclear matter as in the original D1M force, and, in order to have a correct description of 
asymmetric nuclei, the same value of $E_{\rm sym}(0.1)$, i.e., the symmetry energy at density 0.1 fm$^{-3}$.
The last condition is based on the fact that the binding energies of finite nuclei constrain 
the symmetry energy at an average density of nuclei of about 0.1 fm$^{-3}$ more tightly than 
at the saturation density $\rho_0$ \cite{horowitz01a,Centelles09}.
To preserve the pairing properties in the $S=0$, $T=1$ channel,
we demand in the new forces the same value of D1M for the two combinations of parameters $W_i-B_i-H_i+M_i$ ($i=1,2$). 
Thus, we are able to obtain seven of the eight free parameters of D1M$^*$ and of D1M$^{**}$ as a function of a 
single parameter, which we choose to be $B_1$. This parameter is used to modify the slope $L$ of the symmetry energy at 
saturation and, therefore, the behavior of the neutron matter EoS above saturation, which in turn determines the 
maximum mass of NSs by solving the TOV equations [see Eqs.(\ref{eq:TOV}) and (\ref{eq:TOV2}) in Chapter~\ref{chapter2}]. 
We adjust $B_1$ so that the maximum NS mass predicted by
the D1M$^*$ force is 2$M_{\odot}$. With the same strategy, we have also fitted the 
D1M$^{**}$ force, but imposing a constraint on the maximum
NS mass of  1.91$M_{\odot}$, which is close to the lower limit, within the error 
bars, of the heaviest observed masses of NS at the date of the fitting
\cite{Demorest10,Antoniadis13}. 
Finally, in the case of D1M$^*$, we perform a small readjustment of the zero-range strength $t_3$ of about 1 MeV fm$^4$ to optimize 
the results for nuclear masses, which
induces a slight change in the values of the saturation properties of uniform matter~\cite{gonzalez18,gonzalez18a}.

\begin{table}[t!]
\centering
\begin{tabular}{c|rrrrr}
\hline
D1M      & \multicolumn{1}{c}{$W_i$} & \multicolumn{1}{c}{$B_i$} & \multicolumn{1}{c}{$H_i$} & \multicolumn{1}{c}{$M_i$} & \multicolumn{1}{c}{$\mu_i$}  \\ \hline\hline
$i$=1    & -12797.57  & 14048.85   & -15144.43  & 11963.81   & 0.50 \\
$i$=2    &    490.95  &  -752.27   &    675.12  &  -693.57   & 1.00  \\\hline
D1M$^*$     & \multicolumn{1}{c}{$W_i$} & \multicolumn{1}{c}{$B_i$} & \multicolumn{1}{c}{$H_i$} & \multicolumn{1}{c}{$M_i$} & \multicolumn{1}{c}{$\mu_i$}  \\ \hline\hline
$i$=1    & -17242.0144 & 19604.4056  & -20699.9856 & 16408.3344 & 0.50 \\
$i$=2    &    712.2732 &  -982.8150  &    905.6650 &  -878.0060 & 1.00 \\\hline
D1M$^{**}$     & \multicolumn{1}{c}{$W_i$} & \multicolumn{1}{c}{$B_i$} & \multicolumn{1}{c}{$H_i$} & \multicolumn{1}{c}{$M_i$} & \multicolumn{1}{c}{$\mu_i$}  \\ \hline\hline
$i$=1    & -15019.7922 & 16826.6278 & -17922.2078 & 14186.1122 & 0.50 \\
$i$=2    &    583.1680 &  -867.5425  &   790.3925 &  -785.7880 & 1.00 \\\hline
\end{tabular}
\caption{Parameters of the D1M, D1M$^*$ and D1M$^{**}$ Gogny interactions, where $W_i$, $B_i$, $H_i$ and
$M_i$ are in MeV and $\mu_i$ in fm. The coefficients $x_3=1$, $\alpha=1/3$ and $W_{LS}=115.36$ MeV fm$^5$ 
are the same in the three interactions, and $t_3$ has values of $t_3=1562.22$ MeV fm$^4$ for the Gogny D1M and D1M$^{**}$
 forces and $t_3=1561.22$ MeV fm$^4$ for the D1M$^*$ interaction.}
\label{param}
\end{table}
 
\begin{table}[t!]
\centering
\begin{tabular}{lccccccc}
\hline
         &$\rho_0$ & $E_0$ & $K_0$ & $m^*/m$   & $E_{\rm sym}(\rho_0)$ & $E_{\rm sym}(0.1)$ & $L$  \\
         & (fm$^{-3}$) &  (MeV)  &  (MeV) &        & (MeV) &(MeV) & (MeV)\\  \hline \hline
D1M$^*$  &       0.1650          & $-$16.06  & 225.4   &  0.746  & 30.25   & 23.82 & 43.18  \\
D1M$^{**}$  &    0.1647          & $-$16.02  & 225.0   &  0.746  &29.37   & 23.80 & 33.91  \\
D1M      &       0.1647          & $-$16.02  & 225.0   &  0.746  & 28.55   & 23.80 & 24.83  \\
D1N      &       0.1612          & $-$15.96  & 225.7   &  0.697  & 29.60   & 23.80 & 33.58 \\
D1S      &       0.1633          & $-$16.01  & 202.9   &  0.747  & 31.13   & 25.93 & 22.43 \\
D2       &       0.1628          & $-$16.00  & 209.3   &  0.738  & 31.13   & 24.32 & 44.85  \\
SLy4     &       0.1596          & $-$15.98  & 229.9   &  0.695  & 32.00   & 25.15 & 45.96 \\\hline
\end{tabular}
\caption{Nuclear matter properties predicted by the D1M$^{*}$, D1M$^{**}$, D1M, D1N, D1S and D2 Gogny
interactions and the SLy4 Skyrme force.}
\label{inm}
\end{table}
 
The parameters of the new forces D1M$^*$ and D1M$^{**}$ are collected in Table \ref{param} and several nuclear
matter properties predicted by these forces are collected in Table \ref{inm} \cite{gonzalez18, gonzalez18a}. 
Table~\ref{inm} also collects the properties predicted by other Gogny forces, namely the D1M, D1N, D1S, and D2 parametrizations,
and by the Skyrme SLy4 interaction.
As stated previously, the SLy4 force \cite{sly42} is a Skyrme force specially designed to predict results in agreement with experimental
data of finite nuclei as well as with astronomical observations. We have used this force as a benchmark for comparison with the results 
obtained with the D1M$^*$ and D1M$^{**}$ models. 

We observe that the change of the finite-range parameters, as compared with the original
ones of the D1M force, is larger for the D1M$^*$ force than for the D1M$^{**}$ interaction because the variation in the
isovector sector is more important in the former than in the latter force. 
Though the change in the $W_i$, $B_i$, $H_i$, $M_i$ values is relatively large with respect to the D1M values \cite{goriely09}, 
the saturation properties of symmetric nuclear matter obtained with D1M$^*$ and D1M$^{**}$, {\it e.g.}
the saturation density $\rho_0$, the binding energy per nucleon at
saturation $E_0$, the incompressibility $K_0$ and the effective mass $m^*/m$, as well as the symmetry energy at 0.1~fm$^{-3}$  
are basically the same as in D1M (see Table~\ref{inm}). 
The mainly modified property is the density dependence of the symmetry energy, with a change in the slope from $L=24.83$ MeV 
to $L=43.18$ MeV for D1M$^*$ and to $L=33.91$ MeV for D1M$^{**}$, in order to provide a stiffer neutron matter 
EoS and limiting NS masses of $2 M_\odot$ and $1.91 M_\odot$, 
respectively.
The different $L$ value, as we fixed $E_{\rm sym}(0.1)$, implies that 
the symmetry energy $E_{\rm sym}(\rho_0)$ at saturation differs in D1M$^*$ and D1M$^{**}$ from D1M, but in a much less extent.
The D2 interaction, which differs from the other Gogny interactions used in this thesis for its finite-range density 
dependent term (see Section~\ref{Gogny} of Chapter~\ref{chapter1}), has a slope parameter of $L=44.85$ MeV. This value is fairly larger than 
the values predicted by the D1 family and close to $L$ obtained for D1M$^*$. 

\begin{figure}[t]
\centering
\includegraphics[width=0.75\columnwidth,clip=true]{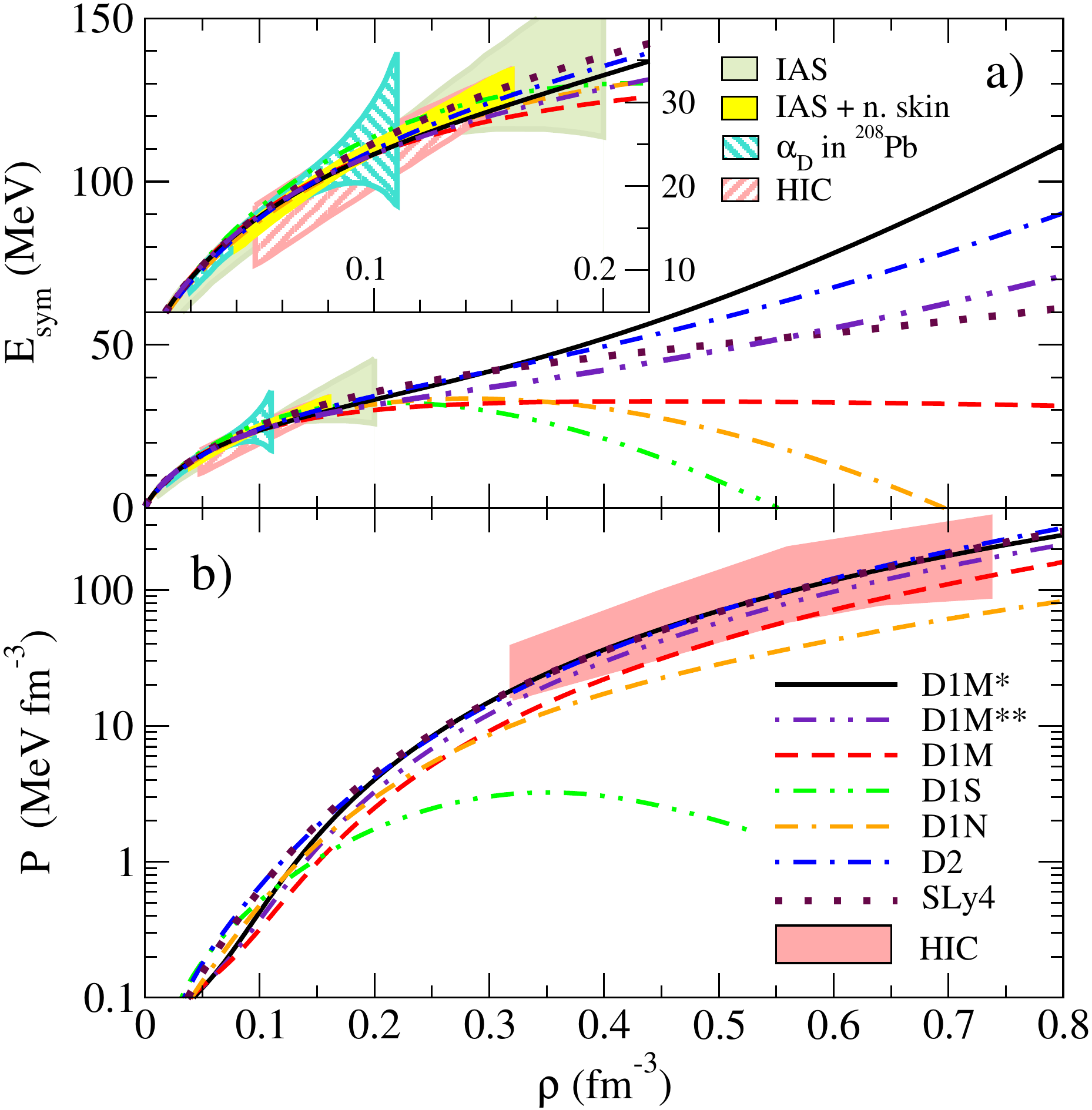}
\caption{a) Symmetry energy versus density from the D1S, D1N, D1M, D1M$^*$, D1M$^{**}$ and D2 Gogny forces 
and from the SLy4 Skyrme force. The inset is a magnified view of the low-density region.
Also plotted are the constraints from isobaric analog states (IAS) and from IAS and neutron skins (IAS+n.skin) \cite{Danielewicz13}, 
from the electric dipole polarizability in lead ($\alpha_D$ in $^{208}$Pb) \cite{Zhang15}
and from transport in heavy-ion collisions (HIC) \cite{Tsang08}.
b) Pressure in $\beta$-stable nuclear matter in logarithmic scale
as a function of density for the same interactions of panel~a). The shaded area depicts the
region compatible with collective flow in HICs~\cite{Danielewicz:2002pu}.\label{fig:essympres}}
\end{figure}

The symmetry energy as a function of density is displayed in Fig.~\ref{fig:essympres}(a) for several
Gogny forces and for the SLy4 Skyrme interaction. 
At subsaturation densities, the symmetry energy of the considered forces displays a similar behavior  
and takes a value of about 30 MeV at saturation (see Table~\ref{inm}).
The subsaturation regime is also the finite nuclei regime, where the parameters of the nuclear forces are fitted to.
Indeed, we observe in Fig.~\ref{fig:essympres}(a) that at subsaturation the present forces fall within, or are very close, to the region 
compatible with recent constraints on $E_{\rm sym} (\rho)$ deduced from several nuclear observables \cite{Danielewicz13,Zhang15,Tsang08}.
In contrast, above saturation density, the behavior of the calculated symmetry energy shows a strong model dependence. 
From this figure, two different patterns can be observed. On the one hand,
as seen in Chapter~\ref{chapter2},
the symmetry energy computed with the  D1S, D1N, and D1M interactions increases till reaching a maximum
value around $30$--$40$ MeV and then bends and decreases with increasing density until vanishing at some
density where the isospin instability starts. 
Although this happens at large densities for terrestrial phenomena, it is critical for NSs, 
where larger densities occur in the star's interior. 
On the other hand, the other forces, namely D1M$^*$, D1M$^{**}$
and D2, predict a symmetry energy with a well defined increasing trend with growing density. This different
behaviour of the symmetry energy strongly influences the EoS of the NS matter as it can be seen in Fig~\ref{fig:essympres}(b). 
In this figure, 
the EoS (total pressure against density) of $\beta$-stable, globally charge-neutral NS matter \cite{Sellahewa14,gonzalez17, gonzalez18} 
calculated with the given functionals is displayed.
The new Gogny forces D1M$^*$ and D1M$^{**}$, and the D2 force predict a high-density EoS with a similar stiffness to 
the SLy4 EoS and they agree well with the region constrained by collective flow in energetic 
heavy-ion collisions (HIC) \cite{Danielewicz:2002pu}, shown as the shaded area in 
Fig.~\ref{fig:essympres}(b).\footnote{Though the constraint of \cite{Danielewicz:2002pu} was proposed for neutron matter,
at these densities the pressures of $\beta$-stable matter and neutron matter are very similar.}
The EoSs from the original D1M parametrization and from D1N are significantly softer, stating that even though
 interactions may have increasing EoSs with the density, they may not be able to provide NSs of $2 M_\odot$.
The D1S force yields a too soft EoS soon after saturation density, which implies it is not 
suitable for describing NSs.

\begin{figure}[t!]
\centering
\includegraphics[width=0.65\columnwidth,clip=true]{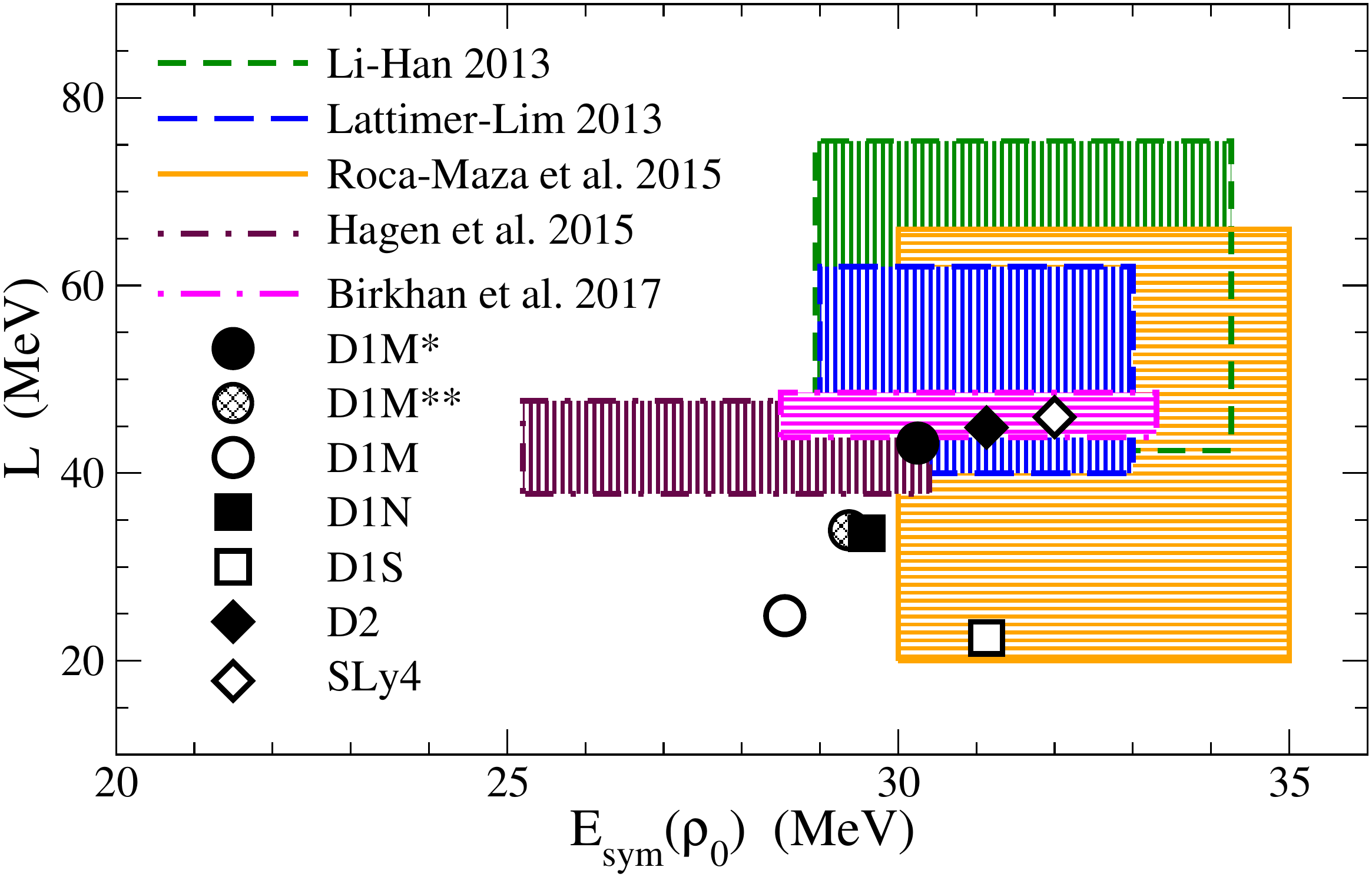}
\caption{Slope $L$ and value $E_{\rm sym}(\rho_0)$ of the symmetry energy at saturation density
for the discussed interactions. The hatched regions are the experimental and theoretical constraints 
derived in \cite{BaoAnLi13,Lattimer2013,Roca-Maza15,Hagen:2015yea,Birkhan16}.}
\label{esym_l}
\end{figure}

A few recent bounds on $E_{\rm sym}(\rho_0)$ and $L$ proposed  
from analyzing different laboratory data and astrophysical observations 
\cite{BaoAnLi13,Lattimer2013,Roca-Maza15} and from ab initio nuclear calculations using
chiral interactions \cite{Hagen:2015yea,Birkhan16}
are represented in Fig.~\ref{esym_l}. The prediction of D1M$^*$ is seen to overlap with the various constraints.
We note this was not incorporated in the fit of D1M$^*$ (nor of D1M$^{**}$).
It follows as a consequence of having tuned the density dependence of the symmetry energy of the interaction  
to be able to reproduce heavy NS masses simultaneously with the properties of nuclear matter and nuclei.
On the other hand, the predictions for D1M$^{**}$ fall in between the ones of D1M and D1M$^*$, 
very near the values for D1N, outside the constraint bands. 
D2 and SLy4 also show good agreement with the constraints of Fig.~\ref{esym_l}. 
We observe that the three interactions D1M$^*$, D2 and SLy4 have $E_{\rm sym}(\rho_0)$ values of 30--32 MeV
and $L$ values of about 45 MeV. A similar feature was recently found in the frame of RMF 
models if the radii of canonical NSs are to be no larger than $\sim$13~km 
\cite{Chen:2014mza,Tolos16}. It seems remarkable that 
mean-field models of different nature (Gogny, Skyrme, and RMF) converge
to specific values $E_{\rm sym}(\rho_0)\sim$30--32 MeV and $L\sim$\,45 MeV for the nuclear symmetry energy
when the models successfully predict the properties of nuclear matter and finite nuclei 
and heavy NSs with small stellar radii. 

\begin{figure}[!b]
 \centering
 \includegraphics[width=0.65\linewidth, clip=true]{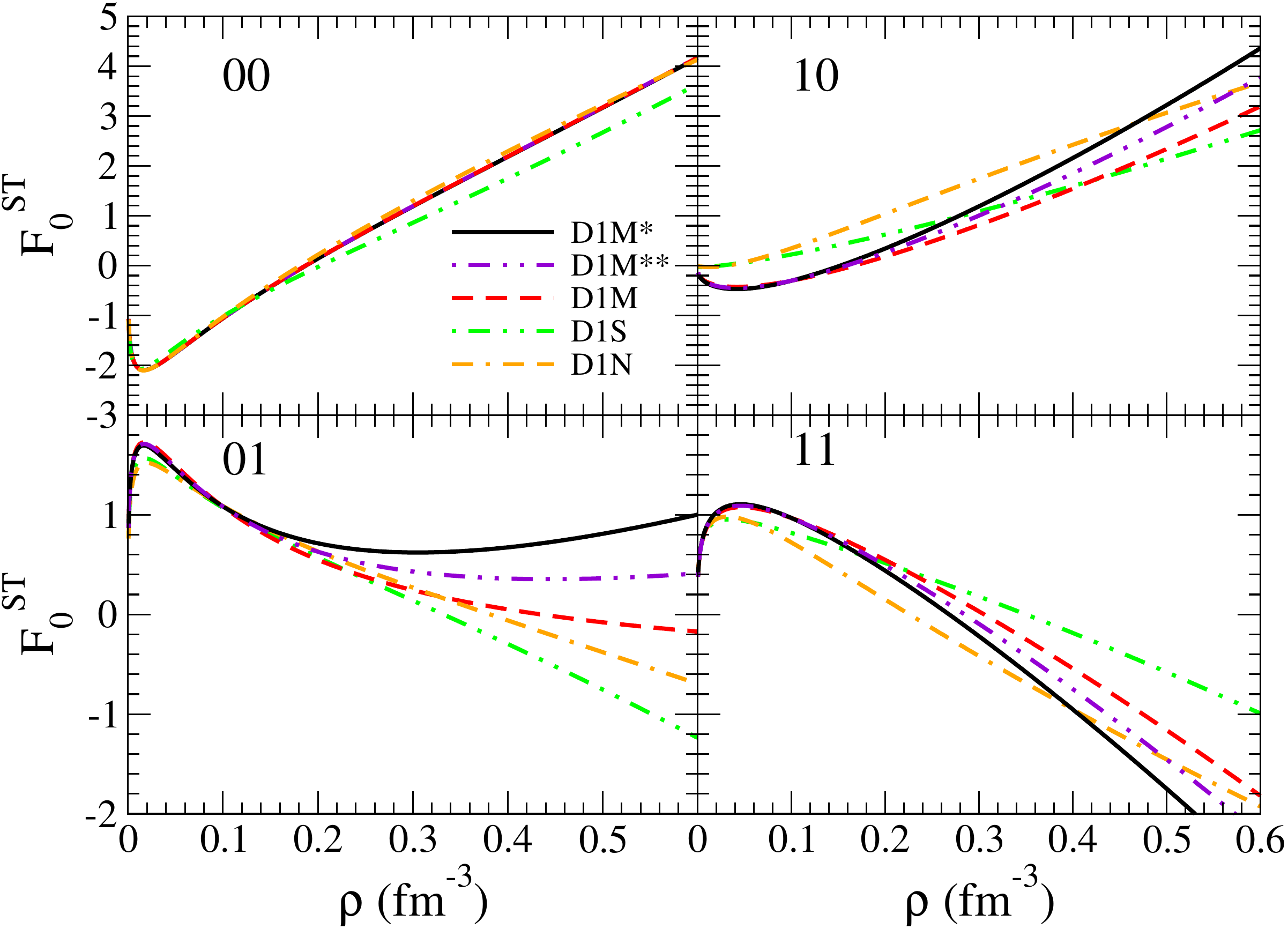}
  \caption{Density dependence of the Landau parameters $F_0^\mathrm{ST}$ for the D1S, D1M, D1N, 
D1M$^*$ and D1M$^{**}$ Gogny interactions.\label{landau0}}
\end{figure}
In order to further study the behavior of the new interactions D1M$^*$ and D1M$^{**}$ we 
analyze their corresponding Landau parameters $F_l^\mathrm{ST}$.
The Landau parameters for Gogny interactions corresponding to the spin-isospin channels ST, 
$f_l^\mathrm{ST}$, of $l=0$ and $l=1$ are the following~\cite{Ventura94}:
\begin{eqnarray}
 f_0^{00} (k_F, k_F, 0) &=& \frac{3}{8} t_3 (\alpha + 1) (\alpha +2) \rho^\alpha + \frac{\pi^{3/2}}{4} \sum_i
 \mu_i^3 \left\{ \left[ 4 W_i + 2 \left( B_i - H_i\right) - M_i\right] \vphantom{\frac{1}{1}}\right. \nonumber\\
 &&\left.- \left[ W_i - 2 \left( B_i - H_i \right) - 4 M_i \right] e^{-\gamma_i} \frac{\sinh (\gamma_i)}{\gamma_i} \right\}
\\
 f_0^{01} (k_F, k_F, 0) &=& -\frac{1}{4} t_3 \rho^\alpha (1+ 2 x_3)- \frac{\pi^{3/2}}{4} \sum_i
 \mu_i^3 \left\{ \left( 2 H_i + M_i \right) \vphantom{\frac{1}{1}}\right. \nonumber\\
 &&\left. +  \left( W_i + 2 B_i \right) e^{-\gamma_i} \frac{\sinh (\gamma_i)}{\gamma_i} \right\}
\\
 f_0^{10} (k_F, k_F, 0) &=& -\frac{1}{4} t_3 \rho^\alpha (1- 2 x_3)+ \frac{\pi^{3/2}}{4} \sum_i
 \mu_i^3 \left\{ \left( 2 B_i - M_i \right) \vphantom{\frac{1}{1}}\right. \nonumber\\
 &&\left. -  \left( W_i - 2 H_i \right) e^{-\gamma_i} \frac{\sinh (\gamma_i)}{\gamma_i} \right\}
\\
 f_0^{11} (k_F, k_F, 0) &=& -\frac{1}{4} t_3 \rho^\alpha - \frac{\pi^{3/2}}{4} \sum_i
 \mu_i^3 \left\{M_i 
 +  W_i e^{-\gamma_i} \frac{\sinh (\gamma_i)}{\gamma_i} \right\}
\\
f_1^{00} (k_F, k_F, 0) &=& - \frac{3\pi^{3/2}}{4} \sum_i
 \mu_i^3 \left[ W_i + 2 \left(  B_i - H_i \right) - 4 M_i \right] e^{-\gamma_i} \nonumber\\
 && \times \left( \frac{ \cosh (\gamma_i)}{\gamma_i} - \frac{\sinh (\gamma_i)}{\gamma_i^2} \right) 
 \\
 f_1^{01} (k_F, k_F, 0) &=& - \frac{3\pi^{3/2}}{4} \sum_i
 \mu_i^3 \left( 2 B_i + W_i\right) e^{-\gamma_i} \left( \frac{ \cosh (\gamma_i)}{\gamma_i} - \frac{\sinh (\gamma_i)}{\gamma_i^2} \right) 
 \\
  f_1^{10} (k_F, k_F, 0) &=& - \frac{3\pi^{3/2}}{4} \sum_i
 \mu_i^3 \left(  W_i- 2 H_i\right) e^{-\gamma_i} \left( \frac{ \cosh (\gamma_i)}{\gamma_i} - \frac{\sinh (\gamma_i)}{\gamma_i^2} \right) 
 \\
 f_1^{11} (k_F, k_F, 0) &=& - \frac{3\pi^{3/2}}{4} \sum_i
 \mu_i^3W_i e^{-\gamma_i} \left( \frac{ \cosh (\gamma_i)}{\gamma_i} - \frac{\sinh (\gamma_i)}{\gamma_i^2} \right) 
\end{eqnarray}
where 

\begin{equation}
 \gamma_i = \frac{\mu_i^2 k_F^2}{2}.
\end{equation}
The dimensionless Landau parameters are given by
\begin{equation}
 F_l^{\mathrm{ST}} = \frac{2 m^* (k_F) k_F}{\pi^2 \hbar^2} f_l^{\mathrm{ST}},
\end{equation}
where $m^*$ is the effective mass which is given by
\begin{eqnarray}
 \frac{m}{m^*} &=& 1- \sum_i \left\{ \vphantom{\frac{1}{1}} \frac{C_{1i}}{\mu_i^3 k_F^3 } \left[ 2 - 
 \mu_i^2 k_F^2 - \left( 2 + \mu_i^2 k_F^2 \right) e^{-\mu_i^2 k_F^2 }\right] \right.\nonumber\\
 && \left.- \frac{C_{2i}}{\mu_i^3 k_F^3 } \left[ \left( 2 + \mu_i^2 k_F^2  \right)  e^{-\mu_i^2 k_F^2 } - 
 \left( 2 - \mu_i^2 k_F^2 \right) \right]\right\}.
\end{eqnarray}
 Moreover, the coefficients $C_{1i}$ and $C_{2i}$ are, respectively,
\begin{equation}
 C_{1i} = \frac{m \mu_i^2}{\hbar^2 \sqrt{\pi}} \left( B_i + \frac{W_i}{2} - M_i -  \frac{H_i}{2} \right)
\end{equation}
and
\begin{equation}
 C_{2i} = \frac{m \mu_i^2}{\hbar^2 \sqrt{\pi}} \left( M_i + \frac{H_i}{2}\right).
\end{equation}

\begin{figure}[t!]
 \centering
 \includegraphics[width=0.65\linewidth, clip=true]{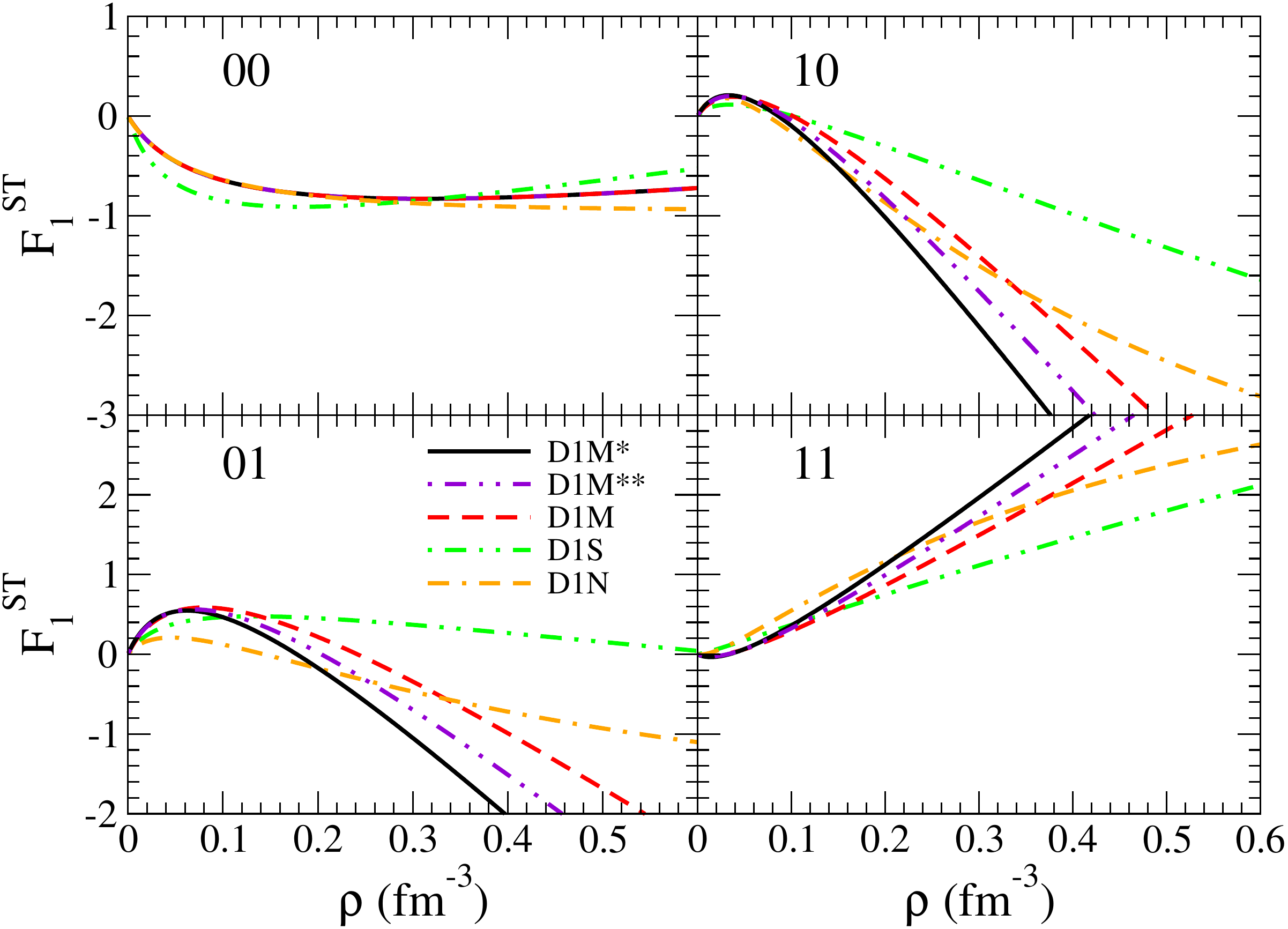}
  \caption{Same as Fig.~\ref{landau0} but for $F_1^\mathrm{ST}$. \label{landau1}}

\end{figure}

We plot in Fig.~\ref{landau0} the density dependence of the Landau parameters $F_0^\mathrm{ST}$ and  in Fig.~\ref{landau1} the density dependence of the 
Landau parameters $F_1^\mathrm{ST}$ 
for the Gogny D1S, D1M, D1N, D1M$^*$ and D1M$^{**}$ interactions. 
As expected, the behaviours of $F_0^\mathrm{00}$ and $F_1^\mathrm{00}$ calculated with D1M$^*$ and D1M$^{**}$ are 
the same of D1M, because in the 
fitting of these new interactions we have not changed the 
symmetric nuclear matter properties. 
In the other spin-isospin (ST) channels we have small differences if we compare D1M$^*$ and D1M$^{**}$ with D1M, 
more prominent beyond $\rho=0.2$ fm$^{-3}$ for $F_0^\mathrm{S1}$ and beyond $\rho=0.1$ fm$^{-3}$ for $F_1^\mathrm{S1}$.
These differences take into account the changes we have applied in the isospin sector when fitting D1M$^*$ and D1M$^{**}$.
Notice that, as we have tried to preserve the properties of the $S=0$ $T=1$ channel, the $F_0^\mathrm{01}$ and $F_1^\mathrm{01}$
parameters obtained with D1M$^*$ and D1M$^{**}$ have similar behaviors as the ones given by D1M.

\section{Neutron star mass-radius relation computed with the D1M$^*$ and D1M$^{**}$ interactions}
\begin{figure}[!b]
\centering
\includegraphics[width=0.8\columnwidth,clip=true]{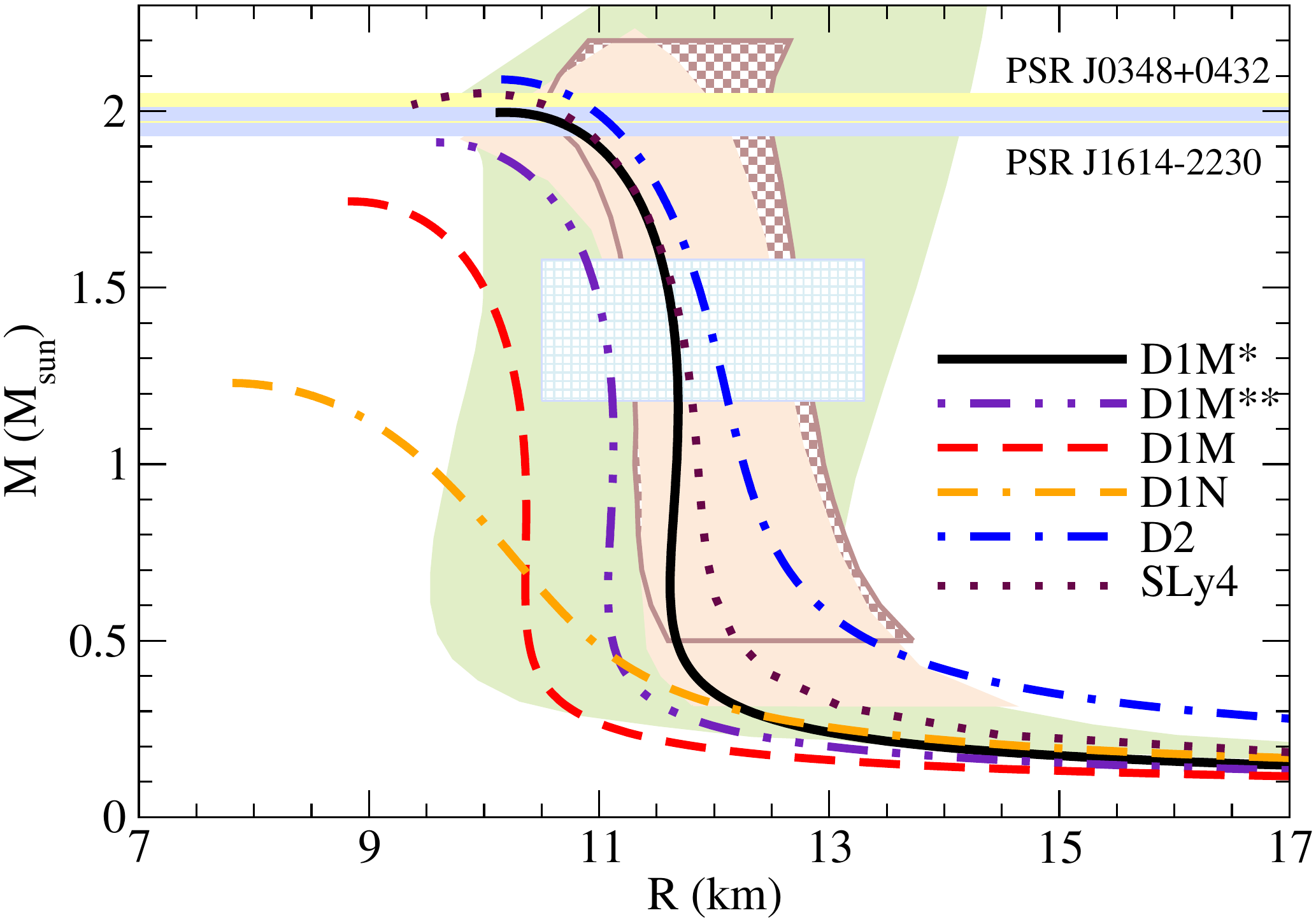}
\caption{Mass-radius relation in NSs from the D1N, D1M, D1M$^*$, D1M$^{**}$, D2 Gogny forces and the SLy4 Skyrme 
force.~The horizontal bands depict the heaviest observed NS masses \cite{Demorest10, Antoniadis13}.~The 
vertical green band shows the \mbox{M-R} region deduced from chiral nuclear interactions up to normal density 
plus astrophysically constrained high-density EoS extrapolations \cite{Hebeler13}. 
The brown dotted band is the zone constrained by the cooling tails of type-I X-ray bursts 
in three low-mass X-ray binaries and a Bayesian analysis \cite{Nattila16}, and the beige 
constraint at the front is from five quiescent low-mass X-ray binaries and five photospheric
radius expansion X-ray bursters after a Bayesian analysis~\cite{Lattimer14}. Finally, the squared blue band accounts for 
a Bayesian analysis of the data coming from the GW170817 detection of gravitational waves from a binary NS merger~\cite{Abbott2018}.}
 \label{fig:MR2}
\end{figure}

The mass-radius (M-R) relation in NSs is dictated by the corresponding EoS, which is the essential
ingredient to solve the TOV equations \cite{shapiro83}. 
In Fig.~\ref{fig:MR2} we display the mass of an NS as a 
function of its size for the D1M$^*$, D1M$^{**}$, D1M, D1N, and D2 Gogny interactions.
As explained in Section~\ref{MRbeta} of Chapter~\ref{chapter2}, to solve the TOV equations for an NS, knowledge of the EoS from the center to the surface 
of the star is needed. At present we do not have microscopic calculations of the EoS of the inner crust with 
Gogny forces.
Following previous literature \cite{Link1999,carriere03,xu09a,Zhang15, gonzalez17}, we interpolate 
the inner-crust EoS by a polytropic form  $P = a + b \epsilon^{4/3}$ ($\epsilon$ is the 
mass-energy density), where the index $4/3$ assumes that the pressure at these densities 
is dominated by the relativistic degenerate electrons [see Section~\ref{MRbeta} of Chapter~\ref{chapter2}]. 
We match this formula continuously to our Gogny EoSs of the homogeneous core
and to the Haensel-Pichon EoS of the outer crust \cite{douchin01}. 
The core-crust transition density is selfconsistently computed for each Gogny force by the thermodynamical method \cite{gonzalez17} [see the following Chapter~\ref{chapter4}]. 
We also plot as a benchmark the M-R curve calculated with the unified NS EoS proposed by Douchin and Haensel \cite{douchin01},
which uses the Skyrme SLy4 force.
It can be seen that standard Gogny forces, such as D1M and D1N, predict too low maximum stellar masses, with D1N being unable to  
generate masses above $1.4M_\odot$. We note that this common failure of conventional
Gogny parametrizations \cite{Loan2011,Sellahewa14,gonzalez17} has been cured in the new D1M$^*$ and D1M$^{**}$ forces, which, as well as D2 and SLy4, 
are successful in reaching the masses around $2M_\odot$ observed in NSs \cite{Demorest10,Antoniadis13}.
This fact is directly related to the behavior of the EoS in $\beta$-stable matter, which we have plotted in Fig.~\ref{fig:essympres}. As can be seen by
looking at Figs.~\ref{fig:essympres}(b) and \ref{fig:MR2}, the stiffer the EoS at high density, the larger the maximum NS mass.
We notice, however, that the maximum mass predicted by a 
given model does not only depend on the value of the parameter $L$ but also on the behaviour of 
the EoS of NS matter at high density. For example, the D1N and D1M$^{**}$ forces have almost 
the same value of $L$ (see Table \ref{inm}), but the maximum mass predicted by D1N is much
smaller than the one predicted by D1M$^{**}$. The maximum mass predicted
by the D1N interaction is around $1.23 M_{\odot}$ and $1.91M_{\odot}$ for the D1M$^{**}$ force.
This fact points out that, in spite of the same
slope of the symmetry energy at saturation, the behaviour of the symmetry energy above saturation (see Fig.~\ref{fig:essympres}(a))
strongly determines the EoS at high density and therefore the maximum NS mass predicted by each force.

As mentioned at the beginning of the chapter, we have included in the fitting procedure of D1M$^*$ 
(D1M$^{**}$) the constraint of a maximum mass of $2 M_{\odot}$ ($1.91 M_{\odot}$). The solution of the TOV equations 
also provides the radius of the NS. As seen in Table~\ref{Table-NSs2}, 
the radii corresponding to the NS of maximum mass are $10.2$ km for D1M$^*$ and $9.6$ km for D1M$^{**}$. Moreover, 
for an NS of a canonical mass of 1.4$M_{\odot}$ the radii obtained are of 11.6 km  for D1M$^*$ and of 11.1 km for D1M$^{**}$.
These values are similar to 
the predictions of the SLy4 model and are in harmony with recent extractions of the NS radius from low-mass X-ray binaries, 
X-ray bursters, and gravitational waves \cite{Nattila16,Lattimer14, Abbott2018}.

\begin{table}[t]
\resizebox{\columnwidth}{!}{%
\begin{tabular}{lcccccccc}
\hline 
\multirow{2}{*}{Force} & $L$    & $M_\mathrm{max}$ & $R(M_\mathrm{max})$ & $\rho_c(M_\mathrm{max})$ & $\epsilon_c(M_\mathrm{max})$ & $R(1.4M_\odot)$ & $\rho_c(1.4M_\odot)$ & $\epsilon_c(1.4M_\odot)$ \\
                       & (MeV)  & ($M_\odot$)      & (km)                & (fm$^{-3}$)              & ($10^{15}$ g cm$^{-3}$)         & (km)            & (fm$^{-3}$)          & ($10^{15}$ g cm$^{-3}$)                                                                       \\\hline \hline
D1M                    & 24.83  & 1.745            & 8.84                & 1.58                     & 3.65                            & 10.14           & 0.80                 & 1.51                                                                                          \\
D1M$^*$                & 43.18  &  1.997           & 10.18                & 1.19                     & 2.73                           & 11.65            & 0.53                 & 0.96                                                                                           \\
D1M$^{**}$             & 33.91  &  1.912           & 9.58               &  1.33                   &  3.09                          &    11.04        &    0.62             &	1.14		\\ \hline                                                                                         
\end{tabular}
}
                                                                              
\caption{Properties (mass $M$, radius $R$, central density $\rho_c$ and central mass-energy density $\epsilon_c$) 
of NS maximum mass and $1.4 M_\odot$ configurations for the D1M Gogny interaction and the new D1M$^*$ and D1M$^{**}$ parametrizations.
\label{Table-NSs2}}
\end{table}

\begin{table}[!b]
\centering
\resizebox{0.5\columnwidth}{!}{%
\begin{tabular}{lccc}
\hline
           & $A \leq 80$    &  $80 < A \leq 160$    & $A > 160$ \\ \hline\hline
D1M$^{*}$  & 1.55        &  1.31                  &   1.26    \\
D1M        & 1.82        &  1.12                &   1.29    \\ \hline
\end{tabular}
}
\caption{Partial rms deviation (in MeV) 
from the experimental binding energies \cite{Audi12} in even-even nuclei, computed in the given mass-number intervals.}
\label{prms}
\end{table}
\section{Properties of nuclei}
One of the goals of the D1M$^*$ and D1M$^{**}$ parametrizations is to reproduce nuclear structure 
properties of finite nuclei with the same global quality as the original D1M force \cite{gonzalez18, gonzalez18a, Vinas19}. 
 We have checked that the basic bulk properties of 
D1M$^*$, such as binding energies or charge radii of even-even nuclei, remain 
globally unaltered as compared to D1M. The finite nuclei 
calculations have been carried out with the \mbox{HFBaxial} code
\cite{robledo02} using an approximate second-order gradient method to 
solve the HFB equations \cite{rob11} in a harmonic oscillator (HO) 
basis. The code preserves axial symmetry but is allowed to break 
reflection symmetry. It has already been used in large-scale 
calculations of nuclear properties with the D1M force, as e.g.\ in Ref.\ \cite{Rob11b}.
Although the calculations of finite nuclei properties with the D1M$^{**}$ 
force have not been performed as extensively as with D1M$^{*}$, looking at 
their parameters reported in Table \ref{param}, it is expected that 
the predictions of D1M$^{**}$ will lie between the ones of D1M and D1M$^{*}$. 
Our preliminary investigations confirm this expectation.

\subsection{Binding energies and neutron and proton radii}
\begin{figure}[b!]
\centering
\includegraphics[width=0.7\columnwidth]{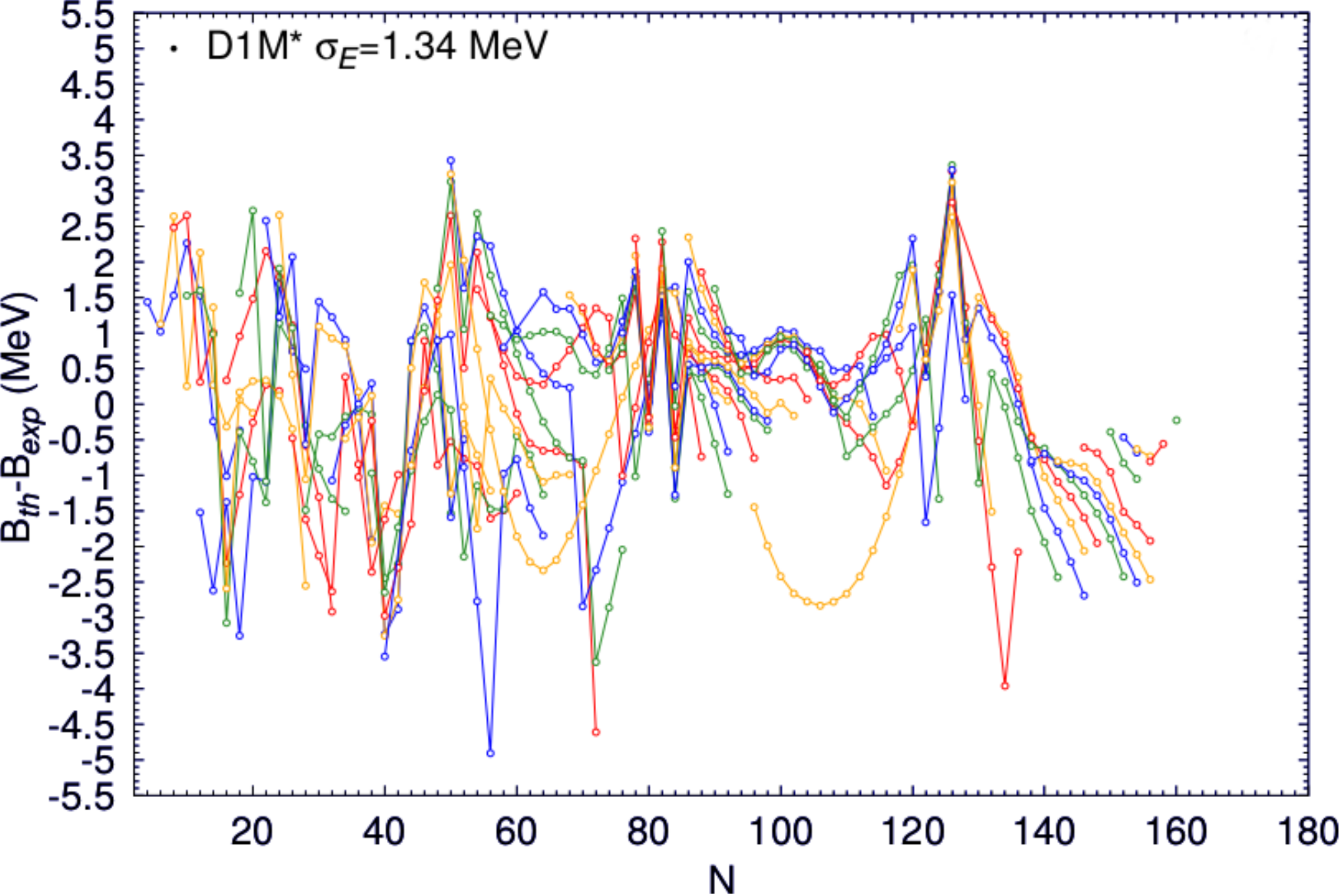}
\caption{Binding energy differences in 620 even-even nuclei between computed and 
experimental values \cite{Audi12}
for the new D1M$^*$ force, as a function of neutron number $N$.\label{Binding}}
\end{figure}
\begin{figure}[t!]
\centering
\includegraphics[width=0.7\columnwidth]{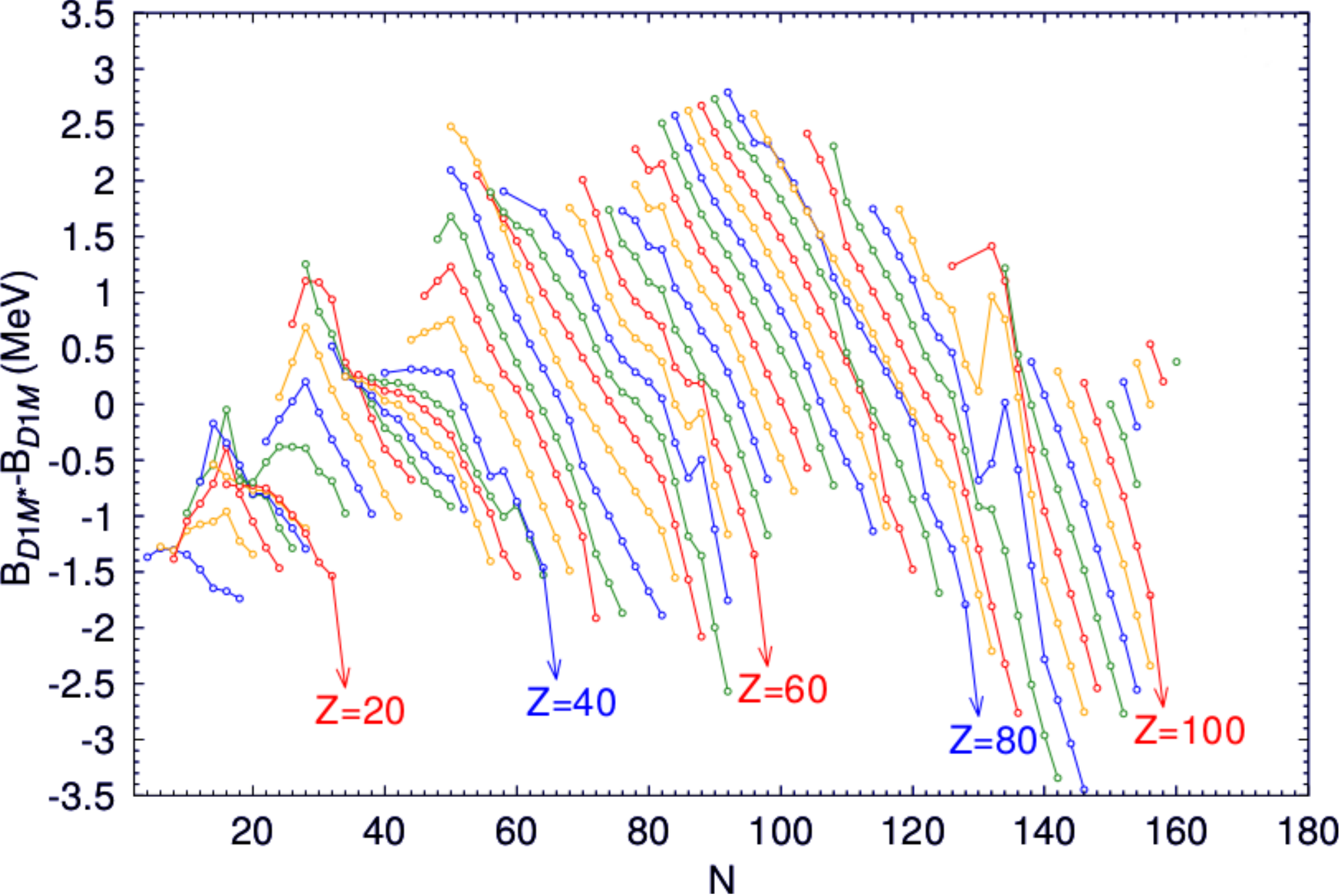}
\caption{Binding energy differences between the theoretical predictions of D1M$^{*}$ and D1M
for 818 even-even nuclei.\label{Bfn}}
\end{figure}

The binding energy is obtained by subtracting to the HFB energy the rotational energy correction,
as given in Ref.\ \cite{RRG00}. The ground-state calculation is repeated with an
enlarged basis containing two more HO major shells and an extrapolation
scheme to an infinite HO basis is used to obtain the final binding energy \cite{Hilaire.07,Baldo13a}.
In this framework, the zero-point energy (ZPE)
of quadrupole motion used in the original fitting of D1M \cite{goriely09} is not taken into account
because it requires considering $\beta$-$\gamma$ potential energy
surfaces (PES) and solving
the five-dimensional collective Hamiltonian for all the nuclei. This is still an enormous task
and we follow a different strategy where the quadrupole ZPE is replaced by a constant binding energy shift. This is 
somehow justified as in general the ZPE shows a weak mass dependence
(see \cite{Rob15} for an example with the octupole degree of freedom). The energy shift is fixed
by minimizing the global rms deviation, $\sigma_{E}$, for the 
known binding energies of 620 even-even nuclei \cite{Audi12}.

With a shift of 2.7 MeV we obtain for D1M a $\sigma_{E}$ of 1.36 MeV,
which is larger than the 798 keV reported for D1M in \cite{goriely09} including also odd-even and odd-odd nuclei. 
The result is still satisfying and gives us confidence in the procedure followed.
Using the same approach for D1M$^{*}$ we obtain a $\sigma_{E}$ of
1.34 MeV (with a shift of 1.1 MeV), which compares favorably with our $\sigma_{E}$ of 1.36 MeV for D1M.
This indicates a similar performance of both parametrizations in the average 
description of binding energies along the periodic table.

The differences between the binding energies of
D1M$^*$ and the experimental values, $\Delta B=B_\textrm{th}-B_\textrm{exp}$, for 620 even-even nuclei belonging to different isotopic 
chains are displayed in Fig.~\ref{Binding} against the neutron number $N$. 
The $\Delta B$ values are scattered around zero and show no drift with increasing $N$.
The agreement between theory and experiment is especially good for medium-mass and heavy
nuclei away from magic numbers and deteriorates for light nuclei, 
as may be seen from the partial $\sigma_{E}$ deviations given in Table~\ref{prms}.
From the partial $\sigma_{E}$ values of Table~\ref{prms}, we also conclude that the
closeness in the total $\sigma_{E}$ of D1M and D1M$^{*}$ involves subtle
cancellations that take place all over the nuclear chart.
We plot the differences in binding energy predictions between D1M$^*$ and D1M in Fig.~\ref{Bfn} 
against $N$ for 818 even-even nuclei.
The differences between the binding
energies computed with the D1M$^*$ and D1M are never larger than $\pm 2.5$ MeV and show a clear shift along
isotopic chains because of the different density dependence of the symmetry energy in both forces. A similar 
behavior can be observed in a recent comparison \cite{Pillet17} between D2 and D1S. It is also interesting to note that
the results for neutron radii show a similar isotopic drift as the
 binding energies. Namely, the difference $r_\mathrm{D1M^{*}}-r_\mathrm{D1M}$
(where $r$ is the rms radius computed from the HFB
wave function) increases linearly with $N$ for the neutron radii, 
whereas it remains essentially constant with $N$ for the proton radii. This is again
a consequence of the larger slope $L$ of the symmetry energy in D1M$^{*}$ \cite{Brown00,Centelles09}. 

\begin{figure}[t!]
\centering
\includegraphics[clip=true,width=0.9\columnwidth]{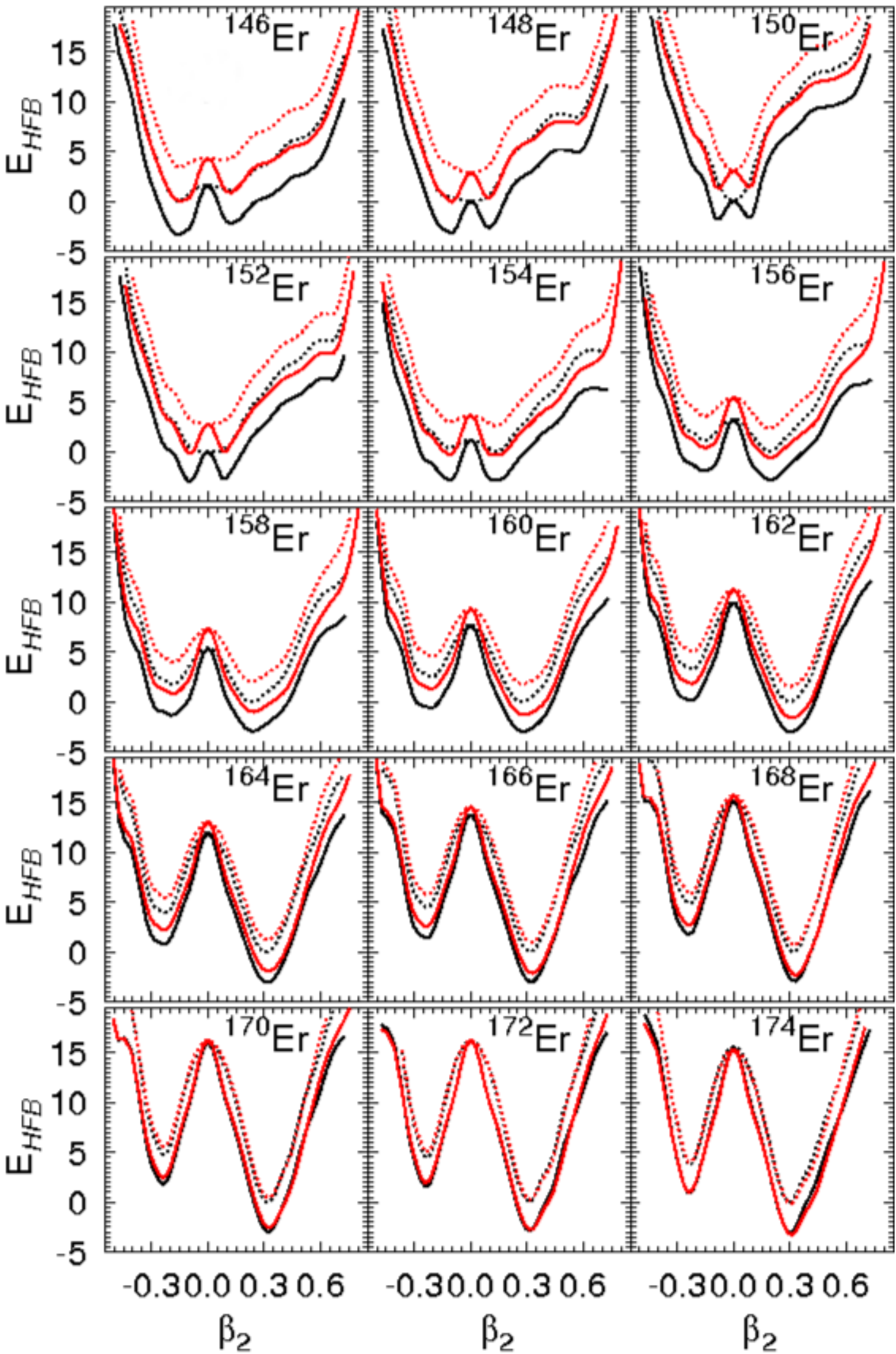}
\caption[]{Potential energy surfaces of the Er isotopic chain computed with the D1M (red) and 
D1M$^*$ (black) interactions as a function of the quadrupole deformation parameter $\beta_2$.\label{Er2}}
\end{figure}

\begin{figure}[b!]
\centering
\includegraphics[clip=true,width=0.65\columnwidth]{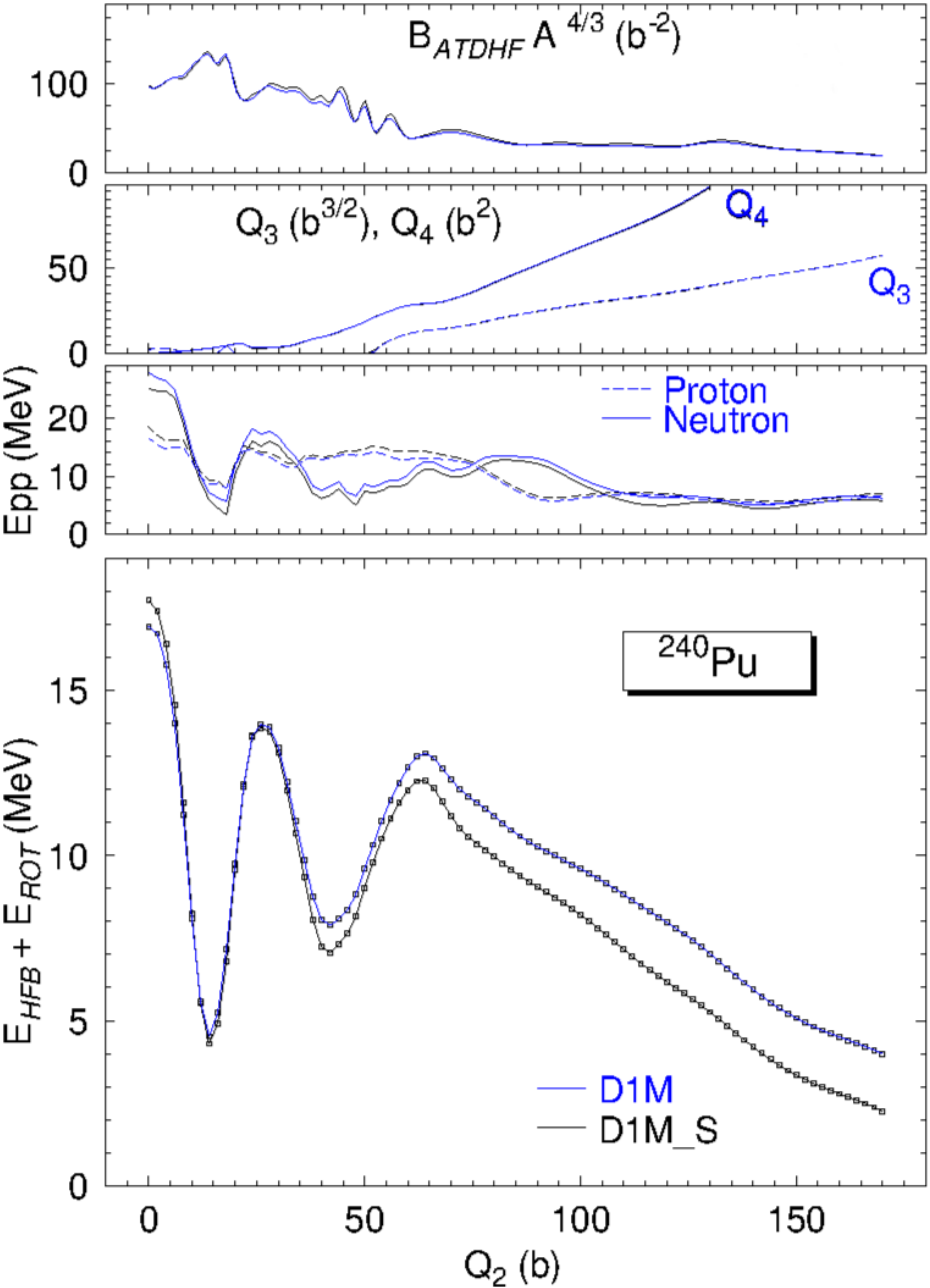}
\caption[]{Fission barrier of the nucleus $^{240}$Pu 
as a function of the quadrupole moment $Q_2$ calculated with the same Gogny forces. The evolution of the mass 
parameter, octupole and hexadecapole moments and neutron and proton pairing energies along the fission path 
are also displayed in the same figure.\label{Fis}}
\end{figure}
\subsection{Potential Energy Surfaces}
An important aspect of any nuclear interaction is the way it determines the response of the
nucleus to shape deformation, in particular to the quadrupole deformation. To know if a nucleus is quadrupole 
deformed or not plays a crucial role in the determination of the low energy-spectrum. To study the 
response of the D1M$^*$ force to the quadrupole deformation, we have performed constrained HFB calculations 
in finite nuclei fixing the quadrupole moment $Q_{20}$ to given values, which allows one to obtain the PES. 
As an example, in Fig.~\ref{Er2} the PES along the Er (Z=68) isotopic chain is displayed as a function of the deformation parameter 
$\beta_2$ for the original D1M interaction and for the modified D1M$^*$ force.
It can be observed that the curves corresponding to the calculations performed with the D1M 
and D1M$^*$ forces follow basically the same trends  with a small displacement of one curve with respect to the 
other~\cite{gonzalez18, gonzalez18a}.

\newpage

\subsection{Fission Barriers}
Finally, we plot in Fig.~\ref{Fis} the fission barrier of the paradigmatic nucleus $^{240}$Pu~\cite{gonzalez18a}.
We see that the inner fission barrier predicted by D1M and D1M$^*$ is the same 
in both models with a value $B_I$=9.5 MeV. This value is a little bit large compared with the experimental 
value of 6.05 MeV. However, it should be pointed out that triaxiality effects, not accounted for in the present 
calculation, might lower the inner barrier by 2-3 MeV. The excitation energy of the fission isomer $E_{II}$ is
3.36 MeV computed with D1M and 2.80 MeV with D1M$^*$. The outer fission barrier $B_{II}$ height are 
8.58 and 8.00 MeV calculated with the D1M and D1M$^*$ forces, respectively. These values clearly 
overestimate the empirical value, which is 5.15 MeV. In the other panels of Fig.~\ref{Fis} we have  
displayed as a function of the quadrupole deformation the neutron and proton pairing energies, the octupole 
moment (responsible for asymmetric fission) and the hexadecapole moment of the mass distributions.
All these quantities take very similar values computed with both interactions.
 
\chapter{Core-crust transition in neutron stars}\label{chapter4}
Our aim in this chapter is to study the properties of the transition between the core and the crust inside neutron stars (NSs). 
The structure of an NS consists of a
solid crust at low densities encompassing a homogeneous core in a liquid phase. The density is maximum at the center, several times
the nuclear matter density, which has a value of $\sim$ 2.3 $\times$ 10$^{14}$ g cm$^{-3}$, and decreases with the distance from the center
reaching a value of the terrestrial
iron around $\sim$ 7.5 g cm$^{-3}$ at the surface of the star.
The external part of the star, i.e., its outer crust, consists of nuclei 
distributed in a solid body-centered-cubic (bcc) lattice permeated by a free electron gas. 
When the density increases, the nuclei in the crust become so neutron-rich that neutrons start to drip from 
them. 
In this scenario, the lattice structure of nuclear clusters is embedded in free 
electron and neutron gases. When the average density reaches a value of about half of the nuclear  
matter saturation density, the lattice structure disappears due to energetic reasons and the system 
changes to a liquid phase. The boundary between the outer and inner crust is determined by the nuclear masses, 
and corresponds to the neutron drip out density around $\sim 4\times 10^{11}$ g cm$^{-3}$ \cite{ruster06,hampel08}.
However, the transition density from the inner crust to
the core is much more uncertain and is strongly model dependent 
\cite{baym70,kubis06,ducoin07,xu09a,xu09b,xu10a,xu10b,Moustakidis10,Moustakidis12,Seif14,routray16,gonzalez17}. 
To determine the core-crust transition density from the crust side it is required to have a precise 
knowledge of the EoS in this region of the star, as the boundary between the liquid core and the 
inhomogeneous solid crust is connected to the isospin dependence of nuclear models below saturation density.
However, this is a very challenging 
task owing to the presence of the neutron gas and the possible existence of complex structures in the
deep layers of the inner crust, where the nuclear clusters may adopt shapes differently
from the spherical one (i.e., the so-called ``pasta phases") in order to minimize the energy
\cite{baym71, lattimer95, shen98a, shen98b, douchin01, sharma15, Carreau19}.
Therefore, it is easier to investigate the core-crust transition from the core side. To this
end, one searches for the violation of the stability conditions of the homogeneous core under
small-amplitude oscillations, which indicates the appearance of nuclear clusters and,  
consequently, the transition to the inner crust. There are different ways to determine the transition 
density from the core side, namely, the thermodynamical method \cite{kubis04,kubis06,xu09a,Moustakidis10,Moustakidis12,Cai2012,Seif14,routray16,gonzalez17}, 
the dynamical method
\cite{baym71,xu09a,xu09b,xu10a,xu10b,pethick95,ducoin07,Tsaloukidis19, gonzalez19}, the random phase approximation 
\cite{horowitz01a, horowitz01b,carriere03} or the Vlasov equation method \cite{chomaz04,providencia06,
ducoin08a,ducoin08b,pais10}. 
 Also, a variety of different functionals (and many-body theories) have been used to 
determine the properties of the core-crust transition, including Skyrme forces 
\cite{xu09a,ducoin07,Pearson12,Newton2014}, 
finite-range functionals \cite{routray16, gonzalez17}, relativistic mean-field (RMF) models 
\cite{horowitz01a,carriere03,Klahn06,Moustakidis10,Fattoyev:2010tb,Cai2012,Newton2014}, 
momentum-dependent interactions \cite{xu09a,Moustakidis12, gonzalez19} and 
Brueckner--Hartree--Fock theory~\cite{Vidana2009,Ducoin11,Li2016}.  

In the low-density regime of the core near the crust, the NS matter is composed of
neutrons, protons, and electrons. 
To find the instabilities in the core, in this thesis we restrict ourselves to two approaches:
the thermodynamical method and the dynamical method.
In the thermodynamical approach, the stability is discussed in terms of the bulk properties of 
the EoS by imposing mechanical and chemical stability conditions, which set the boundaries of
 the core in the homogeneous  case.  
On the other hand, in the dynamical method, one introduces density fluctuations
for neutrons, protons, and electrons, which
can be expanded in plane waves of amplitude $A_i$ ($i=n, p, e$) and of wave-vector ${\bf k}$.
These fluctuations may be induced by collisions, which transfer some momentum to the system.
Following Refs.~\cite{baym71,pethick95,ducoin08a,ducoin08b,xu09a} one writes the variation of 
the energy density generated by these fluctuations in terms of the energy curvature matrix 
$C^f$. 
The system is stable when the curvature matrix is convex, that is, the transition density is 
obtained from the condition $\vert C^f \vert$=0. 
This condition determines the dispersion relation $\omega(k)$ of the collective excitations 
of the system due to the perturbation and the 
instability will appear when this frequency $\omega(k)$ becomes imaginary. 
The dynamical method is more realistic than the thermodynamical approach, as it incorporates 
surface and Coulomb effects in the stability condition that are not taken into account in the 
thermodynamical method.

In Section~\ref{Theory_thermo} we introduce the thermodynamical method to find the core-crust transition and
in Section~\ref{Results_thermo} we obtain the corresponding core-crust transition properties for Skyrme and Gogny interactions. 
In Section~\ref{Theory_dyn} we give the theory for the more sophisticated dynamical method, 
and in Section~\ref{Results_dyn} we present the results of the transition properties obtained with it. 

\section{The thermodynamical method}\label{Theory_thermo}
We will first focus on the search of the transition between the core and the crust of NSs 
using the so-called thermodynamical method, 
which has been widely used in the literature \cite{kubis04,kubis06,xu09a,Moustakidis10,Cai2012,
Moustakidis12,Seif14,routray16}.
Within this approach, the stability of the NS core is discussed in terms of its bulk properties. 
The following mechanical and chemical stability conditions set the boundaries of the homogeneous core:
\begin{eqnarray}\label{cond1}
 -\left( \frac{\partial P}{\partial v} \right)_{\mu_{np}} & > & 0 ,
\\[2mm]
\label{cond2}
 -\left( \frac{\partial \mu_{np}}{\partial q} \right)_v & > & 0 .
\end{eqnarray}
Here, $P$ is the total pressure of $\beta$-stable matter, 
$\mu_{np}= \mu_n-\mu_p$ is the difference between the neutron and proton chemical potentials [related to the $\beta$-equilibrium 
condition in~Eq.~(\ref{betaeq})],
$v=1/\rho$ is the volume per baryon and $q$ is the charge per baryon. 

First, we consider the mechanical stability condition in Eq.~(\ref{cond1}). 
The electron pressure does not contribute to this term, due to the fact that
the derivative is performed at a constant $\mu_{np}$. 
In $\beta$-equilibrium, this involves a constant electron chemical potential $\mu_e$ and, 
because the electron pressure 
in Eq.~(\ref{eq:P_lepton}) is a function only of $\mu_e$, the derivative of $P_e$ with respect to $v$ vanishes.
Equation~(\ref{cond1}) can therefore be rewritten as
\begin{equation}\label{cond1Pb}
 -\left( \frac{\partial P_b}{\partial v} \right)_{\mu_{np}} >0,
\end{equation}
where $P_b$ is the baryon pressure.
Moreover, the isospin asymmetry of the $\beta$-stable system is a function of density, $\delta(\rho)$. 
With $\mu_{np} = 2 \partial E_b(\rho, \delta)/\partial \delta$, and using Eq.~(\ref{eq:pre}) for baryons,
we can express the mechanical stability condition as \cite{xu09a, Cai2012,  Moustakidis10, Moustakidis12}
\begin{eqnarray}\label{cond11}
  -\left( \frac{\partial P_b}{\partial v} \right)_{\mu_{np}} =\rho^2 \left[ 2 \rho 
  \frac{\partial E_b (\rho, \delta)}{\partial \rho} + \rho^2 \frac{\partial^2 E_b (\rho , \delta)}{\partial \rho^2}  \right. 
\left. -\frac{\left( \rho \frac{\partial^2 E_b (\rho , \delta)}{\partial \rho \partial \delta} \right)^2}{\frac{\partial^2 
 E_b (\rho , \delta)}{\partial \delta^2}}\right] >0 .
\end{eqnarray}
In the chemical stability condition of Eq.~(\ref{cond2}), the charge $q$ can be written as $q=x_p - \rho_e/\rho$, 
where \mbox{$x_p = (1- \delta)/2$} is the proton fraction. 
In the ultrarelativistic limit, the electron number density is related to the chemical potential by $\rho_e= \mu_e^3/(3\pi^2)$. 
We can thus recast (\ref{cond2}) as
\begin{equation}\label{cond22}
 -\left( \frac{\partial q}{\partial \mu_{np}} \right)_v= \frac{1}{4} \left[ \frac{\partial^2 E_b (\rho, \delta)}{\partial \delta^2}
 \right]^{-1} + \frac{\mu_e^2}{\pi^2 \rho}>0. 
\end{equation}
In the low-density regime of interest for the core-crust transition, the first term on the right-hand side is positive 
for the Skyrme and Gogny parameterizations studied here. With a second term that is also positive, we conclude that the 
inequality of Eq.~(\ref{cond22}) is fulfilled. 
Hence, the stability condition for $\beta$-stable matter can be expressed in terms of Eq.~(\ref{cond11}) alone, 
with the result 
\cite{kubis04, kubis06,xu09a,Moustakidis10}
\begin{equation}\label{Vthermal}
  V_{\mathrm{ther}} (\rho) = 2 \rho \frac{\partial E_b (\rho, \delta)}{\partial \rho} + \rho^2 
  \frac{\partial^2 E_b (\rho , \delta)}{\partial \rho^2}
 -
 \left( \rho \frac{\partial^2 E_b (\rho , \delta)}{\partial \rho \partial \delta} \right)^2 \left(\frac{\partial^2 
 E_b (\rho , \delta)}{\partial \delta^2} \right)^{-1}>0,
\end{equation}
where we have introduced the so-called thermodynamical potential, $V_\mathrm{ther} (\rho)$. 

If the condition for $V_\mathrm{ther} (\rho)$ is rewritten using the Taylor expansion of 
$E_b (\rho, \delta)$ given in Eq.~(\ref{eq:EOSexpgeneral}), one finds
\begin{eqnarray}\label{eq:Vtherapprox}
 V_{\mathrm{ther}}  (\rho) &=& \rho^2 \frac{\partial^2 E_b (\rho, \delta=0)}{\partial \rho^2} + 2 \rho 
 \frac{\partial E_b (\rho, \delta=0)}{\partial \rho} \nonumber
 \\
&&+ \sum_{k} \delta^{2k} \left( \rho^2 \frac{\partial^2 E_{\mathrm{sym}, 2k}(\rho)}{\partial \rho^2} + 2 \rho 
\frac{\partial E_{\mathrm{sym}, 2k}(\rho)}{\partial \rho}\right) \nonumber
\\
&&-2\rho^2 \delta^2 \left( \sum_{k} k \delta^{2k-2}  \frac{\partial E_{\mathrm{sym}, 2k}(\rho)}{\partial \rho} \right)^2 \nonumber
\\
&&\times \left[ \sum_{k} (2k-1)k\delta^{2k-2} E_{\mathrm{sym}, 2k}(\rho)\right]^{-1} >0 . \nonumber 
\\
\end{eqnarray}
This equation can be solved order by order, together with the $\beta$-equilibrium condition, Eq.~(\ref{betamatter})
(or Eq.~(\ref{betamatter-PA}) in the PA case), to evaluate the influence on
the predictions for the core-crust transition of truncating the Taylor expansion of the EoS of asymmetric nuclear matter.
We collect in Appendix~\ref{appendix_thermal} the expressions for the derivatives of $E_b(\rho, \delta)$ for Skyrme and Gogny forces 
that are needed to 
calculate $V_\mathrm{ther} (\rho)$ in both Eqs.~(\ref{Vthermal}) and (\ref{eq:Vtherapprox}).

\section{Core-crust transition studied within the thermodynamical method}\label{Results_thermo}
\begin{figure}[t!]
 \centering
 \subfigure{\label{fig:vtherskyrme}\includegraphics[width=0.49\linewidth]{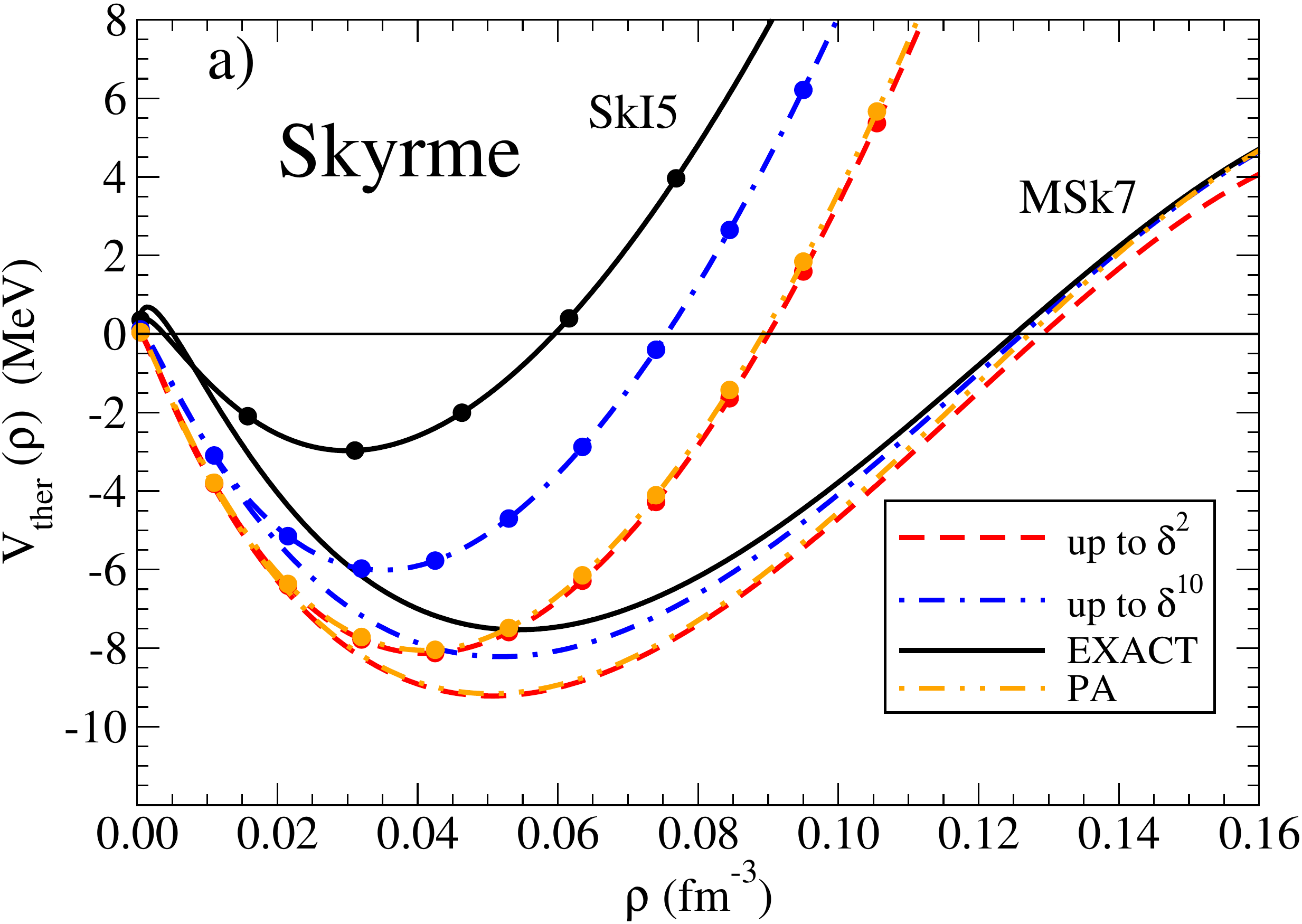}}
\subfigure{\label{fig:vthergogny}\includegraphics[width=0.49 \linewidth]{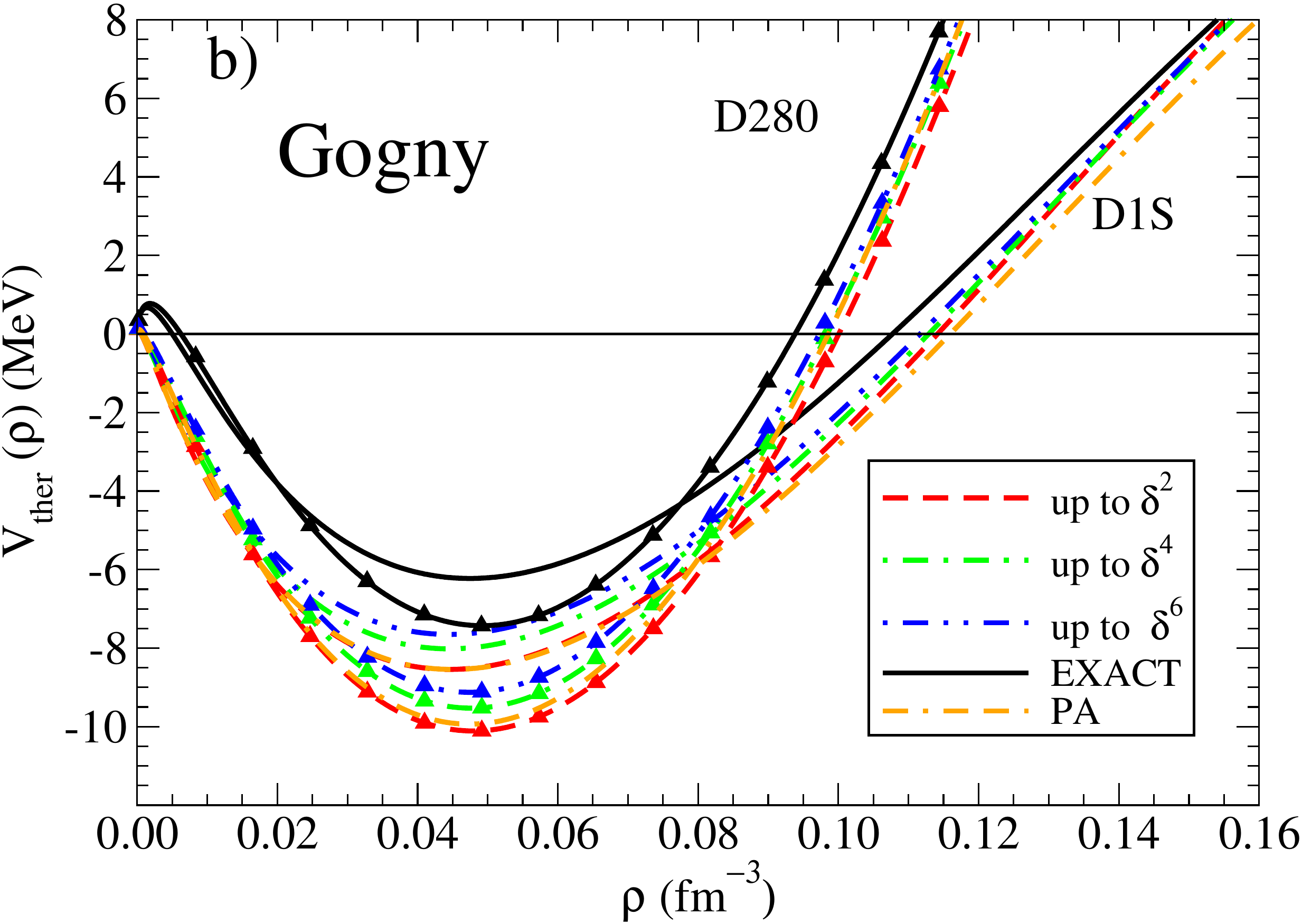}}
  \caption{Density dependence of the thermodynamical potential in $\beta$-stable nuclear matter calculated using the 
  exact expression of the EoS or the expression in Eq. (\ref{eq:EOSexpgeneral}) up to second and 
  tenth order for two Skyrme forces (panel (a)) and up to second, fourth and sixth order for two Gogny forces (panel (b)).
  The results for the parabolic approximation are also included in both panels.\label{fig:vther}}
\end{figure}

We show in Fig.~\ref{fig:vther} the density dependence of the thermodynamical potential $V_\mathrm{ther} (\rho)$ 
in $\beta$-stable matter for Skyrme (Fig.~\ref{fig:vtherskyrme}) and Gogny (Fig.~\ref{fig:vthergogny}) interactions, 
calculated with the exact expression of the EoS (solid 
lines), with its Taylor expansion up to second and tenth order for Skyrme forces and up to 
second, fourth and sixth order for Gogny models, and with the 
PA \cite{gonzalez17}. An instability region characterized by negative $V_\mathrm{ther} (\rho)$ is found 
below $\rho \approx 0.13$ fm$^{-3}$ for both types of interactions. The condition $V_\mathrm{ther}(\rho_t) = 0$ 
defines the density $\rho_t$ of the transition from the homogeneous core to the crust.
We see in Fig.~\ref{fig:vther} that adding more terms to Eq.~(\ref{eq:EOSexpgeneral}) brings the 
results for $V_\mathrm{ther} (\rho)$ closer to the exact values. 
At densities near the core-crust transition, for the Skyrme MSk7 and the Gogny D1S and D280 interactions,
the higher-order results are rather 
similar, but differ significantly from the exact ones. The differences are larger for the SkI5 interaction, which 
is the interaction of the ones considered here that has the 
largest slope of the symmetry energy.
We note that, all in all, the order-by-order convergence of the $\delta^2$ expansion in 
$V_\mathrm{ther} (\rho)$ is slow. This indicates that the non-trivial isospin and 
density dependence arising from exchange terms needs to be considered in a complete 
manner for realistic core-crust transition physics 
\cite{Chen09,Vidana2009,Seif14, gonzalez17}.
If we look at the unstable low-density zone, both the exact and the approximated 
results for $V_\mathrm{ther} (\rho)$ go to zero for vanishing density, but they keep a 
different slope. In this case, we have found that the discrepancies are largely 
explained by the differences in the low-density behaviour of the approximated 
kinetic energy terms, in consonance with the findings of Ref.~\cite{routray16}.

We next analyze more closely the properties of the core-crust transition, using both exact and order-by-order 
predictions. The complete results for the sets of Skyrme and Gogny functionals are provided, respectively, in numerical 
form in Appendix~\ref{app_taules}.

For a better understanding, we discuss each one of the key physical properties of the transition (asymmetry, density, and pressure) 
in separate figures. We plot our predictions as a function of the slope parameter $L$ of each functional, which does not necessarily 
provide a stringent correlation with core-crust properties \cite{Ducoin11}. 
The slope parameter, however, can be constrained in terrestrial experiments and astrophysical observations 
\cite{Tsang2012,Lattimer2013,Lattimer2016,BaoAnLi13,Vinas14,Roca-Maza15}
and is, therefore, an informative parameter in terms of the isovector properties of the functional.

\begin{figure}[t!]
 \centering
 \subfigure{\label{fig:deltatskyrme}\includegraphics[width=0.49\linewidth]{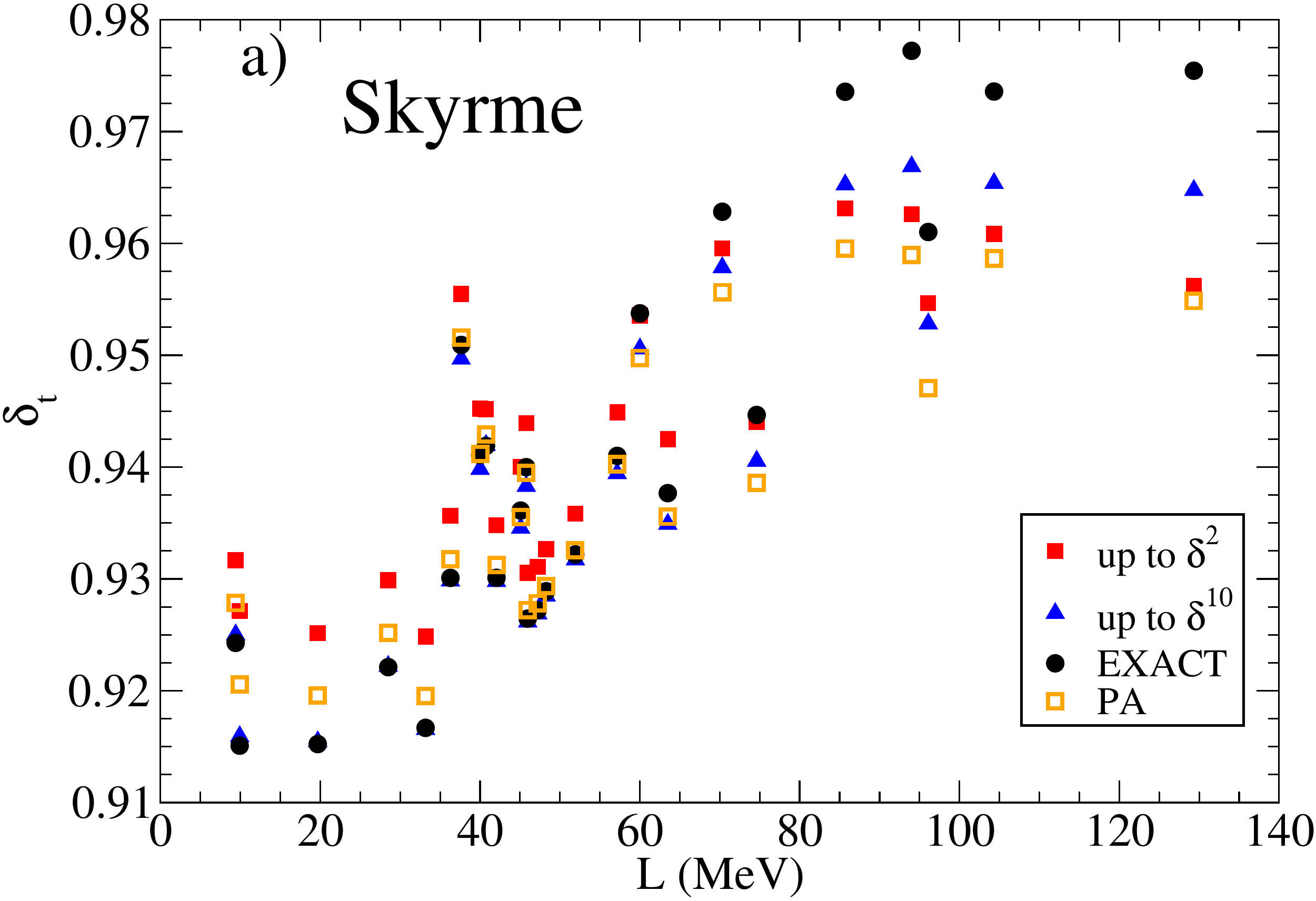}}
\subfigure{\label{fig:deltatgogny}\includegraphics[width=0.49 \linewidth]{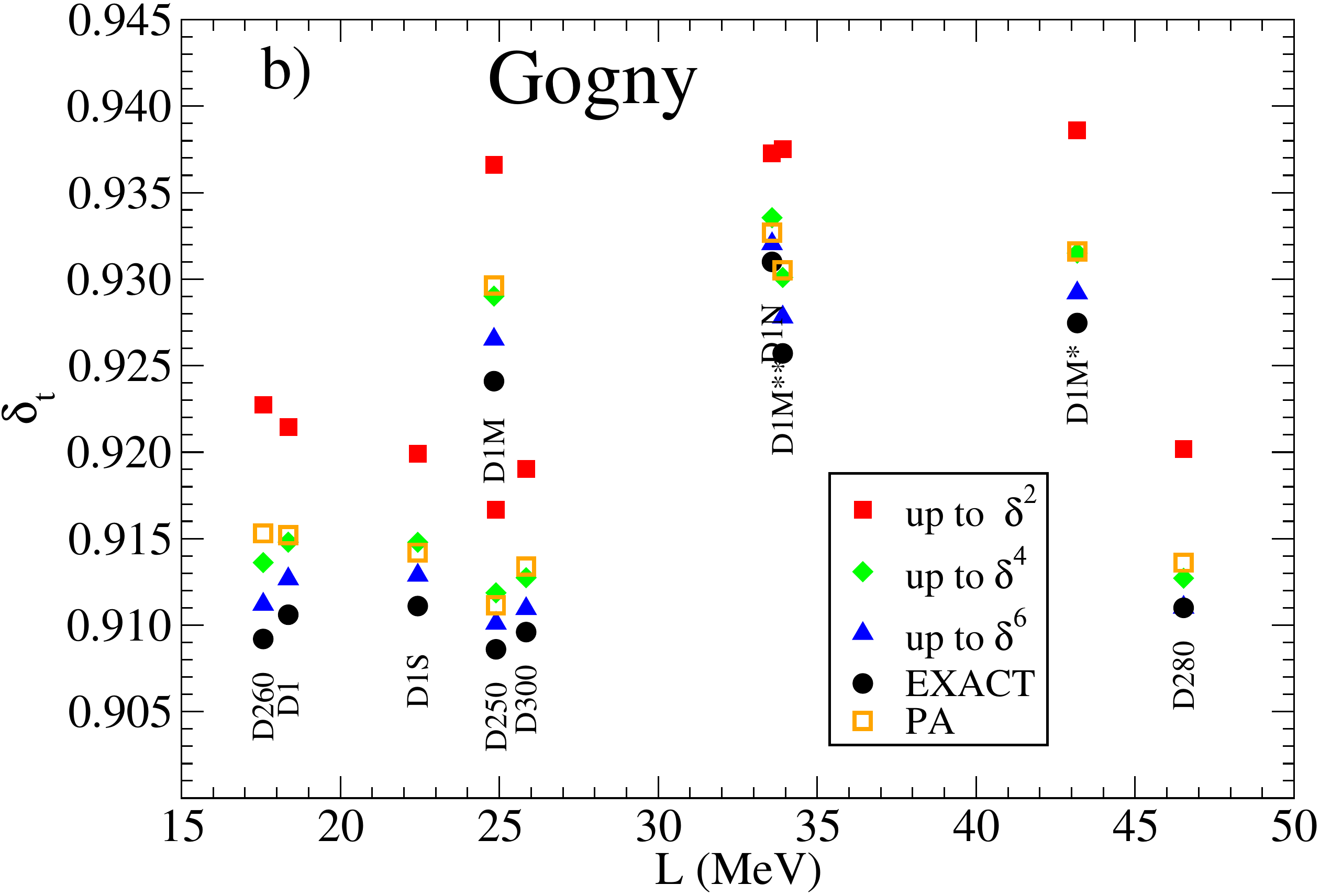}}
   \caption{Panel a: Core-crust transition asymmetry, $\delta_t$, as a function of the slope parameter $L$ for a set of Skyrme forces
   calculated with the thermodynamical approach
   using the exact expression of the EoS (solid circles), 
   and the approximations up to second (solid squares) and tenth order (solid triangles). 
   The parabolic approximation is also included (empty squares).
   Panel b: Same as panel (a) for a set of Gogny forces 
   calculated 
   using the exact expression of the EoS (solid circles), 
   and the approximations up to second (solid squares), fourth (solid diamonds) and sixth order (solid triangles). 
   The parabolic approximation is also included (empty squares).}
 \label{fig:deltat}
\end{figure}

In Fig.~\ref{fig:deltat}, we display the results for the transition asymmetry, $\delta_t$ \cite{gonzalez17}. Black dots 
correspond to the calculations with the exact EoS. We find that the set of Skyrme forces collected in Table~\ref{table:Skyrmeprops} predict a range
of $0.924 \lesssim \delta_t \lesssim 0.978$ for the asymmetry at the transition point (see Fig.~\ref{fig:deltatskyrme}), 
while the set of Gogny forces found in Table~\ref{table:Gognyprops} plus the new D1M$^*$ and D1M$^{**}$ parametrizations predict a range of
$0.909 \lesssim \delta_t \lesssim 0.931$ (see Fig.~\ref{fig:deltatgogny}).
For Skyrme interactions, we see that $\delta_t$ presents an increasing tendency as the slope of the 
symmetry energy of the interaction is larger. Of the interactions we have considered, G$_\sigma$, SkI5, SkI2, and R$_\sigma$
are the ones with higher values, over $\delta_t \gtrsim 0.97$.
In the case of Gogny interactions, the D1N, D1M, D1M$^*$ and D1M$^{**}$ forces are the ones providing
distinctively large transition asymmetries,
whereas the other interactions predict very similar values $\delta_t \approx 0.91$ in spite of having different 
slope parameters.
When using the Taylor expansion of the EoS up to second order (shown by red squares in both panels), 
the predictions 
for $\delta_t$ are generally far from the exact result. For interactions (in both cases of Skyrme and Gogny functionals) 
with the slope of the symmetry energy below $L\lesssim 70$ MeV the results are well above the exact result,
whereas for interactions with higher slope $L$ the results obtained with the Taylor expansion up to second
order remain below the exact results. 
For Skyrme interactions, the results obtained with the tenth-order expansion (blue triangles) are close to the 
ones obtained with the exact EoS for interactions that have a relatively low slope of the symmetry energy, 
while they remain quite far from them when $L$ is larger.
The results obtained with the PA (empty orange squares) provide a very good approximation if the slope $L$ is small, but works 
less well as $L$ becomes larger. Also, if $L\gtrsim 60$ MeV, the values obtained with the PA are farther from the 
exact results than the results obtained with the second-order approximation.
For Gogny interactions, the fourth-order values (green diamonds) are still above the exact 
ones but closer, and the sixth-order 
calculations (blue triangles) produce results that are very close to the exact $\delta_t$. 
The $\delta_t$ values obtained with the PA (empty orange squares) differ from the second-order approximation 
and turn out to be closer to the exact results.

\begin{figure}[t!]
 \centering
 \subfigure{\label{fig:rhotskyrme}\includegraphics[width=0.49\linewidth]{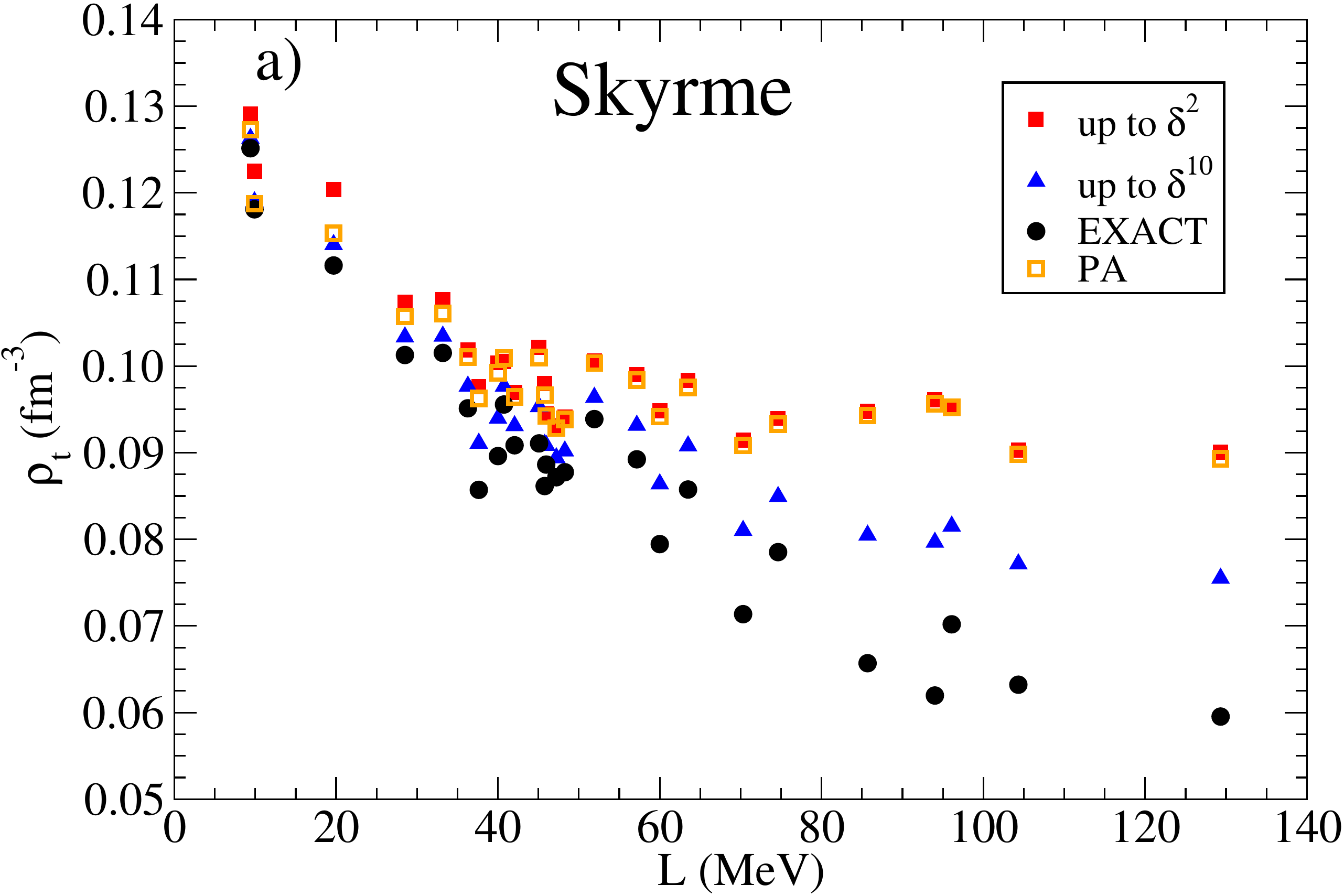}}
\subfigure{\label{fig:rhotgogny}\includegraphics[width=0.49 \linewidth]{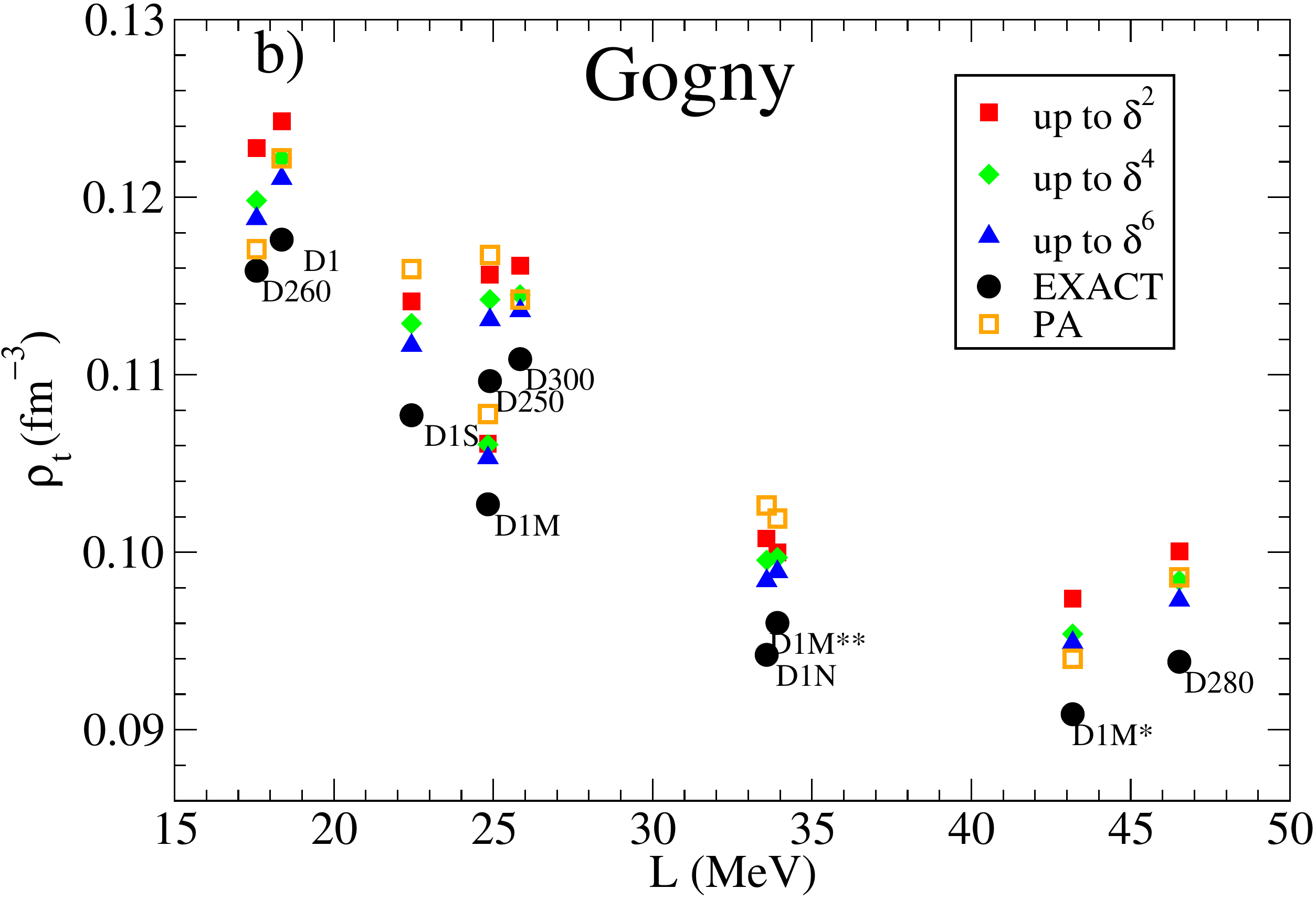}}
  \caption{Core-crust transition density, $\rho_t$, in the thermodynamical approach as a function of the slope parameter $L$
  for a set of Skyrme interactions (panel (a)) and for a set of Gogny forces (panel (b)). Symbols are the same as in Fig.~\ref{fig:deltat}.}
  \label{fig:rhot}
\end{figure}

\begin{figure}[b!]
 \centering
 \subfigure{\label{fig:ptskyrme}\includegraphics[width=0.49\linewidth]{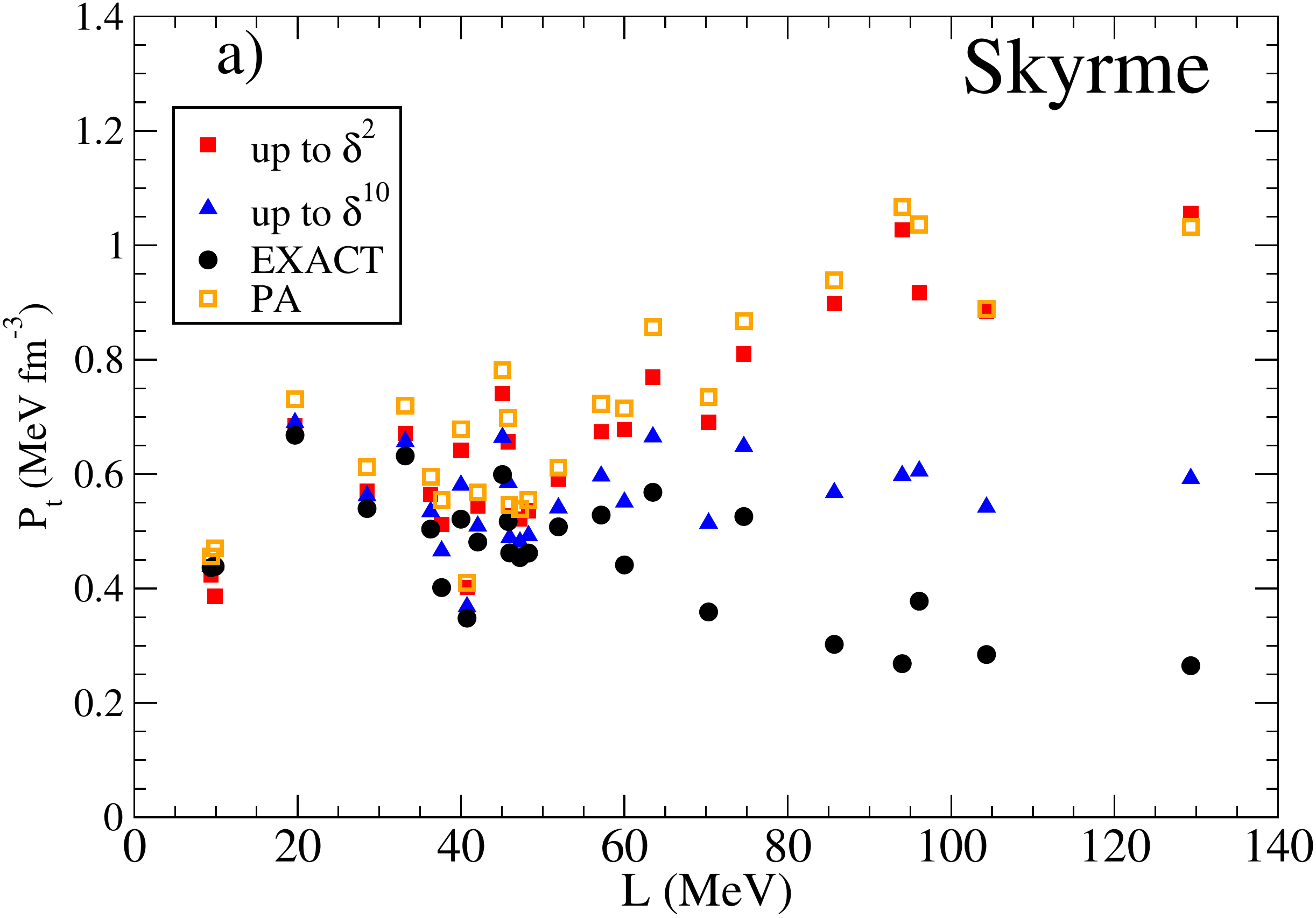}}
\subfigure{\label{fig:ptgogny}\includegraphics[width=0.49 \linewidth]{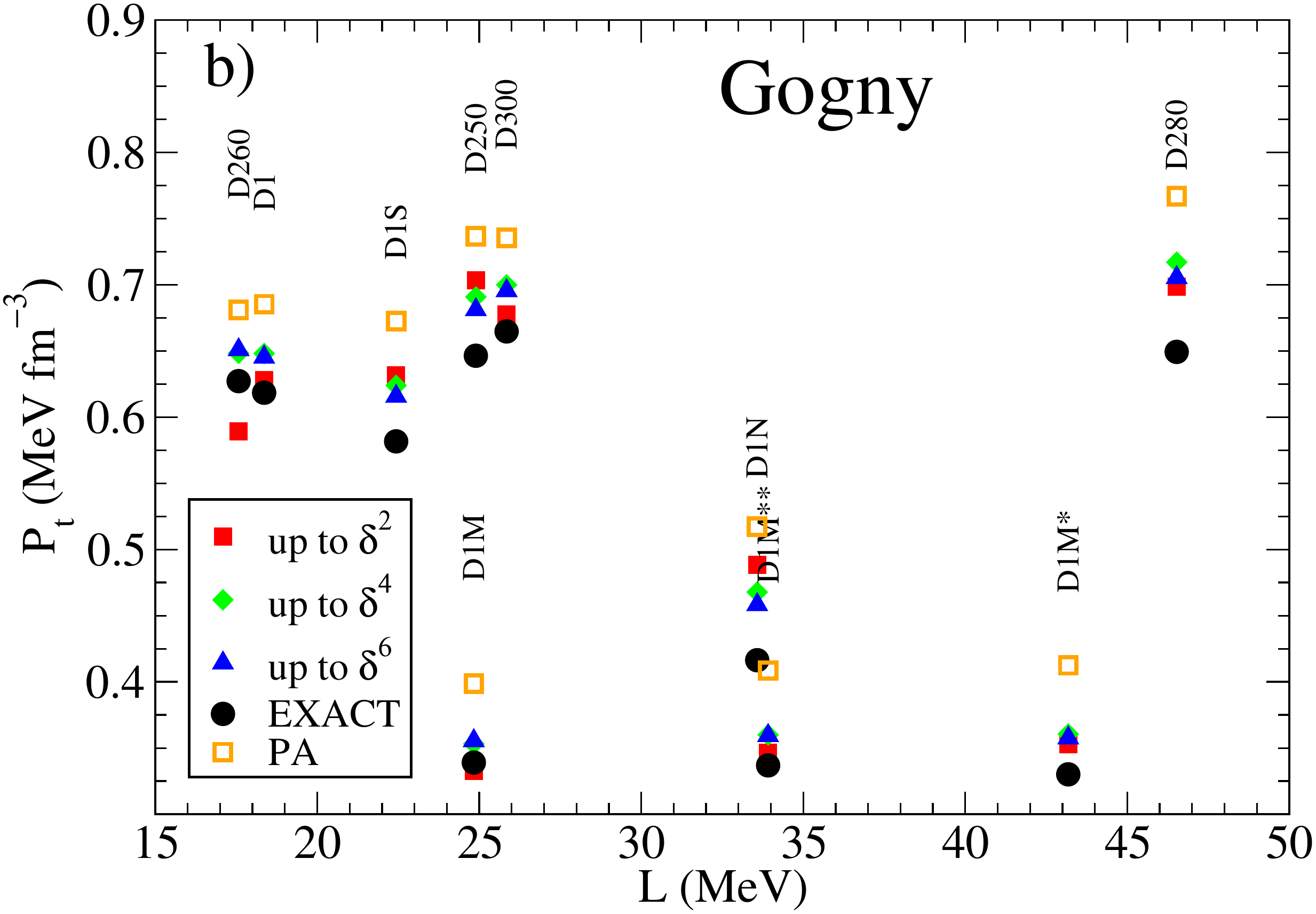}}
   \caption{Core-crust transition pressure, $P_t$, in the thermodynamical approach as a function of the slope parameter $L$
  for a set of Skyrme interactions (panel (a)) and for a set of Gogny forces (panel (b)). Symbols are the same as in Fig.~\ref{fig:deltat}.}
 \label{fig:Pt}
\end{figure}

We show in Fig.~\ref{fig:rhot} the predictions for the density of the core-crust transition, $\rho_t$, for the same of Skyrme 
(Fig.\ref{fig:rhotskyrme}) and Gogny (Fig.~\ref{fig:rhotgogny}) forces~\cite{gonzalez17}. The calculations 
with the exact EoS of the models give a window $0.060 \text{ fm}^{-3} \lesssim \rho_t \lesssim 0.125\text{ fm}^{-3}$  for the Skyrme 
parametrizations and $0.094 \text{ fm}^{-3} \lesssim \rho_t \lesssim 0.118\text{ fm}^{-3}$ for the set of Gogny interactions. 
We find that the approximations of the EoS only provide upper bounds to the exact values. 
For Skyrme interactions, the relative differences between the transition densities predicted using the $\delta^2$ approximation of the EoS
and the exact densities are about $3\%-55 \%$,
and when one uses the EoS expanded up to $\delta^{10}$, the differences are reduced 
to $1\%-30 \%$. This rather large window for the value for the relative 
differences comes from the fact that the approximated values are closer to the ones calculated 
with the full EoS if the slope of the symmetry energy of the interactions is smaller. 
On the other hand, for the Gogny functionals, 
the relative differences between the transition densities predicted using the $\delta^2$ approximation of the EoS
and the exact densities are about $4\%-7 \%$.
When the EoS up to $\delta^4$ is used, the differences are slightly reduced 
to $3\%-6 \%$, and the sixth-order results remain at a similar level of accuracy, within $3\%-5 \%$. 
In other words, the order-by-order convergence for the transition density is very slow.
As mentioned earlier in the discussion of Fig.~\ref{fig:vther}, the non-trivial density and isospin asymmetry 
dependence of the thermodynamical potential arising from the exchange contributions is likely to be the underlying 
cause of this slow convergence pattern. The results for $\rho_t$ of the PA do not exhibit a regular trend with respect 
to the other approximations. In some cases, the PA is the closest approximation to the results obtained with 
the full EoS, like in the cases of the SIII Skyrme interaction or the D260 Gogny parametrization.

We find that there is a decreasing quasi-linear correlation between the transition 
density $\rho_t$ and the slope parameter $L$. In fact, it is known from previous literature that the transition 
densities calculated with Skyrme interactions and RMF models have an anticorrelation with $L$ 
\cite{horowitz01a,xu09a, Moustakidis12,Ducoin11,Providencia14,Fattoyev:2010tb,Pais2016}. We confirm 
this tendency and find that the transition densities calculated with Skyrme and Gogny functionals are in consonance with other 
mean-field models. Moreover, if we take into account the slope parameter of these interactions, the Gogny results are 
within the expected window of values provided by the Skyrme and RMF models \cite{xu09a,Vidana2009}. 

In Fig.~\ref{fig:Pt}, we present the pressure at the transition point, $P_t$, for the same interactions of Figs.~\ref{fig:deltat} and \ref{fig:rhot}. 
The results of the exact Skyrme EoSs (Fig.~\ref{fig:ptskyrme}) lie in the range
$0.269 \text{ MeV fm}^{-3} \lesssim P_t \lesssim 0.668 \text{ MeV fm}^{-3}$ and for the exact 
Gogny EoSs (Fig.~\ref{fig:ptgogny}) they lie in the range $0.339 \text{ MeV fm}^{-3} \lesssim P_t \lesssim 0.665 \text{ MeV fm}^{-3}$ \cite{gonzalez17}.
According to Ref.~\cite{Lattimer01}, in general, the transition pressure for realistic EoSs 
varies over a window $0.25 \text{ MeV fm}^{-3} \lesssim P_t \lesssim 0.65 \text{ MeV fm}^{-3}$. Skyrme and 
Gogny forces,
therefore, seem to deliver reasonable predictions.
If we look at the accuracy of the isospin Taylor expansion of the EoS for predicting $P_t$, we find that the 
second-order approximation gives transition pressures  above the values of the exact EoS in almost 
all of the forces. 
For Skyrme interactions, the differences are of $3\%-54\%$ if the interaction is in the range of $L \lesssim 60$ MeV. 
However, the relative differences can reach up to $\sim 300 \%$ if the slope of the symmetry energy is large. 
These differences reduce to $0.6\%-25\%$ and $\sim 124\%$, respectively, if the EoS expansion up to tenth order is
used.
For Gogny functionals, the differences are of about $2\%-17 \%$ and become $3\%-12 \%$ at fourth order of the expansion, and $4\%-10 \%$ at sixth order. 
On the whole, Fig.~\ref{fig:Pt} shows that the order-by-order convergence for the transition pressure is 
not only slow but actually erratic at times. 
For some parametrizations, like the Gogny D1 or D300 forces, the fourth- and sixth-order predictions for $P_t$ differ more 
from the exact value than if we stop at second order. 
We also see that the PA overestimates the transition pressure for all parametrizations---in fact, the PA 
provides worse predictions for the transition pressure than any of the finite-order approximations.

We note that we do not find a general trend with the slope parameter $L$ in our results for the pressure of 
the transition, i.e., forces with similar $L$ may have quite different pressure values at the border between 
the core and the crust. As in the case of the transition density, the transition pressure has been studied in 
previous literature. However, the predictions on the correlation between the transition pressure and $L$ 
diverge \cite{xu09a,Moustakidis12,Fattoyev:2010tb,PRC90Piekarewicz2014,Moustakidis10}.
In our case, we obtain that the transition pressure is uncorrelated with the slope parameter~$L$. 
The same was concluded in Ref.~\cite{Fattoyev:2010tb} in an analysis with RMF models.

With these values we can conclude that for interactions with soft symmetry energy, if we want to use the
approximations of the EoS, we do not have to go to larger orders than the second, whereas if we are using 
interactions with a stiff symmetry  energy, especially with $L$ larger than $L \gtrsim 60$ MeV, if possible, one should 
perform the calculations using the exact expression of the EoS. 

\section{The dynamical method}\label{Theory_dyn}
We proceed to study the stability of NS matter against the formation of nuclear clusters using the dynamical method. 
As mentioned before, at least at low densities near the transition to the crust, the matter of the NS core is 
composed of neutrons, protons, and electrons. It is globally charge neutral and satisfies the $\beta$-equilibrium 
condition \cite{shapiro83, haensel07}.
Following~\cite{baym71}, we write the particle density as an unperturbed constant part, $\rho_U$, plus a 
position-dependent fluctuating contribution:
\begin{equation} 
\rho({\bf R})= \rho_U + \delta \rho ({\bf R}),
\label{eq6}
\end{equation}
where  ${\bf R} = ({\bf r}+{\bf r'})/2$ is the center of mass coordinate and the small variations are of sinusoidal type, i.e.,
\begin{equation}
\delta \rho_i ({\bf r}) = \int \frac{d{\bf k}}{(2\pi)^3} \delta n_i({\bf k})e^{i {\bf k} \cdot {\bf r}},
\label{density}
\end{equation} 
being $i = {n,p,e}$ and $n({\bf k})$ the density in momentum space.
These fluctuations of the densities fulfill $n_i({\bf k}) = n_i^*(-{\bf k})$ in order to ensure that the variations
 of the particle densities are real. Next, the total energy of the system is expanded up to second order in the variations of 
the densities, which implies
\begin{equation}
E = E_0 + \frac{1}{2} \sum_{i,j} \int \frac{d{\bf k}}{(2\pi)^3}
\frac{\delta^2 E}{\delta n_{i}({\bf k}) \delta n_{j}^*({\bf k})} \delta n_{i}({\bf k}) \delta n_{j}^*({\bf k}),
\label{eq10}
\end{equation}
where $E_0$ is the energy of the uniform phase and the subscripts $i$ and $j$ concern to the different types of particle.
The first-order variation of the energy vanishes due to the particle number conservation for each type of particle.
The second-order variation of the energy, that is, its curvature matrix: 
\begin{equation}
C^f =\frac{\delta^2 E}{\delta n_{i}({\bf k}) \delta n_{j}^*({\bf k})},
\label{eq11}
\end{equation}
can be written as a sum of three matrices as
\begin{eqnarray}
C^f &=& \left( \begin{array}{ccc}
\partial \mu_n / \partial \rho_n & \partial \mu_n / \partial \rho_p & 0 \\
 \partial \mu_p / \partial \rho_n& \partial \mu_p / \partial \rho_p & 0 \\
0 & 0 & \partial \mu_e / \partial \rho_e \end{array} \right)
\nonumber \\ &&
 +  \left( \begin{array}{ccc}
D_{nn}(\rho, k) & D_{np}(\rho, k) & 0 \\
 D_{pn}(\rho,k) & D_{pp}(\rho, k) & 0 \\
0 &  0& 0 \end{array} \right) + \frac{4 \pi e^2}{k^2} \left( \begin{array}{ccc}
0 & 0 &  0 \\
0 & 1 & -1 \\
0 &-1 &  1 \end{array} \right), \nonumber
\\
\label{eq12}
\end{eqnarray}
where it is understood that all quantities are evaluated at the unperturbed density $\rho_U$.
The curvature matrix $C^f$ is composed of three different pieces. The first one, which is the dominant term, corresponds
to the bulk contribution. It defines the stability of uniform NS matter and corresponds to
the equilibrium condition of the thermodynamical method for locating the core-crust transition point. The second piece describes the 
contributions due to the gradient expansion of the energy density functional. Finally, the last piece
 is due to the direct Coulomb interactions of protons and electrons. These last two terms
 tend to stabilize the system reducing the instability region predicted by the bulk contribution alone.

The zero-range Skyrme forces directly provide an energy density functional which 
is expressed as a sum of a homogeneous bulk part, an inhomogeneous term
depending on the gradients of the neutron and proton densities and the direct Coulomb energy.
Therefore, the coefficients $D_{qq'}$ ($q,q'=n,p$) in the surface term, which correspond to the terms of the nuclear energy 
density coming from the momentum-dependent
part of the interaction, which in the case of Skyrme forces reads as \cite{baym71, pethick95,ducoin07}
\begin{equation}
 \mathcal{H}^\nabla= C_{nn} \left( \nabla \rho_n\right)^2 + C_{pp} \left( \nabla \rho_p\right)^2 + 2  C_{np} \nabla \rho_n \cdot  \nabla \rho_p, 
\end{equation}
can be found explicitly from the energy density functional.
For Skyrme interactions, the $D_{qq'}$ coefficients are expressed in terms of the interaction parameters $x_1$, $x_2$, $t_1$ and $t_2$ 
as \cite{sly41}:
 \begin{eqnarray}
     D_{qq'} = k^2 C_{qq'},
 \end{eqnarray}
 where
\begin{eqnarray}
 C_{nn} &=&C_{pp} = \frac{3}{16} \left[ t_1 (1-x_1) - t_2 (1+x_2)\right]\\
 C_{np} &=&C_{pn} = \frac{1}{16} \left[ 3t_1 (2+x_1) - t_2 (2+x_2)\right].
\end{eqnarray} 
Notice that the coefficients $D_{qq'}$ for Skyrme interactions are quadratic functions
of the momentum $k$ with constant coefficients.

With finite-range forces, to study the core-crust transition density using the dynamical method, 
 and obtaining the corresponding $D_{qq'} (\rho, k)$ coefficients, 
 is much more involved than in Skyrme forces. This approach has been discussed, to our 
knowledge, only in the particular case of the MDI interaction in Refs.~\cite{xu09a,xu10b}.
In the dynamical method using finite-range interactions, one needs to extract the gradient 
corrections, which are encoded in the force but do not appear explicitly in their energy 
density functional. 
To solve this problem, the authors in Ref.~\cite{xu09a} adopted the phenomenological approach of approximating the gradient contributions
with constant coefficients whose values are taken as the respective average values of the contributions
provided by 51 Skyrme interactions. 
A step further in the application of the dynamical method to estimate the core-crust 
transition density is discussed in Ref.~\cite{xu10b}, also for the case of MDI
interactions. In this work, the authors use the density matrix (DM) expansion proposed by 
Negele and Vautherin \cite{negele72a, negele72b} to derive a Skyrme-type energy density functional 
including gradient contributions, but with
density-dependent coefficients. Using this functional, the authors study the core-crust transition
through the stability conditions 
provided by the linearized Vlasov equations in NS matter.
 In our case, to generalize the dynamical method to finite-range forces, we perform an expansion of the direct energy in terms of the 
gradients of the nuclear densities which, expressed in momentum space, can be summed up at all orders. 
This exact calculation goes beyond the phenomenological approach of Ref.~\cite{xu09a} to deal with these
corrections. It also goes beyond the calculation of Ref.~\cite{xu10b}, where the gradient expansion expressed
in momentum space was truncated at second order in momentum, therefore remaining in the long-wavelength limit. 
As in Ref.~\cite{xu10b}, we include the effects coming from the exchange energy, not considered explicitly in Ref.~\cite{xu09a},
with the help of a DM expansion. This allows us to obtain a quadratic combination of the gradients of the nuclear
densities, which in momentum space becomes a quadratic function of the momentum. In this thesis we use a DM 
expansion based on the Extended Thomas-Fermi (ETF) approximation introduced in Ref.~\cite{soubbotin00} and applied
to finite nuclei in Refs.~\cite{soubbotin03,krewald06,behera16}. 
This DM expansion allows one to estimate the contribution 
of the kinetic energy due to inhomogeneities of the nuclear densities not taken into consideration in the earlier works.
We have formulated the method in a general form and it can be applied to different effective finite-range interactions, 
such as the Gogny, MDI and SEI interactions used in this work. 
In Appendix~\ref{app_vdyn} we collect the steps of the expansion of the direct part of the interaction in terms of 
the gradients of the nuclear densities. Moreover, we show in the same Appendix~\ref{app_vdyn} the use of the DM expansion 
in the ETF approach when looking for the contributions coming from the exchange and kinetic energies to the 
surface term of the curvature matrix~(\ref{eq12}).

Using Eq.~(\ref{eq6}), the total energy of the system can be expanded as:
\begin{eqnarray}
E &=& E_0 + \int d{\bf R} \left[\frac{\partial {\cal H}}{\partial \rho_{n}}\delta \rho_n +
\frac{\partial {\cal H}}{\partial \rho_{p}}\delta \rho_p + \frac{\partial {\cal H}}{\partial \rho_{e}}\delta \rho_e \right]_{\rho_U}\nonumber\\
&+& \frac{1}{2}\int d{\bf R} \left[\frac{\partial^2 {\cal H}}{\partial \rho_{n}^2}(\delta \rho_n)^2 
+ \frac{\partial^2 {\cal H}}{\partial \rho_{p}^2}(\delta \rho_p)^2 + \frac{\partial^2 {\cal H}}{\partial \rho_{e}^2}
(\delta \rho_e)^2 + 2\frac{\partial^2 {\cal H}}{\partial \rho_{n} \partial \rho_{p}}\delta \rho_n \delta \rho_p \right]_{\rho_U}
\nonumber \\ 
&+& \int d{\bf R}\left[B_{nn}(\rho_{n},\rho_{p})\left({\bf \nabla}\delta \rho_n\right)^2 
+ B_{pp}(\rho_{n},\rho_{p})\left({\bf \nabla}\delta \rho_p\right)^2 \right.\nonumber\\
&+& \left.B_{np}(\rho_{n},\rho_{p}){\bf \nabla}\delta \rho_n\cdot{\bf \nabla}\delta \rho_p 
+ B_{pn}(\rho_{p},\rho_{n}){\bf \nabla}\delta \rho_p\cdot{\bf \nabla}\delta \rho_n \right]_{\rho_U}
\nonumber \\
&+& \int d{\bf R}\left[{\cal H}_{dir}(\delta \rho_n,\delta \rho_p) + {\cal H}_{Coul}(\delta \rho_p,\delta \rho_e)\right],
\label{eq7}
\end{eqnarray}
where $E_0$ contains the contribution to the energy from the unperturbed parts of the neutron, 
proton and electron densities, $\rho_{Un}$, $\rho_{Up}$ and $\rho_{Ue}$, respectively. The subscript 
$\rho_U$ labeling the square brackets in Eq.~(\ref{eq7}) implies that the derivatives of the energy 
density ${\cal H}$ as well as the coefficients $B_{qq}$ are 
evaluated at the unperturbed nucleon and electron densities. The last integral in Eq.~(\ref{eq7}) is the 
contribution from the nuclear direct and Coulomb parts arising out of the fluctuation in 
the particle densities. The two terms of this last integral in Eq.~(\ref{eq7}) are explicitly given by
\begin{equation}\label{Hdir}
 {\cal{H}}_{dir}  (\delta \rho_n, \delta \rho_p) = \frac{1}{2} \sum_q \delta \rho_q ({\bf R}) \int d {\bf s} \left[  \sum_i D_{L,dir}^{i} v_i ({\bf s}) 
 \delta \rho_q ({\bf R} - {\bf s}) + \sum_i D_{U,dir}^i v_i ({\bf s}) \delta \rho_{q'} ({\bf R} - {\bf s}) \right]
\end{equation}
and 
 \begin{equation}
  {\cal{H}}_{Coul}  (\delta \rho_n, \delta \rho_p) = \frac{e^2}{2}  (\delta \rho_p  ({\bf R})- \delta \rho_e  ({\bf R})) \int d {\bf s} 
  \frac{\delta \rho_p ({\bf R} - {\bf s} ) -\delta \rho_e ({\bf R} - {\bf s} ) }{s}.
 \end{equation}
Linear terms in the $\delta\rho$ fluctuation vanish in Eq.~(\ref{eq7}) by the following reason. We are assuming that neutrons, protons and
electrons are in $\beta$-equilibrium. Therefore, the corresponding chemical potentials, defined as
$\mu_i={\partial {\cal H}}/{\partial \rho_{i}}\vert_{\rho_U}$ for each kind of particle ($i=n,p,e$), 
fulfill $\mu_n - \mu_p = \mu_e$. Using this fact, the linear terms in Eq.~(\ref{eq7}) can be written as:
\begin{equation}
\mu_n \delta \rho_n + \mu_p \delta \rho_p + \mu_e \delta \rho_e =
\mu_n( \delta \rho_n + \delta \rho_p ) + \mu_e( \delta \rho_e - \delta \rho_p ).
\label{eq7a}
\end{equation}
The integration of this expression over the space vanishes owing to the charge neutrality of the matter 
(i.e., $\int d{\bf R} (\delta \rho_e - \delta \rho_p)=0$) and to the conservation of the baryon number 
(i.e., $\int d{\bf R} (\delta \rho_n + \delta \rho_p)=0$).

Next, we write the varying particle densities as the Fourier transform of 
the corresponding momentum distributions $\delta n_q({\bf k})$ as \cite{baym71}
\begin{equation}
\delta \rho_q({\bf R}) = \int \frac{d{\bf k}}{(2\pi)^3} \delta n_q({\bf k}) e^{i {\bf k} \cdot {\bf R}}.
\label{eq8}
\end{equation}
One can transform this equation to momentum space due 
to the fact that the fluctuating densities are the only quantities in Eq.~(\ref{eq7}) that depend on the position.
Consider for example the crossed gradient term ${\bf \nabla}\delta \rho_n\cdot{\bf \nabla}\delta \rho_p$
in Eq.~(\ref{eq7}).
Taking into account Eq.~(\ref{eq8}) we can write 
\begin{eqnarray}
\int d{\bf R}{\bf \nabla}\delta \rho_n\cdot{\bf \nabla}\delta \rho_p &=&
-\int \frac{d{\bf k_1}}{(2 \pi)^3}\frac{d{\bf k_2}}{(2 \pi)^3}{\bf k_1}\cdot{\bf k_2} \delta n_n({\bf k_1}) \delta n_p({\bf k_2})
\int d{\bf R}e^{i({\bf k_1}+{\bf k_2})\cdot{\bf R}}\nonumber\\
&=&-\int \frac{d{\bf K}d{\bf k}}{(2 \pi)^3} \left(\frac{{\bf K}}{2}+{\bf k}\right) \cdot \left(\frac{{\bf K}}{2}-{\bf k}\right)
\delta n_n\left(\frac{{\bf K}}{2}+{\bf k}\right) \delta n_p\left(\frac{{\bf K}}{2}-{\bf k}\right)
\delta({\bf K})\nonumber \\
&=& \int \frac{d{\bf k}}{(2 \pi)^3}\delta n_n({\bf k}) \delta n_p(-{\bf k})k^2 =
\int \frac{d{\bf k}}{(2 \pi)^3} \delta n_n({\bf k})\delta n_p^*({\bf k})k^2, 
\label{eqB1}
\end{eqnarray}
where we have used the fact that $\delta \rho_q = \delta\rho^* _q$ and, therefore, due to (\ref{eq8}), 
$\delta n_q(-{\bf k}) = \delta n^*_q({\bf k})$. Similarly,
the other quadratic terms in the fluctuating density in Eq.~(\ref{eq7}) can also be transformed 
into integrals in momentum space of quadratic combinations of fluctuations of the momentum distributions (\ref{eq10}).
After some algebra, one obtains
\begin{eqnarray}
E &=& E_0 +  \frac{1}{2}\int \frac{d{\bf k}}{(2\pi)^3} 
\left\{\left[\frac{\partial \mu_n}{\partial \rho_n}\delta n_n({\bf k})\delta n^*_n({\bf k})
+ \frac{\partial \mu_p}{\partial \rho_p}\delta n_p({\bf k})\delta n^*_p({\bf k})
+ \frac{\partial \mu_n}{\partial \rho_p}\delta n_n({\bf k})\delta n^*_p({\bf k})\right.\right.\nonumber\\
&+&\left. \frac{\partial \mu_p}{\partial \rho_n}\delta n_p({\bf k})\delta n^*_n({\bf k})
+ 
 \frac{\partial \mu_e}{\partial \rho_e}\delta n_e({\bf k})\delta n^*_e({\bf k})\right]_{\rho_0}
\nonumber \\
&+& 2k^2\left[B_{nn}(\rho_{n},\rho_{p})\delta n_n({\bf k})\delta n^*_n({\bf k})
+ B_{pp}(\rho_{n},\rho_{p})\delta n_p({\bf k})\delta n^*_p({\bf k})\right.\nonumber\\
&+&\left. B_{np}(\rho_{n},\rho_{p})\left(\delta n_n({\bf k})\delta n^*_p({\bf k}) 
+ \delta n_p({\bf k})\delta n^*_n({\bf k})\right)\right]_{\rho_0}
\nonumber \\
&+& \sum_i\left[D_{L,dir}^i\left(\delta n_n({\bf k})\delta n^*_n({\bf k}) + \delta n_p({\bf k})\delta n^*_p({\bf k})\right)\right.\nonumber \\
&+&\left. D_{U,dir}^i\left(\delta n_n({\bf k})\delta n^*_p({\bf k})+\delta n_p({\bf k})\delta n^*_n({\bf k})\right)
\right]({\cal F}_i(k)- {\cal F}_i(0)) 
\nonumber \\
&+&\left. \frac{4 \pi e^2}{k^2}\left(\delta n_p({\bf k})\delta n^*_p({\bf k}) + \delta n_e({\bf k})\delta n^*_e({\bf k})
- \delta n_p({\bf k})\delta n^*_e({\bf k}) - \delta n_e({\bf k})\delta n^*_p({\bf k})\right)\right\}.
\label{eq9}
\end{eqnarray}
The factors ${\cal F}_i(k)$ which enter in the contributions of the direct potential 
in Eq.~(\ref{eq9}) are the Fourier transform of the form factors
 $v_i(s)$.
 For Gaussian type interactions, like Gogny forces, the form factor is\footnote{Notice that the notation 
 of the range in the Gaussian form factor for Gogny interactions has changed from $\mu$ in Chapter~\ref{chapter1} to $\alpha$ to not confuse it with the 
 corresponding range parameter of the Yukawa form factors found in the SEI and MDI interactions.}
 \begin{equation}
  v_i(s)= e^{-s^2/\alpha_i^2}
 \end{equation}
 and for Yukawa-type interactions we will have, for SEI forces
 \begin{equation}
  v_i(s) = \frac{e^{-\mu_i s}}{\mu_i s}
 \end{equation}
and for MDI models
 \begin{equation}
  v_i(s) = \frac{e^{-\mu_i s}}{ s}.
 \end{equation}
 The form factors $v_i(s)$ can be expanded in a sum of distributions, which in momentum 
 space can be ressumated as ${\cal F}_i(k)$. They are given by
\begin{equation}
 {\cal F}_i(k)=\pi^{3/2} \alpha_i^3 e^{-\alpha_i^2 k^2/4}
\end{equation}
for Gogny forces and
\begin{equation}
 {\cal F}_i(k)=\frac{4\pi}{\mu_i(\mu_i^2 + k^2)} \hspace{2mm}\mathrm{(SEI)}\hspace{1cm} \mathrm{or} \hspace{1cm} {\cal F}_i(k)=\frac{4\pi}{\mu_i^2 + k^2} \hspace{2mm}\mathrm{(MDI)}
\end{equation}
for Yukawa-type interactions. The full derivation can be found in Appendix~\ref{app_vdyn}. 

The functions $D_{qq'}(\rho,k)$ ($qq'=n,p$) for finite range interactions 
contain the terms of the nuclear energy density coming from the form factor
of the nuclear interaction plus the $\hbar^2$ contributions of the kinetic 
energy and exchange energy densities, see Appendix~\ref{app_vdyn}. They
can be written as:
\begin{small}
\begin{eqnarray}
 D_{nn}(\rho, k)&=& \sum_i D_{L,dir}^i\big({\cal F}_i(k)-{\cal F}_i(0)\big) + 2k^2 B_{nn}(\rho_{n},\rho_{p})\nonumber \\
 D_{pp}(\rho, k)&=& \sum_i D_{L,dir}^i\big({\cal F}_i(k)-{\cal F}_i(0)\big) + 2k^2 B_{pp}(\rho_{n},\rho_{p})\nonumber \\
 D_{np}(\rho, k)&=&  D_{pn}(\rho, k) \nonumber \\
&=&\sum_m D_{U,dir}^m\big({\cal F}_m(k)-{\cal F}_m(0)\big) + 2k^2 B_{np}(\rho_{n},\rho_{p}),\nonumber\\
\label{eq13}
\end{eqnarray}
\end{small}
The stability of the system against small density fluctuations requires that the curvature matrix $C^f$ has to be convex for all values
of $k$, and if this condition is violated, the system becomes unstable. This is guaranteed if the 3$\times$3 determinant of the matrix is positive, provided that 
$\partial \mu_n/\partial \rho_n$ (or $\partial \mu_p/\partial \rho_p$) and the 2$\times$2 minor of the nuclear sector in (\ref{eq12}) 
are also positive \cite{xu09a}:
\begin{eqnarray}
 a_{11}> 0, \hspace{0.3cm} a_{22}> 0,  \hspace{0.3cm}  \left| \begin{array}{cc}
a_{11} & a_{12}  \\
 a_{21}& a_{22}   \end{array} \right| > 0 ,\hspace{0.3cm}\left| \begin{array}{ccc}
a_{11} & a_{12} & a_{13} \\
 a_{21}& a_{22} &a_{23}  \\
a_{31} &a_{32} &a_{33} \end{array} \right| > 0. \nonumber
\\
\label{minors}
\end{eqnarray}
The two first conditions in~(\ref{minors}) demand the system to be stable under fluctuations of proton and neutron densities separately, 
the third minor implies 
the stability against simultaneous modifications of protons and neutrons, and
the last one implies stability under simultaneous proton, neutron and electron density variations.

Therefore, the stability condition against cluster formation, which indicates the transition from
the core to the crust, is given by the condition that the dynamical potential, defined as 
\begin{equation}
 V_{\mathrm{dyn}} (\rho, k) = \left(\frac{\partial \mu_p}{\partial \rho_p} + D_{pp}(\rho, k) + \frac{4 \pi e^2}{k^2}\right) 
 - \frac{\left( \partial \mu_n / \partial \rho_p + D_{np}(\rho, k)\right)^2}{\partial \mu_n / \partial \rho_n + D_{nn}(\rho, k)} 
- \frac{(4\pi e^2 / k^2)^2}{\partial \mu_e / \partial \rho_e + 4 \pi e^2 / k^2},
\label{eq14}
\end{equation}
has to be positive. For a given baryon density $\rho$, the dynamical potential $V_\mathrm{dyn}(\rho, k)$ is calculated at the $k$ value that 
minimizes Eq.~(\ref{eq14}), i.e., that fulfills
$\left.\partial V_\mathrm{dyn}(\rho, k) / \partial k \right|_\rho = 0$ \cite{baym71,ducoin07,xu09a}.
The first term of (\ref{eq14}) gives the stability of protons, and the second and third ones give the
stability when the protons interact with neutrons and electrons, respectively.
From the condition $\left.\partial V_\mathrm{dyn}(\rho, k) / \partial k \right|_\rho =0$ 
one gets the dependence of the momentum on the density, $k (\rho)$,
and therefore Eq.~(\ref{eq14}) becomes a function only depending on the unperturbed density $\rho=\rho_U$. 
The core-crust transition takes place at the density where the dynamical 
potential vanishes, i.e., $V_\mathrm{dyn} (\rho, k(\rho))=0$.

The derivatives of the chemical potentials of neutrons and protons are related with the derivatives of the baryon 
energy per particle by \cite{xu09a}
\begin{eqnarray}
 \frac{\partial \mu_n}{\partial \rho_n} &=& \rho  \frac{\partial^2 E_b}{\partial \rho^2} 
 + 2\frac{\partial E_b}{\partial \rho}
  +  \frac{\partial^2 E_b}{\partial x_p^2}\frac{ x_p^2}{\rho} - 2 x_p  
  \frac{\partial^2 E_b}{\partial \rho \partial x_p}
  \\
 \frac{\partial \mu_n}{\partial \rho_p} &=&  \frac{\partial \mu_p}{\partial \rho_n}=  
 \frac{\partial^2 E_b}{\partial \rho \partial x_p}  (1-2 x_p)
  + \rho \frac{\partial^2 E_b}{\partial \rho^2} + 2 \frac{\partial E_b}{\partial \rho} +
  \frac{\partial^2 E_b}{\partial x_p^2}  \frac{ x_p}{\rho} (x_p -1)\nonumber
  \\
  \frac{\partial \mu_p}{\partial \rho_p} &=& \frac{1}{\rho} \frac{\partial^2 E_b}{\partial x_p^2} 
  (1-x_p)^2 + 2 \frac{\partial^2 E_b}{\partial \rho \partial x_p} (1-x_p) 
   +  \rho  \frac{\partial^2 E_b}{\partial \rho^2} + \frac{\partial E_b}{\partial \rho},
\label{dermu}
 \end{eqnarray}
 where the Coulomb interaction is not considered in nuclear matter and, 
therefore, the cross derivatives are equal $\partial \mu_n / \partial \rho_p =  \partial \mu_p / \partial \rho_n$. 
The explicit expressions of the derivatives of the energy per particle are given in Appendix~\ref{appendix_thermal}.

We conclude this section by mentioning that dropping the Coulomb contributions in Eq.~(\ref{eq14})
and taking the $k\to0$ limit leads to the so-called thermodynamical potential used in the thermodynamical method
for looking for the core-crust transition (see Section~\ref{Theory_thermo}):
\begin{equation}
V_{\mathrm{ther}}(\rho) = \frac{\partial \mu_p}{\partial \rho_p} 
 - \frac{\left( \partial \mu_n / \partial \rho_p \right)^2}{\partial \mu_n / \partial \rho_n } ,
\label{eqB7}
\end{equation}
where the condition of stability of the uniform matter of the core corresponds to a value of the thermodynamical potential 
$V_{\mathrm{ther}}(\rho)~>~0$.

\section{Core-crust transition within the dynamical method}\label{Results_dyn}
\subsection{Core-crust transition properties obtained with Skyrme interactions}
\begin{figure}[t!]
 \centering
 \includegraphics[width=0.8\linewidth, clip=true]{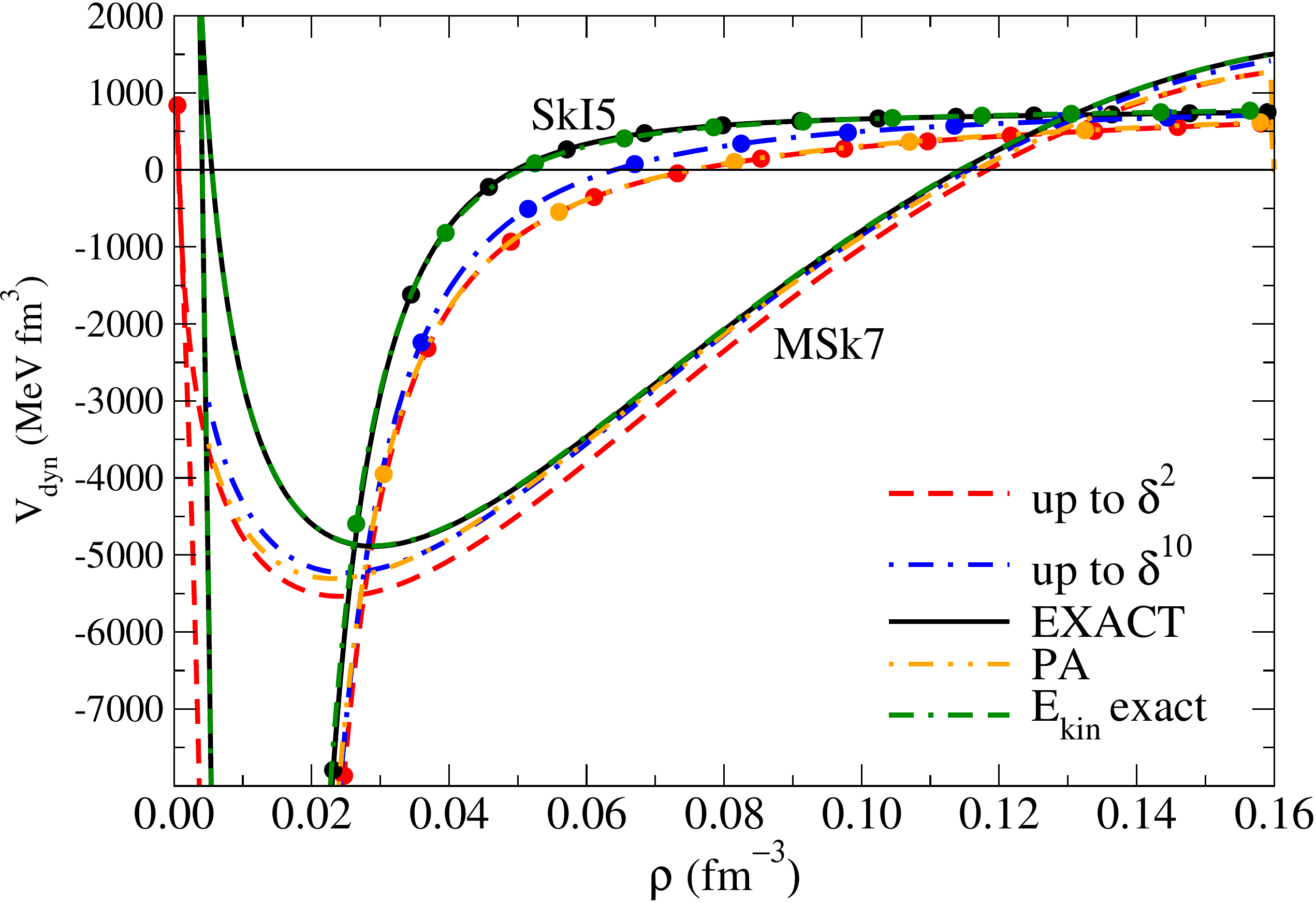}\\
  \caption{ Density dependence of the dynamical potential V$_\mathrm{dyn}$ in $\beta$-stable matter calculated using the exact expression 
of the energy per particle or the expression in Eq.~(\ref{eq:EOSexpgeneral}) up to second and tenth order for two Skyrme forces, MSk7 and SkI5.
The results for the PA and for the case where the expansion up to second order is considered only in the potential part of the interaction 
are also included (label E$_\mathrm{kin}$ exact).}
 \label{fig:dynamic}
\end{figure}
We plot in Fig.~\ref{fig:dynamic}  the density dependence of V$_\mathrm{dyn} (\rho)$ computed with 
the MSk7 ($L=9.41$~MeV) and SkI5 ($L=129.33$ MeV) Skyrme interactions. These two models have the smallest and largest value of the slope parameter of the symmetry energy 
at saturation within the sets of 
Skyrme forces considered, implying that the isovector properties predicted by these two functionals are actually  very different, 
at least around saturation. From Fig.~\ref{fig:dynamic} we see that, in general,  $V_\mathrm{dyn} (\rho)$ as a function of the density shows a 
negative minimum value at $\rho \lesssim 0.04$ fm$^{-3}$, and then increases cutting the $V_\mathrm{dyn} (\rho) = 0$ line at 
a density that corresponds to the transition density, which separates the unstable and stable $npe$ systems.
Taking into account higher-order terms in the expansion~(\ref{eq:EOSexpgeneral}) of the EoS, the corresponding
transition densities approach the value obtained with the full EoS in Eq.~(\ref{eq:Ebanm}). In particular, taking the expansion up to 
$\delta^{10}$, the exact transition density is almost reproduced using the MSk7 interaction. However, this is not the situation 
when the transition density is calculated with the SkI5 force. In this case, although the expansion~(\ref{eq:EOSexpgeneral}) is pushed until 
$\delta^{10}$, the approximate estimate of the transition density overcomes the exact density by more than 0.01 fm$^{-3}$. 
The transition density computed with the PA is similar to the values obtained using the Taylor expansion~(\ref{eq:EOSexpgeneral}) 
up to $\delta^4$ and $\delta^2$ and with the MSk7 and SkI5 forces, respectively.
Fig.~\ref{fig:dynamic} also contains the values of the dynamical potential computed expanding up to 
second order only the potential part of the interaction and using the full expression for the 
kinetic energy.
From Fig.~\ref{fig:dynamic} we notice that $V_\mathrm{dyn} (\rho)$
computed exactly with the  MSk7 force predicts a larger transition density than calculated with the SkI5 interaction. This result suggests
that Skyrme models with small slope parameters tend to cut the  V$_\mathrm{dyn}(\rho) = 0$ line at larger densities than 
the interactions with a larger slope parameter, and, therefore, predict larger core-crust transition densities.  

The exact transition properties, obtained as a solution of V$_\mathrm{dyn} = 0$ with the additional constraint 
$\partial V_\mathrm{dyn}(\rho, k) / \partial k |_\rho = 0$ and calculated through Eqs.~(\ref{eq14})-(\ref{dermu}) using the exact energy 
per particle given by Eq.~(\ref{eq:Ebanm}), are reported in Appendix~\ref{app_taules} for a set of Skyrme interactions available 
in the literature and characterized by a different slope parameter $L$. In the same 
table we also give the isospin asymmetry and the pressure corresponding to the  
transition density calculated assuming $\beta$-stable nuclear matter.

\begin{figure}[t!]
 \centering
 \includegraphics[width=0.8\linewidth, clip=true]{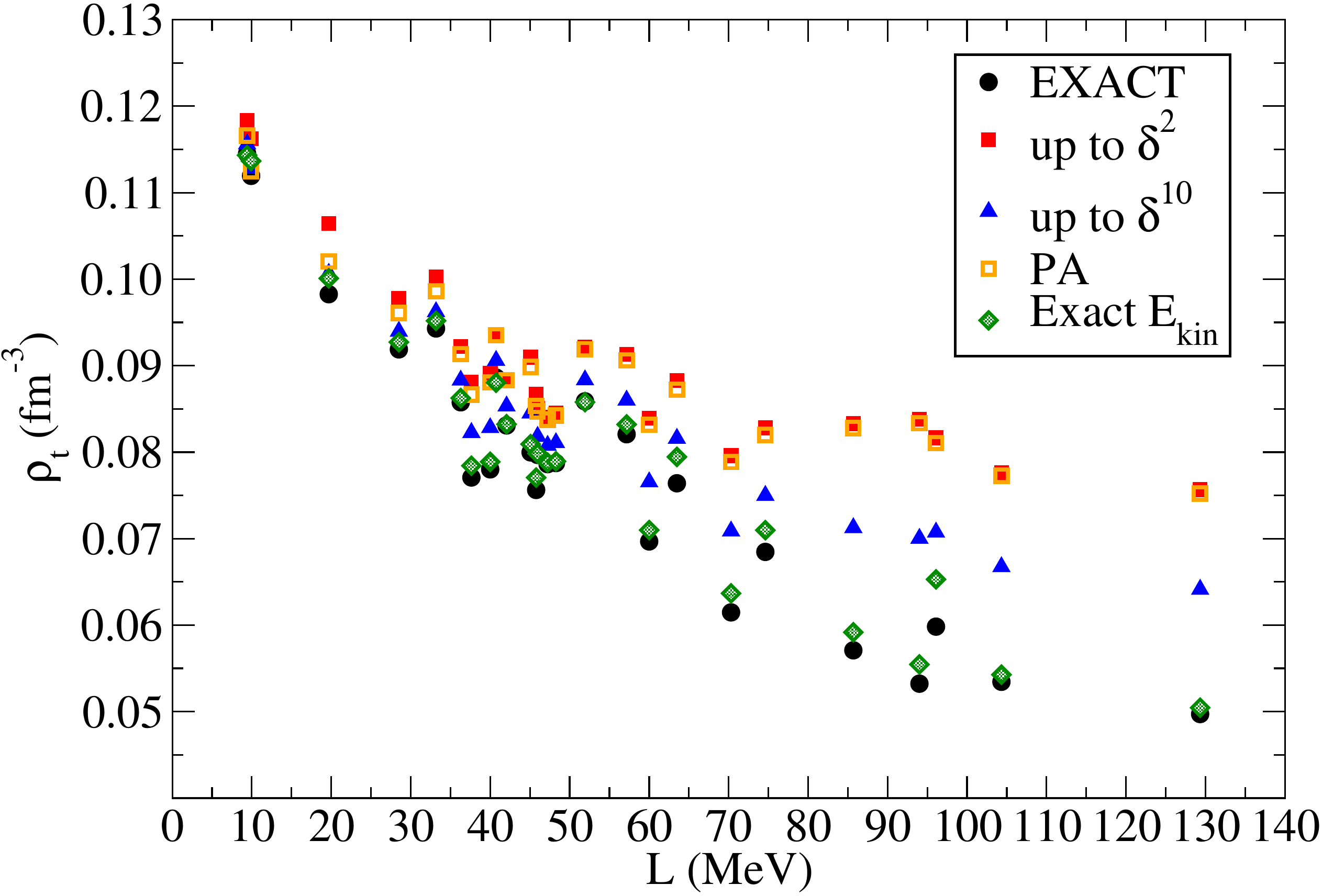}\\
  \caption{Transition density versus the slope parameter $L$ for a set of Skyrme interactions, calculated 
  using the exact expression of the energy per particle and 
  the approximations up to second and tenth order. 
  The results for the PA and for the case where the expansion up to second order is considered only in the potential part of the interaction 
are also included.}
 \label{fig:rhotvdynskyrme}
\end{figure}

Next, we analyze in detail any possible correlation of the transition density, isospin asymmetry and pressure with the slope 
parameter $L$. In Fig.~\ref{fig:rhotvdynskyrme} we plot the transition density as a function of the slope of the symmetry energy for 
the Skyrme interactions reported in Appendix~\ref{app_taules}, calculated using the exact expression of the energy 
per particle as well as with the different approximations of this quantity considered through this work. 
The transition densities follow a quasi-linear decreasing tendency as a function of the slope parameter $L$.
The values of the core-crust transition densities obtained with the full EoS 
lie in the range $0.050$ fm$^{-3}\lesssim \rho_t \lesssim 0.115 $ fm$^{-3}$.
The gap between the asymmetry for the exact results and the approximated ones increases with growing values of the slope parameter $L$. 
We observe that, in general, the exact 
transition densities are better reproduced by the different approximations discussed here if the slope parameter $L$ of the 
interaction is smaller than $L\lesssim60$ MeV. For larger values of $L$ the disagreement between the exact and approximated transition 
densities increases. 
As expected, the agreement between the exact 
and the approximate results improves when more terms of the expansion are considered.   
For example, if we consider the MSk7 interaction ($L$=9.41 MeV), the relative difference between the result calculated with 
the expansion of Eq.~(\ref{eq:EOSexpgeneral}) up to tenth order in $\delta$
and the exact EoS is $\sim 1\%$, while if we consider the SkI5 interaction ($L$ = 129.33 MeV), the value of the relative 
difference is $\sim 30\%$. If the PA is used to estimate the transition densities, the results are quite similar to those calculated 
starting from the second order expansion of the energy per particle. As we have pointed out before, considering 
only the expansion up to send-order only to the potential part of the energy per particle
reproduces very well the exact values of the transition density. 
\begin{figure}[t!]
 \centering
 \includegraphics[width=0.8\linewidth, clip=true]{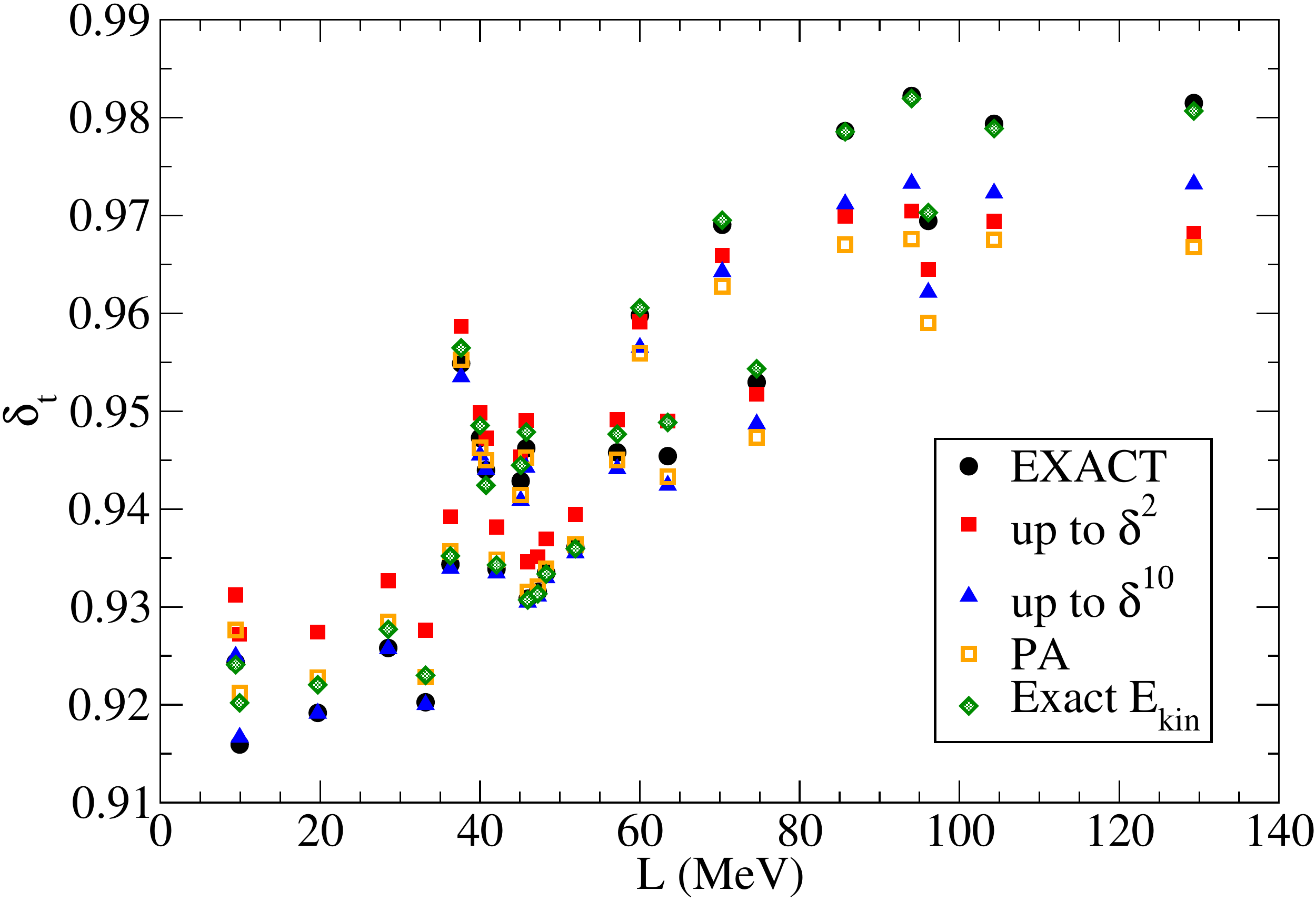}\\
   \caption{Transition asymmetry versus the slope parameter $L$ for a set of Skyrme interactions, calculated 
  using the dynamical method with the exact expression of the energy per particle and 
  the approximations up to second and tenth order. 
  The results for the PA and for the case where the expansion up to second order is considered only in the potential part of the interaction 
are also included.}
 \label{fig:deltatvdynskyrme}
\end{figure}

We have also obtained the asymmetry corresponding to the transition density, i.e., the transition asymmetry, assuming $\beta$-stable matter.
The values calculated from the exact energy per particle with the set of Skyrme forces considered
lie in the range \mbox{$0.916 \lesssim \delta_t \lesssim 0.982$}.
The results of the transition density plotted against the slope of the symmetry energy show
roughly an increasing trend.
We plot in the same figure the transition asymmetries computed with the 
different approximations considered before. 
The relative differences between the values up to second order and the
exact ones are $\lesssim 1\%$ for all the considered Skyrme interactions. Notice that a small difference in the isospin asymmetry can 
develop to large differences in the estimation of other properties, such as the density and the pressure. 
Moreover, the transition asymmetries predicted by the PA are similar to those obtained from the expanded energy per particle
up to the second order in $\delta$. If one expands up to second-order only the potential part of the energy per particle, 
the transition asymmetry obtained with the full EoS is almost reproduced for all the analyzed interactions.   
 \begin{figure}[t!]
 \centering
 \includegraphics[width=0.8\linewidth, clip=true]{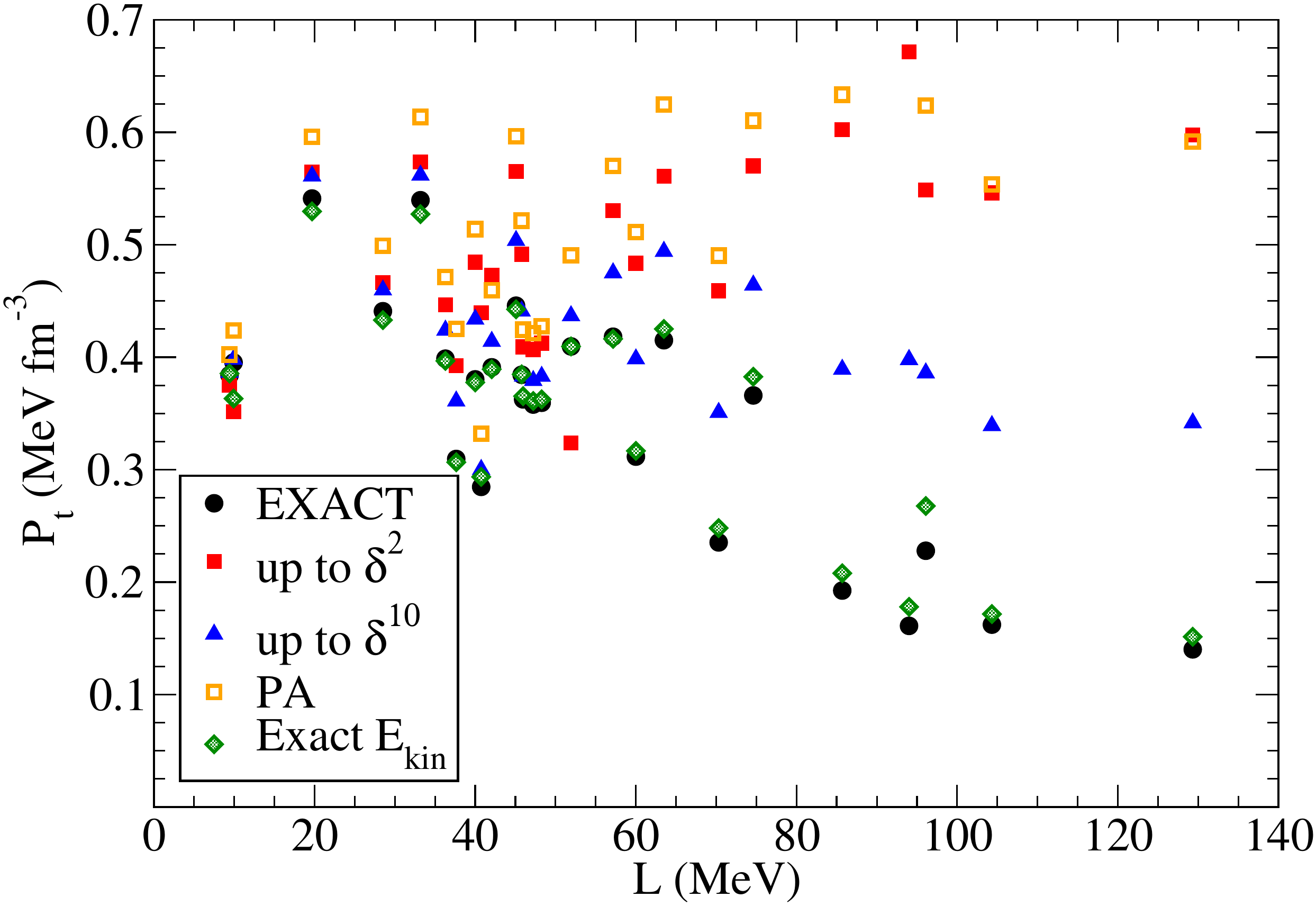}\\
   \caption{Transition pressure versus the slope parameter $L$ for a set of Skyrme interactions, calculated 
  using the dynamical method with the exact expression of the energy per particle and 
  the approximations up to second and tenth order. 
  The results for the PA and for the case where the expansion up to second order is considered only in the potential part of the interaction 
are also included.}
 \label{fig:Ptvdynskyrme}
\end{figure}
In Fig.~\ref{fig:Ptvdynskyrme} we display the core-crust transition pressure as a function of the slope parameter $L$ for the same Skyrme forces.
The transition pressure calculated with the full EoS has a somewhat decreasing trend with $L$.
This correlation is practically destroyed when the pressure is obtained in an approximated way, in particular, if the expansion of the
energy per particle is terminated at second order or the PA are used to this end. As a consequence of the weak correlation between transition pressure
and $L$, forces with similar slope parameters may predict quite different pressures at the transition density. 
The exact values of the transition pressure lie in the range $0.141 $~MeV~fm$^{-3}\lesssim P_t \lesssim 0.541$~MeV~fm$^{-3}$,
decreasing with increasing values of $L$.
As for the transition densities and asymmetries, the differences between the exact pressure and the ones obtained
starting from the different approximations to the energy per particle are nearly inexistent for models with small
slope parameter $L$, but the gaps can become huge for models with $L$ larger than $\sim 60$ MeV.
For example, the relative difference between the exact transition pressure and the estimate obtained using the expansion~(\ref{eq:EOSexpgeneral})
of the energy per particle up to tenth order is only $0.65\%$ computed with the MSk7 interaction, but this relative difference 
becomes $144\%$ calculated with the SkI5 force.
From the same Fig.~\ref{fig:Ptvdynskyrme} we notice that the PA clearly overestimates the transition pressure for all the forces that we 
have considered, giving the largest differences with respect to the exact pressure as compared with the other approximations
to the energy per particle. Once again, the case where only the potential part is expanded 
reproduces very accurately the exact transition pressures
for all the interactions.

\subsection{Core-crust transition properties obtained with finite-range interactions}
We will discuss first the impact of the finite-range terms of the nuclear effective force
on the dynamical potential. Next, we will focus on the study of the core-crust transition density 
and pressure calculated from the dynamical and thermodynamical methods using three different finite-range nuclear 
models, namely Gogny, MDI and SEI interaction. Finally, we will examine the influence of the core-crust transition point on the crustal properties in NSs~\cite{gonzalez19}.

\begin{figure}[t]
\centering
\includegraphics[clip=true,width=0.8\linewidth]{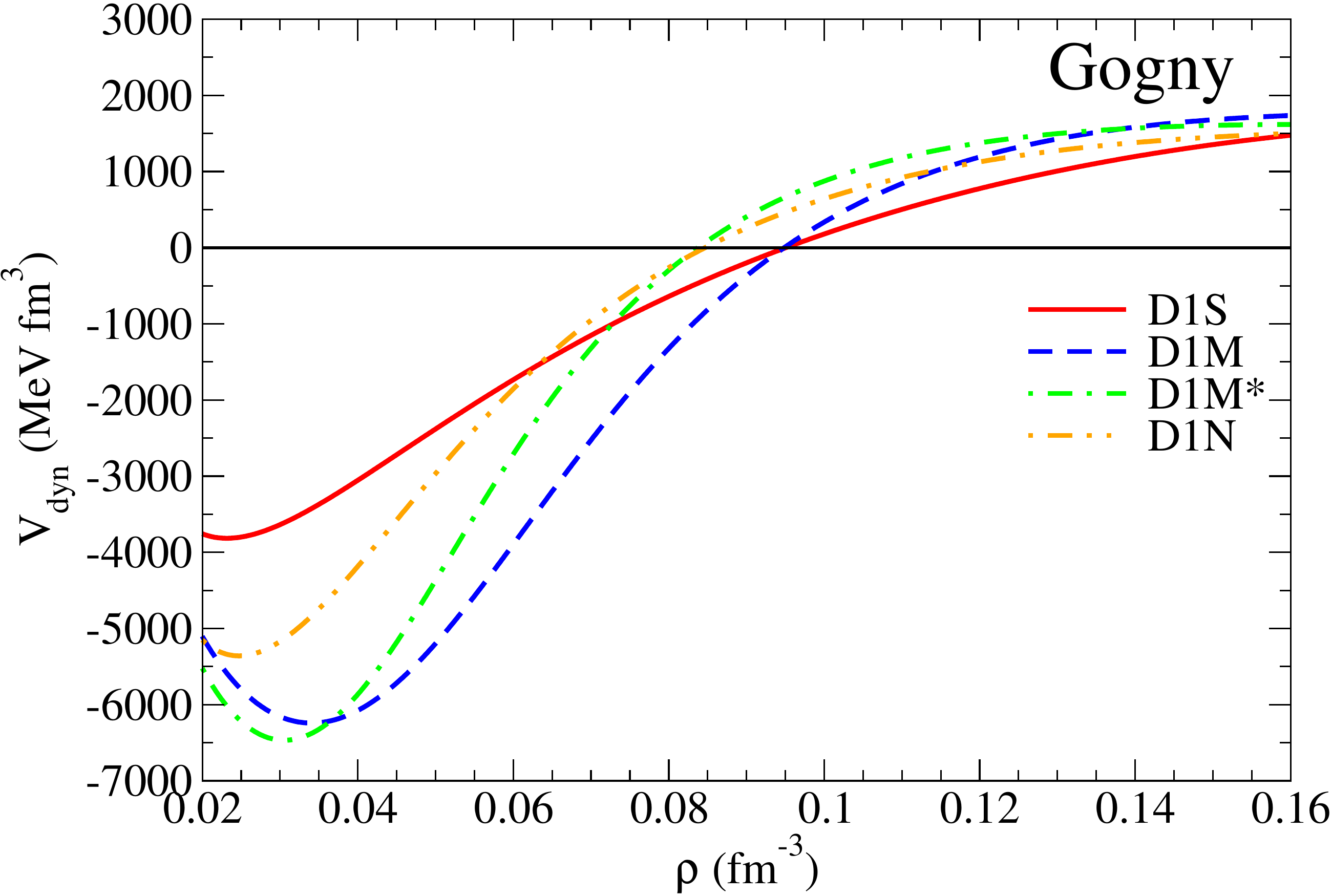}
\caption{Dynamical potential as a function of the density for the D1S, D1M, D1M$^*$ and D1N Gogny interactions. }\label{fig:Vdynfr}
\end{figure}

We show in Fig.~\ref{fig:Vdynfr}~\cite{gonzalez19} the dynamical potential as a function of the unperturbed density computed using the Gogny
forces D1S \cite{berger91}, D1M \cite{goriely09}, D1M$^*$ \cite{gonzalez18} and D1N \cite{chappert08}.
The density at which $V_\mathrm{dyn}(\rho, k(\rho))$ vanishes depicts the transition density. 
Below this point a negative value of $V_\mathrm{dyn}$  implies instability. Above this density, 
the dynamical potential is positive, meaning that the curvature matrix in Eq.~(\ref{eq12}) is convex at all 
values of $k$ and therefore the system is stable against cluster formation.
The Gogny forces D1S and D1M predict larger values of the transition density compared to D1N and D1M$^*$. 
This is due to the different density slope of the symmetry energy predicted 
by these forces.
The values of the slope parameter $L$ of the symmetry energy 
are $22.43$ MeV, $24.83$ MeV, $35.58$ MeV and $43.18$ MeV for the D1S, D1M, D1N and D1M* forces. 
A lower $L$ value implies a softer symmetry energy around the saturation density.
The prediction of higher core-crust
transition density for lower $L$ found in our study is in agreement with the previous 
results reported in the literature (see \cite{xu09a,gonzalez17} and references therein).

As has been done in previous works \cite{baym70,baym71,xu09a}, the dynamical potential (\ref{eq12}) can be
approximated up to order $k^2$ by performing a Taylor expansion of the coefficients $D_{nn}$, $D_{pp}$ and $D_{np}=D_{pn}$ in Eqs.~(\ref{eq13}) \cite{gonzalez19}. 
In this case the dynamical potential can be formally written as \cite{pethick95,ducoin07,xu09a}
\begin{equation}
{\tilde V}_{\mathrm{dyn}}(\rho, k) = V_{\mathrm{ther}}(\rho) + \beta(\rho)k^2 + 
\frac{4\pi e^2}{k^2 + \frac{4\pi e^2}{\partial \mu_e/ \partial \rho_e}},
\label{eq14a}
\end{equation}
where $V_{\mathrm{ther}}(\rho)$ has been defined in Eq.~(\ref{eqB7})
and the expression of $\beta (\rho)$ is given in Eq.~(\ref{eqB8}) of 
Appendix \ref{app_vdyn}. 
The practical advantage of Eq.~(\ref{eq14a}) is that, for finite-range interactions, the $k$-dependence is separated from the $\rho$-dependence,  
whereas in the full expression for the dynamical potential in Eq.~(\ref{eq14}) they are not separated.
\begin{figure}[!t]
\centering
\includegraphics[clip=true, width=0.6\linewidth]{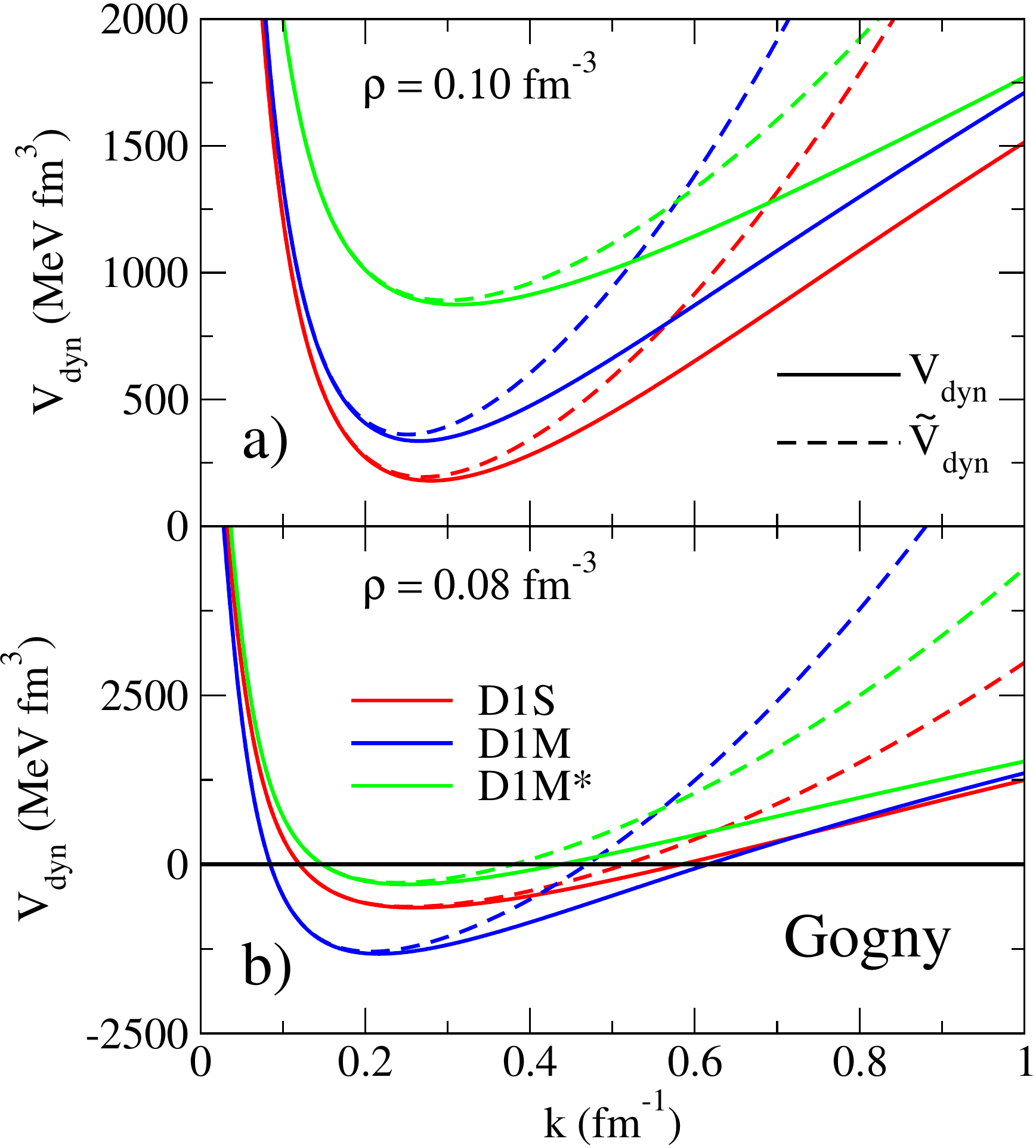}
\caption{Momentum dependence of the dynamical potential at two different densities, $\rho=0.10$ fm$^{-3}$ (panel a)
and $\rho=0.08$ fm$^{-3}$ (panel b), for the D1S, D1M and D1M$^{*}$ Gogny interactions. The results plotted with solid 
lines are obtained using the full expression of the dynamical potential, given in Eq.~(\ref{eq14}), while the dashed lines are 
the results of its $k^2$-approximation, given in Eq.~(\ref{eq14a}).}\label{fig:Vdyncomp}
\end{figure}
\begin{figure}[!t]
\centering
\includegraphics[clip=true, width=0.8\linewidth]{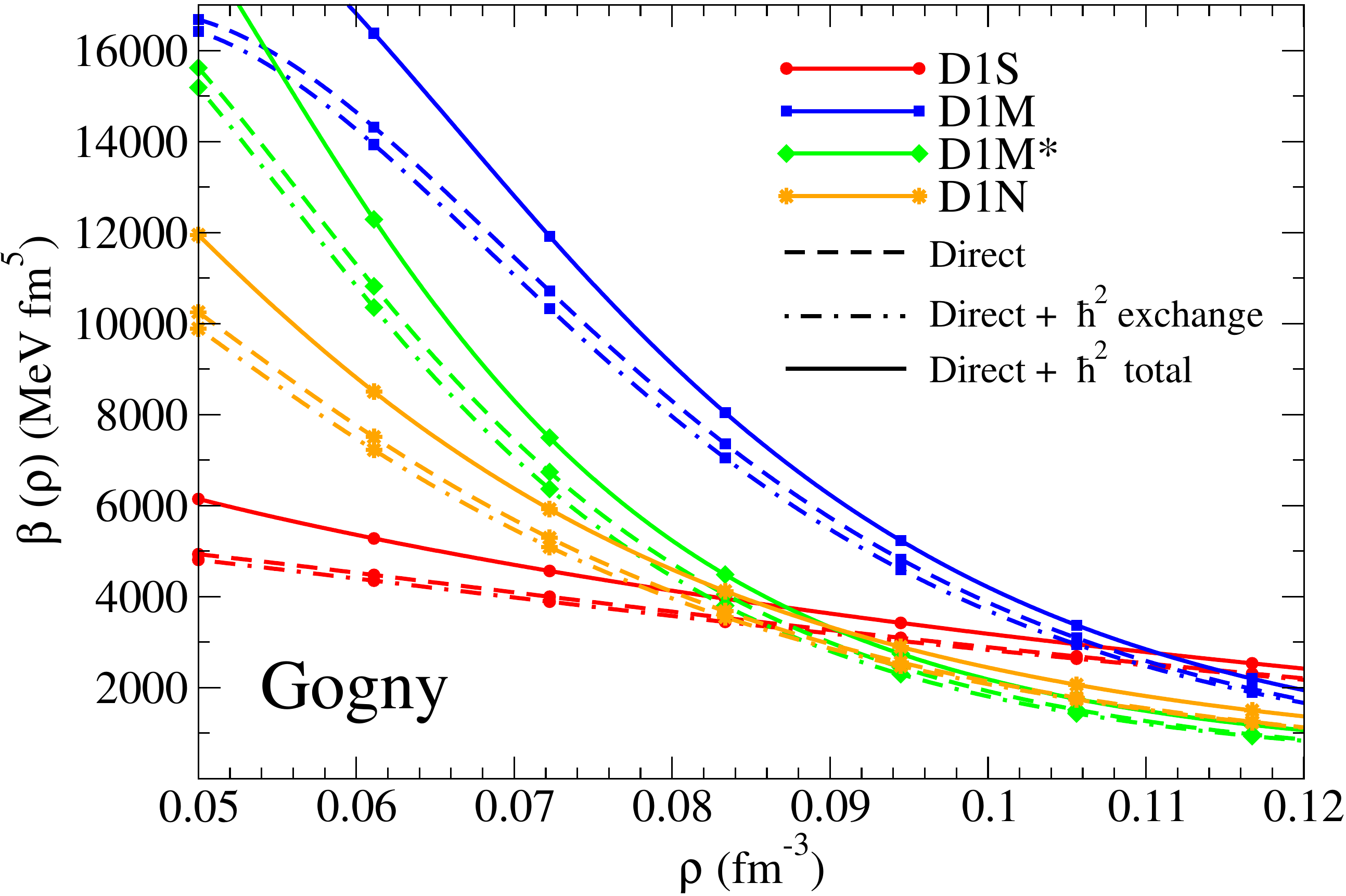}
\caption{Coefficient $\beta(\rho)$ of Eq.~(\ref{eq14a}) as a function of the density for a set of Gogny interactions. The dashed line  
includes only contributions coming from the direct energy, the dash-dotted lines include contributions coming 
from the direct energy and from the from the gradient effects of the exchange energy. 
Finally, the solid lines include all contributions to $\beta(\rho)$.}\label{fig:beta}
\end{figure}
In order to examine the validity of the $k^2$-approximation (long-wavelength limit) of the dynamical potential,
we plot in Fig.~\ref{fig:Vdyncomp} the dynamical potential at a given density as a function of the momentum $k$ 
for both Eq.~(\ref{eq14}) (solid lines)
and its $k^2$-approximation in Eq.~(\ref{eq14a}) (dashed lines) for the Gogny forces D1S, 
D1M and D1M$^*$ \cite{gonzalez19}. 
The momentum dependence 
of the dynamical potential at density $\rho=0.10$ fm$^{-3}$ is shown in the upper panel of Fig.~\ref{fig:Vdyncomp}. 
One sees that at this density the core
has not reached the transition point for any of the considered Gogny forces, as the minima of the 
$ V_{\mathrm{dyn}} (\rho, k)$ curves are positive. This implies that the system is stable against formation of clusters.
In the lower panel of Fig.~\ref{fig:Vdyncomp} the same results but at a density $\rho=0.08$ fm$^{-3}$ are shown.
As the minimum of the dynamical potential for all forces is negative, the matter is unstable against 
cluster formation. From the results in Fig.~\ref{fig:Vdyncomp} it can be seen
that at low values of $k$ the agreement between the results of Eq.~(\ref{eq14}) and its approximation Eq.~(\ref{eq14a}) 
is very good. However, at large momenta beyond $k$ of the minimum of the dynamical potential, 
there are increasingly larger differences between both calculations of the dynamical potential. 
This is consistent with the fact that Eq.~(\ref{eq14a}) is the $k^2$-approximation of Eq.~(\ref{eq14}).

To investigate the impact of the direct and $\hbar^2$ contributions of the finite-range part of the interaction
on the dynamical potential, in Fig.~\ref{fig:beta} we plot for Gogny forces
the behaviour of the coefficient $\beta(\rho)$, which accounts for the finite-size effects
in the stability condition of the core using the dynamical potential ${\tilde V}_{\mathrm{dyn}}(\rho, k)$~\cite{gonzalez19}.
The dashed lines are the result for $\beta (\rho)$ when only the direct contribution from the
finite range is taken into account.
The dash-dotted lines give the value of $\beta(\rho)$ where, along with the direct 
contribution, the gradient effects from the 
exchange energy are included. Finally, solid lines correspond to 
the total value of $\beta(\rho)$, which includes the direct contribution and
the complete $\hbar^2$ corrections, coming from both the exchange and kinetic energies.
 We see that for Gogny forces the gradient correction due to the 
$\hbar^2$ expansion of the exchange energy reduces the result of $\beta (\rho)$ from the direct contribution, at most, by 5\%. When the  
gradient corrections from both the exchange and kinetic energies are taken together, 
the total result for $\beta (\rho)$ increases by 10\% at most with respect to the direct contribution. The effects of the 
kinetic energy $\hbar^2$ corrections are around three times larger than the effects from the exchange energy $\hbar^2$ corrections 
and go in the opposite sense. 
Therefore, we conclude that the coefficient $\beta(\rho)$ is governed dominantly by the gradient expansion of the
direct energy (see Appendix \ref{app_vdyn}), whereas the contributions from the $\hbar^2$ corrections of the 
exchange and kinetic energies are small corrections. 
  
\begin{figure}[!b]
\centering
\includegraphics[clip=true,width=0.8\linewidth]{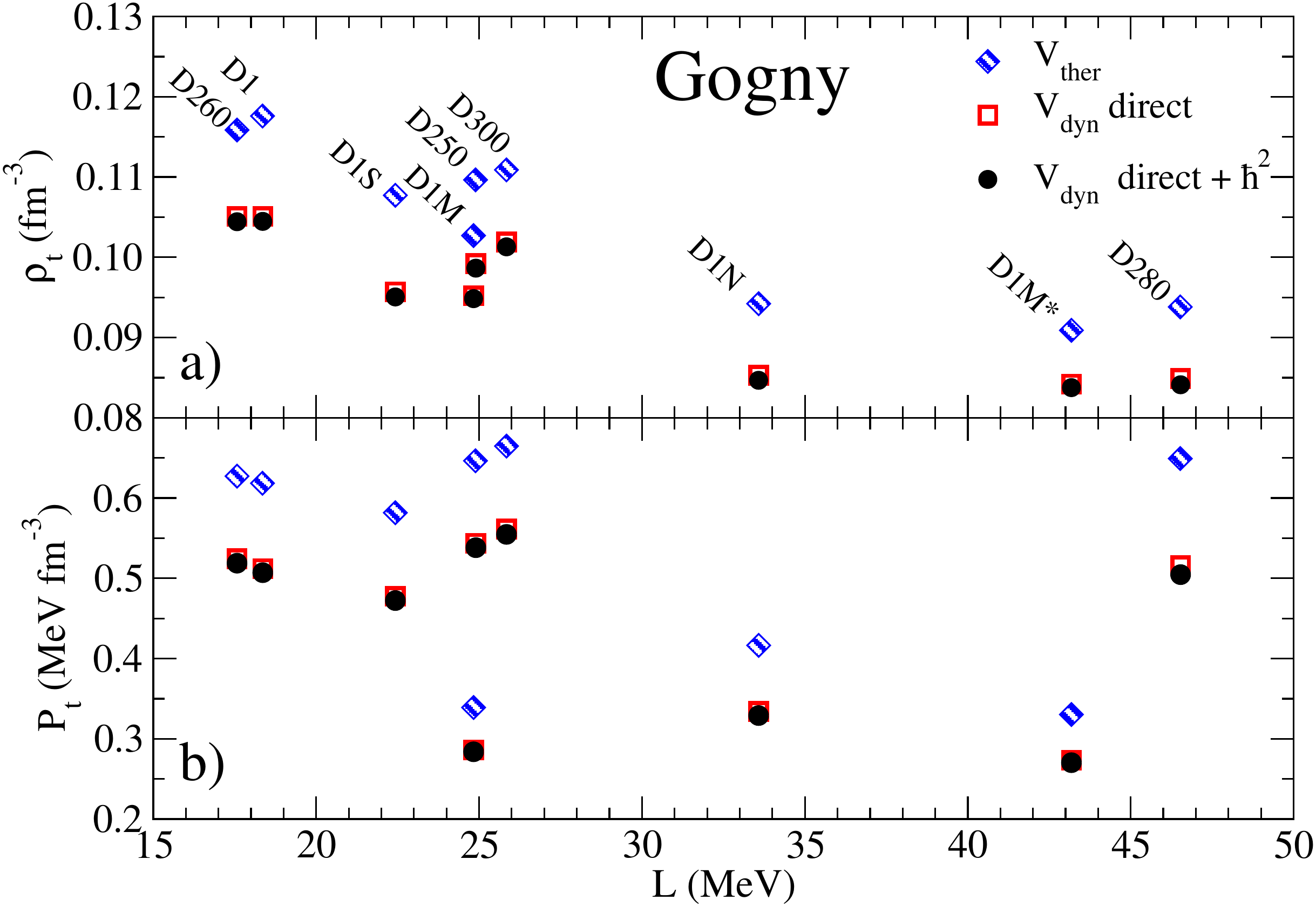}
\caption{Transition density (panel a) and transition pressure (panel b) as a function of the slope of the symmetry 
energy calculated using the thermodynamical and the dynamical methods for 
a set of different Gogny interactions.\label{fig:rhotPtGogny}}
\end{figure}

\begin{figure}[t]
\centering
\includegraphics[clip=true, width=0.8\linewidth]{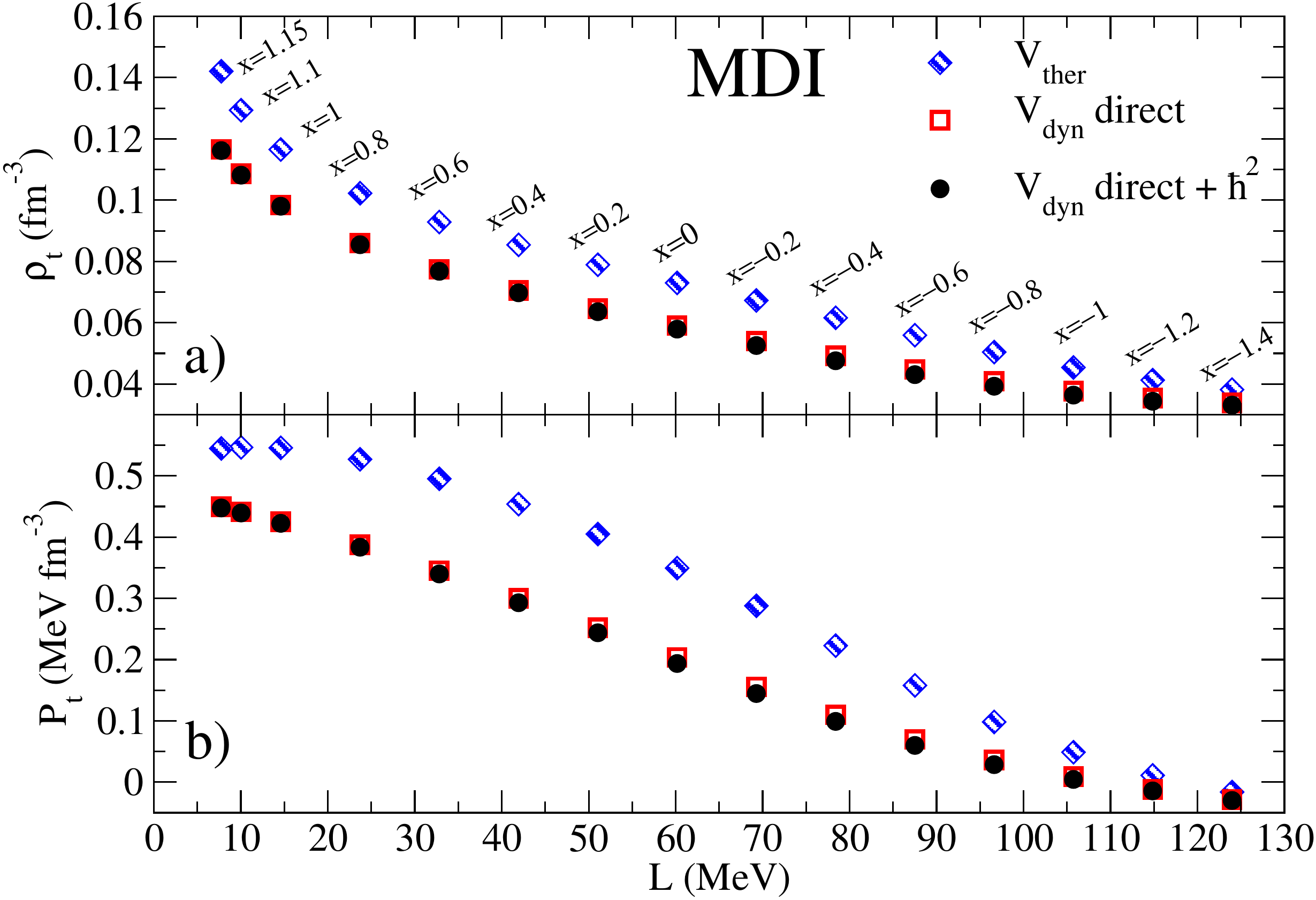}
\caption{Same as Fig.~\ref{fig:rhotPtGogny} but for a family of MDI interactions.\label{fig:rhotPtMDI}}
\end{figure}

\begin{figure}[t]
\centering
\includegraphics[clip=true, width=0.8\linewidth]{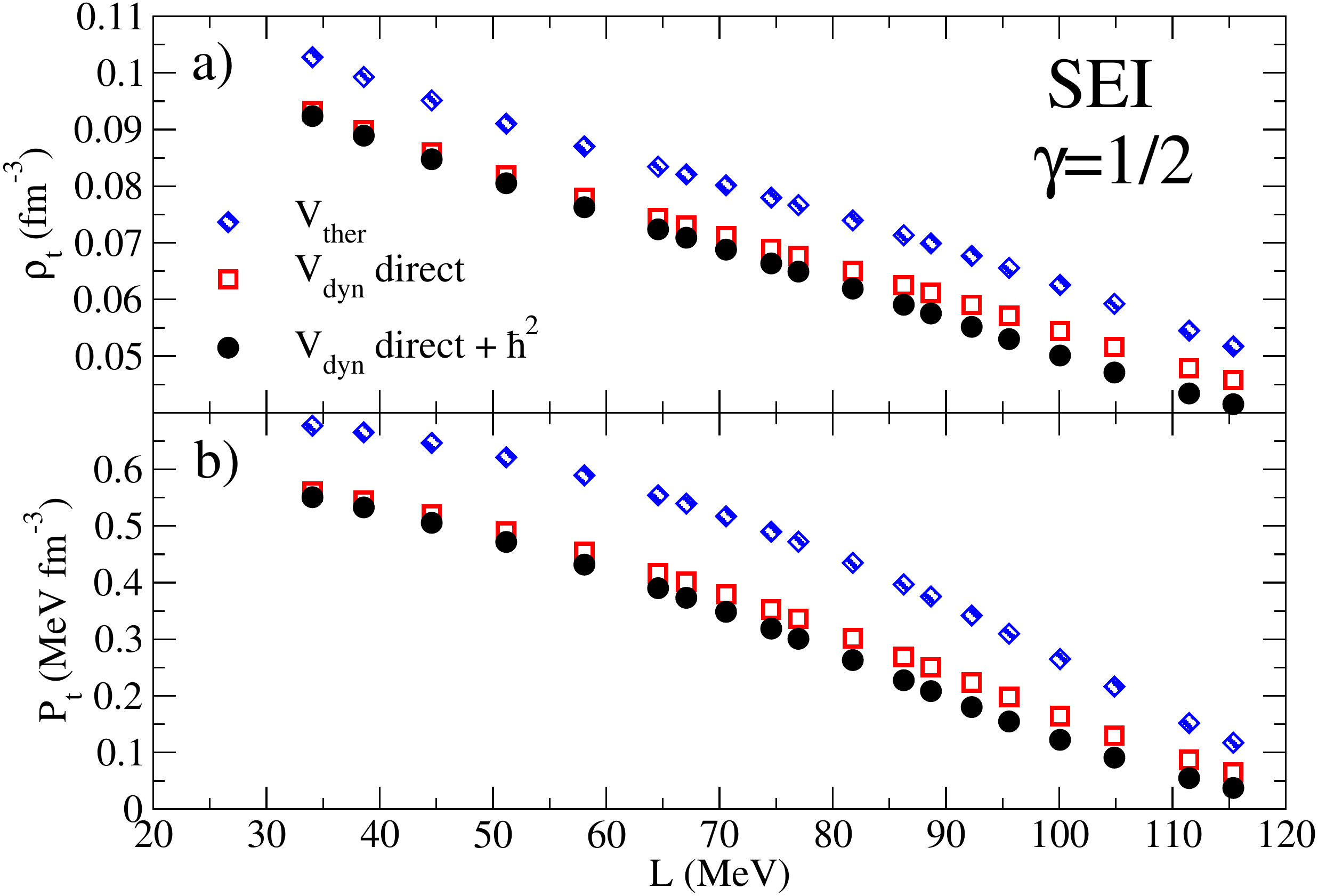}
\caption{Same as Fig.~\ref{fig:rhotPtGogny} but for a family of SEI interactions of $\gamma=1/2$ ($K_0=237.5$ MeV).\label{fig:rhotPtSEI}}
\end{figure}

\begin{figure}[b!]
\centering
\includegraphics[clip=true, width=0.8\linewidth]{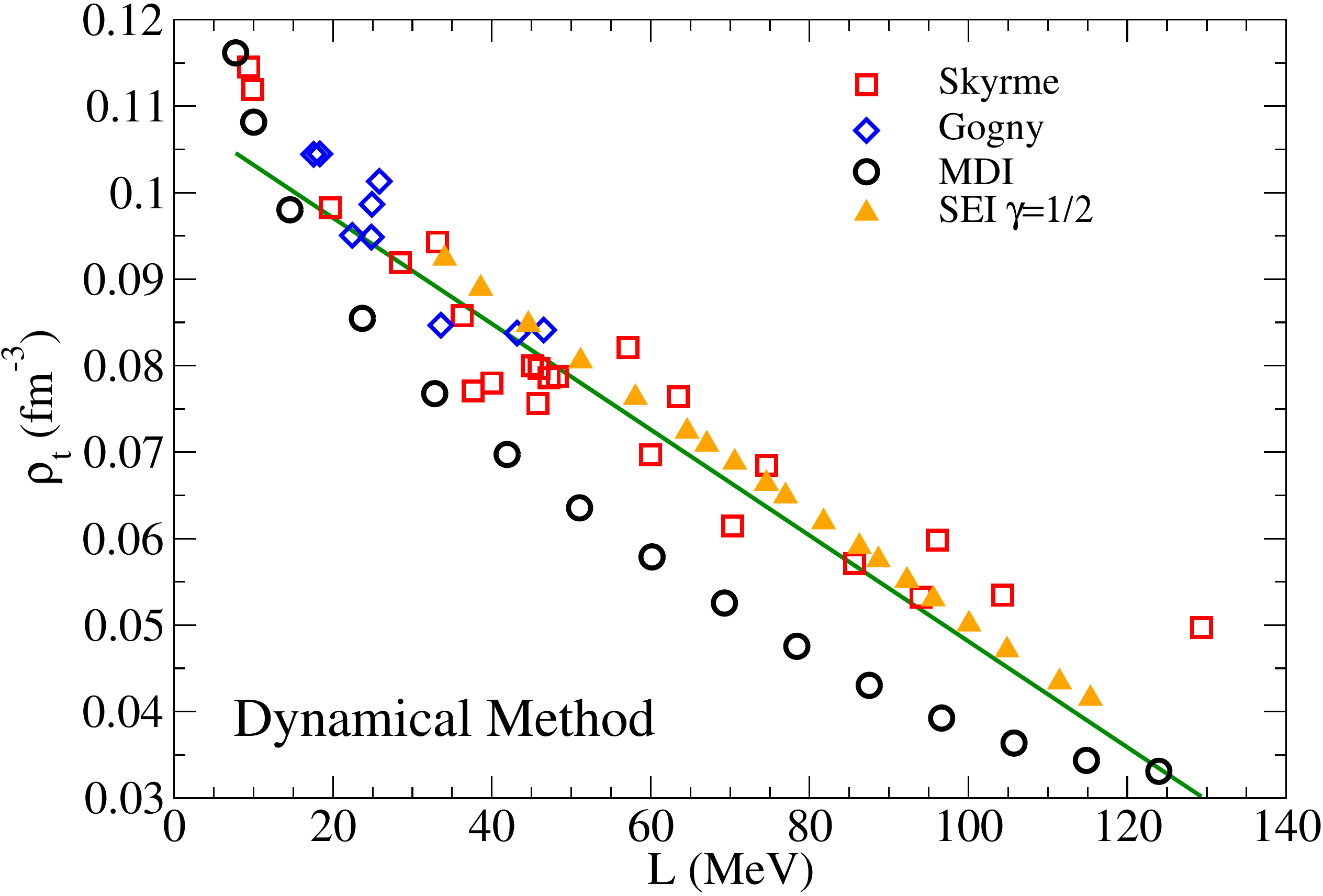}
\caption{Transition density as a function of the slope of the symmetry energy calculated using the dynamical method for 
a set of Skyrme, Gogny, MDI and SEI interactions.\label{fig:rhotall}}
\end{figure}
\begin{figure}[t!]
\centering
\includegraphics[clip=true,width=0.8\linewidth]{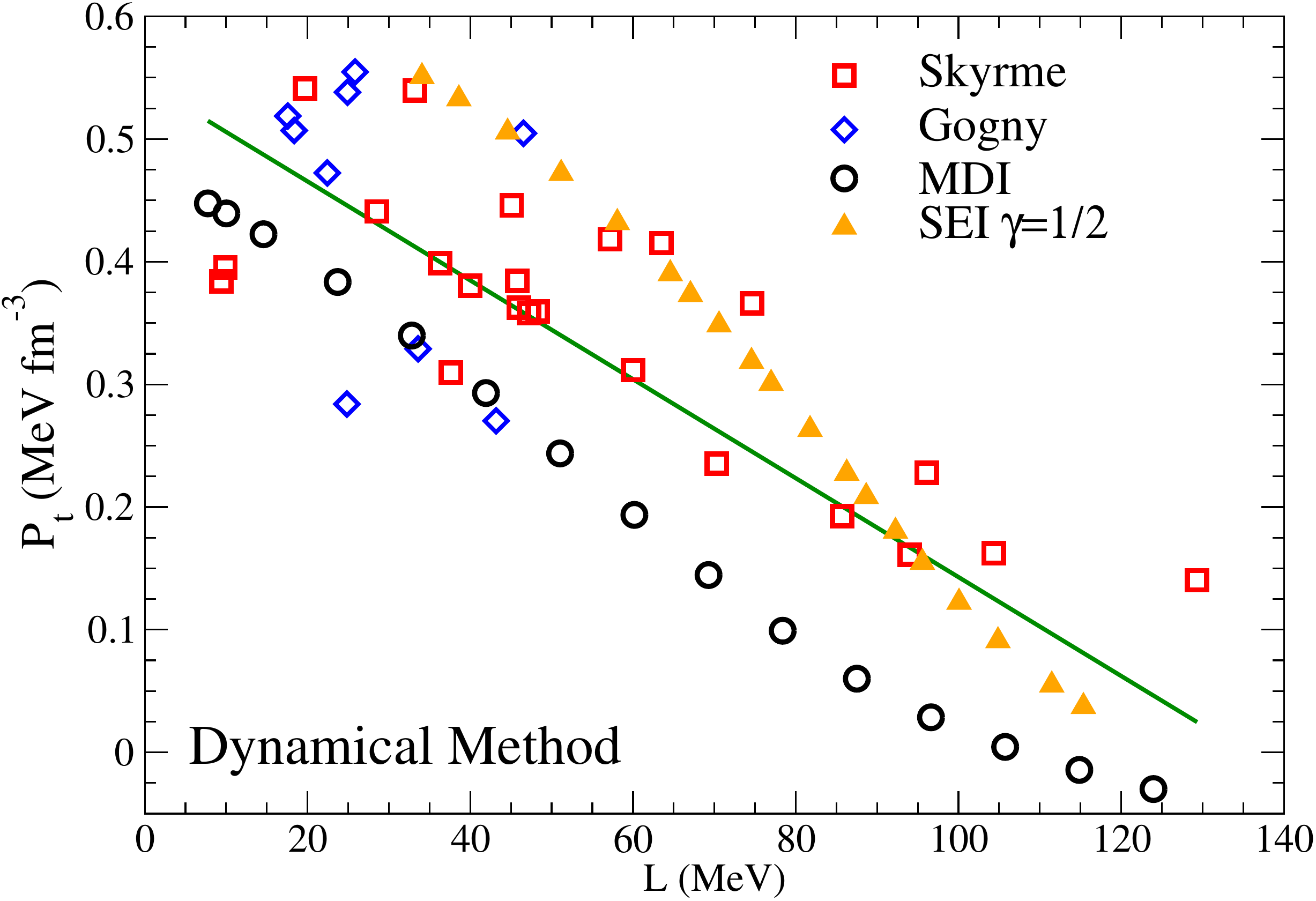}
\caption{Transition density as a function of the slope of the symmetry energy calculated using the dynamical method for 
a set of Skyrme, Gogny, MDI and SEI interactions.\label{fig:Ptall}}
\end{figure}

In Figs.~\ref{fig:rhotPtGogny}, \ref{fig:rhotPtMDI} and \ref{fig:rhotPtSEI} we display, as a 
function of the slope of the symmetry energy $L$,
the transition density (upper panels) and the transition pressure (lower panels) calculated with the 
thermodynamical ($V_\mathrm{ther}$) and the dynamical ($V_\mathrm{dyn}$) methods for three different types of finite-range interactions~\cite{gonzalez19}. 
The expression (\ref{eq14}) with the complete momentum dependence of the dynamical potential has been used for the dynamical calculations.
In the study of the transition properties we have employed several Gogny interactions~\cite{decharge80,Sellahewa14,gonzalez17,gonzalez18} 
(Fig.~\ref{fig:rhotPtGogny}), a 
family of MDI forces \cite{das03,xu10b}, with $L$ from $7.8$ to $124.0$ MeV (corresponding to
values of the $x$ parameter ranging from
$1.15$ to $-1.4$) (Fig.~\ref{fig:rhotPtMDI}), and a family of SEI interactions \cite{behera98} with $\gamma=1/2$ and different 
values of the slope parameter $L$ (Fig.~\ref{fig:rhotPtSEI}). 
Notice that the Gogny interactions have different nuclear matter saturation properties, 
whereas the parametrizations of the MDI or SEI families displayed in Figs.~\ref{fig:rhotPtMDI} and \ref{fig:rhotPtSEI} have the
same nuclear matter saturation properties within each family.
In particular, all parametrizations of the MDI family considered here have the same nuclear matter incompressibility  
$K_0=212.6$ MeV, whereas the SEI parametrizations with $\gamma=1/2$ have $K_0= 237.5$ MeV.
Consistently with previous investigations 
(see e.g. \cite{xu09a, xu10b} and references therein) we find that for a given parametrization, the transition density 
obtained with the dynamical method is 
smaller than the prediction of the thermodynamical approach, as the surface and Coulomb contributions tend to 
further stabilize the uniform matter in the core against the formation of clusters. The relative differences between the 
transition densities obtained with the thermodynamical and dynamical approaches are found to vary within the ranges
$8-14 \%$ for Gogny interactions, 
$15-30 \%$ for MDI, and $10-25 \%$ for SEI $\gamma=1/2$. This clearly points out that important differences arise
in the predicted core-crust transition point between the dynamical method and the simpler thermodynamical method.
For the MDI interactions, the trends of the transition density with $L$ are comparable to those found in Ref.~\cite{xu09a} 
where the finite-size effects in the dynamical calculation
were taken into account phenomenologically through assumed constant values of 132 MeV fm$^{5}$
for the $D_{nn}$, $D_{pp}$ and $D_{np}=D_{pn}$ coefficients.
 The transition densities calculated with $D_{nn}=D_{pp}=D_{np}=132$ MeV fm$^{5}$ lie between the results
 obtained with the thermodynamical method and
those obtained using the dynamical approach in Fig.~\ref{fig:rhotPtMDI}. As far as the results for the transition pressure 
are concerned, from Figs.~\ref{fig:rhotPtGogny}--\ref{fig:rhotPtSEI} it is evident again that the dynamical 
calculation predicts smaller transition pressures than its thermodynamical counterpart, which is also consistent 
with the findings in earlier works (see \cite{xu09a, xu10b} and references therein).

It is to be observed that we have performed the dynamical calculations of the transition properties in two different 
ways. On the one hand, we have considered the contributions to Eqs.~(\ref{eq13}) from the direct energy only 
(the results are labeled as ``$V_\mathrm{dyn}$ direct'' in Figs.~\ref{fig:rhotPtGogny}--\ref{fig:rhotPtSEI}).
This approximation corresponds to neglecting the terms $B_{nn}$, $B_{pp}$ and $B_{np}$ in Eqs.~(\ref{eq13}) that 
define the $D_{qq'}$ coefficients.
Then, we have used the complete expression of Eqs.~(\ref{eq13}) (the results are labeled as ``$V_\mathrm{dyn}$ 
direct$+ \hbar^2$'' in Figs.~\ref{fig:rhotPtGogny}--\ref{fig:rhotPtSEI}).
As with our findings for the dynamical potential in the previous subsection, we see that for the three types of 
forces the finite-range 
effects on $\rho_t$ and $P_t$ coming from the $\hbar^2$ corrections of the exchange and kinetic energies are 
almost negligible
compared to the effects due to the direct energy.

Figs.~\ref{fig:rhotPtGogny}--\ref{fig:rhotPtSEI} also provide information about the 
dependence of the transition properties with the slope of the symmetry energy. 
We can see that within the MDI and SEI families the transition density and pressure
show a clear, nearly linear decreasing trend as a function of the slope parameter $L$. In contrast, the results
of the several Gogny forces in Fig.~\ref{fig:rhotPtGogny} show a weak trend with $L$ for $\rho_t$ and almost 
no trend with $L$ for $P_t$ (as also happened with the different Skyrme forces in the previous subsection).
A more global analysis of the eventual dependence of the transition properties with the slope of 
the symmetry energy is displayed in Figs.~\ref{fig:rhotall} and \ref{fig:Ptall}. These figures include not only 
the results for the core-crust transition 
calculated using the considered sets of finite-range interactions (Gogny, MDI and SEI of $\gamma=1/2$ interactions) 
but also the values for the Skyrme interactions found in the previous section.
Using this 
large set of nuclear models of different nature, it can be seen that the 
decreasing trend of the transition density and pressure with a rising in the slope $L$ of the symmetry energy
is a general feature.
However, the  correlation of the transition properties with the slope parameter $L$ using all interactions is
weaker than within a family of parametrizations where the saturation properties do not change. 
The correlation in the case of the transition density is found to be slightly better than in the case of
the transition pressure. The reason for the weaker correlation obtained in the
different Gogny sets (Fig.~\ref{fig:rhotPtGogny}) and Skyrme sets (Figs.~\ref{fig:rhotall} and \ref{fig:Ptall}) 
is attributed to the fact that these sets have different nuclear matter saturation properties apart from the $L$ values.
Further, we have also examined the correlation of the transition density and transition pressure with the curvature of the 
symmetry energy ($K_\mathrm{sym}$ parameter)
for the models used in this work. The transition density and pressure show a decreasing trend with increasing 
$K_\mathrm{sym}$, in agreement with the findings in earlier works \cite{xu09a, Carreau19}. 
The correlations of the transition density and pressure with the value of $K_\mathrm{sym}$ have a similar quality 
to the case with the $L$ parameter. 

\begin{figure}[!t]
\centering
\includegraphics[clip=true, width=0.8\linewidth]{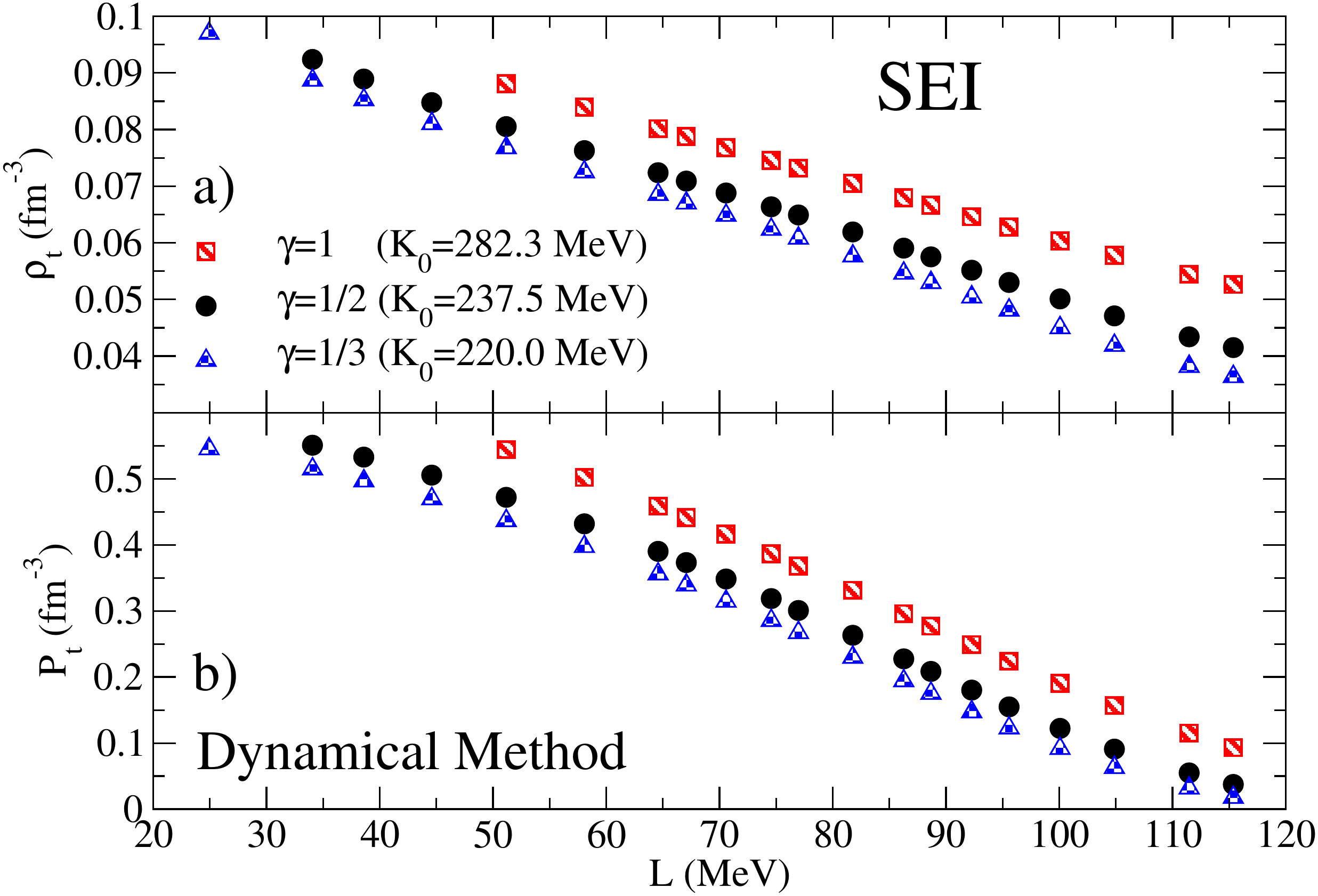}
\caption{Transition density (panel a) and transition pressure (panel b) as a function of the slope of the symmetry energy,
calculated using the dynamical method for three SEI families with different nuclear matter incompressibilities.}\label{fig:SEIK0}
\end{figure}

In order to test the impact of the nuclear matter incompressibility on the correlations
between the core-crust transition properties and the slope of the symmetry energy 
for a given type of finite-range interactions, we plot 
in Fig.~\ref{fig:SEIK0} the transition density (panel a) and the transition pressure (panel b) against the $L$ parameter for 
SEI interactions of $\gamma=1/3$, $\gamma=1/2$ and $\gamma=1$, which correspond to nuclear matter incompressibilities $K_0$ of $220.0$ MeV,
$237.5$ MeV and $282.3$ MeV, respectively, covering an extended 
range of $K_0$ values~\cite{gonzalez19}.
From this figure, we observe a high correlation with $L$ for the different sets of a given SEI force having the same incompressibility
and differing only in the $L$ values. The comparison of the results
obtained with the SEI forces of different incompressibility shows that higher transition density and pressure are predicted for the 
force sets with higher incompressibility.
This also demonstrates, through the example of the nuclear matter incompressibility, a dependence of the features of 
the core-crust transition with the nuclear matter saturation properties of the force,
on top of the dependence on the stiffness of the symmetry energy.

\chapter{Neutron star properties}\label{chapter5}  
This chapter is devoted to the study of different properties of neutron stars (NSs). First of all, we analyze how the choice of the EoS in the inner 
layer of the NS crust can affect the results for the NS mass and radius. 
The EoS in the inner crust is not as widely studied as in the case of the core, due to the 
difficulty of computing the neutron gas and the presence of nuclear clusters that may adopt non-spherical shapes in order to minimize the energy
of the system. 
We discuss possible alternatives one can use when the EoS of the inner crust is not known for the interaction used to describe the core. 

Moreover, we analyze the influence of the core-crust transition point when studying crustal properties, such as the crustal mass, 
crustal thickness and crustal fraction of the moment of inertia. The proper determination of the different properties of the crust 
is important in the understanding of observed phenomena such as pulsar glitches, r-mode oscillations, cooling of isolated NSs, 
etc.~\cite{Link1999,Chamel2008,Fattoyev:2010tb,Chamel2013,PRC90Piekarewicz2014,Newton2015}.

The ratio between the crustal fraction of the moment of inertia and the total moment of inertia is essential when studying 
phenomena such as pulsar glitches. With the same interactions we have used up to now, we integrate the moment of inertia 
and study the performance of these interactions when compared to different theoretical and observational constraints~\cite{Link1999, Andersson2012, Landry18}.

The recent GW170817 detection~\cite{Abbott2017} of gravitational waves from the merger of two NSs by the LIGO and Virgo collaboration (LVC) has helped to 
further constrain the EoS of $\beta$-stable nuclear matter. From the analysis of the data obtained from the detection, specific constraints 
on the mass-weighted tidal deformability at a certain chirp mass of the binary system have been extracted by the LVC~\cite{Abbott2017, Abbott2019}.
Moreover, with further analyses, the LVC was able to provide
constraints on the masses, radii, tidal deformabilities of an NS of $1.4 M_\odot$, etc.~\cite{Abbott2017, Abbott2018, Abbott2019}.
We have applied the interactions we have used in this thesis, paying special attention to the stiffness of their symmetry energy, 
for describing the tidal deformability of NSs and we compare them with the constraints coming from the 
observations.

\section{Crustal properties}
The solution of the TOV equations, see Eqs.~(\ref{eq:TOV}) and (\ref{eq:TOV2}), 
combined with the determination of the core-crust transition, allows one
to separate the crust and the core regions inside an NS~\cite{Chamel2008}. 
In particular, one can compute the thickness ($R_\mathrm{crust}$) and mass ($M_\mathrm{crust}$) of the crust. The 
crustal thickness is defined as the radial coordinate that goes from the core-crust transition to the 
surface of the star, and the mass enclosed in this region is the crustal mass. 
The study of the crustal properties depends on the EoS used to describe the crust, especially in its inner layers. 
Ideally, one should use a unified EoS, which is obtained with the same interaction in all regions of the NS
(outer crust, inner crust, and core). However, the computation of the EoS in the inner crust is tougher 
than the computation in the core due to the presence of the neutron gas and nuclear clusters with shapes different from the 
spherical one. 
Hence, the EoS of the inner crust has been studied with fewer interactions than the core and, therefore, 
the EoS in the crust region is still not computed for many forces. 

There are different alternatives to characterize the inner crust when the used interaction to study the NS core
does not have its equivalent in the crust. 
One alternative is to use a polytropic EoS matching the core EoS with the outer crust EoS. 
The outer crust, which goes from the surface of the star, which has densities of $\sim 10^{-12}$ fm$^{-3}$, to densities around $\sim 10^{-4}$ fm$^{-3}$, 
contributes in a very small fraction to the total mass and thickness of the NS, and it is mostly determined by nuclear masses experimentally known. 
Hence, the EoS that describes the outer crust of the star will have a relatively low influence on the determination of the properties of the NS.
Therefore, to describe the outer crust, one can use an already existing EoS given in the literature. 
Having the EoS for the outer crust and for the core, the inner crust can be obtained as a polytrope of the type $P=a+b \epsilon^{4/3}$, where $P$ is the pressure and $\epsilon$ is the energy density. 
This prescription has been widely used in previous works\cite{Link1999, carriere03,xu09a,Zhang15, gonzalez17}, where one adjusts the coefficients $a$ and $b$
 by demanding continuity at the outer crust-inner crust and core-crust interfaces. 
 The expressions to obtain $a$ and $b$ are given in Eqs.~(\ref{eq:coefa}) and (\ref{eq:coefb}) of Chapter~\ref{chapter2}.
 A polytropic form with an index of average value of about $4/3$ is found to be a good approximation to the EoS in this region~\cite{Link1999,Lattimer01,Lattimer2016}, as 
 the pressure of matter at these densities is largely influenced by the relativistic degenerate electrons. 
 The assumption of a polytropic EoS for the inner crust is the prescription we have been using until now through all this thesis. 

\begin{figure}[!t]
\centering
\includegraphics[clip=true, width=0.8\linewidth]{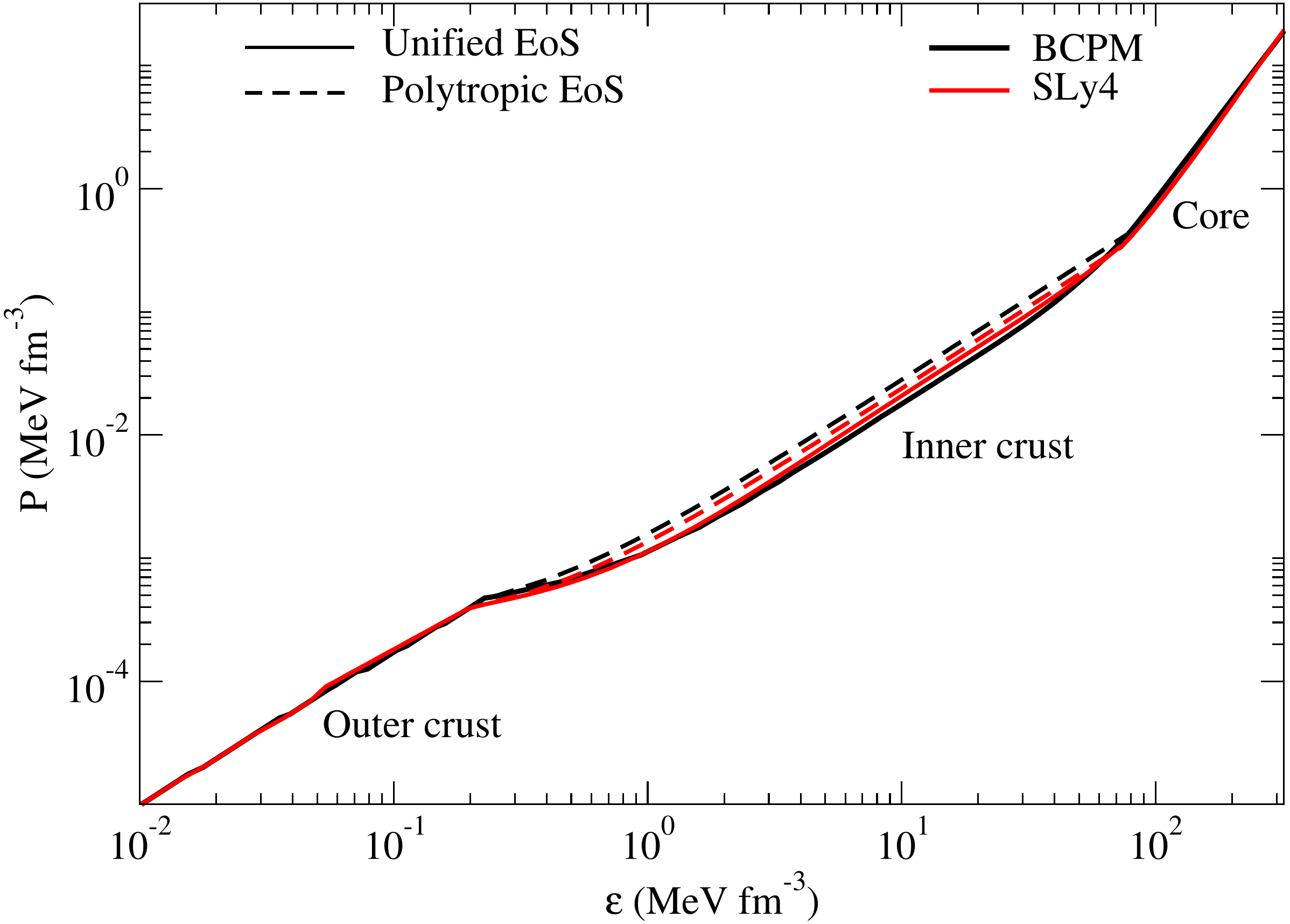}
\caption{Total pressure against the total energy density computed with the SLy4 Skyrme interaction
and the BCPM functional. The results plotted with straight lines account for the respective 
unified EoS, while the results plotted with dashed lines account for the use of a polytropic EoS in the 
inner crust region.}\label{fig:inner}
\end{figure}

 For some interactions, the unified EoS for all the NS has been computed. For these forces, we can fortunately compare the 
 behaviour of the polytropic EoS for the inner crust and their respective unified EoS computed taking into 
 account the physics of the inner crust with the same interaction. 
 We plot in Fig.~\ref{fig:inner} in double logarithmic scale the EoS (total pressure against the total energy density) 
 of two interactions for which their unified EoSs exist in the literature. These two models are the SLy4~\cite{douchin01} Skyrme 
 interaction and the BCPM~\cite{sharma15} density functional. For each one of them, we plot with straight lines the unified EoS, which
 has all regions of the NS (outer crust, inner crust, core) computed with the same interaction. 
 Moreover, in the same figure, we plot with dashed lines the EoS resulting from using a polytrope in the inner crust. 
 We see that the unified EoSs of BCPM and the SLy4 forces are very similar, and that their respective polytropic approximations are quite accurate with the unified ones.
 On the other hand, it may happen that the agreement between the unified EoS and the polytropic approximation for the inner crust is not 
 as good as in the cases of the BCPM and the SLy4 functionals. Thus, some differences between the microscopic calculations of the EoS and the 
 polytropic prescription may lead to differences in the computation of NS properties, in particular in the case of the crustal properties, such
 as the crustal mass and crustal thickness.

Another way to circumvent the lack of the crust EoS for a specific nuclear force is to use the energy density and pressure 
given by a different interaction in this region of the star and, at the core-crust transition interface, match the 
EoS of the core with the one coming from the crust. 
The choice of a polytropic EoS has the advantage of providing a continuous EoS along all the NS, while, if using an already computed EoS for the inner 
crust layers which is different from the one used in the core region, the values of the energy density and pressure
at the core-crust transition may not coincide. 
On the other hand, already computed EoSs for the inner crust have the advantage of having considered the physics of the inner 
crust and, in a sense, are more realistic.

The choice of the inner crust EoS may not have large consequences when studying
global properties of the star, such as the total mass and the total radius, but can influence the values obtained for the 
crustal properties.
A method recently proposed by Zdunik, Fortin, and Haensel~\cite{Zdunik17} allows one to accurately
estimate the total and crustal masses and radii of the NSs without the explicit knowledge of the EoS of the crust. 
These authors have presented an approximate description of the NS crust structure in terms of the function relating the chemical potential and the pressure. 
It only requires the knowledge of the chemical potential at the boundaries of a given layer in the crust. In particular, if one wants to 
study the thickness of the whole crust, one needs the chemical potential at the core-crust transition and at the surface of the star. 
One integrates the TOV 
equations from the center of the NS up to the core-crust transition and, 
with this approximation, using the fact that the crustal mass is small compared to the total mass of the NS, $M_\mathrm{crust} << M$, and that 
$4 \pi r^3 P /mc^2 << 1$, the crustal thickness
 of an NS of mass $M$ is then given by~\cite{Zdunik17}:
\begin{equation}\label{rcrustz}
 R_\mathrm{crust} = \Phi R_\mathrm{core} \frac{1-2 G M / R_\mathrm{core} c^2}{1- \Phi \left( 1- 2 G M /R_\mathrm{core} c^2\right)},
\end{equation}
with   
\begin{equation}
 \Phi \equiv \frac{(\alpha-1) R_\mathrm{core} c^2}{2 G M}
\end{equation}
and 
\begin{equation}
 \alpha = \left( \frac{\mu_t}{\mu_0}\right)^2,
\end{equation}
where $\mu_t = (P_t + {\cal H}_t c^2)/\rho_t$ is the chemical potential at the core-crust transition, 
$\mu_0$ is the chemical potential at the surface of the NS and $R_\mathrm{core}$ is
the thickness of the core. The crustal mass is obtained as 
\begin{equation}\label{mcrustz}
 M_\mathrm{crust} = \frac{4 \pi P_t R_\mathrm{core}^4}{G M_\mathrm{core}} \left(1- \frac{2 G M_\mathrm{core}}{R_\mathrm{core}c^2} \right),
\end{equation}
where $M_\mathrm{core}$ is the mass of the NS core. The total mass of the NS is $M=M_\mathrm{crust}+M_\mathrm{core}$.
This approximate approach predicts the radius and mass of the crust with
accuracy better than $\sim 1\%$ in $R_\mathrm{crust}$ and $\sim 5\%$ in $M_\mathrm{crust}$ for typical
NS masses larger than $1 M_\odot$~\cite{Zdunik17}.

\begin{figure}[!b]
\centering
\includegraphics[clip=true, width=0.9\linewidth]{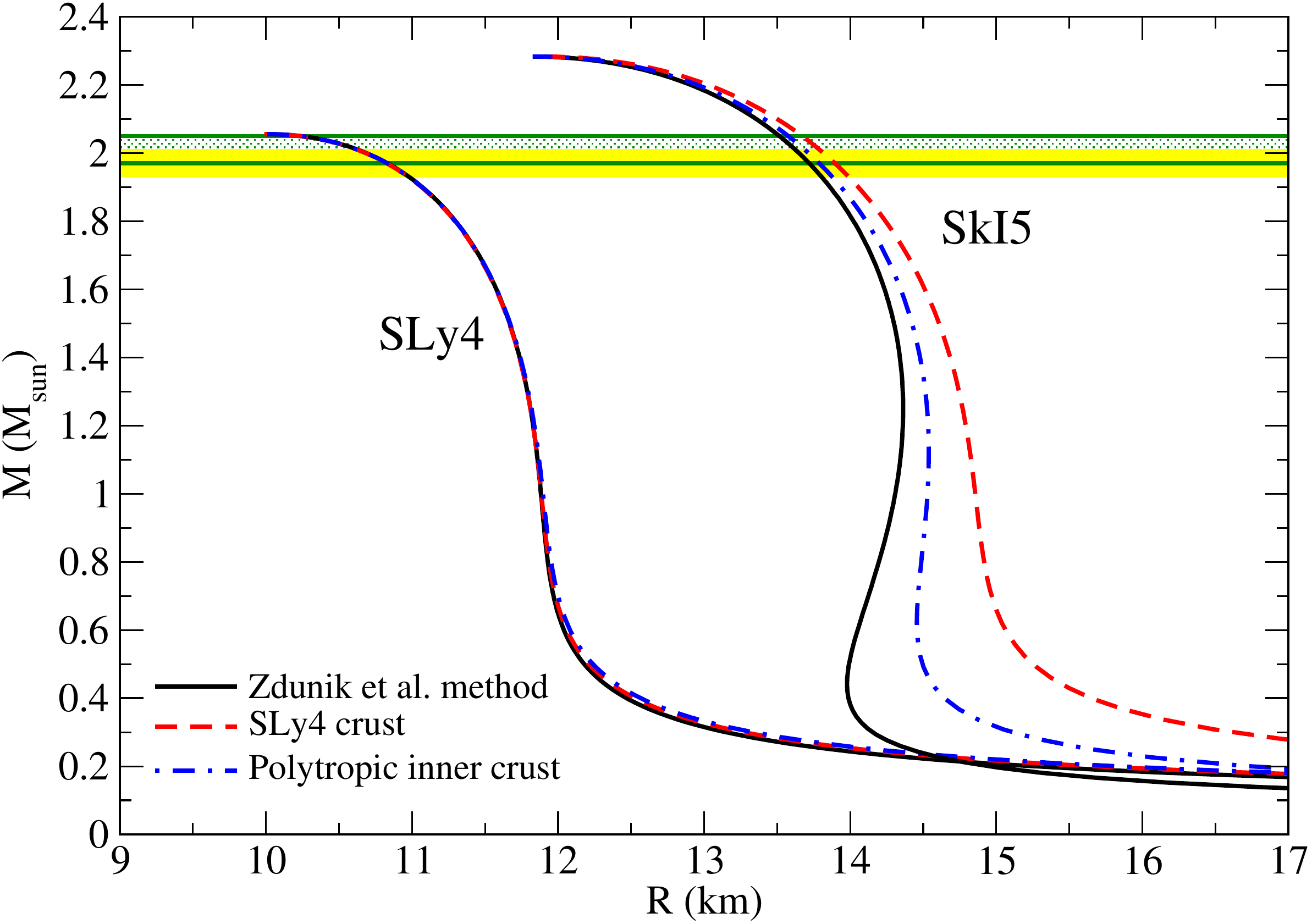}
\caption{Mass-radius relation calculated using the SLy4 and SkI5 Skyrme interactions for the NS core. 
With black solid lines we plot the results if using the method proposed by Zdunik et al. to describe the crust, 
with red dashed lines we plot the results if the SLy4 EoS~\cite{douchin01} for the inner crust is used
and with blue dashed-dotted lines we include the values for the M-R relation obtained 
using a polytropic approximation for the inner crust. The constraints for the maximum value of the mass
coming from Refs.~\cite{Demorest10, Antoniadis13}
are also included.\label{fig:MR_crust}} 
\end{figure}
We plot in Fig.~\ref{fig:MR_crust} the results for the mass-radius (M-R) relation computed with the SLy4 and SkI5 Skyrme interactions.
We use these two interactions as SLy4 has a moderate slope of the symmetry energy $L=46$ MeV, while the SkI5 force 
produces a stiffer EoS with slope parameter $L=129$ MeV. We do not plot the results for the BCPM functional as they 
have a very similar behaviour compared to the ones predicted by SLy4.
We represent with black solid lines the results obtained using the Zdunik et al. method~\cite{Zdunik17} to compute the crustal thickness and crustal mass. 
Moreover, we include in the same Fig.~\ref{fig:MR_crust} with red dashed lines the results obtained if using the crust of the SLy4 EoS. 
This case corresponds to the use of the SLy4 unified EoS while for the SkI5 interaction, the crust EoS will be microscopically calculated with a different interaction.
Finally, in the same figure, we plot with blue dashed-dotted lines the results for the M-R relation obtained using a polytropic EoS for the inner crust.
In the three approaches, the core-crust transition has been obtained using the dynamical method, and the outer crust is the one of Haensel-Pichon computed
with the SLy4 force~\cite{HaenselPichon}.
We use as a benchmark the results obtained with the Zdunik et al. method, as in Ref.~\cite{Zdunik17}.
In the case of the SLy4 interaction, the three prescriptions for the crust EoS provide almost the same results for the NS M-R relation. Let us remind that 
in this case, the line labeled as ``SLy4 crust'' in Fig.~\ref{fig:MR_crust} is obtained with the SLy4 unified EoS. The results we find 
for this interaction are in agreement with the ones found in Ref.~\cite{Zdunik17}.
On the other hand, the SkI5 interaction, with a stiff EoS, presents more differences in the results computed with different
descriptions of the inner crust. We see that, in the case of the SkI5 force, the use of a crust computed with a different interaction, 
in this case the SLy4 force, predicts results that may be quite far from the ones obtained using the Zdunik et al. method for the crust. 
This behavior may come from the fact that the symmetry energy of the SLy4 interaction is much softer than in the case of the SkI5 force. 
In the case of the SkI5 parametrization, the M-R values given by the TOV equations if using the polytropic approximation for the inner crust EoS are closer 
to the ones of Zdunik et al. than if the SLy4 crust is used.  
The differences between the results using different definitions of the inner crust EoS are more prominent in the low-mass regime about $M\lesssim 1 M_\odot$, 
and for interactions with  stiff symmetry energy.

Let us remark that, as we have seen, for interactions such as the SLy4 force, which has a moderate value of the slope parameter,
the differences between the M-R relation obtained using 
either of the three prescriptions of the crust are minimal.
Very recently we have been able to calculate the EoS of the crust for Gogny forces. In particular, we have obtained firsts results for the EoS of the inner crust calculated with
the D1M$^*$ interaction, which, similarly to the SLy4 interaction, has a moderate value of the slope parameter of the symmetry energy ($L=43$ MeV). 
The preliminary results for the inner crust EoS provide M-R relations very similar to the ones obtained if the 
inner crust is reproduced with the polytropic approximation, which we have used to fit the interaction.

\begin{figure}[!t]
\centering
\includegraphics[clip=true, width=1\linewidth]{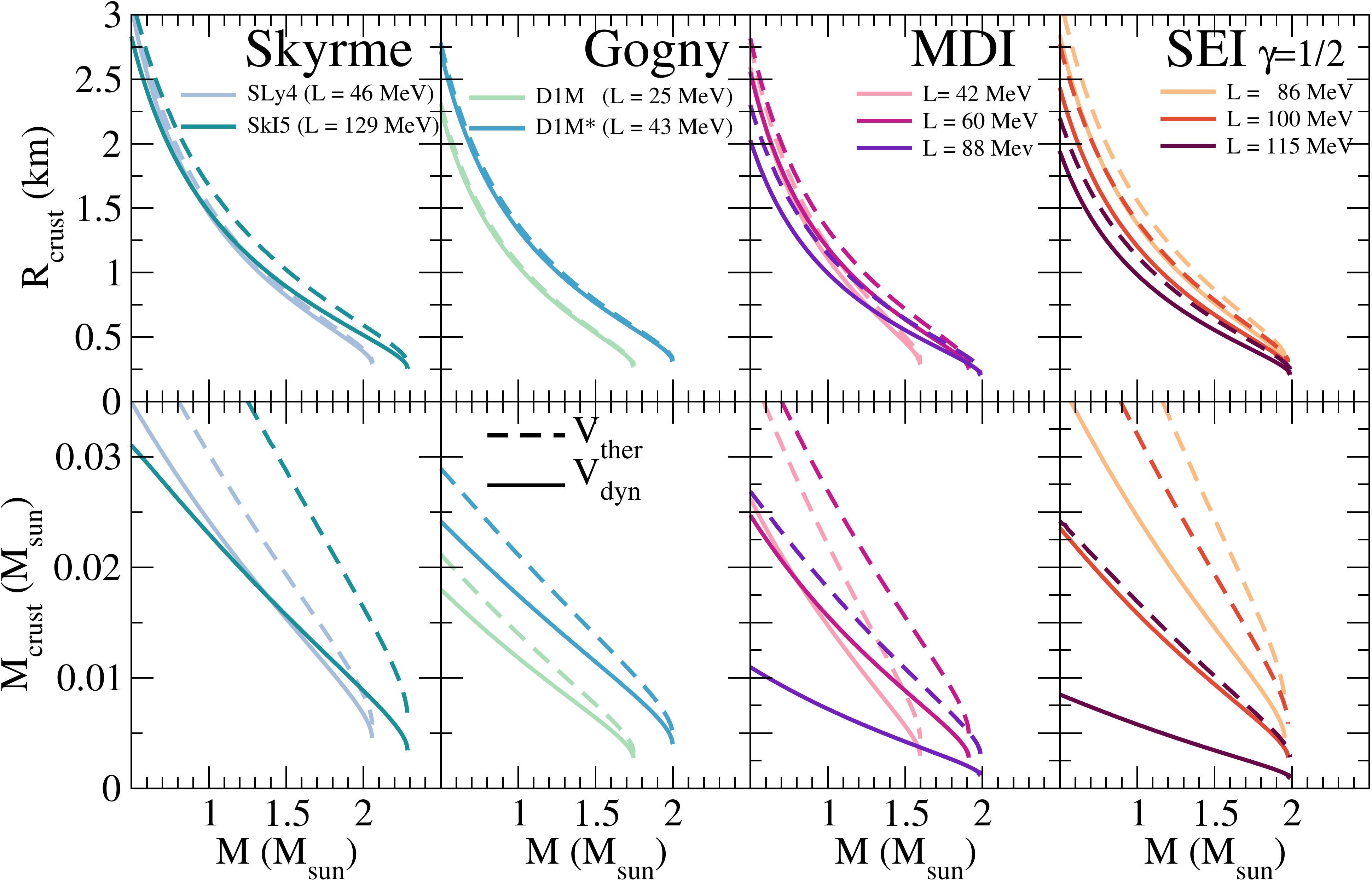}
\caption{Neutron star crustal thickness (upper panels) and crustal mass (lower panels) 
against the total mass of the NS for Skyrme (left), Gogny (center-left), MDI (center-right) and SEI (right) interactions. The core-crust 
transition has been determined using the thermodynamical potential (dashed lines) and the dynamical potential 
(solid lines).}\label{fig:crust_MR}
\end{figure}
On the whole, we can conclude that the global properties of an NS depend, up to a certain extend, on the treatment of the crust EoS, and this is more relevant 
in the low-mass regime of the M-R relation plot, where the 
crust of the NS is more prominent than its core.  

We proceed to study the crustal properties of NSs, such as the crustal thickness and crustal mass. 
To obtain the results, we use the method proposed by Zdunik et al. as it does not require any prescription 
for the crust EoS and it only uses information related to the core. 
We compute the results by solving the TOV equations from the center of the star up to the core-crust transition, and 
we use Eqs.~(\ref{rcrustz}) and (\ref{mcrustz}) to get the values of the crustal thickness and crustal mass. 
We plot in Fig.~\ref{fig:crust_MR} the crustal thickness and crustal mass against the total mass of 
the NS. As representative examples, we show the results provided by the SLy4 and SkI5 Skyrme forces, the D1M and D1M$^{*}$ Gogny interactions, 
three MDI and three SEI forces with different $L$ values. 
All considered forces in this figure, except the Gogny D1M and the MDI force with $L=42$ MeV,
predict NSs of mass about $2 M_\odot$, in agreement with astronomical observations \cite{Demorest10, Antoniadis13}.
The whole set of interactions is located in a wide range of the values of the slope of the symmetry energy.
As stated in Chapter~\ref{chapter4}, one can obtain the core-crust transition interface using the thermodynamical method
or the more sophisticated dynamical method.
For each interaction, we have obtained the crustal properties using the transition point given by the thermodynamical and the dynamical approaches.
In general, the global behaviour of these properties is similar for all four types of interactions. 
The values of both the crustal thickness $R_\mathrm{crust}$ and crustal mass $M_\mathrm{crust}$
decrease as the NS is more massive, being the contribution of the crust smaller. This is true for either the case
where one uses the core-crust thermodynamical transition density and the case
where one uses the dynamical method. The first case, that is, when using the thermodynamical transition density, 
predicts higher values of the crustal properties than the case where the dynamical transition density is used, as the 
values predicted for the transition density and transition pressure are higher in the first case than in the second one (see Chapter~\ref{chapter4}). 
The influence of the transition point is smaller on the crustal radius, and has a larger impact on 
the calculation of the crustal mass.
Moreover, we see that the differences between the two calculations are larger for interactions with a large $L$ parameter and stiff EoS.
The crustal properties play a crucial role in the description of several phenomena such as 
pulsar glitches, r-mode oscillations, cooling of isolated NSs, etc.~\cite{Link1999,Chamel2008,Fattoyev:2010tb,Chamel2013,PRC90Piekarewicz2014,Newton2015}. 
Hence, one should use the core-crust transition estimated using the dynamical method when looking for the results of NS
crustal properties because, as said in Chapter~\ref{chapter4}, the dynamical method includes surface and Coulomb 
contributions when one studies the stability of the NS core. 

Finally, we present in Table~\ref{tablecrust} some quantitative values of the crustal properties 
for the same interactions appearing in Fig.~\ref{fig:crust_MR}. We provide the results computed 
either using the core-crust transition estimated with the thermodynamical approach ($V_\mathrm{ther}$)
and with the dynamical approach ($V_\mathrm{dyn}$), for the NS configurations of maximum mass ($M_\mathrm{max}$) and for a 
canonical NS of $1.4 M_\odot$.
As already seen in Fig.~\ref{fig:crust_MR}, the values for the crustal properties are higher for the $V_\mathrm{ther}$
case than for the $V_\mathrm{dyn}$ case. 
Moreover, the values for a canonical NS are higher than in the cases of maximum mass configurations.
The differences between the results obtained using the thermodynamical core-crust transition
 and the ones obtained using the dynamical core-crust transition are larger as stiffer is the EoS describing the core. 
For example, for the SLy4 interaction, one obtains differences of $0.0012 M_\odot$ and $0.01$ km for the crustal 
mass and thickness in the maximum mass configuration. On the other hand, for the SEI $L=115$ MeV parametrization these differences are of 
$0.0067 M_\odot$ and $0.03$ km, respectively. For a canonical NS, the differences for the crustal mass and 
thickness are of $0.005 M_\odot$ and $0.04$ km in the case of SLy4 and 
$0.008 M_\odot$ and $0.09$ km in the case of the  SEI $L=115$ MeV force.
\begin{table}[t!]
\centering
\resizebox{\columnwidth}{!}{
\begin{tabular}{c|lcccccccc}
\hline
\multicolumn{2}{c|}{\multirow{2}{*}{Force}} & \multicolumn{2}{c|}{$M_\mathrm{crust}(M_\mathrm{max})$}                        & \multicolumn{2}{c|}{$R_\mathrm{crust}(M_\mathrm{max})$}                         & \multicolumn{2}{c|}{$M_\mathrm{crust}(1.4M_\odot)$}       & \multicolumn{2}{c|}{$R_\mathrm{crust}(1.4M_\odot)$}       \\ \cline{3-10} 
\multicolumn{2}{c|}{}                       & \multicolumn{1}{c|}{$V_\mathrm{ther}$} & \multicolumn{1}{c|}{$V_\mathrm{dyn}$} & \multicolumn{1}{c|}{$V_\mathrm{ther}$} & \multicolumn{1}{c|}{$V_\mathrm{dyn}$} & \multicolumn{1}{c|}{$V_\mathrm{ther}$} & \multicolumn{1}{c|}{$V_\mathrm{dyn}$} & \multicolumn{1}{c|}{$V_\mathrm{ther}$} & \multicolumn{1}{c|}{$V_\mathrm{dyn}$} \\ \hline\hline
\multirow{2}{*}{Skyrme}    & SLy4 ($L=46$ MeV)         & 0.0058                                 & 0.0046                                & 0.31                                   & 0.30                                  & 0.022                                  & 0.017            & 0.96                                   & 0.92             \\
                           & SkI5 ($L=129$ MeV)         & 0.0066                                 & 0.0035                                & 0.30                                   & 0.26                                  & 0.032                                  & 0.017            & 1.13                                   & 0.99             \\ \hline
\multirow{2}{*}{Gogny}     & D1M  ($L=25$ MeV)         & 0.0033                                 & 0.0027                                & 0.28                                   & 0.27                                  & 0.009                                  & 0.008            & 0.63                                   & 0.62             \\
                           & D1M$^*$ ($L=43$ MeV)       & 0.0049                                 & 0.0040                                & 0.32                                   & 0.31                                  & 0.013                                  & 0.015            & 0.87                                   & 0.85             \\ \hline
\multirow{3}{*}{MDI}       & $L=42$ MeV     & 0.0045                                 & 0.0029                                & 0.32                                   & 0.29                                  & 0.011                                  & 0.008            & 0.62                                   & 0.57             \\
                           & $L=60$ MeV     & 0.0050                                 & 0.0028                                & 0.29                                   & 0.26                                  & 0.018                                  & 0.010            & 0.81                                   & 0.73             \\
                           & $L=88$ MeV     & 0.0031                                 & 0.0012                                & 0.24                                   & 0.21                                  & 0.012                                  & 0.005            & 0.72                                   & 0.63             \\ \hline
\multirow{3}{*}{SEI}       & $L=86$ MeV     & 0.0078                                 & 0.0046                                & 0.35                                   & 0.31                                  & 0.028                                  & 0.017            & 0.98                                   & 0.86             \\
                           & $L=100$ MeV    & 0.0059                                 & 0.0028                                & 0.31                                   & 0.26                                  & 0.022                                  & 0.011            & 0.89                                   & 0.76             \\
                           & $L=115$ MeV    & 0.0096                                 & 0.0029                                & 0.24                                   & 0.21                                  & 0.012                                  & 0.004            & 0.72                                   & 0.63            \\ \hline
\end{tabular}
}
\caption{Crustal mass ($M_\mathrm{crust}$) and crustal thickness ($R_\mathrm{crust}$) for an
NS of maximum mass ($M_\mathrm{max}$) and for a canonical NS mass of $1.4 M_\odot$ evaluated
taking into account the core-crust transition obtained using the thermodynamical method ($V_\mathrm{ther}$)
or the dynamical method ($V_\mathrm{dyn}$) for a set of mean-field models.
The values of the mass are given in solar masses and the results for the radii in km. \label{tablecrust}}
\end{table}

\section{Moment of inertia}
One property of interest, due to potential observational evidence in pulsar glitches as well as 
the connection to the core-crust transition, is the ratio 
between the fraction of the moment of inertia enclosed in the NS crust, $\Delta I_\mathrm{crust}$, and the star's total moment of inertia, 
$I$~\cite{Ravenhall1994,haensel07,Lattimer2005,Lattimer2016}.
To lowest order in angular velocity, the moment of inertia of the star can be computed 
from the static mass distribution and gravitational potentials encoded in the TOV equations 
\cite{Hartle1967}. 
In the slow-rotation limit, the moment of inertia of a spherically symmetric NS is given by~\cite{Fattoyev:2010tb}
\begin{equation}\label{inertia1}
 I\equiv \frac{J}{\Omega} = \frac{8 \pi}{3} \int_0^R r^4 e^{-\nu(r)} \frac{\bar{\omega} (r)}{\Omega}
 \frac{(\varepsilon(r) + P(r))}{\sqrt{1-2Gm(r)/rc^2}}dr,
\end{equation}
where $J$ is the angular momentum, $\Omega$ is the stellar rotational frequency, $\nu(r)$ and $ \bar{\omega}$ are 
radially dependent metric functions and $m(r)$, $\varepsilon(r)$ and $P(r)$ are, respectively, the NS mass, energy density
and total pressure enclosed in a radius $r$.

The metric function $\nu(r)$ satisfies
\begin{equation}
 \nu(r) = \frac{1}{2} \mathrm{ln} \left(1-\frac{2 G M}{R c^2} \right) - 
 \frac{G}{c^2} \int_r^R \frac{(M(x) + 4 \pi x^3 P(x)}{x^2 (1-2 G M(x)/xc^2)}dx,
\end{equation}
and the relative frequency $\bar{\omega}(r)$, which is defined as 
\begin{equation}
 \bar{\omega}(r) \equiv \Omega - \omega(r),
\end{equation}
represents the angular velocity of the fluid as measured in a local reference frame~\cite{Fattoyev:2010tb}.
The frequency $\omega(r)$ is the frequency appearing due to the slow rotation. 
One also can define the relative frequency $\tilde{\omega}(r) \equiv \bar{\omega}(r)/\Omega$, 
which is obtained solving 
\begin{equation}\label{inertia2}
 \frac{d}{dr} \left( r^4 j(r) \frac{d\tilde{\omega}(r)}{dr}\right) + 4 r^3 \frac{dj(r)}{dr} \tilde{\omega}(r)=0,
\end{equation}
with
\begin{equation}
 j(r) =\left \{ \begin{matrix} e^{\nu(r)} \sqrt{1-2Gm(r)/rc^2} & \mbox{if } r \leq R
\\ 1 & \mbox{if }r >R\end{matrix}\right.  .
\end{equation}
The boundary conditions defining the relative frequency $\tilde{\omega}(r)$ are:
\begin{equation}\label{constraintI}
 \tilde{\omega}'(r) (0)=0 \hspace{1cm} \mathrm{and} \hspace{1cm} \tilde{\omega}(r) + \frac{R}{3} \tilde{\omega}'(r) =1.
\end{equation}
One integrates~Eq.~(\ref{inertia2}) considering an arbitrary value of $\tilde{\omega}(0)$. The second 
boundary condition at the surface of the NS usually will not be satisfied for the given $\tilde{\omega}(0)$. 
Hence, one must rescale the solution of (\ref{inertia2}) with an appropriate constant in order to fulfill~(\ref{constraintI}).
Notice that in this slow-rotation regime the solution of the moment of inertia does not depend on the stellar 
frequency $\Omega$.
A further check one can implement to ensure the accuracy of the full calculation is that the equation
\begin{equation}
 \tilde{\omega}'(R) =\frac{6GI}{R^4c^2}
\end{equation}
is fulfilled~\cite{Fattoyev:2010tb}.

\begin{figure}[t!]
\centering
\includegraphics[clip=true, width=0.9\linewidth]{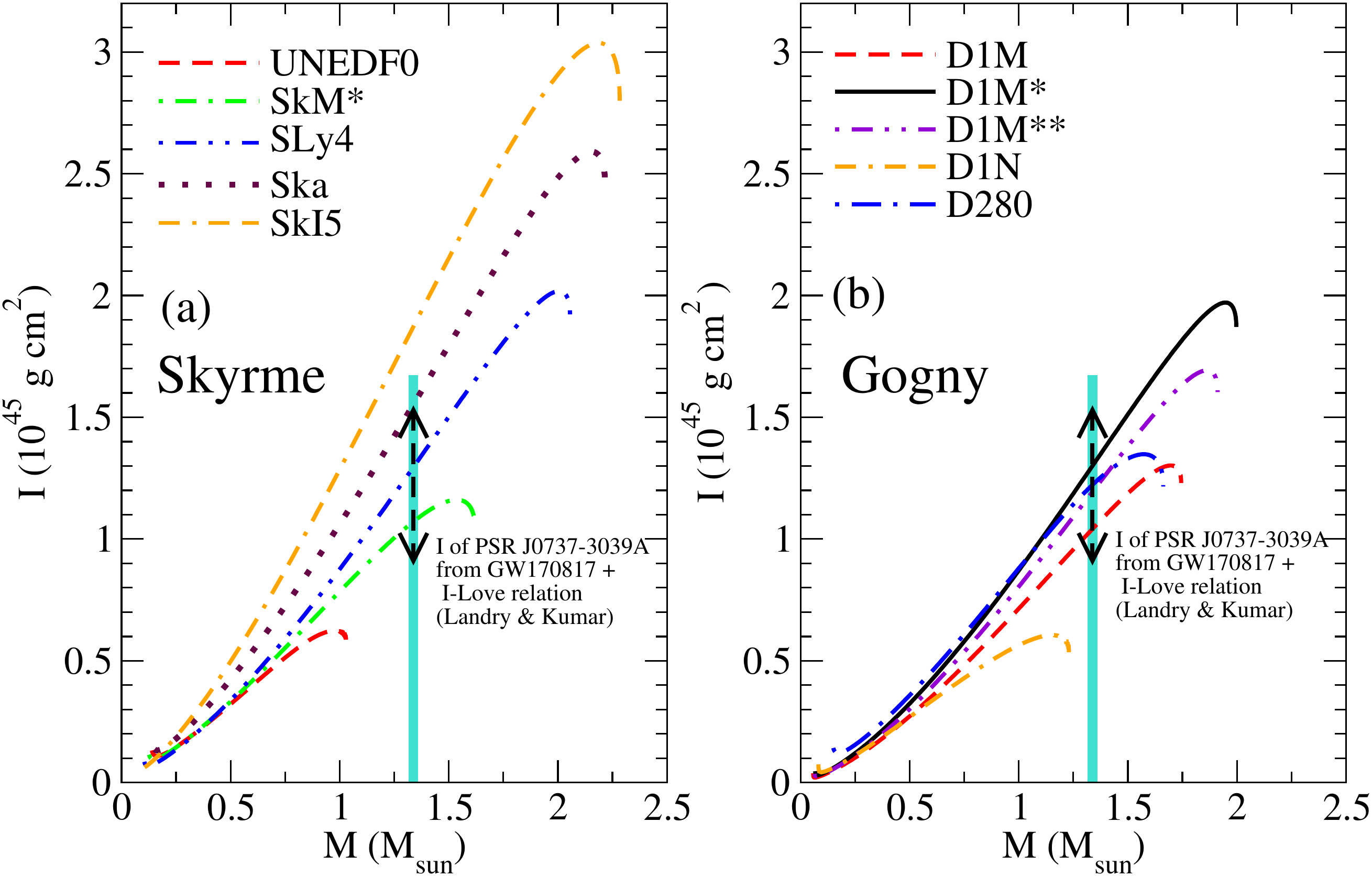}
\caption{Total NS moment of inertia against the total mass for a set of Skyrme (panel (a)) and Gogny (panel (b)) interactions. 
The two constraints from Ref.~\cite{Landry18} at $1.338 M_\odot$ for the primary component of the double pulsar 
PSR J0737-303 are also included.}\label{fig:momI}
\end{figure}

We show in Fig.~\ref{fig:momI} \cite{gonzalez17} the results for 
the moment of inertia against the total mass of the NS for a set of Skyrme (panel (a)) and Gogny (panel (b)) parametrizations of interest.
To integrate the TOV equations and to find the total moment of inertia we have used the outer crust of Haensel-Pichon and a polytropic EoS 
for the inner crust. One cannot use the method of Zdunik et al. to describe the crust, as one needs the EoS in all the NS to integrate the moment of inertia. 
In agreement with the findings of the behaviour of the EoS, the moments of inertia are larger as stiffer are their respective EoSs.
In particular, we find that the moments of inertia of the Gogny parametrizations are below 
the predictions of the SLy4 Skyrme interaction. 
As expected, the maximum value of $I$ is reached slightly below the maximum 
mass configuration for all forces \cite{haensel07}. 
For Gogny interactions, we observe~\cite{gonzalez17} that the D1M and D280 interactions give maximum values of 
$I_\text{max} \approx 1.3-1.4 \times 10^{45}$ g cm$^2$, below the typical 
maximum values of $\approx 2 \times 10^{45}$ g cm$^2$ obtained with stiffer EoSs~\cite{haensel07}. 
Our results for D1N are commensurate with those of Ref.~\cite{Loan2011}.
The new recently fitted D1M$^*$ and D1M$^{**}$ forces provide maximum values of 
$I_\text{max} = 1.97 \times 10^{45}$ g cm$^2$ and $I_\text{max} = 1.70 \times 10^{45}$ g cm$^2$ respectively.

Binary pulsar observations may provide in the future information on the moment of inertia of NSs, 
giving new constraints to the EoS of NS matter~\cite{Lattimer2005}. 
The binary PSR J0737-3039 is the only double-pulsar system known to date, and it is expected a 
precise measurement in the near future of the moment of inertia of its primary component PSR J0737-3039A (or pulsar A) from
radio observations~\cite{burgay03, Lyne04}.
In Ref.~\cite{Landry18}, Landry and Kumar estimate a range of $I=1.15^{+0.38}_{-0.24} \times 10^{45}$ g cm$^{2}$
for pulsar A, which has a mass of $M=1.338 M_\odot$. 
To obtain these constraints, Landry and Kumar have combined the values for the tidal deformability of a $1.4 M_\odot$
NS reported by the LVC obtained from the GW170817 event~\cite{Abbott2019} (see next Section~\ref{seclambda}) with 
approximately universal relations
among NS observables~\cite{Yagi13a, Yagi13b}, known
as the binary-Love and I-Love relations.
In the same Ref.~\cite{Landry18}, a wider range of \mbox{$I \leq 1.67 \times 10^{45}$ g cm$^2$} is given for the moment of inertia of pulsar A, 
obtained from the less restrictive limit on the tidal deformability, $\tilde{\Lambda} \leq 800$,  obtained from the first analysis
of the GW170817 event~\cite{Abbott2017}.
We have plotted in Fig.~\ref{fig:momI} the two constraints for pulsar A of the PSR J0737-3039 binary system.
We observe that not all Skyrme forces that provide NSs above the $2M_\odot$ constraint limit may fit inside these boundaries.
This is the case of for example the SkI5 Skyrme force, where the EoS is so stiff that it gives too large values for the 
moment of inertia. Interactions with smaller $L$ value, such as the new D1M$^*$
and D1M$^{**}$ Gogny interactions that we have constructed, fit inside the constraints from Landry and Kumar
for the moment of inertia of pulsar PSR J0737-3039 A. 
We also observe that the SLy4 Skyrme interaction fits well inside both boundary limits.
A useful comparison with the systematics of other NS EoSs is provided by the 
dimensionless quantity $\frac{I}{MR^2}$. This has been found to scale with the NS 
compactness $\chi=GM/Rc^2$.
In fact, in a relatively wide region of $\chi$ values, the dimensionless ratio $\frac{I}{MR^2}$ 
for the mass and radius combinations of several EoSs can be fitted by universal 
relations~\cite{Ravenhall1994,Lattimer2005,Breu2016}.

\begin{figure}[t!]
\centering
\includegraphics[clip=true, width=0.9\linewidth]{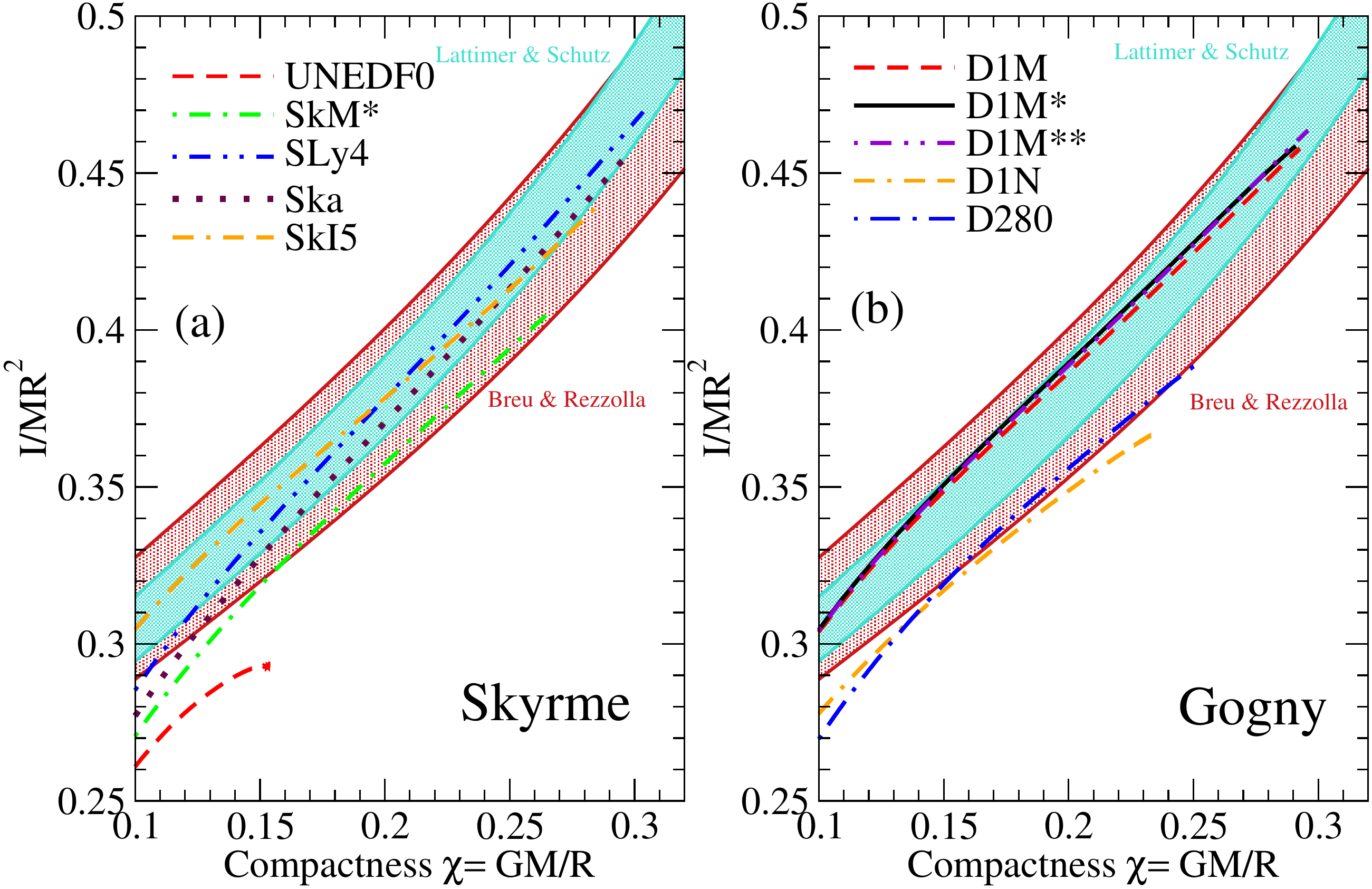}
\caption{Dimensionless quantity $\frac{I}{MR^2}$ against the compactness $\chi$ of an NS obtained
for a set of Skyrme (panel (a)) and Gogny (panel (b)) forces. The constraining bands from Refs.~\cite{Lattimer2005, Breu2016} are also included.}\label{fig:momI2}
\end{figure}

We show in Fig.~\ref{fig:momI2}~\cite{gonzalez17}
the dimensionless ratio $\frac{I}{MR^2}$ as a function of the compactness $\chi$ for the same set of Skyrme and
Gogny forces as in the previous Fig.~\ref{fig:momI}. Our results are compared to the recent fits from Breu and 
Rezzolla~\cite{Breu2016} (red shaded area) and the older results from
Lattimer and Schutz~\cite{Lattimer2005} (blue shaded region).
 These fits have been obtained from a very wide range of different theoretical EoS predictions.
For compactness $\chi > 0.1$, in the case of Skyrme interactions, only the SLy4, Ska and SkI5 parametrizations 
fit inside both bands. On the other hand, the SkM$^*$ and UNEDF0, not being able to provide large enough moments 
of inertia, lie outside the constraint from Lattimer and Schutz in the case of SkM$^*$ and outside
both constraints in the case of UNEDF0. 
For Gogny interactions, we see that the D1M, D1M$^*$ and D1M$^{**}$ forces give practically identical values of 
the $I/MR^2$ value if plotted against the NS compactness. In the three cases, the results fit inside both bands. 
On the other hand, D280 is close to the lower limit of this 
fit, but well below the lower bounds of the fit in Lattimer and Schutz~\cite{Lattimer2005}. We 
find that, in spite of the significant differences in their absolute moments of inertia, 
D1M, D1M$^*$, D1M$^{**}$ and SLy4 produce dimensionless ratios that agree well with each other. In contrast, and as 
expected, D1N produces too small moments of inertia for a given mass and radius, 
and systematically falls below the fits. 

When studying the ratio between the crustal fraction of the moment of inertia and the 
total moment of inertia $\Delta I_\mathrm{crust}/I$, one way to circumvent the problem of 
not having a good definition of the EoS in the inner crust is to use the 
the approximation given by \cite{Lattimer00, Lattimer01, Lattimer07},
which allows one to express this quantity as
\begin{eqnarray}\label{eq:Iaprox}
 \frac{\Delta I_\mathrm{crust}}{I} = \frac{28 \pi P_t R^3}{3 M c^2} \frac{\left(1-1.67 \chi-0.6 \chi^2 \right)}{\chi} 
 \times \left[ 1+ \frac{2 P_t \left( 1+ 5 \chi -14 \chi^2\right)}{\rho_t m c^2 \chi^2}\right]^{-1}, 
\end{eqnarray}
where $m$ is the baryon mass. 
The reason for this choice is that, as stated previously, we do not have unified EoSs for all the interactions we are considering 
in these studies. The approximate expression for the crustal fraction of the moment of inertia given in Eq.~(\ref{eq:Iaprox}),
combined with the Zdunik et al. method~\cite{Zdunik17} to obtain the total mass and radius of an NS, requires only of the core EoS
to find the crustal fraction of the moment of inertia. 

\begin{figure}[b!]
\centering
\includegraphics[clip=true, width=0.9\linewidth]{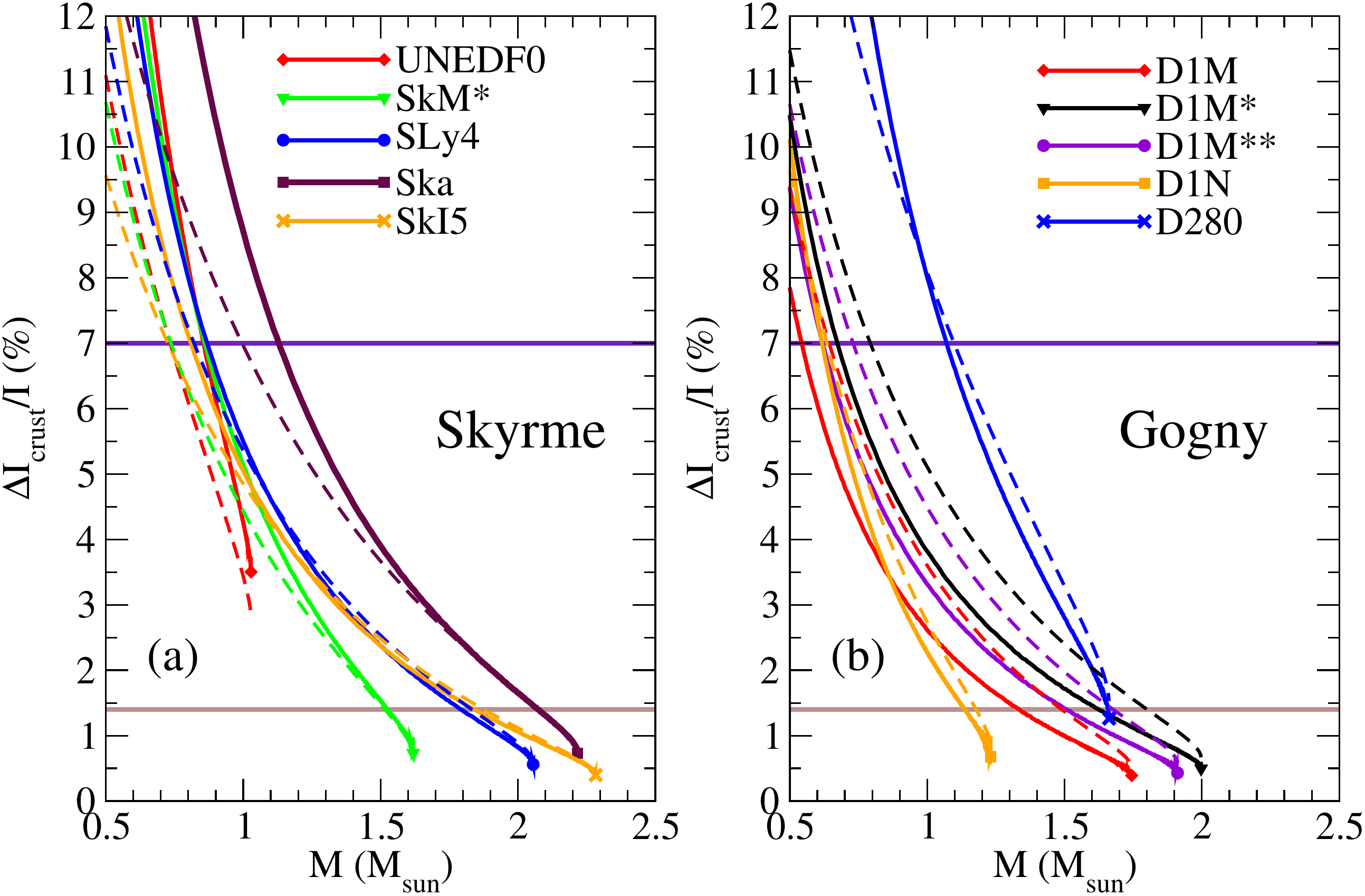}
\caption{Crustal fraction of the moment of inertia against the total NS mass for a set of Skyrme (panel (a)) and Gogny (panel (b))
interactions computed using a polytropic inner crust EoS (straight lines) or the approximation for 
$\Delta I_\mathrm{crust}/I$ in Eq.~(\ref{eq:Iaprox}) (dashed lines). The core-crust transition has been 
obtained using the dynamical approach.
The constraints from the Vela pulsar in Refs.~\cite{Link1999, Andersson2012} are also included.}\label{fig:Icrust2}
\end{figure}
We compare in Fig.~\ref{fig:Icrust2}, for a set of Skyrme (panel (a)) and Gogny (panel (b)) interactions, the results 
of the crustal fraction moment of inertia obtained if using a polytropic 
EoS for the inner crust (straight lines) or obtained if using the combination of 
Eq.~(\ref{eq:Iaprox}) and the method proposed by Zdunik et al. to find the total thickness and mass of an NS (dashed lines).
In both cases, the core-crust transition has been obtained for each particular interaction using the dynamical method.
We find a very good agreement between the approximated formula (\ref{eq:Iaprox})
and the full results if using a polytropic EoS above $1-1.2 M_\odot$, and the agreement improves for both types of interactions
as the mass of the 
pulsar increases. This is in keeping with the findings of Ref.~\cite{xu09a, gonzalez17}. 
\begin{figure}[!t]
\centering
\includegraphics[clip=true, width=1\linewidth]{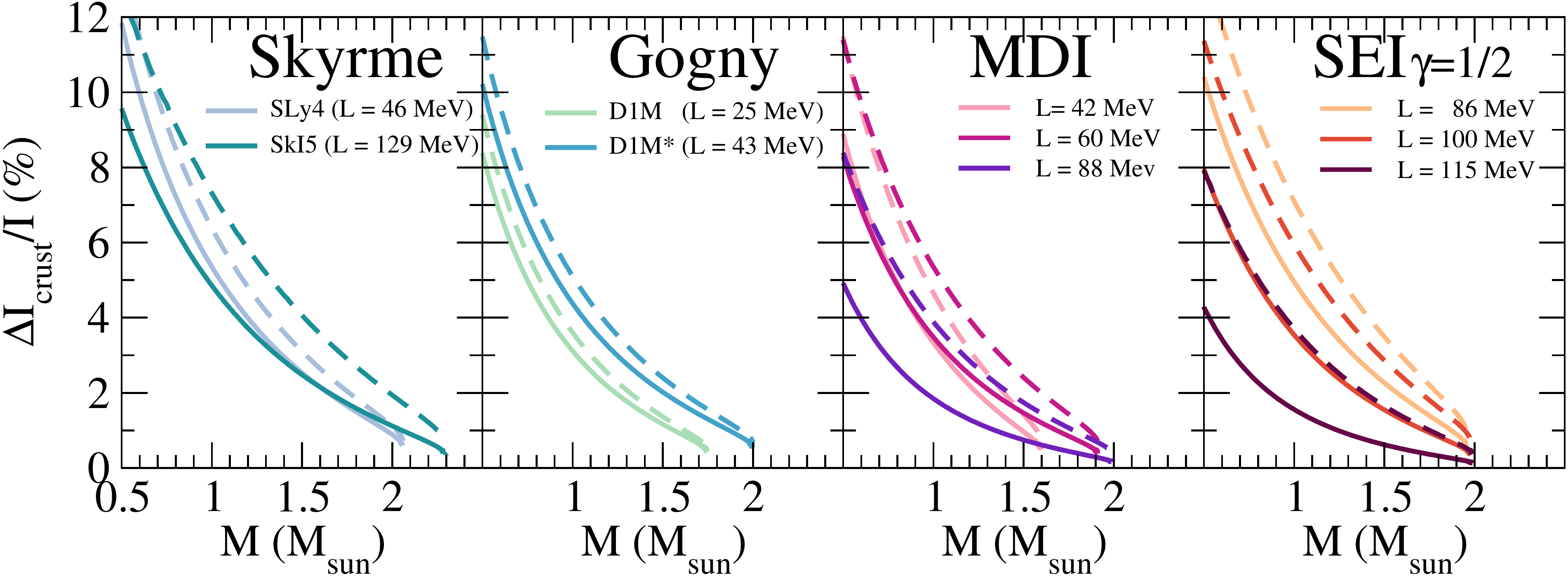}
\caption{Neutron star crustal fraction of the moment of inertia
against the total mass of the NS for Skyrme (left), Gogny (center-left), MDI (center-right) and SEI (right) interactions. The core-crust 
transition has been determined using the thermodynamical potential (dashed lines) and the dynamical potential 
(solid lines).}\label{fig:Icrust1}
\end{figure}

\begin{table}[!b]
\centering
\begin{tabular}{c|lcccc}
\hline
\multicolumn{2}{c|}{\multirow{2}{*}{Force}} & \multicolumn{2}{c|}{$\Delta I_\mathrm{crust}/I (M_\mathrm{max})$ ($\%$)}                                        & \multicolumn{2}{c}{$\Delta I_\mathrm{crust}/I(1.4M_\odot)$ $\%$}            \\ \cline{3-6} 
\multicolumn{2}{c|}{}                       & \multicolumn{1}{c|}{$V_\mathrm{ther}$} & \multicolumn{1}{c|}{$V_\mathrm{dyn}$} & \multicolumn{1}{c|}{$V_\mathrm{ther}$} & \multicolumn{1}{c}{$V_\mathrm{dyn}$}  \\ \hline\hline
\multirow{2}{*}{Skyrme}    & SLy4  ($L=46$ MeV)          &                0.76   &             0.60       &                  3.61     &   2.96                                              \\
                           & SkI5  ($L=129$ MeV)         &                    0.73      &        0.41       &                   4.60   &     2.85                        \\ \hline
\multirow{2}{*}{Gogny}     & D1M     ($L=25$ MeV)       &                     0.52     &         0.44    &                        1.70  &     1.45                                   \\
                           & D1M$^*$  ($L=43$ MeV)      &                   0.67        &      0.56       &                     2.81   &     2.37                                                  \\ \hline
\multirow{3}{*}{MDI}       & $L=42$ MeV     &                    0.73        &    0.49       &                      1.95   &     1.34                                               \\
                           & $L=60$ MeV     &                    0.67    &        0.39        &                         2.86   &     1.76                                                 \\
                           & $L=88$ MeV     &                       0.41   &        0.16      &                      2.07      &       0.90                                               \\ \hline
\multirow{3}{*}{SEI}       & $L=86$ MeV     &                     0.99      &        0.60       &                    4.08     &     2.68                                                       \\
                           & $L=100$ MeV    &                      0.75      &           0.37   &                   3.36      &    1.84                                                 \\
                           & $L=115$ MeV    &                       0.38      &    0.13       &                     1.96      &    0.78                                                    \\ \hline
\end{tabular}
\caption{Crustal fraction of the moment of inertia (in \%) for an
NS of maximum mass ($M_\mathrm{max}$) and for a canonical NS mass of ($1.4 M_\odot$) evaluated
taking into account the core-crust transition obtained using the thermodynamical method ($V_\mathrm{ther}$)
or the dynamical method ($V_\mathrm{dyn}$) for a set of mean-field models. \label{tablecrustI}}
\end{table}
To account for the sizes of observed glitches, the widely used pinning model requires that a certain amount of angular momentum is 
carried by the crust. This can be translated into constraints on the crustal fraction of the moment of inertia. 
Initial estimates suggested that $\Delta I_\text{crust}/I>1.4 \, \%$ to explain  Vela and other glitching sources 
\cite{Link1999}. We show this value as the bottom horizontal line in Fig.~\ref{fig:Icrust2}.
We note that this does not pose 
mass limitations on the UNEDF0 and D280 interactions, which have minimum values of $\Delta I_\text{crust}/I$ above that limit. 
For other interactions that cross the limit, in contrast, glitching sources that satisfy this constraint should have below a certain mass. 
For the SkM$^*$, SLy4, Ska and SkI5 interactions, these masses are, respectively, $M< 1.5$, $1.8$, $2.1$ and $1.8$ $M_\odot$, 
and for Gogny interactions the limits are $M< 1.4$, $1.7$, $1.6$ and $1.2$ $M_\odot$ for the D1M, D1M$^*$, D1M$^{**}$ and D1N interactions, respectively.
More recently, a more stringent constraint has 
been obtained by accounting for the entrainment of neutrons in the crust \cite{Andersson2012}. With entrained neutrons, a 
larger crustal fraction of moment of inertia, $\Delta I_\text{crust}/I>7 \, \%$ (top horizontal line in the same figure), is needed 
to explain glitches.  Of course, a more realistic account 
of nuclear structure and superfluidity in the crust will modify the estimates. 
Then, all interactions would need 
significantly lower masses to account for glitching phenomena.

We present in Fig.~\ref{fig:Icrust1} the results for the 
crustal fraction of the moment of inertia against the total mass of the NS computed with the same interactions 
as in Fig.~\ref{fig:crust_MR}, namely the SLy4 and SkI5 Skyrme forces, the D1M and D1M$^{*}$ Gogny interactions, 
three MDI and three SEI forces with different $L$ values. 
The results enclosed in these figures have been obtained using the Zdunik et al. method to find the core properties and 
employing Eq.~(\ref{eq:Iaprox}) to find the ratio $\Delta I_\mathrm{crust}/I$.
Similarly to the analysis made for the crustal mass and thickness of an NS, 
we have obtained for each interaction the crustal fraction of the moment of inertia
using the transition point given by the thermodynamical and the dynamical approaches.
Moreover, we include in Table~\ref{tablecrustI} the values of the crustal fraction
of the moment of inertia $\Delta I_\mathrm{crust}/I$  for an NS of maximum mass $M_\mathrm{max}$ given by the interaction
and for a canonical NS of $1.4 M_\odot$, obtained either using the thermodynamical ($V_\mathrm{ther}$) or the dynamical ($V_\mathrm{dyn}$)
approaches.

The global behaviour of the moment of inertia is, again, akin to all four types of interactions,
$\Delta I_\text{crust}/I$ decreasing with the NS mass.
The location of the transition point has a large impact on 
the calculation of the crustal fraction of the moment of inertia, similar to the case when one studies the crustal mass.
In the case of the crustal fraction of the moment of inertia, the differences 
between the predictions using the core-crust transition found with the thermodynamical method 
or with the more realistic dynamical method are larger for interactions with a larger value of $L$. 
These differences are very prominent in the typical NS mass region and could influence 
the properties where the crust has an important role, such as pulsar glitches~\cite{Link1999,Fattoyev:2010tb,Chamel2013,PRC90Piekarewicz2014,Newton2015, gonzalez17}.

\section{Tidal deformability}\label{seclambda}

In a binary system composed of two NSs, each component star induces a
 perturbing gravitational tidal field on its companion, leading to a mass-quadrupole 
deformation in each star. 
To linear order, the tidal deformation of each component of the binary system is 
described by the so-called tidal deformability $\Lambda$, which is defined as 
the ratio between the 
induced quadrupole moment and the external tidal field~\cite{Flanagan08, Hinderer08}.
For a single NS, the tidal deformability can be written in terms of the dimensionless tidal 
Love number $k_2$, and the mass and radius of the NS~\cite{Flanagan08, Hinderer08, Hinderer2010}:
\begin{equation}
 \Lambda= \frac{2}{3}k_2 \left(\frac{R c^2}{G M} \right)^5,
\end{equation}
where $G$ is the gravitational constant and $c$ the speed of the light.
The mass and radius of an NS are determined by the solution of the TOV 
equations, and the Love number $k_2$ is given by
\begin{eqnarray}
 k_2&=& \frac{8\chi^5}{5} \left(1-2\chi \right)^2 \left[ 2+2\chi (y-1) -y\right]
 \times \left\{ 2\chi \left[6-3y + 3\chi(5y-8) \right] \right.\nonumber \\
 &+&  4 \chi^3 \left[ 13-11y+\chi (3 y -2) + 2 \chi^2 (1+y)\right]\nonumber\\
 &+&3 \left.(1-2\chi)^2 \left[2-y+2\chi(y-1) \right] \mathrm{ln}(1-2\chi)\right\}^{-1} ,
\end{eqnarray}
where 
\begin{equation}\label{compactness}
\chi=\frac{GM}{Rc^2},
\end{equation} 
is the compactness of the star and 
\begin{equation}
y=\frac{R \beta(R)}{H (R)}.
\end{equation}
The values of the functions $\beta(R)$ and $H(R)$ can be obtained by solving 
the following set of coupled differential equations~\cite{Hinderer08, Hinderer2010}:
\begin{eqnarray}
 \frac{dH(r)}{dr} &=& \beta (r) \\
 \frac{d\beta (r)}{d r} &=& \frac{2G}{c^2} \left(1-\frac{2Gm(r)}{r c^2} \right)^{-1} H (r)
 \left\{ -2 \pi \left[ 5 \epsilon + 9p + \frac{d \epsilon}{d p} 
(\epsilon+p) \right] + \frac{3 c^2}{r^2 G} \right.\nonumber \\
 &+&\left. \frac{2G}{c^2} \left(1-\frac{2Gm(r)}{r c^2} \right)^{-1} \left( \frac{m(r)}{r^2} 
+ 4 \pi r p\right)^2 \right\}\nonumber \\
 &+& \frac{2 \beta (r)}{r} \left(1-\frac{2Gm(r)}{r c^2} \right)^{-1} \left\{-1+\frac{Gm(r)}{r c^2} 
+ \frac{2 \pi r^2 G}{c^2} \left( \epsilon -p\right) \right\},
\label{eq.14a}
\end{eqnarray}
where $m(r)$ is the mass enclosed inside a radius $r$, and $\epsilon$ and $p$ are the corresponding 
energy density and pressure. 
One solves this system, together with the TOV 
equations given in Eqs.~(\ref{eq:TOV}) and (\ref{eq:TOV2}) integrating outwards and considering as boundary conditions
$H(r)= a_0 r^2$ and $\beta(r)= 2 a_0 r$
as $r \rightarrow 0$. The constant $a_0$ is arbitrary, as it cancels in the expression for 
the Love number~\cite{Hinderer2010}.

For a binary NS system, the mass weighted tidal deformability $\tilde{\Lambda}$, 
defined as
\begin{equation}\label{eq:wLambda}
 \tilde{\Lambda} = \frac{16}{13} \frac{(M_1 + 12M_2)M_1^4 \Lambda_1 +(M_2 + 12M_1)M_2^4 
\Lambda_2 }{(M_1+M_2)^5},
\end{equation}
takes into account the contribution from the tidal effects to the phase evolution of the 
gravitational wave spectrum of the inspiraling NS binary. 
In the definition of $\tilde{\Lambda}$ in Eq.~(\ref{eq:wLambda}), $\Lambda_1$ and $\Lambda_2$
refer to the tidal deformabilities of each NS in the system and $M_1$ and $M_2$ are their corresponding masses. 
The definition of $\tilde{\Lambda}$ fulfills $\tilde{\Lambda}=\Lambda_1=\Lambda_2$ 
when $M_1=M_2$.

The recent event GW170817 accounting for the detection of GWs coming 
from the merger of an NS 
binary system has allowed the LVC to 
obtain constraints on the mass-weighted 
tidal deformability $\tilde{\Lambda}$ and on the chirp mass $\mathcal{M}$, 
which for a binary NS system conformed of two stars of masses $M_1$ and $M_2$ is defined as 
\begin{equation}
 \mathcal{M}= \frac{(M_1 M_2)^{3/5}}{(M_1+M_2)^{1/5}}.
\end{equation}
In the first data analysis of GW170817 by the LIGO and Virgo collaboration, values of 
$\tilde{\Lambda} \leq 800$ and $\mathcal{M} = 1.188^{+0.004}_{-0.005} M_\odot$ were 
reported~\cite{Abbott2017}. Moreover, they estimated the masses of the two NSs to be in the 
range $M_1 \in (1.36,1.60) M_\odot$ and  $M_2 \in (1.17,1.36) M_\odot$. In a recent 
 reanalysis of the data~\cite{Abbott2019},
the values have been further constrained to $\tilde{\Lambda} =300^{+420}_{-230}$, 
$\mathcal{M} = 1.186^{+0.001}_{-0.001} M_\odot$,
$M_1 \in (1.36,1.60) M_\odot$ and  $M_2 \in (1.16,1.36) M_\odot$.

\begin{figure}[t!]
\centering
\includegraphics[clip=true, width=0.9\linewidth]{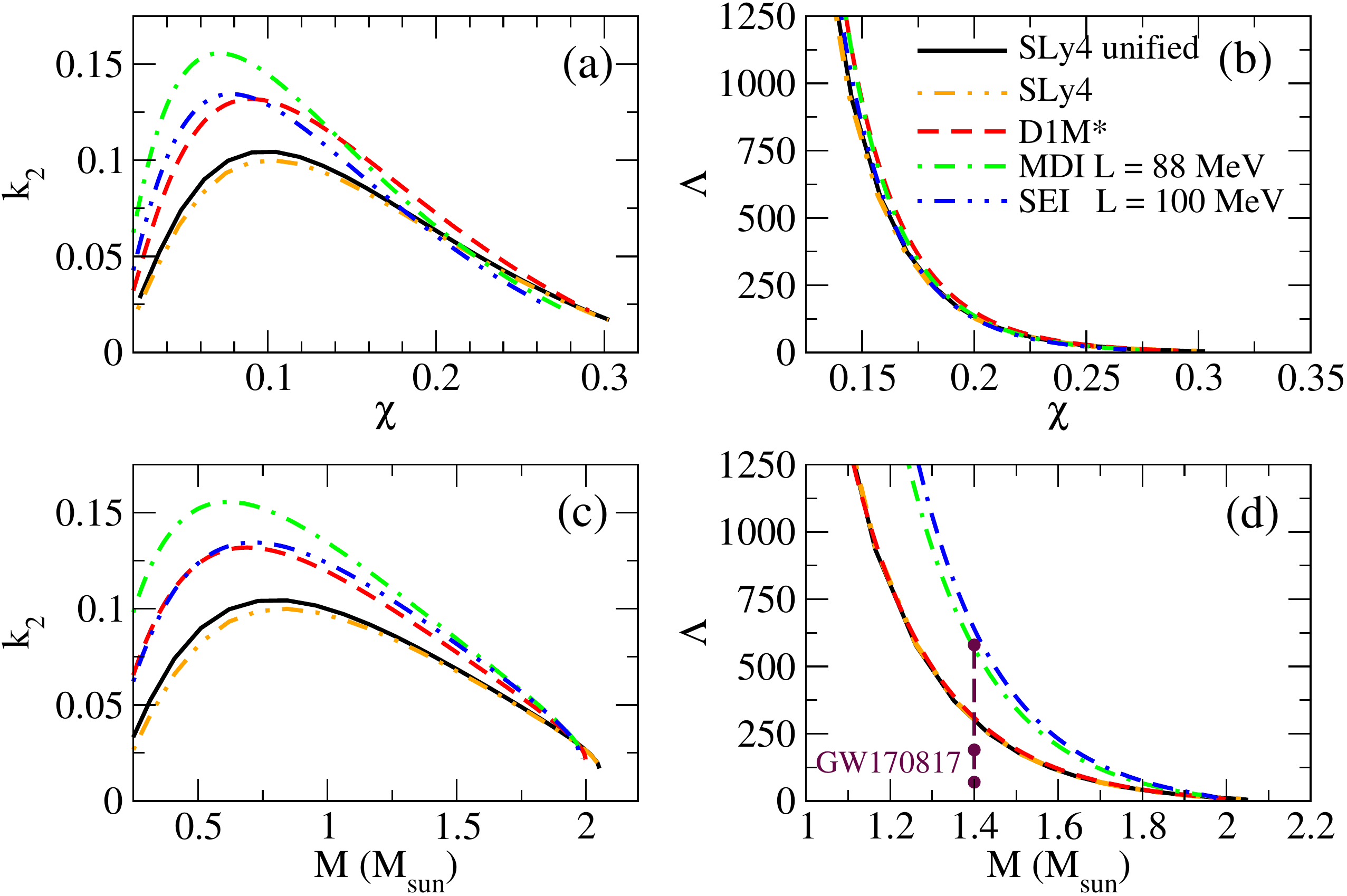}
\caption{Tidal Love number $k_2$ and dimensionless tidal deformability $\Lambda$
against the NS compactness $\chi=\frac{GM}{Rc^2}$ and the total NS mass ($M$) for the SLy4 Skyrme force, 
the D1M$^*$ interaction, the MDI with $L=88$ MeV model and the SEI with $L=100$ MeV parametrization. For these cases, the inner 
crust has been computed using a polytropic EoS for the inner crust. Moreover, the 
results using the SLy4 unified EoS are also included.
The constraint for the tidal deformability at $M=1.4 M_\odot$ obtained from the GW170817 event detection~\cite{Abbott2019} is also shown in panel (d).}\label{fig:tidal1}
\end{figure}

We plot in Fig.~\ref{fig:tidal1} the tidal Love number $k_2$ against the star compactness $\chi=\frac{GM}{Rc^2}$ (panel (a)) and against the 
total NS mass (panel (c)) for a set of representative mean-field interactions.
Moreover, we plot in the same Fig.~\ref{fig:tidal1} the values for the dimensionless 
tidal deformability $\Lambda$ against the star compactness (panel (b)) and against the total NS mass (panel (d)).
The results have been obtained considering the outer crust EoS of Haensel-Pichon computed with the SLy4 interaction~\cite{HaenselPichon}
and a polytropic EoS for the 
inner region. In this case, one cannot use the method proposed by Zdunik et al., as the EoS along all the NS is needed
when calculating the tidal deformability. Moreover, the core-crust transition has been obtained using the dynamical method. 
We first center our discussion on the behaviour of the Love number $k_2$. We see an increase of its value up to
$k_2 \sim 0.10-0.15$ at around compactness $\chi \sim 0.05-0.10$ or mass $M \sim 0.5-1 M_\odot$. Afterwards, it decreases
until vanishing at the black hole compactness $\chi=0.5$ for all EoSs~\cite{Krastev19}.
This can be understood  knowing that $k_2$ gives an estimation of how easy is for the bulk NS matter to deform. 
Therefore, more centrally condensed stellar models will have smaller $k_2$ and smaller $\Lambda$.
Also, for low values of the compactness or the mass, the crust part of the EoS is more prominent than the core. Hence, 
the NS becomes more centrally condensed and $k_2$ becomes smaller~\cite{Krastev19}.

We proceed to study the dependence of the dimensionless tidal deformability $\Lambda$ with the compactness and the NS mass.
We see in panel (b) of Fig.~\ref{fig:tidal1} that the dependence of the tidal deformability with the compactness
is almost the same for all EoSs, having a decreasing tendency as larger is $\chi$. 
On the other hand, if one plots $\Lambda$ against the total mass $M$, the results separate between them. 
For a given mass, the $\Lambda$ results obtained with a stiffer EoS are larger than if obtained with softer equations of state.
We plot in panel (d) of Fig.~\ref{fig:tidal1} the constraint for the tidal deformability of a canonical NS of $1.4 M_\odot$, 
$\Lambda_{1.4}= 190^{+390}_{-120}$ obtained from the analysis of the GW170817 data in Ref.~\cite{Abbott2018}.
We see that the SLy4 and D1M$^*$ interactions fit very well inside the constraint, while models with stiffer EoSs, 
like the MDI with $L=88$ MeV, reach the upper limit of the GW170817 data, and the SEI with $L=100$ MeV does not provide 
values of the tidal deformability inside the observed bands. 

We have to mention that the results found for the Love number $k_2$ are rather sensitive to the EoS of the crust. 
We see that the results of $k_2$ computed with the SLy4 unified EoS are a little bit different with respect to the ones
obtained using the polytropic inner crust. The relative differences between them reach up to values of $5\%$ in this case. 
On the other hand, the results for $\Lambda$ do not depend on this choice, as the changes in $k_2$ are compensated by the 
possible differences between the values of the NS radius~\cite{Piekarewicz19}. Hence, 
one can compare the results for the tidal deformability obtained with the polytropic inner crust, 
as it is our case, with the constraint for $\Lambda_{1.4}$ coming from the data analysis of the 
GW170817 event.

\begin{figure}[t!]
\centering
\includegraphics[clip=true, width=1\linewidth]{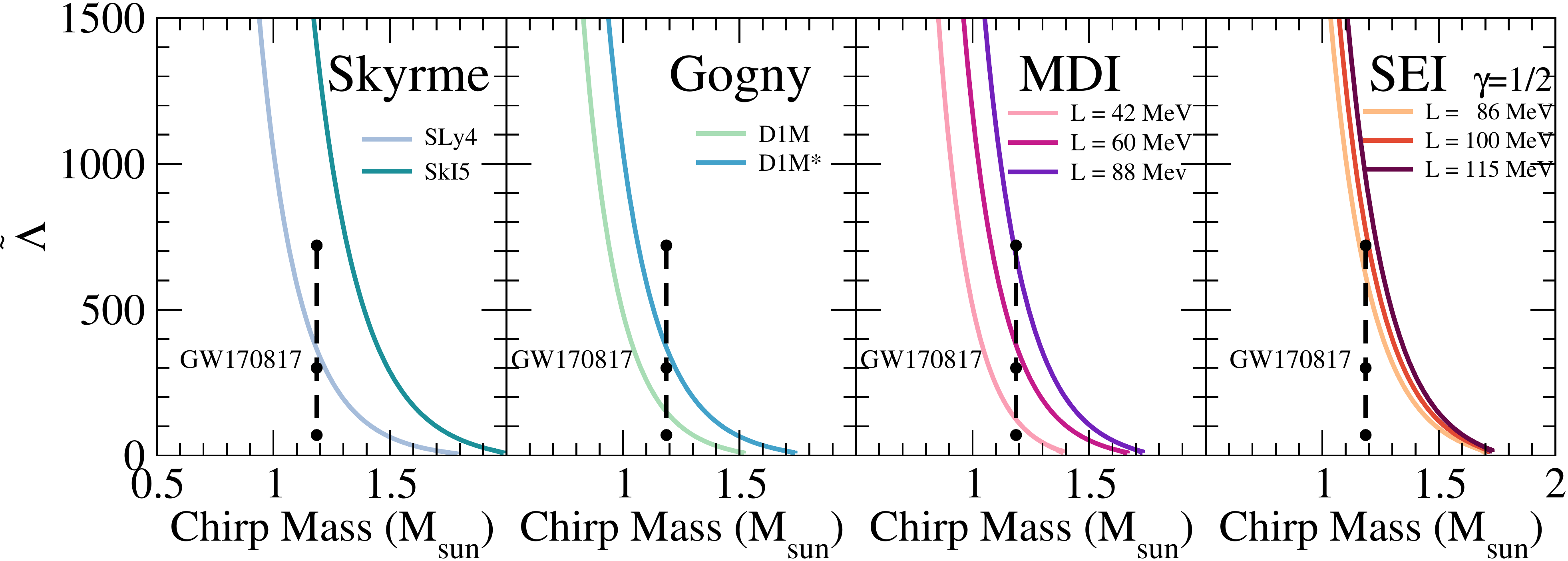}
\caption{Dimensionless mass-weighted tidal deformability plotted against the chirp mass $\mathcal{M}$
of a binary NS system
for a set of mean-field interactions. The constraint for $\tilde{\Lambda}$ at $\mathcal{M} = 1.186 M_\odot$~\cite{Abbott2019}
is also included.}\label{fig:tidal2}
\end{figure}

We plot in Fig.~\ref{fig:tidal2} the mass-weighted tidal deformability 
of a binary NS system against the chirp mass $\mathcal{M}$ for a set of Skyrme (left),
Gogny (center-left), MDI (center-right) and SEI (right) interactions. 
In all cases, we see that $\tilde{\Lambda}$ has a decreasing behaviour as the chirp mass becomes larger. 
We have followed the same prescription for the crust as the results obtained in Fig.~\ref{fig:tidal1}. In this case, the changes for 
$\tilde{\Lambda}$
should not differ taking into account other descriptions for the crust~\cite{Piekarewicz19}.
We have also plotted in each panel the constraint for $\tilde{\Lambda}$ at $\mathcal{M}=1.186 M_\odot$ coming from the detection 
of the GW170817 event~\cite{Abbott2019}. 
Most interactions that have been considered here lay 
inside the constraint. However, the SkI5 and the SEI with respective slopes $L=100$ MeV and $L=115$ MeV, and very stiff EoSs are not able to 
provide results for $\tilde{\Lambda}$ inside the observational bounds. 
The GW constraint gives some upper and lower bounds for the slope of the symmetry energy. 
At the same time, one can use bounds to $L$ to constraint other NS properties, such as the 
radius of a $1.4 M_\odot$ NS~\cite{Lourenco19, Lourenco19a}. 
Some main properties of NSs such as the total mass and radius for the maximum mass configuration and for a 
$1.4 M_\odot$ NS are contained in Table~\ref{tabletidal}. 
Notice that minor changes with the values in Table~\ref{Table-NSs} of Chapter~\ref{chapter2} are due to the 
prescription used for the core-crust transition.
We include in the same table the 
values of the dimensionless tidal deformability for a canonical NS and the mass-weighted tidal deformability 
of a binary NS at a chirp mass $\mathcal{M}= 1.186 M_\odot$.
We find in the lower limit of the constraint the results obtained with the D1M and MDI with $L=42$ MeV interactions, which 
provide radius $R_{1.4}$ for a $1.4 M_\odot$ NS of $10.05$ km and $10.11$ km, respectively.
On the other hand, the MDI with $L=88$ MeV and the SEI with $L=100$ MeV lay close to the upper limit of the constraint, 
providing, respectively, $R_{1.4}$ of $12.76$ km and $13.18$ km.
Therefore, the predictions using these few interactions for the radius of a canonical star lay in the 
range $10$ km $\lesssim R_{1.4} \lesssim 13$ km, which is in consonance with the values obtained by the LIGO and Virgo 
collaboration for the value of the radius of a canonical NS of $1.4 M_\odot$, i.e., $R_{1.4} = 11.9^{+1.4}_{-1.4}$ km~\cite{Abbott2018}.

\begin{table}[t!]
\centering
\resizebox{\columnwidth}{!}{
\begin{tabular}{c|lccccc}
\hline
\multicolumn{2}{c|}{Force}     & $M_\mathrm{max}$  &$R (M_\mathrm{max})$       & $R (1.4M_\odot)$       & $\Lambda (1.4M_\odot)$    &  $\tilde{\Lambda} (\mathcal{M}=1.186 M_\odot)$  \\  \hline\hline
\multirow{2}{*}{Skyrme}    & SLy4 ($L=46$ MeV)      &     2.06   &     10.00    &   11.74     &          304       &        365                \\
                           & SkI5 ($L=129$ MeV)       &     2.28   &     11.85    &   14.48     &          1185      &        1402      \\ \hline
\multirow{2}{*}{Gogny}     & D1M   ($L=25$ MeV)     &    1.74   &     8.80     &   10.05      &         121       &         149      \\
                           & D1M$^*$ ($L=43$ MeV)      &     2.00  &      10.13   &    11.52     &           310     &         370       \\ \hline
\multirow{3}{*}{MDI}       & $L=42$ MeV    &     1.60   &     8.62    &    10.11     &           94      &          124    \\
                           & $L=60$ MeV    &       1.91  &      9.91  &      11.85   &            312     &           380 \\
                           & $L=88$ MeV    &       1.99  &      10.59 &      12.76   &            567      &          686  \\ \hline
\multirow{3}{*}{SEI}       & $L=86$ MeV   &       1.95  &      10.67 &      12.90    &           511      &          618    \\
                           & $L=100$ MeV  &        1.98 &       10.89 &      13.18   &            640     &             773      \\
                           & $L=115$ MeV    &       1.99 &       11.05 &      13.38   &            789     &          954   \\ \hline
\end{tabular}
}

\caption{Neutron star maximum mass $M_\mathrm{max}$, radius at the maximum mass $R (M_\mathrm{max})$ and for a $1.4 M_\odot$
NS. The table also includes the values of the dimensionless tidal deformability $\Lambda$ for a canonical $1.4M_\odot$ NS
and of the mass-weighted tidal deformability $\tilde{\Lambda}$ at a chirp mass of $\mathcal{M}=1.186 M_\odot$.
The mass results are given in units of solar masses and the radii in units of km. \label{tabletidal}}
\end{table}

\chapter{Summary and Conclusions}\label{conclusions}
\fancyhead[RE, LO]{Chapter 7}
In this thesis we have further extended the analysis of the properties of neutron stars (NSs) and of finite
nuclei through relating them to microphysical predictions associated to the isospin dependence of the equation of state (EoS)
used to characterize both the nuclei and the NS core. 
 
We have recalled in Chapter~\ref{chapter1} the basic idea of the mean-field approximation, 
where the system is described as a set of non-interacting quasiparticles moving independently inside an effective 
mean-field potential. 
We have summarized the main features of the Hartree-Fock method, and we have reminded the concept
of phenomenological potentials. The definition of some properties of the EoS of symmetric nuclear matter 
and of asymmetric nuclear matter that we have used in the following chapters are also collected there.
In our work, we have used the zero-range
density-dependent Skyrme~\cite{skyrme56, vautherin72,sly41} interactions, and the finite-range 
Gogny~\cite{decharge80, berger91}, MDI~\cite{das03,li08} and SEI~\cite{behera98, Behera05} models. We provide for each one of them 
the explicit interaction, as well as the corresponding expressions for different isoscalar and isovector nuclear matter 
properties. All these interactions perform fairly well in the finite-nuclei density regime, and we 
use them to study nuclear matter at larger densities, like the ones found in systems such as NSs. 
Skyrme interactions have been already widely used to 
study stellar objects as, for example, the celebrated SLy4 force, as well as some MDI and SEI interactions. On the other hand, our study with 
Gogny forces has been one of the few to date where these interactions have been applied to NSs. 

We have studied some properties of asymmetric nuclear matter, and we have presented our results in 
Chapter~\ref{chapter2} for a large set of Skyrme models with different nuclear matter properties and a set of 
different Gogny interactions. 
We first analyze in detail the impact on different nuclear and NS properties of the Taylor expansion of the energy per particle in asymmetric nuclear matter in
even powers of the isospin asymmetry $\delta$. 
The lowest order is the contribution in 
symmetric nuclear matter and the next term, quadratic in $\delta$, corresponds to the usual symmetry energy coefficient. 
This truncation of the EoS at second order is widely used in the literature when performing microscopic calculations. Terms of a higher order than two
 in the Taylor expansion provide additional corrections that account for the departure of the energy from a quadratic law in $\delta$. 
In our work, we have expanded the energy per particle up to tenth order when working with Skyrme interactions and up to 
sixth order when working with Gogny forces.
 
From our study of the symmetry energy, we notice that the considered Skyrme forces can be separated into two different groups. In one of them
the symmetry energy, defined as the second-order coefficient of the expansion of the energy per particle, vanishes at some suprasaturation 
density several times the saturation one, which implies that for larger densities the asymmetric nuclear matter obtained with these forces 
becomes isospin unstable. This is a general trend exhibited by the Skyrme models with a slope parameter smaller than about $46$ MeV. The other group of 
Skyrme interactions have larger slope parameters and the corresponding symmetry energy has an increasing behaviour as a function of the density.
Terms of order higher than two in the expansion will contain contributions coming only from the kinetic and from non-local terms.
The coefficients in the expansion from fourth to tenth order in $\delta$ do not show any well defined common trend as a function of the density 
and are strongly model-dependent. 
 
On the other hand, in the analyzed Gogny interactions, the second-order symmetry energy coefficient shows an
isospin instability at large values of the density, above $0.4 - 0.5$~fm$^{-3}$~\cite{gonzalez17}.
The fourth- and sixth-order symmetry energy coefficients contain contributions from the kinetic and
exchange terms exclusively. The results indicate that Gogny parametrizations also fall into two different 
groups according to the density behaviour of these coefficients above saturation.
In the first group (D1S, D1M, D1N, and D250), the fourth- and sixth-order coefficients reach a maximum 
and then decrease with growing density. In the second group (D1, D260, D280, and D300), these coefficients 
are always increasing functions of density in the range analyzed. The different behaviour
of the two groups can be traced back to the density dependence of the exchange terms, which  
add to the kinetic part of the fourth- and sixth-order coefficients. 

At saturation density, the higher-order symmetry energy coefficients
are relatively small for both Skyrme and Gogny interactions. This supports the accuracy 
of the Taylor expansion at second order in calculations of the energy in asymmetric nuclear matter 
around the saturation density.  
 
An alternative definition of the symmetry energy is provided by the difference  between the energy per
particle in neutron matter and in symmetric matter, and we have called it parabolic symmetry energy.
This difference also coincides with the infinite sum of all the coefficients of the Taylor expansion
of the energy per particle in powers of the isospin asymmetry if one considers isospin asymmetry equal to one~\cite{gonzalez17}.
We find that around saturation, the difference between the PA estimate $E_{\mathrm{sym}}^{PA} (\rho)$ 
and the $E_{\mathrm{sym},2} (\rho)$ coefficient is largely accounted by the sum of higher-order contributions.
Another important quantity in studies of the symmetry energy is the slope parameter $L$, which is 
commonly used to characterize the density dependence of the symmetry energy near saturation.
We find, either using Skyrme or Gogny models, that large discrepancies of several MeV can arise between the $L$ value calculated with 
$E_{\mathrm{sym},2} (\rho)$ or with $E_{\mathrm{sym}}^{PA} (\rho)$. 
Again, adding higher-order contributions accounts for most of these differences.

To study several properties of NSs one needs to consider $\beta$-stable stellar 
matter first. To proceed with this study, we take into account neutrons, protons, and leptons in chemical equilibrium.  
To test the accuracy of this approach at high densities, we have performed a systematic study with Skyrme and Gogny functionals~\cite{gonzalez17}. 
By 
solving the equations with the exact EoS and with the Taylor expansion of Eq.~(\ref{eq:EOSexpgeneral}) at increasing 
orders in $\delta$, we are able to analyze the convergence of the solutions with the expansion. 
 The agreement between the $\beta$-equilibrium asymmetries obtained using the exact EoS 
and the truncated Taylor expansion improves order by order.
However, the convergence of this expansion is rather slow, in particular for forces with large slope parameter~$L$~\cite{gonzalez17}. 
In this scenario, to reproduce the exact asymmetry of $\beta$-stable matter with approximations based on the Taylor expansion of
the energy per particle requires to include contributions of higher-order than two.
The parabolic approximation case, in where the energy per particle for a given asymmetry and density comes 
from a $\delta^2$ interpolation between the 
corresponding values in neutron matter and in symmetric matter, is also unable to reproduce the exact asymmetry of 
$\beta$-stable matter in the range of densities considered. 
These differences will make an impact when one tries to obtain other NS properties, and remark on the importance of the knowledge of the exact
EoS. 
Another approximation, used sometimes in the past, 
for example in cases when it is complicated to obtain the full expression of the EoS, 
consists in considering the exact
kinetic energy part and performing the Taylor expansion only in the potential contribution to the energy per particle. This approach
works very well, reproducing closely the exact asymmetry in $\beta$-stable matter for all 
the range of densities considered. 

We have also studied in Chapter~\ref{chapter2} the mass-radius relation of NSs by solving the TOV equations and
using Skyrme and Gogny interactions~\cite{gonzalez17}. In this case, we have used a polytropic EoS for the inner crust
and the transition to the core is obtained by the thermodynamical method~\cite{xu09a, Moustakidis12,Ducoin11,Providencia14,Fattoyev:2010tb,Pais2016,routray16,gonzalez17}.
We find that interactions with soft symmetry energy are not able to 
provide numerically stable solutions of the TOV equations, and that only stiff enough 
EoSs may give an NS of $2 M_\odot$~\cite{Demorest10, Antoniadis13}.
We have considered a large set of Skyrme interactions, and we have closely analyzed a sample of 5 of them, namely the 
MSk7, UNEDF0, SkM$^*$, SLy4 and SkI5 parameterizations. Of this subset,  
only the SLy4 and the SkI5 interactions fit inside the astrophysical constraints of $2 M_\odot$
Moreover, the SLy4 force is the only one that, giving a radius of $\sim 10$ km at the maximum mass configuration 
and a radius of $11.8$ km for a canonical NS of $1.4M_\odot$ fits inside the
constraints for the radii coming from low-mass X-ray binaries and
X-ray bursters~\cite{Nattila16, Lattimer14}.
Of the considered Gogny interactions, i.e., D1, D1S, D1M, D1N, D250, D260, D280 and D300,
only the D1M and D280 parameterizations are able to generate  
NSs above the $1.4 M_\odot$ value, reaching maximum masses of $M=1.74 M_\odot$ and $1.66 M_\odot$, respectively. 
However, one has to take into account that these interactions have not been fitted to 
reproduce highly asymmetric nuclear matter like the one found in the interior of NSs,
even though in the fitting of D1N and D1M it was imposed the reproduction of the equation of state of neutron matter of 
Friedman and Pandharipande~\cite{Friedman81}.
Finally, in Chapter~\ref{chapter2}, we have analyzed the convergence of the mass-radius relation if the 
Taylor expansion of the EoS is used instead of its full form. The prescription for the inner crust and the transition is 
the same as before. In general, if the second-order expansion of the EoS is used, one finds results quite far from the 
exact ones. If higher-orders are used, the results approach the ones obtained using the full expression of the 
EoS. However, for interactions with very stiff EoSs, the convergence of the solution is slower than for softer interactions. 
This again points out the necessity of using the full expression of the EoS or, in its absence, its Taylor expansion cut at an order 
higher than $\delta^2$.

The non-existence of Gogny interactions of the D1 type that are able to provide 
NSs with large masses around $2M_\odot$ because of their soft symmetry energies~\cite{Sellahewa14, gonzalez17} leads us to
introduce in Chapter~\ref{chapter3} a new Gogny parametrization, dubbed D1M$^*$ 
that, while preserving 
the description of nuclei similar to the one obtained with D1M, the slope of 
the symmetry energy is
modified to make the EoS of $\beta$-stable matter stiff enough to 
obtain NS masses of $2 M_\odot$ inside the observational constraints~\cite{gonzalez18}. 
We have also introduced a second new parametrization, which we name D1M$^{**}$, following the 
same fitting procedure as D1M$^{*}$ and which is able to provide a $1.91 M_\odot$ NS in the lower 
region of the observational bounds~\cite{gonzalez18a, Vinas19}.
The D1M force \cite{goriely09} is susceptible to being used in this procedure,
but not D1S and D1N, as they are too far from the $2M_\odot$ target. 
We find that the new sets of parameters also perform at the same level 
as D1M in all aspects of finite nuclei analyzed in this work. Stellar properties from 
D1M$^*$, such as the \mbox{M-R} relation and the moment of inertia,
are in good agreement with the results from the Douchin-Haensel SLy4 EoS~\cite{douchin01}, which is 
designed especially for working in the astrophysical scenario.

With these modified interactions we also study 
some ground-state properties of finite nuclei, such as binding energies, neutron and proton radii, response to quadrupole 
deformation and fission barriers. We find that both D1M$^*$ and D1M$^{**}$ interactions perform as well as D1M in all the 
concerned properties of finite nuclei~\cite{gonzalez18, gonzalez18a}. On the whole, we can say
that the D1M$^*$ and D1M$^{**}$ forces presented in this work are a good alternative to describe simultaneously finite nuclei and NSs providing 
very good results
in harmony with the experimental and observational data. Moreover, 
these two interactions are much less demanding than the D2 force in
terms of computational resources \cite{chappert15}.

The correct determination of the transition between the core and the crust in NSs is essential in the study 
of different NS properties, such as the properties of the crust, or pulsar glitches. 
In Chapter~\ref{chapter4} we have calculated the core-crust transition searching
for the density where the uniform $\beta$-stable matter becomes unstable 
against small fluctuations of the neutron, proton and electron distributions. 
The instabilities in the core have been determined following two methods, namely the thermodynamical method~\cite{kubis04,kubis06,xu09a,Moustakidis10,Cai2012,
Moustakidis12,Seif14,routray16, gonzalez17} and the dynamical method~\cite{baym71, pethick95,ducoin07, gonzalez19}. 
First, we evaluate the core-crust transition using the thermodynamical method, 
where one requires the mechanical and chemical stability of the NS core.
We have obtained the results for Skyrme and Gogny interactions~\cite{gonzalez17}. As 
noted in earlier literature, 
the core-crust transition density is anticorrelated with the slope parameter $L$ of the models for both types of interactions. 
On the other hand, in contrast to the transition density, the transition pressure is not seen 
to have a high correlation with~$L$.
Moreover, we have studied the convergence of the core-crust transition properties, i.e., density, pressure, and isospin 
asymmetry at the core-crust boundary, when the Taylor expansion of the EoS is used instead of the full EoS. 
In general, adding more terms to the Taylor expansion of the EoS brings the 
transition density closer to the exact values.
However, there can be still significant differences even when the Taylor expansion is pushed 
to tenth (Skyrme) or sixth (Gogny) order. This points out that the convergence for the transition properties is slow.
In fact, the order-by-order convergence for the transition
pressure is  sometimes not only slow, but actually erratic~\cite{gonzalez17}. 

We also compute the transition density using the so-called dynamical 
method, where the stable densities correspond to the values for which the curvature matrix, i.e., the second variation of the total energy, 
including bulk, finite size and Coulomb effects, is convex with respect to the density.
In this case, we have performed the calculations using Skyrme interactions and three types of finite-range interactions, Gogny, MDI and SEI~\cite{gonzalez19}. 
The values of the transition properties show the same global trends as exhibited when they are obtained using the thermodynamical method, 
but they are shifted to lower values, as the addition of the surface and Coulomb terms help to stabilize more the system. 

We have first obtained the results for Skyrme interactions, where we have analyzed the convergence of the results when one 
uses the Taylor expansion of the EoS. 
As seen before, the agreement of these core-crust transition
properties estimated using the energy per particle expanded as a Taylor series with the exact predictions improves if more terms are included
in the truncated sum. However, the quality of this agreement is model dependent as far as it depends on the slope parameter $L$.
Retaining up to $\delta^{10}$-terms in the expansion, the gap between the exact and approximated transition densities, asymmetries, and pressures
for interactions with $L$ below $\sim 60$ MeV is nearly inexistent. However, for forces with $L$ above $\sim 60$ MeV, these gaps are not negligible and
may be quite large. Moreover, the parabolic approach, very useful for estimating the energy per particle in asymmetric nuclear matter, does not predict
very accurate results for the crust-core properties compared with the exact results, in particular for the transition pressures.   
On the other hand, we have tested the case where one considers the full expression for the kinetic part of the interaction and 
the Taylor expansion up to second order is performed only in its potential part. 
In this case, one almost recovers the results obtained with the exact EoS.

A global conclusion of our analysis is that the use of 
the Taylor expansion up to quadratic terms in isospin asymmetry of the energy per particle 
may be relatively sufficient to study the crust-core transition for models with soft symmetry energy. However, models with slope
parameter $L$ above $\sim 60$ MeV require the use of the exact energy per particle or its improved delta expansion to describe properly the 
relevant physical quantities in the crust-core transition.

In this Chapter~\ref{chapter4} we have also analyzed the core-crust transition in NSs with the dynamical method
using several finite-range interactions of Gogny, MDI and SEI types. Contrary to the case of Skyrme interactions that have the contributions   
to the dynamical potential explicitly differentiated, we have to derive the energy curvature matrix in momentum space.
Our study extends previous works available
in the literature for finite-range forces, which were basically performed using MDI interactions and where the 
surface effect was considered by the gradient contributions coming from the interaction part only. 
We have taken into account contributions to the surface from both the interaction part and the kinetic 
energy part. In this context, it can be considered
as more self-contained compared to the earlier studies on the subject. 
We have obtained the contribution coming from the direct energy through the
expansion of its finite-range form factors in terms of distributions, which allows one to write the direct contribution beyond
the long-wavelength limit (expansion till $k^2$-order). We have used the ETF approximation 
to write the kinetic and exchange energies as the sum of a bulk term plus a $\hbar^2$ correction. This $\hbar^2$
term can be written as a linear combination of the square of the gradients of the neutron and proton distributions, with density-dependent coefficients,
which in turn provide a $k^2$-dependence in the energy curvature matrix.

We have found that the effects of the finite-range part of the nuclear interaction on the energy curvature matrix mainly
arise from the direct part of the energy, where the $\hbar^2$-contributions 
from the kinetic and exchange part of the finite-range interaction can be considered as a small correction.
We have concluded that in the application of the dynamical method with finite-range forces it is an
accurate approximation, at least in the many cases we have studied here, to compute the $D_{qq'}$ coefficients 
of the dynamical potential by considering only
the contributions from the direct energy, namely~\cite{gonzalez19}:
\begin{small}
\begin{eqnarray}
 D_{nn}(k)&=&  D_{pp}(k) \nonumber \\
&=&\sum_m \bigg[W_m + \frac{B_m}{2} - H_m - \frac{M_m}{2}\bigg] \big({\cal F}_m(k)-{\cal F}_m(0)\big) , \nonumber \\
 D_{np}(k)&=&  D_{pn}(k) \nonumber \\
&=&\sum_m \bigg[W_m + \frac{B_m}{2}\bigg] \big({\cal F}_m(k)-{\cal F}_m(0)\big) .
\end{eqnarray}
\end{small}
Therefore, as the ${\cal F}_m(k)$ factors are just the Fourier transforms of the form factors
 $v_m(s)$, i.e., ${\cal F}_m(k)=\pi^{3/2} \alpha_m^3 e^{-\alpha_m^2 k^2/4}$ for Gaussians and ${\cal F}_m(k)= 4\pi \mu_m^{-1}(\mu_m^2 + k^2)^{-1}$ for Yukawians,
the dynamical calculation of the core-crust transition with the dynamical method using finite-range
forces becomes almost as simple as with Skyrme forces.

We have also analyzed the global behaviour of the core-crust transition density and pressure as a function of the slope of
the symmetry energy at saturation for a large set of nuclear models, which include finite-range
interactions as well as Skyrme forces. 
The results for MDI sets found under the present formulation are in good agreement with the earlier ones 
reported by Xu and Ko~\cite{xu10b}, where they use a different
density matrix expansion to deal with the exchange energy and adopted the Vlasov equation method to obtain the curvature
energy matrix.
We also observe that within the MDI and SEI families of interactions, 
the transition density and pressure are highly correlated with the slope of the symmetry
energy at saturation. However, when nuclear models with different saturation properties are considered, these correlations are deteriorated,
in particular the one related to the transition pressure. 
From our analysis, we have also found noticeable differences in the transition density and pressure calculated 
with the thermodynamical method and the dynamical method, and later we point out its consequences when studying properties of the 
NS crust.

We have devoted Chapter~\ref{chapter5} to study global NS properties. First, we study stellar masses and radii 
paying special attention to the properties related to the crust, that is, the  crustal thickness and crustal mass, which can influence
observational properties. We calculate them using different sets of mean-field models.
These properties are directly related to the core-crust transition and to the choice of the inner crust. Many mean-field interactions used to 
compute the core EoS have not been used to find the corresponding EoS in the inner crust
as this region is particularly difficult to describe owing to the presence of the neutron gas and the possible existence of nuclear clusters 
with non-spherical shapes.
As it is usual in the literature, we use either a polytropic EoS in the inner crust 
or an EoS computed with a different model in the crust region. 
The use of different inner crust prescriptions may not have a large impact when studying global properties of the NS such as the total mass or radius, if the 
star is massive enough. On the other hand, if the NS is light with masses $M \lesssim 1 M_\odot$, the 
 crust region has a noticeable influence on the determination of these properties. If one studies
the crustal properties of the NS, the choice of the inner crust may lead to large uncertainties. 
Because of that, we have studied the mass and thickness of the crust using the recent approximation derived by Zdunik et al.~\cite{Zdunik17} who solved the 
TOV equations from the center of the star to the point corresponding to the transition density. 
This points out again the importance of an accurate determination of the core-crust transition. Until now, in previous literature, the 
thermodynamical method has been widely used in all types of interactions. These calculations are relatively easy to perform because only the derivatives of the energy 
per particle are needed, and it takes into account neither surface nor Coulomb effects. On the other hand, to perform the calculation with the dynamical method is 
more complicated, specifically when using finite-range forces, where the exchange contributions in the surface terms are also considered. 
We analyze how the determination of the core-crust transition affects the results obtained for the crustal mass and crustal thickness, and it is found 
that the differences in the transition point determined with the thermodynamical and dynamical methods have a relevant impact on the 
considered crustal properties, in particular for models with stiff symmetry energy~\cite{gonzalez19}.
These crustal 
properties play a crucial role in the description of several observed phenomena, like glitches, r-mode oscillation, etc.
Therefore, the core-crust transition density needs to be ascertained as precisely as possible by 
taking into account the associated physical conditions in order to have a realistic estimation of the observed phenomena.

The detection of gravitational waves has opened a new era in astrophysics, cosmology, and nuclear physics. The GW170817 detection 
by the LIGO and Virgo collaboration~\cite{Abbott2017, Abbott2018, Abbott2019}, accounting for the merger of a binary NS system, led to 
a new set of constraints in both the astrophysical and nuclear sector. 
One of these constraints, directly measured from the GW signal of the merger, is the dimensionless mass-weighted tidal deformability $\tilde{\Lambda}$
at a certain chirp mass $\mathcal{M}$. Other constraints coming from the data analysis of the detection are on the dimensionless tidal 
deformability of a single NS, $\Lambda$, for a canonical NS of $1.4 M_\odot$, on the masses of the two NSs, on the radii, etc.~\cite{Abbott2017, Abbott2018, Abbott2019}.
In our work, we have calculated the tidal deformability $\Lambda$ using different mean-field models. 
We observe that very stiff EoSs will provide $\Lambda$ at $1.4 M_\odot$ that are not inside the boundary limits. 
If one computes $\tilde{\Lambda}$ for an NS binary system one finds the same behaviour: very stiff EoSs with $L \gtrsim 90-100$ MeV
are not able to provide a system that gives small enough $\tilde{\Lambda}$.
With this limitation on $L$ coming from the constraints on the tidal deformability, 
one can also restrict other NS properties, such as the radius of a canonical NS. In our case, with the interactions we have used, 
we find that this radius should be in the range $10 \lesssim R_{1.4} \lesssim 13$ km, which is in consonance with the values 
obtained by the LIGO and Virgo collaboration of $R_{1.4} = 11.9 \pm 1.4$ km.

In this work, the moment of inertia of NSs has also been studied in Chapter~\ref{chapter5}. 
We find large values of the moment of inertia when the considered EoS to describe the core is stiff~\cite{gonzalez17}. 
Using universal relations between $\Lambda$ and the moment of inertia, and using the data extracted from the GW170817 event, 
Landry and Kumar extracted some constraints for the moment of inertia of the double binary pulsar PSR J0737-3039. 
With these constraints, we find that very stiff interactions are not suitable to describe the moment of inertia of these observables. 
In particular, the two new interactions D1M$^*$ and D1M$^{**}$ that we have formulated fit inside the constraint, as well as inside constraining bands from Refs.~\cite{Lattimer2005, Breu2016}
for the plots of the dimensionless 
quantity $I/MR^2$ against the compactness. 
This points out again the good performance of these interactions when studying NS systems. 
Finally, the fraction of the moment of inertia enclosed in the crust is analyzed. Again, we see the importance of the determination 
of the core-crust transition, as the results calculated if considering the core-crust transition given by the dynamical method are much lower than the ones obtained if using the transition 
density given by the 
transition density given by the
thermodynamical approach. This may play a crucial role when predicting other phenomena like glitches or r-mode oscillations. 
 
To conclude, we mention some future prospects:

\begin{itemize}
 \item A more extensive and systematical application to the study of the properties of finite nuclei with the newly proposed D1M$^*$ and D1M$^{**}$ interactions is 
 to be performed. 
 This work is in progress and will be presented in the near future. 
 \item When studying NS properties, we have emphasized the necessity of having a unified EoS describing with the same interaction all regions of the NS. 
 Because of that, there is work under progress in our group aimed to obtain the inner crust and outer crust EoSs using 
 Gogny interactions and, in particular, with the new D1M$^*$ and D1M$^{**}$ interactions.
 \item As stated previously, the detection of gravitational waves has opened a new window in physics. New observations on mergers of binary NS
 systems or binary NS+black hole systems will provide new constraints on NS observables, such as the tidal deformability, the stellar radii, the moment of inertia, etc. 
 This will help to better determine the EoS of highly dense and asymmetric nuclear matter. Moreover, the NICER mission will aim to establish better boundaries on the 
 radii of NSs. Universal relations, for example between $\Lambda$ and the moment of inertia, may also be of great help to further constraint the nuclear 
 matter EoS. 
\end{itemize}

\fancyhead[RE, LO]{Chapter 7}

\begin{appendices}
\addtocontents{toc}{\protect\setcounter{tocdepth}{0}}
\fancyhead[RE, LO]{Chapter \thechapter}
\chapter{Derivatives of the equation of state to study the stability of uniform nuclear matter}\label{appendix_thermal}
The stability conditions for the thermodynamical potential $V_\mathrm{ther} (\rho)$ and the dynamical potential $V_\mathrm{dyn} (\rho)$ 
discussed in Chapter~\ref{chapter4} require the calculation of
the first and second derivatives of the energy per baryon $E_b (\rho, \delta)$ with respect to the density $\rho$ and the isospin asymmetry $\delta$. 
In this Appendix, we provide the corresponding expressions obtained with the exact EoS and with the Taylor expansion of the EoS up to order $\delta^{10}$
for Skyrme forces and up to order $\delta^6$ for Gogny interactions.

\section{Derivatives using the exact expression of the EoS} 
We collect here the derivatives of $E_b (\rho, \delta)$ involved in the study of the stability of homogeneous matter, obtained using 
the full expression of the EoS. 
The derivative $\partial E_b (\rho, \delta)/\partial \rho$ is immediately obtained from the expression for the pressure $P_b(\rho,\delta)$ we 
have given in Eq.~(\ref{eq:press_skyrme}) for Skyrme and in Eq.~(\ref{eq:pressure_bars}) for Gogny interactions, 
taking into account that $\partial E_b (\rho, \delta)/\partial \rho = P_b(\rho,\delta)/\rho^2$. 
The other derivatives that are needed to compute the thermodynamical and dynamical potentials for Skyrme interactions are the following:
\begin{eqnarray}
\frac{\partial^2 E_b (\rho, \delta)}{\partial \rho^2}&=& -\frac{1}{15} \frac{\hbar^2}{2m} 
\left(\frac{3 \pi^2}{2} \right)^{2/3} \rho^{-4/3}\left[ (1+\delta)^{5/3} + (1-\delta)^{5/3} \right] \nonumber\\
&+& \frac{(\sigma + 1) \sigma}{48} t_3 \rho^{\sigma-1} \left[ 2(x_3+2) - \frac{1}{2} (2x_3 + 1) 
\left[ (1+\delta)^2 + (1-\delta)^2\right]\right]\nonumber\\
&+& \frac{1}{24} \left( \frac{3 \pi^2}{2}\right)^{2/3} \rho^{-1/3}\left\{  \vphantom{\frac{1}{2}}
\left[ t_1(x_1+2) + t_2(x_2+2)\right] \left[ (1+\delta)^{5/3} + (1-\delta)^{5/3}\right] \right. \nonumber\\ 
&+&\left.  \frac{1}{2} \left[ t_2(2x_2 + 1) - t_1 (2x_1 + 1)\right] \left[ (1+\delta)^{8/3} + (1-\delta)^{8/3}\right] \right\}
\end{eqnarray}

\begin{eqnarray}
\frac{\partial^2 E_b (\rho, \delta)}{\partial \rho \partial \delta}&=& \frac{\hbar^2}{6m} 
\left(\frac{3 \pi^2}{2} \right)^{2/3} \rho^{-1/3}\left[ (1+\delta)^{2/3} - (1-\delta)^{2/3} \right]\nonumber\\
&-& \frac{1}{4} t_0 (2x_0+1) \delta - \frac{(\sigma +1)}{24} t_3 \rho^{\sigma} (2x_3+1) \delta  \nonumber\\
&+& \frac{1}{48} \left( \frac{3 \pi^2}{2}\right)^{2/3} \rho^{2/3}\left\{5 \left[ t_1(x_1+2) + t_2(x_2+2)\right] 
\left[ (1+\delta)^{2/3} - (1-\delta)^{2/3}\right] \right.\nonumber \\ 
&+&\left. 4 \left[ t_2(2x_2 + 1) - t_1 (2x_1 + 1)\right] \left[ (1+\delta)^{5/3} - (1-\delta)^{5/3}\right] \right\}
\end{eqnarray}

\begin{eqnarray}
\frac{\partial^2 E_b (\rho, \delta)}{\partial \delta^2}&=& \frac{\hbar^2}{6m}
\left(\frac{3 \pi^2}{2} \right)^{2/3} \rho^{2/3}\left[ (1+\delta)^{-1/3} + (1-\delta)^{-1/3} \right]\nonumber\\
&-&  \frac{t_0}{4} \rho (2x_0+1) - \frac{1}{24} t_3 \rho^{\sigma+1} (2x_3+1) \nonumber \\
&+& \frac{1}{24} \left( \frac{3 \pi^2}{2}\right)^{2/3} \rho^{5/3}\left\{ \left[ t_1(x_1+2) + 
t_2(x_2+2)\right] \left[ (1+\delta)^{-1/3} + (1-\delta)^{-1/3}\right] \right.\nonumber \\ 
&+&\left.  2 \left[ t_2(2x_2 + 1) - t_1 (2x_1 + 1)\right] \left[ (1+\delta)^{2/3} + (1-\delta)^{2/3}\right] \right\}.
\end{eqnarray}

On the other hand, for Gogny interactions we have:

\begin{eqnarray}
\frac{\partial^2 E_b (\rho, \delta)}{\partial \rho^2}&=& - \frac{\hbar^2}{30m} \left(\frac{3 \pi^2}{2} 
\right)^{2/3} \rho^{-4/3}\left[ (1+\delta)^{5/3} + (1-\delta)^{5/3} \right]\nonumber
\\
&+& \frac{(\alpha + 1) \alpha}{8} t_3 \rho^{\alpha-1} \left[ 3 - \left( 2 x_3 + 1 \right) \delta^2 \right] 
\nonumber
\\
&+&\sum_{i=1,2}\frac{1}{6 \rho^2 k_F^{3} \mu_i^3}  \Bigg\{  {\cal C}_i \left[ 
\vphantom{\frac{1}{2}}2 \left( -6 + 
k_{Fn}^2\mu_i^2 + k_{Fp}^2\mu_i^2\right) +  e^{- k_{Fn}^2\mu_i^2} \left( 6  + 4  k_{Fn}^2\mu_i^2 + 
k_{Fn}^4\mu_i^4\right)  \right.  \nonumber
\\
&+& \left.
  e^{- k_{Fp}^2\mu_i^2} \left( 6  + 4  k_{Fp}^2\mu_i^2 + k_{Fp}^4\mu_i^4\right) \vphantom{\frac{1}{2}} \right]
 + {D}_i e^{-\frac{1}{4} \left( k_{Fn}^2 + k_{Fp}^2\right)\mu_i^2} \nonumber
\\
&\times& \left[  \left( -12  k_{Fn} k_{Fp}\mu_i^2- k_{Fn}^3 k_{Fp}\mu_i^4 - k_{Fn} k_{Fp}^3\mu_i^4 \right) 
\mathrm{cosh} \left[ \frac{k_{Fn} k_{Fp}\mu_i^2}{2}  \right] \right.  \nonumber
\\
&+&  2 \left( 12 + k_{Fn}^2\mu_i^2  +  k_{Fp}^2\mu_i^2 + k_{Fn}^2 k_{Fp}^2\mu_i^4\right) 
\left.  \mathrm{sinh} \left[ \frac{k_{Fn} k_{Fp}\mu_i^2}{2} \right] \right] \Bigg\} ,
\end{eqnarray}
\begin{eqnarray}
\frac{\partial^2 E_b (\rho, 
\delta)}{\partial \rho \partial \delta} &=&  \frac{\hbar^2}{6m} \left(\frac{3 \pi^2}{2} \right)^{2/3} 
\rho^{-1/3}\left[ (1+\delta)^{2/3} - (1-\delta)^{2/3} \right]\nonumber
\\
&-& \frac{(\alpha +1)}{4} t_3 \rho^{\alpha} (2x_3+1) \delta  +  \sum_{i=1,2} \mu_i^3 \pi^{3/2} {\cal B}_i\delta \nonumber
\\
&-& \sum_{i=1,2}\frac{1}{6 \rho} \left\{{\cal C}_i \left[ \frac{-1 + e^{-k_{Fp}^2 \mu_i^2} 
\left( 1+ k_{Fp}^2 \mu_i^2\right)}{k_{Fp} \mu_i} - \frac{-1 + e^{-k_{Fn}^2 \mu_i^2} 
\left( 1+ k_{Fn}^2 \mu_i^2\right)}{k_{Fn} \mu_i} \right]\nonumber \right. \nonumber
\\
&-& {\cal D}_i e^{-\frac{1}{4} \left(k_{Fn}^2+ k_{Fp}^2\right)\mu_i^2} 
\left[ \left(k_{Fn} \mu_i - k_{Fp} \mu_i \right)
\cosh \left[ \frac{k_{Fn} k_{Fp}\mu_i^2}{2} \right] \right. \nonumber
\\
&-& \left.  \frac{2}{k_{Fn} k_{Fp} \mu_i^2}  \left( k_{Fn} \mu_i - k_{Fp} \mu_i + \delta k_F^3 \mu_i^3\right)
\sinh \left[  \frac{k_{Fn} k_{Fp}\mu_i^2}{2} \right] \right] \Bigg\} ,
\end{eqnarray}
\begin{eqnarray}
\frac{\partial^2 E_b (\rho, \delta)}{ \partial \delta^2} &=& 
\frac{\hbar^2}{6m} \left(\frac{3 \pi^2}{2} \right)^{2/3} \rho^{2/3}\left[ (1+\delta)^{-1/3} + (1-\delta)^{-1/3} \right]
\nonumber\\
&-&  \frac{t_3}{4} \rho^{\alpha+1} (2x_3+1) + \frac{1}{4}\sum_{i=1,2} \mu_i^3 \pi^{3/2}  {\cal B}_i \rho \nonumber
\\
&-& \frac{1}{6} \sum_{i=1,2} \left\{ {\cal C}_i \left [  \frac{1 - e^{- k_{Fp}^2 \mu_i^2}
\left( 1 + k_{Fp}^2 \mu_i^2 \right)}{(1-\delta) k_{Fp} \mu_i} + 
\frac{1 - e^{- k_{Fn}^2 \mu_i^2} 
\left( 1 + k_{Fn}^2 \mu_i^2 \right)}{(1+\delta) k_{Fn} \mu_i} \right] \right.\nonumber
\\
&+&  {\cal D}_i  e^{-\frac{1}{4} \left( k_{Fp}^2 +k_{Fn}^2 \right) \mu_i^2}
 \left[ \left(  k_{Fn} \mu_i\left(1-\delta \right)^{-1} +  k_{Fp} \mu_i\left(1+\delta \right)^{-1}\right)  
\cosh \left[\frac{k_{Fn} k_{Fp}\mu_i^2}{2}\right] \right. \nonumber
\\
&-&   \frac{2}{\left( 1- \delta^2 \right) k_{Fn} k_{Fp} \mu_i^2} \left( k_{Fn} \mu_i + k_{Fp} \mu_i 
- k_F^3 \mu_i^3  \right. \nonumber
\\
&+&\left.   \left. \delta \left(  k_{Fn} \mu_i - k_{Fp} \mu_i   
+ \delta k_F^3 \mu_i^3 \right) \right)\sinh \left[\frac{k_{Fn} k_{Fp}\mu_i^2}{2}\right] \right] \Bigg\} .
\end{eqnarray}

\section{Derivatives using the Taylor expansion of the EoS}
The derivatives of the energy per particle taking into account the Taylor expansion of the EoS can be rewritten as 
\begin{eqnarray}
 \frac{\partial E_b (\rho, \delta)}{\partial \rho}&=& \frac{\partial E_b(\rho, \delta=0)}{\partial \rho}+
 \sum_k \frac{\partial E_{\mathrm{sym}, 2k}}{\partial \rho} \delta^{2k}\\
 \frac{\partial^2 E_b (\rho, \delta)}{\partial \rho \partial \delta}&=& \frac{\partial^2 E_b(\rho, \delta=0)}{\partial \rho^2}+
 \sum_k \frac{\partial^2 E_{\mathrm{sym}, 2k}}{\partial \rho^2} \delta^{2k}
 \\
 \frac{\partial^2 E_b (\rho, \delta)}{\partial \rho^2}&=& 2 \sum_k k \delta^{2k-1} \frac{\partial E_{\mathrm{sym}, 2k}}{\partial \rho} 
 \\
 \frac{\partial^2 E_b (\rho, \delta)}{ \partial \delta^2} &=& 2 \sum_k k (2k-1) \delta^{2k-2} E_{\mathrm{sym}, 2k}
\end{eqnarray}

For the Skyrme interaction, the density derivatives of the energy per baryon in symmetric 
nuclear matter $E_b(\rho, \delta=0)$ given by

\begin{eqnarray}
\frac{\partial E_b(\rho, \delta=0)}{\partial \rho} &=& \frac{\hbar^2}{5m} \left(  \frac{3 \pi^2}{2}\right)^{2/3} \rho^{-1/3} + 
\frac{3}{8} t_0  + \frac{(\sigma +1)}{16} t_3 \rho^{\sigma}   \nonumber\\
 &+& \frac{1}{16} \left(  \frac{3 \pi^2}{2}\right)^{2/3} \rho^{2/3} \left[ 3 t_1 + t_2 (4 x_2 +5) \right]
\end{eqnarray}

\begin{eqnarray}
\frac{\partial^2 E_b (\rho, \delta=0)}{\partial \rho^2} &=& -\frac{\hbar^2}{15m} \left(  \frac{3 \pi^2}{2}\right)^{2/3} \rho^{-4/3} 
+ \frac{(\sigma +1) \sigma}{16} t_3 \rho^{\sigma-1}   \nonumber\\
 &+& \frac{1}{24} \left(  \frac{3 \pi^2}{2}\right)^{2/3} \rho^{-1/3} \left[ 3 t_1 + t_2 (4 x_2 +5) \right]
\end{eqnarray}

and for the Gogny interaction they read as
\begin{eqnarray}
\frac{\partial E_b(\rho, \delta=0)}{\partial \rho} &=& \frac{\hbar^2}{5m} \left(  \frac{3 \pi^2}{2}\right)^{2/3} \rho^{-1/3} 
+ \frac{3(\alpha +1)}{8} t_3 \rho^{\alpha}   +
 \frac{1}{2} \sum_{i=1,2} \mu_i^3 \pi^{3/2}  {\cal A}_i \nonumber
\\
 &-&\sum_{i=1,2}\frac{1}{2 \rho k_F^3 \mu_i^3}  \left( {\cal C}_i - {\cal D}_i \right)
 \left[ -2 + k_F^2 \mu_i^2 + e^{-k_F^2 \mu_i^2} \left(2 + k_F^2 \mu_i^2 \right) \right] ,
\\
\frac{\partial^2 E_b(\rho, \delta=0)}{\partial \rho^2} &=& -\frac{\hbar^2}{15m} \left(  \frac{3 \pi^2}{2}\right)^{2/3} \rho^{-4/3} 
+ \frac{3(\alpha +1) \alpha}{8} t_3 \rho^{\alpha-1}
\\  
&-& \sum_{i=1,2}\frac{1}{3 \rho^2 k_F^3 \mu_i^3}  \left( {\cal C}_i - {\cal D}_i \right)
\left[ 6 -2 k_F^2 \mu_i^2 - e^{-k_F^2 \mu_i^2} \left(6 +4 k_F^2 \mu_i^2 +k_F^4 \mu_i^4\right) \right] .\nonumber
\end{eqnarray}
The first and second derivatives with respect to the density of the symmetry energy coefficients 
can be readily computed from Eqs.~(\ref{eq:esym2skyrme2})---(\ref{eq:esym10skyrme}) 
for Skyrme interactions and from Eqs.~(\ref{eq:esym2gog})---(\ref{eq:esym6gog}) for Gogny interactions,
by 
taking derivatives of the $G_n (\eta)$ functions defined in Eqs.~(\ref{G1}), (\ref{G2}) and (\ref{G3})---(\ref{G6}), and using 
$\displaystyle \frac{\partial G_n (\eta)}{\partial \rho} = \frac{\partial G_n (\eta)}{\partial \eta} \, \frac{\partial \eta}{\partial \rho}$,
where $\displaystyle \frac{\partial \eta}{\partial \rho} = \frac{\pi^2 \mu_i}{2 k_F^2}$ for $\eta = \mu_i k_F$.
As this is relatively straightforward, we omit here the explicit results for these derivatives.

\chapter{Core-crust transition properties}\label{app_taules}
In this Appendix we collect the values of the transition density, pressure and asymmetry 
found using the thermodynamical and the dynamical methods (Chapter~\ref{chapter4}) for a set of different Skyrme 
parametrizations. The results have been obtained if using the full expression of the EoS,
its Taylor expansion to different orders, and the parabolic approximation. 
The results for Gogny interactions when using the thermodynamical method are also included, obtained 
using both the exact and Taylor expanded EoSs. Finally, the values of the transition properties 
obtained using the exact expression of the EoS for the finite-range Gogny, MDI and SEI interactions 
are also collected. 

\begin{table}[htb]
\begin{tabular}{cccccccccc}
\hline
\multicolumn{9}{c}{THERMODYNAMICAL METHOD} \\ \hline
Force & $L$ & \begin{tabular}[c]{@{}c@{}}$\delta_t$ \\ ($\delta^2$)\end{tabular} & 
\begin{tabular}[c]{@{}c@{}}$\delta_t$\\ ($\delta^4$)\end{tabular} & \begin{tabular}[c]{@{}c@{}}$\delta_t$\\ 
($\delta^6$)\end{tabular} & \begin{tabular}[c]{@{}c@{}}$\delta_t$\\ ($\delta^8$)\end{tabular} & 
\begin{tabular}[c]{@{}c@{}}$\delta_t$\\ ($\delta^{10}$)\end{tabular} & \begin{tabular}[c]{@{}c@{}}$\delta_t$\\ 
(Exact)\end{tabular} & \begin{tabular}[c]{@{}c@{}}$\delta_t$\\ (PA)\end{tabular} \\\hline \hline
MSk7 & 9.41 & 0.9317 & 0.9275 & 0.9260 & 0.9254 & 0.9250 & 0.9243 & 0.9278 \\
SIII & 9.91 & 0.9271 & 0.9195 & 0.9173 & 0.9164 & 0.9159 & 0.9151 & 0.9206 \\
SkP & 19.68 & 0.9252 & 0.9182 & 0.9164 & 0.9158 & 0.9155 & 0.9152 & 0.9196 \\
HFB-27 & 28.50 & 0.9299 & 0.9245 & 0.9230 & 0.9224 & 0.9222 & 0.9221 & 0.9252 \\
SKX & 33.19 & 0.9248 & 0.9187 & 0.9172 & 0.9167 & 0.9166 & 0.9167 & 0.9195 \\
HFB-17 & 36.29 & 0.9357 & 0.9315 & 0.9304 & 0.9300 & 0.9298 & 0.9301 & 0.9318 \\
SGII & 37.63 & 0.9555 & 0.9510 & 0.9500 & 0.9497 & 0.9497 & 0.9509 & 0.9516 \\
UNEDF1 & 40.01 & 0.9452 & 0.9408 & 0.9400 & 0.9397 & 0.9398 & 0.9411 & 0.9412 \\
Sk$\chi$500 & 40.74 & 0.9452 & 0.9432 & 0.9424 & 0.9421 & 0.9420 & 0.9419 & 0.9429 \\
Sk$\chi$450 & 42.06 & 0.9348 & 0.9311 & 0.9302 & 0.9299 & 0.9298 & 0.9301 & 0.9312 \\
UNEDF0 & 45.08 & 0.9400 & 0.9353 & 0.9345 & 0.9344 & 0.9346 & 0.9361 & 0.9355 \\
SkM* & 45.78 & 0.9440 & 0.9392 & 0.9383 & 0.9382 & 0.9383 & 0.9400 & 0.9395 \\
SLy4 & 45.96 & 0.9305 & 0.9275 & 0.9266 & 0.9263 & 0.9262 & 0.9265 & 0.9272 \\
SLy7 & 47.22 & 0.9311 & 0.9282 & 0.9273 & 0.9270 & 0.9269 & 0.9272 & 0.9278 \\
SLy5& 48.27 & 0.9327 & 0.9297 & 0.9289 & 0.9286 & 0.9285 & 0.9289 & 0.9295 \\
Sk$\chi$414 &51.92  & 0.9358 & 0.9328 & 0.9320 & 0.9318 & 0.9317 & 0.9322 & 0.9326 \\
MSka & 57.17 & 0.9449 & 0.9403 & 0.9395 & 0.9393 & 0.9394 & 0.9410 & 0.9403 \\
MSL0 & 60.00 & 0.9536 & 0.9503 & 0.9500 & 0.9503 & 0.9506 & 0.9538 & 0.9490 \\
SIV & 63.50 & 0.9425 & 0.9358 & 0.9347 & 0.9347 & 0.9349 & 0.9377 & 0.9356 \\
SkMP & 70.31 & 0.9596 & 0.9568 & 0.9568 & 0.9573 & 0.9578 & 0.9628 & 0.9556 \\
SKa & 74.62 & 0.9440 & 0.9397 & 0.9395 & 0.9400 & 0.9405 & 0.9447 & 0.9386 \\
R$_\sigma$ & 85.69 & 0.9631 & 0.9622 & 0.9632 & 0.9643 & 0.9653 & 0.9736 & 0.9596 \\
G$_\sigma$ & 94.01 & 0.9626 & 0.9626 & 0.9643 & 0.9656 & 0.9669 & 0.9772 & 0.9590 \\
SV & 96.09 & 0.9547 & 0.9502 & 0.9507 & 0.9518 & 0.9528 & 0.9610 & 0.9471 \\
SkI2 & 104.33 & 0.9609 & 0.9618 & 0.9631 & 0.9643 & 0.9654 & 0.9736 & 0.9587 \\
SkI5 & 129.33 & 0.9562 & 0.9591 & 0.9614 & 0.9633 & 0.9648 & 0.9754 & 0.9549 \\ \hline
\end{tabular}
\caption{Values of the core-crust transition asymmetry $\delta_t$ for Skyrme interactions 
calculated within the thermodynamical approach. The results have been computed with the exact expression of the EoS (Exact), the parabolic approximation (PA), 
or the approximations of the full EoS with Eq.~(\ref{eq:EOSexpgeneral}) up to second ($\delta^2$), fourth ($\delta^4$), sixth ($\delta^6$) 
eighth ($\delta^8$) and tenth ($\delta^{10}$) order. The value for each interaction of the slope parameter of the symmetry energy $L$ is also included in units of MeV.}
\end{table}

\begin{table}[htb]
\begin{tabular}{ccccccccc}
\hline
\multicolumn{9}{c}{THERMODYNAMICAL METHOD} \\ \hline
Force & $L$ & \begin{tabular}[c]{@{}c@{}}$\rho_t$ \\ ($\delta^2$)\end{tabular} & 
\begin{tabular}[c]{@{}c@{}}$\rho_t$\\ ($\delta^4$)\end{tabular} & \begin{tabular}[c]{@{}c@{}}$\rho_t$\\ 
($\delta^6$)\end{tabular} & \begin{tabular}[c]{@{}c@{}}$\rho_t$\\ ($\delta^8$)\end{tabular} & 
\begin{tabular}[c]{@{}c@{}}$\rho_t$\\ ($\delta^{10}$)\end{tabular} & \begin{tabular}[c]{@{}c@{}}$\rho_t$\\
(Exact)\end{tabular} & \begin{tabular}[c]{@{}c@{}}$\rho_t$\\ (PA)\end{tabular} \\ \hline\hline
MSk7 & 9.41 & 0.1291 & 0.1276 & 0.1270 & 0.1266 & 0.1263 & 0.1251 & 0.1273 \\
SIII & 9.91 & 0.1225 & 0.1202 & 0.1196 & 0.1192 & 0.1190 & 0.1181 & 0.1186 \\
SkP & 19.68 & 0.1204 & 0.1170 & 0.1156 & 0.1147 & 0.1140 & 0.1116 & 0.1153 \\
HFB-27 & 28.50 & 0.1074 & 0.1055 & 0.1046 & 0.1039 & 0.1034 & 0.1013 & 0.1057 \\
SKX & 33.19 & 0.1076 & 0.1058 & 0.1047 & 0.1040 & 0.1034 & 0.1015 & 0.1061 \\
HFB-17 & 36.29 & 0.1019 & 0.1002 & 0.0991 & 0.0983 & 0.0977 & 0.0951 & 0.1011 \\
SGII & 37.63 & 0.0976 & 0.0951 & 0.0934 & 0.0921 & 0.0911 & 0.0857 & 0.0963 \\
UNEDF1 & 40.01 & 0.1004 & 0.0978 & 0.0961 & 0.0949 & 0.0940 & 0.0896 & 0.0992 \\
Sk$\chi$500 & 40.74 & 0.1005 & 0.0995 & 0.0987 & 0.0981 & 0.0977 & 0.0956 & 0.1009 \\
Sk$\chi$450 & 42.06 & 0.0969 & 0.0954 & 0.0944 & 0.0937 & 0.0931 & 0.0909 & 0.0965 \\
UNEDF0 & 45.08 & 0.1021 & 0.0994 & 0.0976 & 0.0963 & 0.0953 & 0.0911 & 0.1010 \\
SkM* & 45.78 & 0.0980 & 0.0952 & 0.0934 & 0.0920 & 0.0910 & 0.0861 & 0.0967 \\
SLy4 & 45.96 & 0.0945 & 0.0931 & 0.0921 & 0.0914 & 0.0908 & 0.0886 & 0.0942 \\
SLy7 & 47.22 & 0.0931 & 0.0917 & 0.0907 & 0.0897 & 0.0894 & 0.0872 & 0.0928 \\
SLy5 & 48.27 & 0.0941 & 0.0926 & 0.0916 & 0.0908 & 0.0902 & 0.0877 & 0.0938 \\
Sk$\chi$414 & 51.92 & 0.1006 & 0.0989 & 0.0978 & 0.0970 & 0.0964 & 0.0939 & 0.1004 \\
MSka & 57.17 & 0.0990 & 0.0967 & 0.0951 & 0.0940 & 0.0932 & 0.0892 & 0.0984 \\
MSL0 & 60.00 & 0.0949 & 0.0916 & 0.0893 & 0.0877 & 0.0864 & 0.0795 & 0.0942 \\
SIV & 63.50 & 0.0984 & 0.0954 & 0.0934 & 0.0919 & 0.0908 & 0.0858 & 0.0975 \\
SkMP & 70.31 & 0.0915 & 0.0874 & 0.0846 & 0.0826 & 0.0810 & 0.0714 & 0.0908 \\
SKa & 74.62 & 0.0940 & 0.0904 & 0.0880 & 0.0862 & 0.0849 & 0.0785 & 0.0933 \\
R$_\sigma$ & 85.69 & 0.0948 & 0.0890 & 0.0852 & 0.0825 & 0.0805 & 0.0657 & 0.0943 \\
G$_\sigma$ & 94.01 & 0.0961 & 0.0893 & 0.0851 & 0.0820 & 0.0797 & 0.0620 & 0.0957 \\
SV & 96.09 & 0.0954 & 0.0898 & 0.0862 & 0.0835 & 0.0815 & 0.0702 & 0.0952 \\
SkI2 & 104.33 & 0.0903 & 0.0851 & 0.0817 & 0.0791 & 0.0772 & 0.0632 & 0.0898 \\
SkI5 & 129.33 & 0.0901 & 0.0846 & 0.0807 & 0.0778 & 0.0755 & 0.0595 & 0.0893 \\ \hline
\end{tabular}
\caption{Values of the core-crust transition density $\rho_t$ (fm$^{-3}$) for Skyrme interactions 
calculated within the thermodynamical approach. The results have been computed with the exact expression of the EoS (Exact), the parabolic approximation (PA), 
or the approximations of the full EoS with Eq.~(\ref{eq:EOSexpgeneral}) up to second ($\delta^2$), fourth ($\delta^4$), sixth ($\delta^6$) 
eighth ($\delta^8$) and tenth ($\delta^{10}$) order. The value for each interaction of the slope parameter of the symmetry energy $L$ is also included in units of MeV.}
\end{table}

\begin{table}[htb]
\begin{tabular}{ccccccccc}
\hline
\multicolumn{9}{c}{THERMODYNAMICAL METHOD} \\ \hline
Force & $L$ & \begin{tabular}[c]{@{}c@{}}P$_t$\\ ($\delta^2$)\end{tabular} & 
\begin{tabular}[c]{@{}c@{}}P$_t$\\ ($\delta^4$)\end{tabular} & \begin{tabular}[c]{@{}c@{}}P$_t$\\
($\delta^6$)\end{tabular} & \begin{tabular}[c]{@{}c@{}}P$_t$\\ ($\delta^8$)\end{tabular} & \begin{tabular}[c]{@{}c@{}}P$_t$\\
($\delta^{10}$)\end{tabular} & \begin{tabular}[c]{@{}c@{}}P$_t$\\ (Exact)\end{tabular} 
& \begin{tabular}[c]{@{}c@{}}P$_t$\\ (PA)\end{tabular} \\ \hline\hline
MSk7 & 9.41 & 0.4270 & 0.4404 & 0.4424 & 0.4398 & 0.4391 & 0.4366 & 0.4563 \\
SIII & 9.91 & 0.3864 & 0.4376 & 0.4438 & 0.4424 & 0.4418 & 0.4386 & 0.4703 \\
SkP & 19.68 & 0.6854 & 0.7106 & 0.7048 & 0.6958 & 0.6898 & 0.6681 & 0.7309 \\
HFB-27 & 28.50 & 0.5702 & 0.5833 & 0.5766 & 0.5679 & 0.5620 & 0.5398 & 0.6119 \\
SKX & 33.19 & 0.6708 & 0.6847 & 0.6748 & 0.6640 & 0.6568 & 0.6318 & 0.7198 \\
HFB-17 & 36.29 & 0.5649 & 0.5640 & 0.5531 & 0.5419 & 0.5343 & 0.5039 & 0.5957 \\
SGII & 37.63 & 0.5118 & 0.5128 & 0.4953 & 0.4781 & 0.4657 & 0.4016 & 0.5545 \\
UNEDF1 & 40.01 & 0.6413 & 0.6335 & 0.6126 & 0.5939 & 0.5806 & 0.5212 & 0.6785 \\
Sk$\chi$500 & 40.74 & 0.4018 & 0.3888 & 0.3807 & 0.3728 & 0.3681 & 0.3481 & 0.4099 \\
Sk$\chi$450 & 42.06 & 0.5440 & 0.5384 & 0.5270 & 0.5162 & 0.5090 & 0.4812 & 0.5680 \\
UNEDF0 & 45.08 & 0.7409 & 0.7277 & 0.7017 & 0.6797 & 0.6641 & 0.5991 & 0.7820 \\
SkM* & 45.78 & 0.6567 & 0.6469 & 0.6226 & 0.6012 & 0.5857 & 0.5173 & 0.6982 \\
SLy4 & 45.96 & 0.5274 & 0.5170 & 0.5061 & 0.4958 & 0.4888 & 0.4623 & 0.5459 \\
SLy7 & 47.22 & 0.5212 & 0.5099 & 0.4987 & 0.4883 & 0.4812 & 0.4540 & 0.5388 \\
SLy5 & 48.27 & 0.5361 & 0.5237 & 0.5114 & 0.5000 & 0.4923 & 0.4620 & 0.5546 \\
Sk$\chi$414 & 51.92 & 0.5909 & 0.5771 & 0.5624 & 0.5495 & 0.5410 & 0.5078 & 0.6112 \\
MSka & 57.17 & 0.6741 & 0.6600 & 0.6342 & 0.6122 & 0.5965 & 0.5284 & 0.7229 \\
MSL0 & 60.00 & 0.6774 & 0.6408 & 0.6031 & 0.5729 & 0.5510 & 0.4412 & 0.7145 \\
SIV & 63.50 & 0.7695 & 0.7569 & 0.7195 & 0.6880 & 0.6651 & 0.5682 & 0.8568 \\
SkMP & 70.31 & 0.6904 & 0.6339 & 0.5827 & 0.5432 & 0.5144 & 0.3590 & 0.7344 \\
SKa & 74.62 & 0.8102 & 0.7631 & 0.7139 & 0.6758 & 0.6483 & 0.5259 & 0.8671 \\
R$_\sigma$ & 85.69 & 0.8977 & 0.7663 & 0.6769 & 0.6133 & 0.5676 & 0.3024 & 0.9385 \\
G$_\sigma$ & 94.01 & 1.0270 & 0.8463 & 0.7323 & 0.6535 & 0.5974 & 0.2686 & 1.0671 \\
SV & 96.09 & 0.9172 & 0.8098 & 0.7181 & 0.6523 & 0.6054 & 0.3779 & 1.0363 \\
SkI2 & 104.33 & 0.8850 & 0.7410 & 0.6511 & 0.5877 & 0.5423 & 0.2845 & 0.8885 \\
SkI5 & 129.33 & 1.0561 & 0.8589 & 0.7367 & 0.6522 & 0.5921 & 0.2651 & 1.0319 \\ \hline
\end{tabular}
\caption{Values of the core-crust transition pressure $P_t$ (MeV fm$^{-3}$) for Skyrme interactions 
calculated within the thermodynamical approach. The results have been computed with the exact expression of the EoS (Exact), the parabolic approximation (PA), 
or the approximations of the full EoS with Eq.~(\ref{eq:EOSexpgeneral}) up to second ($\delta^2$), fourth ($\delta^4$), sixth ($\delta^6$) 
eighth ($\delta^8$) and tenth ($\delta^{10}$) order. The value for each interaction of the slope parameter of the symmetry energy $L$ is also included in units of MeV.}
\end{table}

\begin{table}[htb]
\centering
\resizebox{\columnwidth}{!}{
 \begin{tabular}{ccccccccccc}
 \hline
 \multicolumn{11}{c}{THERMODYNAMICAL METHOD}  \\    \hline
Force                       & D1     & D1S    & D1M    & D1N     & D250   & D260   & D280    & D300  &D1M$^*$&D1M$^{**}$ \\ \hline\hline
$L$                         &18.36   & 22.43  & 24.83  & 33.58   & 24.90  &  17.57 & 46.53   & 25.84  & 43.18 & 33.91\\\hline
$\delta_t^{\delta^2}$       & 0.9215 & 0.9199 & 0.9366 & 0.9373  & 0.9167 & 0.9227 & 0.9202  & 0.9190 &0.9386 &0.9375\\
$\delta_t^{\delta^4}$       & 0.9148 & 0.9148 & 0.9290 & 0.9336  & 0.9119 & 0.9136 & 0.9127  & 0.9128 &0.9315  & 0.9301 \\
$\delta_t^{\delta^6}$       & 0.9127 & 0.9129 & 0.9265 & 0.9321  & 0.9101 & 0.9112 & 0.9110  & 0.9110 &  0.9292&0.9278\\
$\delta_t^{\mathrm{exact}}$ & 0.9106 & 0.9111 & 0.9241 & 0.9310  & 0.9086 & 0.9092 & 0.9110  & 0.9096 & 0.9275 &0.9257\\
$\delta_t^\mathrm{PA}$      & 0.9152 & 0.9142 & 0.9296 & 0.9327  & 0.9111 & 0.9153 & 0.9136  & 0.9134 & 0.9316 &0.9305 \\ \hline

$\rho_t^{\delta^2}$         & 0.1243 & 0.1141 & 0.1061 & 0.1008  & 0.1156 & 0.1228 & 0.1001  & 0.1161 & 0.0974 &0.1000\\
$\rho_t^{\delta^4}$         & 0.1222 & 0.1129 & 0.1061 & 0.0996  & 0.1143 & 0.1198 & 0.0984  & 0.1145 &  0.095&0.0997\\
$\rho_t^{\delta^6}$         & 0.1211 & 0.1117 & 0.1053 & 0.0984  & 0.1131 & 0.1188 & 0.0973  & 0.1136 & 0.0949 &0.0989\\
$\rho_t^{\mathrm{exact}}$   & 0.1176 & 0.1077 & 0.1027 & 0.0942 & 0.1097 & 0.1159 & 0.0938 & 0.1109 & 0.0909 &0.0960\\
$\rho_t^\mathrm{PA}$               & 0.1222 & 0.1160 & 0.1078 & 0.1027  & 0.1168 & 0.1171 & 0.0986  & 0.1142& 0.0940 & 0.1019\\ \hline

$P_t^{\delta^2}$            & 0.6279 & 0.6316 & 0.3326 & 0.4882  & 0.7034 & 0.5892 & 0.6984  & 0.6776 & 0.3528 &0.3464 \\
$P_t^{\delta^4}$            & 0.6479 & 0.6239 & 0.3531 & 0.4676  & 0.6908 & 0.6483 & 0.7170  & 0.6998 & 0.3605 &0.3599\\
$P_t^{\delta^6}$            & 0.6452 & 0.6156 & 0.3554 & 0.4582  & 0.6811 & 0.6509 & 0.7053  & 0.6955 & 0.3575 &0.359\\
$P_t^{\mathrm{exact}}$      & 0.6184 & 0.5817 & 0.3390 & 0.4164  & 0.6464 & 0.6272 & 0.6493  & 0.6647 &0.3301  &0.3368\\
$P_t^\mathrm{PA}$                  & 0.6853 & 0.6725 & 0.3986 & 0.5173  & 0.7368 & 0.6809 & 0.7668  & 0.7356 & 0.4125 &0.4085\\\hline
\end{tabular}
}
\caption{Values of the core-crust transition density $\rho_t$ (in fm$^{-3}$) for Gogny forces
calculated within the thermodynamical approach and using exact expression of the EoS ($\rho_t^\mathrm{exact}$), the parabolic approximation ($\rho_t^{PA}$), 
or the approximations of the full EoS with Eq.~(\ref{eq:EOSexpgeneral}) up to second ($\rho_t^{\delta^2}$), fourth ($\rho_t^{\delta^4}$) and sixth ($\rho_t^{\delta^6}$) order.
The table includes the corresponding values of the transition pressure $P_t$ (in MeV fm$^{-3}$) and isospin asymmetry~$\delta_t$.}
\label{Table-transition}
\end{table}

\begin{table}[htb]
\begin{tabular}{ccccccccc}
\hline
\multicolumn{9}{c}{DYNAMICAL METHOD} \\ \hline
Force & $L$& \begin{tabular}[c]{@{}c@{}}$\delta_t$ \\ ($\delta^2$)\end{tabular} & \begin{tabular}[c]{@{}c@{}}$\delta_t$\\ 
($\delta^4$)\end{tabular} & \begin{tabular}[c]{@{}c@{}}$\delta_t$\\ ($\delta^6$)\end{tabular} & \begin{tabular}[c]{@{}c@{}}$\delta_t$\\ 
($\delta^8$)\end{tabular} & \begin{tabular}[c]{@{}c@{}}$\delta_t$\\ ($\delta^{10}$)\end{tabular} & \begin{tabular}[c]{@{}c@{}}$\delta_t$\\ 
(Exact)\end{tabular} & \begin{tabular}[c]{@{}c@{}}$\delta_t$\\ (PA)\end{tabular} \\ \hline \hline
MSk7 & 9.41 & 0.9312 & 0.9272 & 0.9259 & 0.9253 & 0.9250 & 0.9243 & 0.9276 \\
SIII & 9.91 & 0.9272 & 0.9201 & 0.9180 & 0.9171 & 0.9167 & 0.9159 & 0.9212 \\
SkP & 19.68 & 0.9274 & 0.9214 & 0.9199 & 0.9194 & 0.9191 & 0.9192 & 0.9228 \\
HFB-27 & 28.50 & 0.9327 & 0.9278 & 0.9264 & 0.9259 & 0.9257 & 0.9258 & 0.9285 \\
SKX & 33.19 & 0.9276 & 0.9219 & 0.9206 & 0.9202 & 0.9200 & 0.9203 & 0.9228 \\
HFB-17 & 36.29 & 0.9392 & 0.9354 & 0.9344 & 0.9341 & 0.9339 & 0.9344 & 0.9357 \\
SGII & 37.63 & 0.9587 & 0.9547 & 0.9538 & 0.9535 & 0.9535 & 0.9549 & 0.9553 \\
UNEDF1 & 40.01 & 0.9498 & 0.9461 & 0.9455 & 0.9454 & 0.9455 & 0.9473 & 0.9463 \\
Sk$\chi$500 & 40.74 & 0.9473 & 0.9452 & 0.9445 & 0.9442 & 0.9440 & 0.9440 & 0.9450 \\
Sk$\chi$450 & 42.06 & 0.9381 & 0.9347 & 0.9338 & 0.9336 & 0.9335 & 0.9339 & 0.9348 \\
UNEDF0 & 45.08 & 0.9453 & 0.9413 & 0.9407 & 0.9407 & 0.9409 & 0.9429 & 0.9414 \\
SkM* & 45.78 & 0.9490 & 0.9449 & 0.9442 & 0.9441 & 0.9443 & 0.9462 & 0.9453 \\
SLy4 & 45.96 & 0.9346 & 0.9318 & 0.9309 & 0.9306 & 0.9305 & 0.9308 & 0.9315 \\
SLy7 & 47.22 & 0.9351 & 0.9323 & 0.9315 & 0.9312 & 0.9311 & 0.9315 & 0.9321 \\
SLy5 & 48.27 & 0.9370 & 0.9342 & 0.9333 & 0.9331 & 0.9330 & 0.9335 & 0.9339 \\
Sk$\chi$414 & 51.92 & 0.9394 & 0.9365 & 0.9358 & 0.9356 & 0.9355 & 0.9361 & 0.9364 \\
MSka & 57.17 & 0.9491 & 0.9449 & 0.9442 & 0.9440 & 0.9441 & 0.9458 & 0.9450 \\
MSL0 & 60.00 & 0.9592 & 0.9563 & 0.9560 & 0.9562 & 0.9565 & 0.9598 & 0.9559 \\
SIV & 63.50 & 0.9490 & 0.9431 & 0.9422 & 0.9422 & 0.9424 & 0.9454 & 0.9433 \\
SkMP & 70.31 & 0.9659 & 0.9635 & 0.9634 & 0.9638 & 0.9643 & 0.9691 & 0.9628 \\
SKa & 74.62 & 0.9517 & 0.9480 & 0.9478 & 0.9482 & 0.9487 & 0.9530 & 0.9473 \\
R$_\sigma$ & 85.69 & 0.9699 & 0.9689 & 0.9696 & 0.9704 & 0.9712 & 0.9786 & 0.9670 \\
G$_\sigma$ & 94.01 & 0.9705 & 0.9701 & 0.9712 & 0.9723 & 0.9733 & 0.9822 & 0.9676 \\
SV & 96.09 & 0.9645 & 0.9605 & 0.9607 & 0.9614 & 0.9622 & 0.9695 & 0.9590 \\
SkI2 & 104.33 & 0.9694 & 0.9697 & 0.9706 & 0.9715 & 0.9723 & 0.9794 & 0.9675 \\
SkI5 & 129.33 & 0.9682 & 0.9697 & 0.9711 & 0.9723 & 0.9732 & 0.9815 & 0.9668 \\ \hline
\end{tabular}
\caption{Values of the core-crust transition asymmetry $\delta_t$ for Skyrme interactions 
calculated within the dynamical approach. The results have been computed with the exact expression of the EoS (Exact), the parabolic approximation (PA), 
or the approximations of the full EoS with Eq.~(\ref{eq:EOSexpgeneral}) up to second ($\delta^2$), fourth ($\delta^4$), sixth ($\delta^6$) 
eighth ($\delta^8$) and tenth ($\delta^{10}$) order. The value for each interaction of the slope parameter of the symmetry energy $L$ is also included in units of MeV.}
\end{table}

\begin{table}[htb]
\begin{tabular}{ccccccccc}
\hline
\multicolumn{9}{c}{DYNAMICAL METHOD} \\ \hline
Force & $L$  & \begin{tabular}[c]{@{}c@{}}$\rho_t$ \\ ($\delta^2$)\end{tabular} & 
\begin{tabular}[c]{@{}c@{}}$\rho_t$\\ ($\delta^4$)\end{tabular} & \begin{tabular}[c]{@{}c@{}}$\rho_t$\\
($\delta^6$)\end{tabular} & \begin{tabular}[c]{@{}c@{}}$\rho_t$\\ ($\delta^8$)\end{tabular} & \begin{tabular}[c]{@{}c@{}}$\rho_t$\\
($\delta^{10}$)\end{tabular} & \begin{tabular}[c]{@{}c@{}}$\rho_t$\\ (Exact)\end{tabular} & \begin{tabular}[c]{@{}c@{}}$\rho_t$\\ (PA)\end{tabular} \\ \hline\hline
MSk7 & 9.41 & 0.1184 & 0.1170 & 0.1163 & 0.1159 & 0.1157 & 0.1145 & 0.1166 \\
SIII & 9.91 & 0.1163 & 0.1140 & 0.1134 & 0.1131 & 0.1128 & 0.1120 & 0.1125 \\
SkP & 19.68 & 0.1065 & 0.1034 & 0.1022 & 0.1013 & 0.1007 & 0.0983 & 0.1021 \\
HFB-27 & 28.50 & 0.0978 & 0.0961 & 0.0952 & 0.0945 & 0.0940 & 0.0919 & 0.0961 \\
SKX & 33.19 & 0.1002 & 0.0985 & 0.0975 & 0.0968 & 0.0962 & 0.0943 & 0.0986 \\
HFB-17 & 36.29 & 0.0922 & 0.0907 & 0.0897 & 0.0889 & 0.0883 & 0.0858 & 0.0914 \\
SGII & 37.63 & 0.0881 & 0.0858 & 0.0843 & 0.0832 & 0.0823 & 0.0771 & 0.0867 \\
UNEDF1 & 40.01 & 0.0891 & 0.0866 & 0.0850 & 0.0838 & 0.0828 & 0.0780 & 0.0881 \\
Sk$\chi$500 & 40.74 & 0.0932 & 0.0923 & 0.0916 & 0.0910 & 0.0906 & 0.0886 & 0.0935 \\
Sk$\chi$450 & 42.06 & 0.0889 & 0.0875 & 0.0865 & 0.0858 & 0.0853 & 0.0831 & 0.0883 \\
UNEDF0 & 45.08 & 0.0910 & 0.0884 & 0.0867 & 0.0855 & 0.0845 & 0.0800 & 0.0899 \\
SkM* & 45.78 & 0.0867 & 0.0843 & 0.0827 & 0.0814 & 0.0805 & 0.0756 & 0.0854 \\
SLy4 & 45.96 & 0.0851 & 0.0838 & 0.0830 & 0.0823 & 0.0818 & 0.0797 & 0.0847 \\
SLy7 & 47.22 & 0.0840 & 0.0828 & 0.0820 & 0.0813 & 0.0808 & 0.0786 & 0.0838 \\
SLy5 & 48.27 & 0.0845 & 0.0833 & 0.0823 & 0.0816 & 0.0811 & 0.0788 & 0.0842 \\
Sk$\chi$414 & 51.92 & 0.0922 & 0.0907 & 0.0897 & 0.0889 & 0.0884 & 0.0859 & 0.0919 \\
MSka & 57.17 & 0.0913 & 0.0893 & 0.0879 & 0.0868 & 0.0860 & 0.0821 & 0.0906 \\
MSL0 & 60.00 & 0.0839 & 0.0811 & 0.0792 & 0.0777 & 0.0766 & 0.0697 & 0.0832 \\
SIV & 63.50 & 0.0883 & 0.0857 & 0.0840 & 0.0826 & 0.0816 & 0.0764 & 0.0872 \\
SkMP & 70.31 & 0.0796 & 0.0763 & 0.0740 & 0.0723 & 0.0709 & 0.0615 & 0.0789 \\
SKa & 74.62 & 0.0828 & 0.0798 & 0.0777 & 0.0762 & 0.0750 & 0.0685 & 0.0820 \\
R$_\sigma$ & 85.69 & 0.0833 & 0.0785 & 0.0754 & 0.0731 & 0.0713 & 0.0571 & 0.0828 \\
G$_\sigma$ & 94.01 & 0.0838 & 0.0782 & 0.0747 & 0.0721 & 0.0700 & 0.0532 & 0.0833 \\
SV & 96.09 & 0.0817 & 0.0774 & 0.0745 & 0.0724 & 0.0708 & 0.0598 & 0.0811 \\
SkI2 & 104.33 & 0.0777 & 0.0735 & 0.0706 & 0.0684 & 0.0668 & 0.0535 & 0.0773 \\
SkI5 & 129.33 & 0.0757 & 0.0713 & 0.0682 & 0.0659 & 0.0641 & 0.0497 & 0.0752 \\ \hline
\end{tabular}
\caption{Values of the core-crust transition density $\rho_t$ (fm$^{-3}$) for Skyrme interactions 
calculated within the dynamical approach. The results have been computed with the exact expression of the EoS (Exact), the parabolic approximation (PA), 
or the approximations of the full EoS with Eq.~(\ref{eq:EOSexpgeneral}) up to second ($\delta^2$), fourth ($\delta^4$), sixth ($\delta^6$) 
eighth ($\delta^8$) and tenth ($\delta^{10}$) order. The value for each interaction of the slope parameter of the symmetry energy $L$ is also included in units of MeV.}
\end{table}

\begin{table}[htb]
\begin{tabular}{ccccccccc}
\hline
\multicolumn{9}{c}{DYNAMICAL METHOD} \\ \hline
Force & $L$ & \begin{tabular}[c]{@{}c@{}}P$_t$\\ ($\delta^2$)\end{tabular} & 
\begin{tabular}[c]{@{}c@{}}P$_t$\\ ($\delta^4$)\end{tabular} & \begin{tabular}[c]{@{}c@{}}P$_t$\\ 
($\delta^6$)\end{tabular} & \begin{tabular}[c]{@{}c@{}}P$_t$\\ ($\delta^8$)\end{tabular} & \begin{tabular}[c]{@{}c@{}}P$_t$\\ 
($\delta^{10}$)\end{tabular} & \begin{tabular}[c]{@{}c@{}}P$_t$\\ (Exact)\end{tabular} &
\begin{tabular}[c]{@{}c@{}}P$_t$\\ (PA)\end{tabular} \\ \hline \hline
MSk7 & 9.41 & 0.3753 & 0.3886 & 0.3898 & 0.3873 & 0.3866 & 0.3840 & 0.4024 \\
SIII & 9.91 & 0.3515 & 0.3950 & 0.4001 & 0.3986 & 0.3981 & 0.3953 & 0.4237 \\
SkP & 19.68 & 0.5645 & 0.5802 & 0.5744 & 0.5665 & 0.5611 & 0.5411 & 0.5959 \\
HFB-27 & 28.50 & 0.4664 & 0.4769 & 0.4717 & 0.4645 & 0.4598 & 0.4410 & 0.4992 \\
SKX & 33.19 & 0.5737 & 0.5854 & 0.5772 & 0.5680 & 0.5618 & 0.5396 & 0.6138 \\
HFB-17 & 36.29 & 0.4465 & 0.4469 & 0.4387 & 0.4299 & 0.4240 & 0.3988 & 0.4712 \\
SGII & 37.63 & 0.3925 & 0.3955 & 0.3831 & 0.3703 & 0.3610 & 0.3095 & 0.4252 \\
UNEDF1 & 40.01 & 0.4845 & 0.4774 & 0.4603 & 0.4448 & 0.4338 & 0.3803 & 0.5139 \\
Sk$\chi$500 & 40.74 & 0.3237 & 0.3153 & 0.3097 & 0.3036 & 0.3001 & 0.2848 & 0.3320 \\
Sk$\chi$450 & 42.06 & 0.4397 & 0.4368 & 0.4282 & 0.4197 & 0.4140 & 0.3910 & 0.4597 \\
UNEDF0 & 45.08 & 0.5651 & 0.5549 & 0.5343 & 0.5165 & 0.5037 & 0.4459 & 0.5966 \\
SkM* & 45.78 & 0.4914 & 0.4860 & 0.4685 & 0.4526 & 0.4409 & 0.3844 & 0.5214 \\
SLy4 & 45.96 & 0.4092 & 0.4033 & 0.3958 & 0.3882 & 0.3831 & 0.3625 & 0.4245 \\
SLy7 & 47.22 & 0.4068 & 0.4001 & 0.3922 & 0.3845 & 0.3792 & 0.3580 & 0.4215 \\
SLy5 & 48.27 & 0.4124 & 0.4053 & 0.3968 & 0.3885 & 0.3829 & 0.3595 & 0.4277 \\
Sk$\chi$414 & 51.92 & 0.4731 & 0.4644 & 0.4535 & 0.4435 & 0.4368 & 0.4095 & 0.4906 \\
MSka & 57.17 & 0.5304 & 0.5233 & 0.5042 & 0.4872 & 0.4750 & 0.4182 & 0.5701 \\
MSL0 & 60.00 & 0.4834 & 0.4614 & 0.4357 & 0.4142 & 0.3985 & 0.3117 & 0.5112 \\
SIV & 63.50 & 0.5609 & 0.5587 & 0.5335 & 0.5109 & 0.4941 & 0.4151 & 0.6247 \\
SkMP & 70.31 & 0.4592 & 0.4283 & 0.3961 & 0.3700 & 0.3508 & 0.2353 & 0.4903 \\
SKa & 74.62 & 0.5700 & 0.5434 & 0.5104 & 0.4837 & 0.4641 & 0.3661 & 0.6103 \\
R$_\sigma$ & 85.69 & 0.6025 & 0.5221 & 0.4634 & 0.4203 & 0.3892 & 0.1925 & 0.6334 \\
G$_\sigma$ & 94.01 & 0.6715 & 0.5613 & 0.4876 & 0.4353 & 0.3979 & 0.1611 & 0.7020 \\
SV & 96.09 & 0.5487 & 0.5044 & 0.4536 & 0.4145 & 0.3860 & 0.2279 & 0.6238 \\
SkI2 & 104.33 & 0.5462 & 0.4616 & 0.4071 & 0.3676 & 0.3391 & 0.1623 & 0.5536 \\
SkI5 & 129.33 & 0.5978 & 0.4891 & 0.4224 & 0.3752 & 0.3416 & 0.1403 & 0.5917 \\ \hline
\end{tabular}
\caption{Values of the core-crust transition pressure $P_t$ (MeV fm$^{-3}$) for Skyrme interactions 
calculated within the dynamical approach. The results have been computed with the exact expression of the EoS (Exact), the parabolic approximation (PA), 
or the approximations of the full EoS with Eq.~(\ref{eq:EOSexpgeneral}) up to second ($\delta^2$), fourth ($\delta^4$), sixth ($\delta^6$) 
eighth ($\delta^8$) and tenth ($\delta^{10}$) order. The value for each interaction of the slope parameter of the symmetry energy $L$ is also included in units of MeV.}
\end{table}

\begin{table}[htb]
\centering
\begin{tabular}{llcccc}
\hline
\multicolumn{6}{c}{DYNAMICAL METHOD} \\ \hline
\multicolumn{2}{c}{Force} & $L$ & $\delta_t$ & $\rho_t$ & $P_t$ \\ \hline\hline
\multicolumn{1}{c|}{\multirow{10}{*}{Gogny}} & D1 & 18.36 & 0.9137 & 0.1045 & 0.5070 \\
\multicolumn{1}{c|}{} & D1S & 22.43 & 0.9145 & 0.0951 & 0.4723 \\
\multicolumn{1}{c|}{} & D1M & 24.83 & 0.9257 & 0.0949 & 0.2839 \\
\multicolumn{1}{c|}{} & D1N & 33.58 & 0.9345 & 0.0847 & 0.3280 \\
\multicolumn{1}{c|}{} & D250 & 24.90 & 0.9121 & 0.0987 & 0.5382 \\
\multicolumn{1}{c|}{} & D260 & 17.57 & 0.9126 & 0.1044 & 0.5188 \\
\multicolumn{1}{c|}{} & D280 & 46.53 & 0.9181 & 0.0841 & 0.5046 \\
\multicolumn{1}{c|}{} & D300 & 25.84 & 0.9135 & 0.1013 & 0.5547 \\
\multicolumn{1}{c|}{} & D1M* & 43.18 & 0.9300 & 0.0838 & 0.2702 \\
\multicolumn{1}{c|}{} & D1M** & 33.91 & 0.9279 & 0.0886 & 0.2786 \\ \hline
\multicolumn{1}{c|}{\multirow{15}{*}{MDI}} & $x=-1.4$ & 123.98 & 0.9967 & 0.0331 & -0.0299 \\
\multicolumn{1}{c|}{} & $x=-1.2$ & 114.86 & 0.9950 & 0.0344 & -0.0144 \\
\multicolumn{1}{c|}{} & $x=-1$ & 105.75 & 0.9925 & 0.0364 & 0.0045 \\
\multicolumn{1}{c|}{} & $x=-0.8$ & 96.63 & 0.9889 & 0.0393 & 0.0286 \\
\multicolumn{1}{c|}{} & $x=-0.6$ & 87.51 & 0.9841 & 0.0430 & 0.0600 \\
\multicolumn{1}{c|}{} & $x=-0.4$ & 78.40 & 0.9778 & 0.0476 & 0.0991 \\
\multicolumn{1}{c|}{} & $x=-0.2$ & 69.28 & 0.9701 & 0.0526 & 0.1445 \\
\multicolumn{1}{c|}{} & $x=0$ & 60.17 & 0.9612 & 0.0579 & 0.1936 \\
\multicolumn{1}{c|}{} & $x=0.2$ & 51.05 & 0.9515 & 0.0636 & 0.0467 \\
\multicolumn{1}{c|}{} & $x=0.4$ & 41.94 & 0.9411 & 0.0697 & 0.0480 \\
\multicolumn{1}{c|}{} & $x=0.6$ & 32.82 & 0.9303 & 0.0768 & 0.0483 \\
\multicolumn{1}{c|}{} & $x=0.8$ & 23.70 & 0.9194 & 0.0855 & 0.0472 \\
\multicolumn{1}{c|}{} & $x=1$ & 14.59 & 0.9089 & 0.0980 & 0.0442 \\
\multicolumn{1}{c|}{} & $x=1.1$ & 10.03 & 0.9044 & 0.1082 & 0.0410 \\
\multicolumn{1}{c|}{} & $x=1.15$ & 7.75 & 0.9019 & 0.1160 & 0.0382 \\ \hline
\end{tabular}
\caption{Values of the core-crust transition asymmetry $\delta_t$, transition density $\rho_t$ (fm$^{-3}$) and transition pressure $P_t$ (MeV fm$^{-3}$) for a set of 
Gogny and MDI interactions 
calculated within the dynamical approach. The results have been computed with the exact expression of the EoS.
The value for each interaction of the slope parameter of the symmetry energy $L$ is also included in units of MeV.}
\end{table}

\begin{table}[htb]
\centering
\begin{tabular}{lccc}
\hline
\multicolumn{4}{c}{DYNAMICAL METHOD} \\ \hline
Force & $\delta_t$ & $\rho_t$ & $P_t$ \\ \hline\hline
SEI $L=34$ MeV & 0.9132 & 0.0923 & 0.5508 \\
SEI $L=39$ MeV & 0.9183 & 0.0889 & 0.5327 \\
SEI $L=45$ MeV& 0.9251 & 0.0848 & 0.5056 \\
SEI $L=51$ MeV& 0.9326 & 0.0805 & 0.4719 \\
SEI $L=58$ MeV& 0.9404 & 0.0763 & 0.4319 \\
SEI $L=65$ MeV& 0.9477 & 0.0724 & 0.3902 \\
SEI $L=67$ MeV& 0.9505 & 0.0709 & 0.3732 \\
SEI $L=71$ MeV& 0.9543 & 0.0688 & 0.3484 \\
SEI $L=75$ MeV& 0.9586 & 0.0664 & 0.3187 \\
SEI $L=77$ MeV& 0.9611 & 0.0649 & 0.3007 \\
SEI $L=82$ MeV& 0.9661 & 0.0619 & 0.2632 \\
SEI $L=86$ MeV& 0.9705 & 0.0591 & 0.2276 \\
SEI $L=89$ MeV& 0.9727 & 0.0575 & 0.2085 \\
SEI $L=92$ MeV& 0.9760 & 0.0552 & 0.1802 \\
SEI $L=96$ MeV& 0.9789 & 0.0530 & 0.1548 \\
SEI $L=100$ MeV& 0.9824 & 0.0501 & 0.1223 \\
SEI $L=105$ MeV& 0.9857 & 0.0471 & 0.0909 \\
SEI $L=111$ MeV& 0.9895 & 0.0434 & 0.0549 \\
SEI $L=115$ MeV& 0.9913 & 0.0415 & 0.0373 \\ \hline
\end{tabular}
\caption{Values of the core-crust transition asymmetry $\delta_t$, transition density $\rho_t$ (fm$^{-3}$) and transition pressure $P_t$ (MeV fm$^{-3}$) for a set of 
SEI interactions of $\gamma=1/2$ and different slope of the symmetry energy
calculated within the dynamical approach. The results have been computed with the exact expression of the EoS.
The value for each interaction of the slope parameter of the symmetry energy $L$ is also included in units of MeV.}
\end{table}

\chapter{Extended Thomas-Fermi approximation with finite-range forces}\label{app_vdyn}
In this Appendix we find the contribution to the surface part of the curvature matrix in the dynamical method (Chapter~\ref{chapter4}) to find the core-crust transition. 
The contribution coming from the exchange and kinetic parts of the interaction are found using a density matrix (DM) expansion in the 
Extended-Thomas-Fermi (ETF) approximation. To find the contributions coming from the direct part of the interaction we perform 
an expansion of the direct energy in terms of the gradients of the nuclear derivatives, which 
in momentum space can be summed at all orders.

The total energy density provided by a finite-range density-dependent effective nucleon-nucleon 
interaction can be decomposed as 
\begin{equation}
 \mathcal{H} = \mathcal{H}_{kin} +  \mathcal{H}_{zr} + \mathcal{H}_{dir} + \mathcal{H}_{exch}+ 
 \mathcal{H}_{Coul}+\mathcal{H}_{LS},
 \label{eqendens}
\end{equation} 
where $\mathcal{H}_{kin}$, $\mathcal{H}_{zr}$, $\mathcal{H}_{dir}$, $\mathcal{H}_{exch}$, 
$\mathcal{H}_{Coul}$ and $\mathcal{H}_{LS}$ are the
 kinetic, zero-range, finite-range direct, finite-range exchange, Coulomb and spin-orbit contributions.
The finite-range part of the interaction can be written in a general way as 
\begin{equation}\label{eqVfin}
V({\bf r},{\bf r'}) = \sum_i \left(W_i + B_i P_{\sigma} - 
H_i P_{\tau} -+
M_i P_{\sigma} P_{\tau}\right) v_i({\bf r} , {\bf r'}),\\
\end{equation}
where $P^{\sigma}$ and $P^{\tau}$ are the spin and isospin exchange operators and $v_i({\bf r} , {\bf r'})$ 
are the form factors of the force. 
For Gaussian-type interactions the form factor is\footnote{Notice that the notation 
 of the range in the Gaussian form factor for Gogny interactions has changed from $\mu$ in Chapter~\ref{chapter1} to $\alpha$ to not confuse it with the 
 corresponding range parameter of the Yukawa form factors found in the SEI and MDI interactions.}
\begin{equation}
 v_i({\bf r} , {\bf r'})=e^{-|{\bf r} - {\bf r'}|^2/\alpha_i^2} ,
\end{equation}
 while for a Yukawa force one has
 \begin{equation}
  v_i({\bf r} , {\bf r'})=\frac{e^{-\mu_i |{\bf r} - {\bf r'}|}}{\mu_i |{\bf r} - {\bf r'}|}.
 \end{equation}
 Let us remind that the form factor of the SEI interaction, which we have used to present all of 
 the expressions of Yukawa type functionals, has an additional parameter $\mu_i$ in the denominator 
 compared with the usual expression of the MDI interaction in the literature.
 Hence, the Yukawa form factor for MDI interactions in the literature is usually written as
 \begin{equation}
  v_i({\bf r} , {\bf r'})=\frac{e^{-\mu_i |{\bf r}   - {\bf r'}|}}{|{\bf r} - {\bf r'}|}.
 \end{equation}
 It is important to take into account this fact when using our expressions for the MDI interactions. 
 
The finite-range term provides the direct ($\mathcal{H}_{dir}$) and exchange ($\mathcal{H}_{exch}$) 
contributions to the total energy density in Eq.~(\ref{eqendens}).
The HF energy due to the finite-range part of the interaction, neglecting zero-range, Coulomb and spin-orbit contributions, reads 
\begin{eqnarray}
E_{HF} &=& \sum_q \int d{\bf r} \left[\frac{\hbar^2}{2m}\tau({\bf r})
+ {\cal H}_{dir} + {\cal H}_{exch} \right]_q= \sum_q \int d{\bf r}\left[\frac{\hbar^2}{2m}\tau({\bf r})\right. \nonumber\\
&+& \left.\frac{1}{2}\rho({\bf r})V^{H}({\bf r})
+ \frac{1}{2}\int d{\bf r'} V^{F}({\bf r},{\bf r'})\rho \left({\bf r} ,{\bf r'}\right)
\right]_q,
\label{eq2}
\end{eqnarray}
where the subscript $q$ refers to each kind of nucleon. In this equation $\rho({\bf r})$ and $\tau({\bf r})$ 
are the particle and kinetic energy densities, respectively, and $\rho \left({\bf r} ,{\bf r'}\right)$ 
is the one-body density matrix.
The direct ($V^H$) and exchange ($V^F$) contributions to the HF potential due to the finite-range interaction (\ref{eqVfin}) are 
given by 
\begin{equation}\label{eq3}
V_q^H ({\bf r}) = \sum_i \int d{\bf r'} v_i ({\bf r} , {\bf r'}) \left[ D_{L, dir}^i \rho_q ({\bf r'}) + D_{U,dir}^i \rho_{q'} ({\bf r'})\right]
\end{equation}
and
\begin{eqnarray}
V_q^F({\bf r}, {\bf r'})&=& - \sum_i v_i ({\bf r} , {\bf r'}) \left[ D_{L, exch}^i \rho_q ({\bf r} ,{\bf r'})
+ D_{U,exch}^i \rho_{q'} ({\bf r},{\bf r'}) \right],
\label{eq4}
\end{eqnarray}
respectively. In Eqs.~(\ref{eq3}) and (\ref{eq4}) the coefficients $D_{L, dir}^i$, $D_{U,dir}^i$, $D_{L, exch}^i$ and $D_{U,exch}^i$ 
 are the usual contributions of the spin
and isospin strengths for the like and unlike nucleons in the direct and exchange potentials:
\begin{eqnarray}
&&D_{L,dir}^i = W_i + \frac{B_i}{2} - H_i - \frac{M_i}{2} \nonumber \\
&&D_{U,dir}^i= W_i + \frac{B_i}{2}\nonumber\\
&&D_{L,exch}^i=M_i + \frac{H_i}{2} - B_i - \frac{W_i}{2}\nonumber\\
&&D_{U,exch}^i=M_i + \frac{H_i}{2}.
\end{eqnarray}

The non-local one-body DM $ \displaystyle\rho({\bf r},{\bf r'})=\sum_k \varphi_k^*({\bf r})\varphi_k ({\bf r'})$
plays an essential role in Hartree-Fock (HF) calculations using effective finite-range forces,
where its full knowledge is needed. 
The HFB theory with finite-range forces is well
established from a theoretical point of view \cite{decharge80,nakada03} and calculations in
finite nuclei are feasible nowadays with a reasonable computing time. However, these calculations 
are still complicated and usually require specific codes, as for example the one provided by Ref.~\cite{robledo02}. 
Therefore, approximate methods based on the expansion of the DM in terms of local densities and their gradients usually allow one to
reduce the non-local energy density to a local form.

The simplest approximation to the DM is to replace locally its quantal value by its expression 
in nuclear matter, i.e., the so-called Slater or Thomas-Fermi (TF) approximation. A more elaborated
treatment was developed by Negele and Vautherin \cite{negele72a, negele72b}, which expanded the DM into a
bulk term (Slater) plus a corrective contribution that takes into account the finite-size effects.
Campi and Bouyssy \cite{campi78a,campi78b} proposed another approximation consisting of a Slater term alone but 
with an effective Fermi momentum, which partially resummates the finite-size corrective terms.
Later on, Soubbotin and Vi\~nas developed the Extended Thomas-Fermi (ETF) approximation 
of the one-body DM in the case of a non-local single-particle Hamiltonian \cite{soubbotin00}. 
The ETF DM includes, on top of the Slater part, corrections of $\hbar^2$ order, which are
expressed through second-order derivatives of the neutron and proton densities. 
In the same Ref.~\cite{soubbotin00} the similarities and differences
with previous DM expansions \cite{negele72a, negele72b,campi78a,campi78b} are discussed in detail.
The ETF approximation to the HF energy for non-local potentials
consists of replacing the quantal density matrix by the semiclassical one that contains, in
addition to the bulk (Slater) term, corrective terms depending on the second-order derivatives
of the proton and neutron densities, which account for contributions up to $\hbar^2$-order. 
The angle-averaged semiclassical ETF density matrix, derived in Refs.~\cite{centelles98,gridnev98,soubbotin00}, 
for each kind of nucleon reads 
\begin{eqnarray}
{\tilde \rho}({\bf R},s) &=&=\rho_U ({\bf R},s) +\rho_2 ({\bf R},s) =  \frac{3j_1(k_F s)}{k_f s}\rho \nonumber\\
&+& \frac{s^2}{216}\left\{\left[\left(9 - 2k_F\frac{f_k}{f}
- 2k_F^2\frac{f_{kk}}{f} +  k_F^2\frac{f_k^2}{f}\right)\frac{j_1(k_F s)}{k_F s} -4 j_0(k_F s)\right]
\frac{({\bf \nabla}\rho)^2}{\rho}\right. \nonumber \\
&-&\left.\left[\left(18 +  6k_F\frac{f_k}{f}\right) \frac{j_1(k_F s)}{k_F s} - 3j_0(k_F s)\right]\Delta \rho 
-\left[18\rho \frac{\Delta f}{f} \right.\right.\nonumber \\
&+& \left.\left. \left(18 - 6k_F \frac{f_k}{f}\right)\frac{{\bf \nabla}\rho \cdot {\bf \nabla} f}{f}
+12 k_F \frac{{\bf \nabla}\rho\cdot {\bf \nabla} f_k}{f} - 9\rho \frac{({\bf \nabla} f)^2}{f} \right]
\frac{j_1(k_F s)}{k_F s} \right\}, \nonumber \\
\label{eqA1}
\end{eqnarray}
where $\rho_U$ is the uniform Slater bulk term, $\rho_2$ contains the ETF gradient corrections of order $\hbar^2$ 
to the DM, and ${\bf R}=({\bf r}+{\bf r'})/2$ and ${\bf s}={\bf r}-{\bf r'}$ are the center of mass and relative 
coordinates of the two nucleons, respectively.
Moreover, in the r.h.s. of~(\ref{eqA1}), $\rho=\rho ({\bf R})$ is the local density, $k_F=(3 \pi^2 \rho({\bf R}))^{1/3}$ is the corresponding Fermi momentum for each type of nucleon, and $j_1(k_F s)$
is the $l=1$ spherical Bessel function. In Eq.~(\ref{eqA1}), $f= \left.f({\bf R},k)\right|_{k=k_F}$ is the inverse of the position and momentum
dependent effective mass, defined for each type of nucleon as
\begin{equation}
f({\bf R},k) = \frac{m}{m^{*}({\bf R},k)}= 1 + \frac{m}{\hbar^2k}\frac{\partial V^{F}_{0}({\bf R},k)}{\partial k}.
\label{eqA2}
\end{equation} 
The value of $f$ in Eq.~(\ref{eqA1}) is computed at $k=k_{F}$ (see below), and $f_{k}= \left.f_{k}({\bf R},k)\right|_{k=k_F}$ and 
$f_{kk}= \left.f_{kk}({\bf R},k)\right|_{k=k_F}$ denote its first and second
derivatives with respect to $k$.
In Eq.~(\ref{eqA2}), $V^{F}_{0}({\bf R},k)$ is the Wigner transform of the exchange potential 
(\ref{eq4}) given by
\begin{eqnarray}\label{eqA3} 
V^{F}_0({\bf R},k) = \int d{\bf s} V^{F}_0({\bf R},s)e^{-i{\bf k}\cdot{\bf s}} 
= \int d{\bf s} V_0^{F} ({\bf R},s) j_0 (k,s),
\end{eqnarray}
At TF level, computed with
Eq.~(\ref{eq4}) using the Slater DM (\ref{eqA1}), it can be rewritten as
\begin{eqnarray}
V^{F}_{0}({\bf R},k) &=& - \sum_i \left[ D_{L,exch}^i \int d{\bf s}v(s)\frac{3j_1(k_{Fq} s)}{k_{Fq} s}\rho_q j_0(k s)\right.\nonumber\\
&+& \left.D_{U,exch}^i \int d{\bf s}v(s)\frac{3j_1(k_{Fq'} s)}{k_{Fq'} s}\rho_{q'} j_0(k s)\right].
\label{eqA9}
  \end{eqnarray}
Notice that due to the structure of the exchange potential (\ref{eqA9}), the space dependence of the effective mass for each 
kind of nucleon is through the Fermi momenta of both type of nucleon, neutrons and protons,
 i.e., $ f_q=f_q(k,k_{Fn}({\bf R}),k_{Fp}({\bf R}))$ ($q=n,p$).
When the inverse 
effective mass and its derivatives with respect to $k$ are used in (\ref{eqA1}), an additional space
dependence arises from the replacement of the momentum $k$ by the local Fermi momentum
$k_F({\bf R})$.
 
Using the DM (\ref{eqA1}), the explicit form of the semiclassical kinetic energy at the ETF level for either neutrons or protons 
can be written as:
\begin{eqnarray}
\tau_{ETF}({\bf R}) &=& \left.\left(\frac{1}{4}\Delta_R - \Delta_s\right) {\tilde \rho}({\bf R},s)\right|_{s=0} =
\tau_{0} + \tau_{2},
\label{eqA4}
\end{eqnarray}
which consists of the well-known TF term
\begin{equation}
 \tau_{0}=\frac{3}{5}k_{F}^2\rho,
\end{equation}
plus the $\hbar^2$ contribution
\begin{eqnarray}
\tau_{2}({\bf R}) &=& 
\frac{1}{36}\frac{({\bf \nabla} \rho)^2}{\rho} \left[ 1 + \frac{2}{3}k_F \frac{f_k}{f} +
\frac{2}{3}k_F^2 \frac{f_{kk}}{f}- \frac{1}{3}k_F^2 \frac{f_k^2}{f^2} \right] + 
\frac{1}{12}\Delta \rho \left[4 + \frac{2}{3}k_F \frac{f_k}{f} \right]\nonumber \\
&+&  \frac{1}{6}\rho \frac{\Delta f}{f}+ \frac{1}{6}\frac{{\bf \nabla}\rho \cdot {\bf \nabla}f}{f}
\left[1 - \frac{1}{3}k_F \frac{f_k}{f}\right] + \frac{1}{9}\frac{{\bf \nabla}\rho \cdot {\bf \nabla}f_k}{f}
- \frac{1}{12}\rho \frac{({\bf \nabla}f)^2}{f^2}.
\label{eqA41}
\end{eqnarray}
This $\hbar^2$ contribution reduces to the standard $\hbar^2$
expression for local forces \cite{skms} if the effective mass depends only on the 
position and not on the momentum.

Collecting all pieces together, the HF energy~(\ref{eq2}) at ETF
level calculated using the DM given by Eq.~(\ref{eqA1}) reads
\begin{eqnarray}
E_{HF}^{ETF} &=& \sum_{q} \int d{\bf R}\left[\frac{\hbar^2}{2m}\frac{3}{5}(3\pi^2)^{2/3}\rho^{5/3}\right. \nonumber\\
&+& \left.\frac{1}{2}\rho({\bf R})V^{H}({\bf R}) + \frac{1}{2} \int d{\bf s}  \rho_U ({\bf R}, {\bf s}) V_0^F ({\bf R}, {\bf s}) 
+ \frac{\hbar^2 \tau_2({\bf R})}{2m} + {\cal H}_{exch,2}({\bf R})\right]_q,\nonumber\\
\label{eq5}
\end{eqnarray}
where ${\cal H}_{exch,2}({\bf R})$ 
is the $\hbar^2$ contribution to the exchange energy. The contributions from the exchange and kinetic energies 
to the surface term of the curvature matrix~(\ref{eq12}) in Chapter~\ref{chapter4}
will come from the $\tau_2$ and ${\cal H}_{exch,2}({\bf R})$, respectively.

The exchange energy density, which is local within the ETF approximation, is obtained from the exchange potential (\ref{eqA3}) and
is given by
\begin{eqnarray}
{\cal H}_{exch}({\bf R}) = 
\frac{1}{2 }\int d{\bf s} \rho_0({\bf R},s) V_0^F ({\bf R},s) +
\int d{\bf s} \rho_2({\bf R},s) V^{F}_{0}({\bf R},s),
\label{eqA5}
\end{eqnarray}
from where, and after some algebra explained in detail in Ref.~\cite{soubbotin00}, one can recast the $\hbar^2$ 
contribution to the exchange energy for each kind of nucleon as
\begin{equation}
{\cal H}_{exch,2}({\bf R}) = 
 \frac{\hbar^2}{2m} \left[(f-1)\left(\tau_{ETF}-\frac{3}{5}k_F^2 \rho - \frac{1}{4}\Delta \rho \right) 
+ k_F f_k \left(\frac{1}{27}\frac{({\bf \nabla} \rho)^2}{\rho} - \frac{1}{36}\Delta \rho\right)\right].
\label{eqA6}
\end{equation}
Notice that $\tau_{ETF}-\frac{3}{5}k_{F}^2\rho=\tau_{2}$ and that 
$\Delta \rho$ vanishes under integral sign if spherical symmetry in coordinate space is assumed. 

From Eq.~(\ref{eq5}) we see that the energy in the ETF approximation for finite-range forces consists of a 
pure TF part, which depends only on the local densities of each type of particles, plus additional $\hbar^2$ 
corrections coming from the $\hbar$-expansion of the kinetic and exchange energy densities, which depend 
on the local Fermi momentum $k_F$ and on second-order derivatives of the nuclear density:
\begin{eqnarray}
\int d{\bf R}\left[{\cal H}_{kin}({\bf R}) +  {\cal H}_{exch,2}({\bf R})\right] &=& 
\frac{\hbar^2}{2m}\int d{\bf R}\left\{\tau_{0} +  
 \left[f\tau_{2} - \frac{1}{4}f \Delta \rho \right.\right.\nonumber\\
&+& \left.\left. k_F f_k \left(\frac{1}{27}
\frac{({\bf \nabla} \rho)^2}{\rho} - \frac{1}{36}\Delta \rho\right)\right]\right\}.
\label{eqA7}
\end{eqnarray}
The full $\hbar^2$ contribution to the total energy corresponding to the finite-range central interaction 
(\ref{eqVfin}), given by Eq.~(\ref{eqA7}) for each kind of nucleon, can be written, after partial integration, as:
\begin{eqnarray}
&&\int d{\bf R}\bigg[\frac{\hbar^2}{2m}\tau_2({\bf R}) + {\cal H}_{exch,2}({\bf R})\bigg]_q =\\
&&\int d{\bf R}\bigg[B_{nn}(\rho_{n},\rho_{p})\big({\bf \nabla}\rho_n\big)^2
+ B_{pp}(\rho_{n},\rho_{p})\big({\bf \nabla}\rho_p\big)^2
+ 2 B_{np}(\rho_{n},\rho_{p}){\bf \nabla}\rho_n\cdot{\bf \nabla}\rho_p\bigg],\nonumber
\label{eqA10}
\end{eqnarray}
where the like coefficients of the gradients of the densities are
\begin{eqnarray}
B_{nn}(\rho_n,\rho_p) &=& \frac{\hbar^2}{2m} \frac{1}{108}\left\{\left[3 f_n + k_{Fn}(2f_{nk} - 3 f_{nk_{Fn}})
+  k^2_{Fn}(5f_{nkk} + 3 f_{nkk_{Fn}})  \right. \right.
\nonumber \\
&-& \left. \left. k^2_{Fn}\frac{(2f_{nk} +  f_{nk_{Fn}})^2}{f_n}\right]\frac{1}{\rho_n}
- \frac{\rho_p}{\rho_n^2}k^2_{Fn}\frac{f^2_{pk_{Fn}}}{f_p}\right\}
\label{eqA11}
\end{eqnarray}
and a similar expression for $B_{pp}(\rho_n,\rho_p)$ obtained by exchanging $n$ by $p$ in Eq.~(\ref{eqA11}). The 
unlike coefficient  $B_{np}(\rho_n,\rho_p)$ of Eq.~(\ref{eqA10}) reads:
\begin{eqnarray}
B_{np}(\rho_n,\rho_p)&=& B_{pn}(\rho_p,\rho_n)= - \frac{\hbar^2}{2m} \frac{1}{316}\left\{\left[ \vphantom{ \frac{2 k_{Fn} k_{Fp}(2f_{nk} + f_{nk_{Fn}})f_{nk_{Fp}}}{f_n}} 3 k_{Fp}f_{nk_{Fp}}
- 3 k_{Fn} k_{Fp}f_{nkk_{Fp}} \right.\right.
\nonumber\\
&+&  \left.\left.\frac{2 k_{Fn} k_{Fp}(2f_{nk} + f_{nk_{Fn}})f_{nk_{Fp}}}{f_n}\right]
\frac{1}{\rho_p}\right.
\nonumber \\ 
&+& \left.\left[3 k_{Fn}f_{pk_{Fn}} - 3 k_{Fp} k_{Fn}f_{pkk_{Fn}} +  
\frac{2 k_{Fp} k_{Fn}(2f_{pk} + f_{pk_{Fp}})f_{pk_{Fn}}}{f_p}\right]\frac{1}{\rho_n}\right\}.\nonumber\\
\label{eqA12}
\end{eqnarray}
As stated before, all derivatives of the neutron (proton) inverse effective mass $f_q (k, k_{Fq}, k_{Fq'})$
with respect to the momentum $k$, $f_{qk}( k_{Fq}, k_{Fq'})$, are 
evaluated at the neutron (proton) Fermi momentum $k_{Fq}$, i.e.
$f_{qk}  (k_{Fq}, k_{Fq'}) =\left.\frac{\partial f_q (k, k_{Fq}, k_{Fq'})}{\partial k}\right|_{k=k_{Fq}}$,
$f_{qkk}  (k_{Fq}, k_{Fq'}) =\left.\frac{\partial^2 f_q (k, k_{Fq}, k_{Fq'})}{\partial k^2}\right|_{k=k_{Fq}}$,
$f_{qkk_{Fq'}}  (k_{Fq}, k_{Fq'}) =\left.\frac{\partial^2 f_q (k, k_{Fq}, k_{Fq'})}{\partial k \partial k_{Fq'}}\right|_{k=k_{Fq}}$, etc.

We proceed to derive the direct term in Eq.~(\ref{eq9}) in Section~\ref{Theory_dyn} of Chapter~\ref{chapter4} due to the fluctuating density.
Let us first obtain the gradient expansion of the direct energy coming from the finite-range part of the force. For the sake of 
simplicity, we consider a single Wigner term. The result for the case including spin and isospin exchange operators can be obtained analogously.
In the case of a Wigner term we have 
\begin{equation}\label{eq:Edir}
 E_{dir} = \frac{1}{2}\int d{\bf R} d{\bf s} \rho ({\bf R}) \rho({\bf R}-{\bf s}) v ({\bf s}).
\end{equation}
Following the procedure of 
Ref.~\cite{durand93}, a central finite-range interaction can be expanded in a series of distributions as follows:  
\begin{equation}
v({s}) = \sum_{n=0}^{\infty} c_{2n}\nabla^{2n} \delta ({\bf s}),
\label{eqB2}
\end{equation}
where the coefficients $c_{2n}$ are chosen in such a way 
that the expansion (\ref{eqB2}) gives the same moments of the interaction $v$($s$). 
This implies that \cite{durand93}
\begin{equation}
c_{2n} = \frac{1}{(2n+1)!} \int d{\bf s} s^{2n} v({s}),
\label{eqB3}
\end{equation}
which allows one to determine the values of the coefficients $c_{2n}$ for any value of $n$.
Using this expansion, the direct energy (\ref{eq:Edir}) can be written as 
\begin{eqnarray}
E_{dir}= \sum_{n=0}^{\infty} \frac{c_{2n}}{2} \int d{\bf R} d{\bf s} \rho({\bf R}) 
\rho({\bf R}- {\bf s}) \nabla^{2n} \delta({\bf s})
=\sum_{n=0}^{\infty}  \frac{c_{2n}}{2} \int d{\bf R} \rho({\bf R})\nabla^{2n} \rho({\bf R}).
\label{eqB4}
\end{eqnarray}      
Expanding now the density $\rho({\bf R})$ in its uniform and varying contributions,
the direct energy due to the fluctuating part of the density
becomes:
\begin{equation}
\delta E_{dir} =  \sum_{n=0}^{\infty} \frac{c_{2n}}{2} 
\int d{\bf R} \delta \rho({\bf R})\nabla^{2n} \delta \rho({\bf R}).
\label{eqB5}
\end{equation}
Notice that the splitting of the density (\ref{eq6}) also provides a contribution to the
non-fluctuating energy $E_0({\rho_U})$ through the constant density $\rho_U$.
Linear terms in $\delta \rho ({\bf R})$ do not contribute 
to the direct energy by the reasons discussed previously.
Proceeding as explained in the following lines, to transform integrals in coordinate space into
integrals in momentum space (see Eqs.~(\ref{eq8}) and (\ref{eqB1})) after some algebra the fluctuating correction to 
the direct energy can be written as follows:
\begin{equation}
\delta E_{dir} = \frac{1}{2} \int \frac{d{\bf k}}{(2\pi)^3} \delta n({\bf k})\delta  n^*({\bf k}) 
{\cal F}(k),
\label{eqB6}
\end{equation}
where 
\begin{equation}
 {\cal F}(k)= \sum_{n=0}^{\infty} c_{2n}k^{2n}
\end{equation}
is a series encoding the 
response of the direct energy to the perturbation induced by the varying density.
This series is the Taylor expansion of the Fourier transform of the form factor $v(s)$. In the case of 
Gaussian ($e^{-s^2/\alpha^2}$) or Yukawian ($e^{-\mu s}/\mu s$) form factors one obtains, respectively, 
\begin{equation}\label{Fgauss}
 {\cal F}(k)=\pi^{3/2} \alpha^3 e^{-\alpha^2 k^2/4}
\end{equation}
and
\begin{equation}\label{Fyuk}
 {\cal F}(k)=\frac{4\pi}{\mu(\mu^2 + k^2)}.
\end{equation}

For the nuclear direct potential the first 
 term ($c_0$) of the series for ${\cal F}(k)$, which can be written as 
${\cal F}(0)$, corresponds to the bulk contribution associated to the fluctuating density 
$\delta \rho({\bf R})$, i.e. it also contributes to $\mu_n$ and $\mu_p$ in Eq.~(\ref{eq10}). 
Then, the fluctuating correction to the direct energy in Eq.~(\ref{eq10}) is given by   
\begin{eqnarray}
 \delta E_{dir} &=& \frac{1}{2} \int \frac{d{\bf k}}{(2\pi)^3} \sum_i\left[D_{L,dir}^i\left(\delta n_n({\bf k})\delta n^*_n({\bf k})
 + \delta n_p({\bf k})\delta n^*_p({\bf k})\right) \right. \nonumber \\
&+& \left. D_{U,dir}^i\left(\delta n_n({\bf k})\delta n^*_p({\bf k})
+\delta n_p({\bf k})\delta n^*_n({\bf k})\right)
\right]({\cal F}_i(k)- {\cal F}_i(0)) .
\end{eqnarray}
Let us also point out that 
if the series ${\cal F}(k)$ is cut at first order, i.e. taking only the $n=1$ term of the series, $c_2$, one recovers 
the typical $k^2$ dependence corresponding to square gradient terms in the energy density functional. 
If this expansion up to quadratic terms in $k$ is used, the dynamical potential can be written 
as Eq.~(\ref{eq14a}), where the coefficient $\beta (\rho)$ reads
\begin{eqnarray}
\beta (\rho) &=& \left[\sum_i D^i_{L,dir}c^i_2 + 2B_{pp} \left(\rho_n, \rho_p\right)\right] + \frac{\left(\frac{\partial \mu_p}{\partial \rho_n}\right)^2}
{\left(\frac{\partial \mu_n}{\partial \rho_n}\right)^2}\left[ \sum_i D^i_{L,dir}c^i_2 + 2B_{nn}\left(\rho_n, \rho_p\right)\right] \nonumber\\
&-&2 \frac{\frac{\partial \mu_p}{\partial \rho_n}}
{\frac{\partial \mu_n}{\partial \rho_n}}\left[ \sum_i D^i_{U,dir}c^i_2 + 2B_{np}\left(\rho_n, \rho_p\right)\right],
\label{eqB8}
\end{eqnarray}
and the $B_{qq}\left(\rho_n, \rho_p\right)$ and $B_{qq'}\left(\rho_n, \rho_p\right)$ functions have been given in 
Eqs.~(\ref{eqA11}) and (\ref{eqA12}), corresponding to the $\hbar^2$ contributions coming from the expansion of the energy
density functional. Moreover, for Gaussian form factors one has 
\begin{equation}
 c_2^i = -\frac{\pi^{3/2} \alpha_i^5}{4},
 \end{equation}
whereas for Yukawian form factors one has 
\begin{equation}
 c_2^i= -\frac{4 \pi}{\mu_i^5}.
\end{equation}

\fancyhead[RE, LO]{Chapter C}
\end{appendices}
\newpage
{\small
\fancyhead[RE, LO]{Bibliography}
\bibliography{bibtex}

\begin{thebibliography}{100}

\bibitem{skyrme56}
T.~Skyrme, ``{CVII}. {The} nuclear surface,'' {\em The Philosophical Magazine:
  A Journal of Theoretical Experimental and Applied Physics}, vol.~1, p.~1043,
  1956.

\bibitem{vautherin72}
D.~Vautherin and D.~M. Brink, ``Hartree-{Fock} {Calculations} with {Skyrme}'s
  {Interaction}. {I}. {Spherical} {Nuclei},'' {\em Physical Review C}, vol.~5,
  p.~626, 1972.

\bibitem{sly41}
E.~Chabanat, P.~Bonche, P.~Haensel, J.~Meyer, and R.~Schaeffer, ``A {Skyrme}
  parametrization from subnuclear to neutron star densities,'' {\em Nuclear
  Physics A}, vol.~627, p.~710, 1997.

\bibitem{decharge80}
J.~Decharg\'{e} and D.~Gogny, ``{Hartree-{Fock}-{Bogolyubov} calculations with
  the {D1} effective interaction on spherical nuclei},'' {\em Physical Review
  C}, vol.~21, p.~1568, 1980.

\bibitem{berger91}
J.~F. Berger, M.~Girod, and D.~Gogny, ``{Time-dependent quantum collective
  dynamics applied to nuclear fission},'' {\em Computer Physics
  Communications}, vol.~63, p.~365, 1991.

\bibitem{das03}
C.~B. Das, S.~Das~Gupta, C.~Gale, and B.-A. Li, ``Momentum dependence of
  symmetry potential in asymmetric nuclear matter for transport model
  calculations,'' {\em Physical Review C}, vol.~67, p.~034611, 2003.

\bibitem{li08}
B.-A. Li, L.-W. Chen, and C.~M. Ko, ``Recent progress and new challenges in
  isospin physics with heavy-ion reactions,'' {\em Physics Reports}, vol.~464,
  p.~113, 2008.

\bibitem{behera98}
B.~Behera, T.~R. Routray, and R.~K. Satpathy, ``Momentum and density dependence
  of the mean field in nuclear matter,'' {\em Journal of Physics G: Nuclear and
  Particle Physics}, vol.~24, p.~2073, 1998.

\bibitem{Behera05}
B.~Behera, T.~Routray, A.~Pradhan, S.~Patra, and P.~Sahu, ``Momentum and
  density dependence of the isospin part of nuclear mean field and equation of
  state of asymmetric nuclear matter,'' {\em Nuclear Physics A}, vol.~753,
  p.~367, 2005.

\bibitem{gonzalez17}
C.~Gonzalez-Boquera, M.~Centelles, X.~Vi{\~n}as, and A.~Rios, ``Higher-order
  symmetry energy and neutron star core-crust transition with {Gogny} forces,''
  {\em Physical Review C}, vol.~96, p.~065806, 2017.

\bibitem{Sellahewa14}
R.~Sellahewa and A.~Rios, ``Isovector properties of the {Gogny} interaction,''
  {\em Physical Review C}, vol.~90, p.~054327, 2014.

\bibitem{gonzalez18}
C.~Gonzalez-Boquera, M.~Centelles, X.~Viñas, and L.~Robledo, ``New {Gogny}
  interaction suitable for astrophysical applications,'' {\em Physics Letters
  B}, vol.~779, p.~195, 2018.

\bibitem{gonzalez18a}
X.~Vi{\~n}as, C.~Gonzalez-Boquera, M.~Centelles, L.~Robledo, and C.~Mondal,
  ``Gogny forces in the {Astrophysical} context,'' {\em Bulgarian Journal of
  Physics, Nuclear Theory}, vol.~37, p.~68, 2018.

\bibitem{Vinas19}
X.~Vi{\~n}as, C.~Gonzalez-Boquera, M.~Centelles, C.~Mondal, and L.~Robledo,
  ``{The modified D1M interactions: new Gogny forces adapted for neutron star
  calculations},'' {\em Acta Physica Polonica B Proceedings Supplement},
  vol.~12, p.~705, 2019.

\bibitem{douchin01}
F.~Douchin and P.~Haensel, ``{A unified equation of state of dense matter and
  neutron star structure},'' {\em Astronomy and Astrophysics}, vol.~380,
  p.~151, 2001.

\bibitem{Link1999}
B.~Link, R.~I. Epstein, and J.~M. Lattimer, ``Pulsar {Constraints} on {Neutron}
  {Star} {Structure} and {Equation} of {State},'' {\em Physical Review
  Letters}, vol.~83, 1999.

\bibitem{Fattoyev:2010tb}
F.~J. {Fattoyev} and J.~{Piekarewicz}, ``{Sensitivity of the moment of inertia
  of neutron stars to the equation of state of neutron-rich matter},'' {\em
  Physical Review C}, vol.~82, p.~025810, 2010.

\bibitem{Chamel2013}
N.~Chamel, ``{Crustal {Entrainment} and {Pulsar} {Glitches}},'' {\em Physical
  Review Letters}, vol.~110, p.~011101, 2013.

\bibitem{PRC90Piekarewicz2014}
J.~Piekarewicz, F.~J. Fattoyev, and C.~J. Horowitz, ``Pulsar glitches: {The}
  crust may be enough,'' {\em Physical Review C}, vol.~90, p.~015803, 2014.

\bibitem{Newton2015}
W.~G. Newton, S.~Berger, and B.~Haskell, ``{Observational constraints on
  neutron star crust-core coupling during glitches},'' {\em Monthly Notices of
  the Royal Astronomical Society}, vol.~454, p.~4400, 2015.

\bibitem{gonzalez19}
C.~Gonzalez-Boquera, M.~Centelles, X.~Vi\~nas, and T.~R. Routray, ``{Core-crust
  transition in neutron stars with finite-range interactions: The dynamical
  method},'' {\em Physical Review C}, vol.~100, p.~015806, 2019.

\bibitem{xu10b}
J.~Xu and C.~M. Ko, ``Density matrix expansion for the isospin- and
  momentum-dependent {MDI} interaction,'' {\em Physical Review C}, vol.~82,
  p.~044311, 2010.

\bibitem{Abbott2017}
B.~Abbott {\em et~al.}, ``{GW170817: Observation of Gravitational Waves from a
  Binary Neutron Star Inspiral},'' {\em Physical Review Letters}, vol.~119,
  p.~161101, 2017.

\bibitem{Abbott2018}
B.~Abbott {\em et~al.}, ``{GW170817: Measurements of Neutron Star Radii and
  Equation of State},'' {\em Physical Review Letters}, vol.~121, p.~161101,
  2018.

\bibitem{Abbott2019}
B.~Abbott {\em et~al.}, ``{Properties of the Binary Neutron Star Merger
  GW170817},'' {\em Physical Review X}, vol.~9, p.~011001, 2019.

\bibitem{Landry18}
P.~Landry and B.~Kumar, ``{Constraints on the Moment of Inertia of {PSR}
  J0737-3039A from {GW}170817},'' {\em The Astrophysical Journal}, vol.~868,
  p.~L22, 2018.

\bibitem{Baade34}
W.~Baade and W.~Zwicky {\em Physical Review}, vol.~45, p.~138, 1934.

\bibitem{Tolman39}
R.~C. Tolman, ``{Static Solutions of Einstein's Field Equations for Spheres of
  Fluid},'' {\em Physical Review}, vol.~55, p.~364, 1939.

\bibitem{Oppenheimer39}
J.~R. Oppenheimer and G.~M. Volkoff, ``{On Massive Neutron Cores},'' {\em
  Physical Review}, vol.~55, p.~374, 1939.

\bibitem{Wheeler58}
B.~Harrison, M.~Wakano, and J.~Wheeler, {\em {Matter-energy at high-density:
  end point of thermonuclear evolution , in La Structure et \'Evolution de
  l'Universe}}.
\newblock R. Stoops, Brussels, 1958.

\bibitem{Cameron59}
A.~G.~W. Cameron, ``{Pycnonuclear Reations and Nova Explosions},'' {\em The
  Atrophysical Journal}, vol.~130, p.~916, 1959.

\bibitem{Vidana18}
I.~Vida{\~{n}}a, ``A short walk through the physics of neutron stars,'' {\em
  The European Physical Journal Plus}, vol.~133, p.~445, 2018.

\bibitem{Hewish68}
A.~Hewish, S.~J. Bell, J.~D.~H. Pilkington, P.~F. Scott, and R.~A. Collins,
  ``Observation of a rapidly pulsating radio source,'' {\em Nature}, vol.~217,
  p.~709, 1968.

\bibitem{shapiro83}
S.~L. Shapiro and S.~A. Teukolsky, {\em {Black holes, white dwarfs, and neutron
  stars: {The} physics of compact objects}}, vol.~20.
\newblock John Wiley {\&} Sons, 1983.

\bibitem{Glendenning2000}
N.~K. Glendenning, {\em {Compact stars : nuclear physics, particle physics, and
  general relativity}}, vol.~242 of {\em Astonomy and Astrophysics Library}.
\newblock Springer (New York), second~ed., 2000.

\bibitem{haensel07}
P.~Haensel, A.~Y. Potekhin, and D.~G. Yakovlev, {\em {Neutron Stars 1:
  {Equation} of {State} and {Structure}}}.
\newblock Springer, 2007.

\bibitem{Lattimer2004}
J.~M. Lattimer and M.~Prakash, ``{The Physics of Neutron Stars},'' {\em
  Science}, vol.~304, p.~536, 2004.

\bibitem{baym71}
G.~Baym, H.~A. Bethe, and C.~J. Pethick, ``Neutron star matter,'' {\em Nuclear
  Physics A}, vol.~175, p.~225, 1971.

\bibitem{Duflo95}
J.~Duflo and A.~Zuker, ``Microscopic mass formulas,'' {\em Physical Review C},
  vol.~52, p.~R23, 1995.

\bibitem{Moller95}
P.~Moller, J.~Nix, W.~Myers, and W.~Swiatecki, ``{Nuclear Ground-State Masses
  and Deformations},'' {\em Atomic Data and Nuclear Data Tables}, vol.~59,
  p.~185, 1995.

\bibitem{ruster06}
S.~B. R\"uster, M.~Hempel, and J.~Schaffner-Bielich, ``Outer crust of
  nonaccreting cold neutron stars,'' {\em Physical Review C}, vol.~73,
  p.~035804, 2006.

\bibitem{XaviRoca08}
X.~Roca-Maza and J.~Piekarewicz, ``Impact of the symmetry energy on the outer
  crust of nonaccreting neutron stars,'' {\em Physical Review C}, vol.~78,
  p.~025807, 2008.

\bibitem{Pearson11}
J.~M. Pearson, S.~Goriely, and N.~Chamel, ``Properties of the outer crust of
  neutron stars from {Hartree-Fock-Bogoliubov} mass models,'' {\em Physical
  Review C}, vol.~83, p.~065810, 2011.

\bibitem{sharma15}
B.~K. Sharma, M.~Centelles, X.~Vi{\~n}as, M.~Baldo, and G.~F. Burgio,
  ``{Unified equation of state for neutron stars on a microscopic basis},''
  {\em Astronomy and Astrophysics}, vol.~584, p.~A103, 2015.

\bibitem{Lattimer19}
J.~M. Lattimer, ``Neutron star masses and radii,'' {\em AIP Conference
  Proceedings}, vol.~2127, p.~020001, 2019.

\bibitem{Wiringa95}
R.~B. Wiringa, V.~G.~J. Stoks, and R.~Schiavilla, ``Accurate nucleon-nucleon
  potential with charge-independence breaking,'' {\em Physical Review C},
  vol.~51, p.~38, 1995.

\bibitem{Vidana2009}
I.~Vida{\~{n}}a, C.~Providencia, A.~Polls, and A.~Rios, ``{Density dependence
  of the nuclear symmetry energy: a microscopic perspective},'' {\em Physical
  Review C}, vol.~80, p.~045806, 2009.

\bibitem{Ducoin11}
C.~Ducoin, J.~Margueron, C.~Provid\^encia, and I.~Vida\~na, ``Core-crust
  transition in neutron stars: Predictivity of density developments,'' {\em
  Physical Review C}, vol.~83, p.~045810, 2011.

\bibitem{Li2016}
A.~Li, J.~M. Dong, J.~B. Wang, and R.~X. Xu, ``{Structures of the Vela Pulsar
  and the Glitch Crisis From the Brueckner Theory},'' {\em The Astrophysical
  Journal Supplement Series}, vol.~223, p.~16, 2016.

\bibitem{xu09a}
J.~Xu, L.-W. Chen, B.-A. Li, and H.-R. Ma, ``Nuclear {Constraints} on
  {Properties} of {Neutron} {Star} {Crusts},'' {\em The Astrophysical Journal},
  vol.~697, p.~1549, 2009.

\bibitem{ducoin07}
C.~Ducoin, P.~Chomaz, and F.~Gulminelli, ``Isospin-dependent clusterization of
  neutron-star matter,'' {\em Nuclear Physics A}, vol.~789, p.~403, 2007.

\bibitem{Pearson12}
J.~M. Pearson, N.~Chamel, S.~Goriely, and C.~Ducoin, ``Inner crust of neutron
  stars with mass-fitted skyrme functionals,'' {\em Physical Review C},
  vol.~85, 2012.

\bibitem{Newton2014}
W.~G. Newton, J.~Hooker, M.~Gearheart, K.~Murphy, D.~H. Wen, F.~J. Fattoyev,
  and B.~A. Li, ``{Constraints on the symmetry energy from observational probes
  of the neutron star crust},'' {\em The European Physical Journal A}, vol.~50,
  p.~1, 2014.

\bibitem{routray16}
T.~R. Routray, X.~Vi{\~{n}}as, D.~N. Basu, S.~P. Pattnaik, M.~Centelles, L.~M.
  Robledo, and B.~Behera, ``{Exact versus {Taylor-expanded} energy density in
  the study of the neutron star crust-core transition},'' {\em Journal of
  Physics G: Nuclear and Particle Physics}, vol.~43, p.~105101, 2016.

\bibitem{horowitz01a}
C.~J. Horowitz and J.~Piekarewicz, ``{Neutron star structure and the neutron
  radius of {${}^{208}\mathrm{Pb}$}},'' {\em Physical Review Letters}, vol.~86,
  p.~5647, 2001.

\bibitem{carriere03}
J.~{Carriere}, C.~J. {Horowitz}, and J.~{Piekarewicz}, ``{{Low-Mass} {Neutron}
  {Stars} and the {Equation} of {State} of {Dense} {Matter}},'' {\em The
  Astrophysical Journal}, vol.~593, p.~463, 2003.

\bibitem{Klahn06}
T.~Kl\"ahn, D.~Blaschke, S.~Typel, E.~N.~E. van Dalen, A.~Faessler, C.~Fuchs,
  T.~Gaitanos, H.~Grigorian, A.~Ho, E.~E. Kolomeitsev, M.~C. Miller,
  G.~R\"opke, J.~Tr\"umper, D.~N. Voskresensky, F.~Weber, and H.~H. Wolter,
  ``Constraints on the high-density nuclear equation of state from the
  phenomenology of compact stars and heavy-ion collisions,'' {\em Physical
  Review C}, vol.~74, p.~035802, 2006.

\bibitem{Moustakidis10}
C.~C. Moustakidis, T.~Nik\ifmmode \check{s}\else
  \v{s}\fi{}i\ifmmode~\acute{c}\else \'{c}\fi{}, G.~A. Lalazissis, D.~Vretenar,
  and P.~Ring, ``Constraints on the inner edge of neutron star crusts from
  relativistic nuclear energy density functionals,'' {\em Physical Review C},
  vol.~81, p.~065803, 2010.

\bibitem{Cai2012}
B.-J. Cai and L.-W. Chen, ``Nuclear matter fourth-order symmetry energy in the
  relativistic mean field models,'' {\em Physical Review C}, vol.~85,
  p.~024302, 2012.

\bibitem{Moustakidis12}
C.~C. Moustakidis, ``Effect of the symmetry energy on the location of the inner
  edge of the neutron star crust,'' {\em Physical Review C}, vol.~86,
  p.~015801, 2012.

\bibitem{Demorest10}
P.~B. Demorest, T.~Pennucci, S.~M. Ransom, M.~S.~E. Roberts, and J.~W.~T.
  Hessels, ``A two-solar-mass neutron star measured using {Shapiro} delay,''
  {\em Nature}, vol.~467, p.~1081, 2010.

\bibitem{Antoniadis13}
J.~Antoniadis, P.~C.~C. Freire, N.~Wex, T.~M. Tauris, R.~S. Lynch, M.~H. van
  Kerkwijk, M.~Kramer, C.~Bassa, V.~S. Dhillon, T.~Driebe, J.~W.~T. Hessels,
  V.~M. Kaspi, V.~I. Kondratiev, N.~Langer, T.~R. Marsh, M.~a. McLaughlin,
  T.~T. Pennucci, S.~M. Ransom, I.~H. Stairs, J.~van Leeuwen, J.~P.~W.
  Verbiest, and D.~G. Whelan, ``A {Massive} {Pulsar} in a {Compact}
  {Relativistic} {Binary},'' {\em Science}, vol.~340, p.~448, 2013.

\bibitem{Fonseca_2016}
E.~Fonseca, T.~T. Pennucci, J.~A. Ellis, I.~H. Stairs, D.~J. Nice, S.~M.
  Ransom, P.~B. Demorest, Z.~Arzoumanian, K.~Crowter, T.~Dolch, R.~D. Ferdman,
  M.~E. Gonzalez, G.~Jones, M.~L. Jones, M.~T. Lam, L.~Levin, M.~A. McLaughlin,
  K.~Stovall, J.~K. Swiggum, and W.~Zhu, ``{The nanograv nine-year data set:
  mass and geometric meassurements of binary millisecond pulsars},'' {\em The
  Astrophysical Journal}, vol.~832, p.~167, 2016.

\bibitem{Cromartie19}
H.~T. Cromartie, E.~Fonseca, S.~M. Ransom, P.~B. Demorest, Z.~Arzoumanian,
  H.~Blumer, P.~R. Brook, M.~E. DeCesar, T.~Dolch, J.~A. Ellis, R.~D. Ferdman,
  E.~C. Ferrara, N.~Garver-Daniels, P.~A. Gentile, M.~L. Jones, M.~T. Lam,
  D.~R. Lorimer, R.~S. Lynch, M.~A. McLaughlin, C.~Ng, D.~J. Nice, T.~T.
  Pennucci, R.~Spiewak, I.~H. Stairs, K.~Stovall, J.~K. Swiggum, and W.~W. Zhu,
  ``Relativistic shapiro delay measurements of an extremely massive millisecond
  pulsar,'' {\em Nature Astronomy}, 2019.

\bibitem{Zhang_2019}
N.-B. Zhang and B.-A. Li, ``{Implications of the Mass $M=2.17^{+0.11}_{-0.10}$
  M$_\odot$ of {PSR} J07409+6620 on the Equation of State of Super-dense
  Neutron-rich Nuclear Matter},'' {\em The Astrophysical Journal}, vol.~879,
  p.~99, 2019.

\bibitem{Oertel:2016bki}
M.~Oertel, M.~Hempel, T.~Kl{\"a}hn, and S.~Typel, ``{Equations of state for
  supernovae and compact stars},'' {\em Reviews of Modern Physics}, vol.~89,
  p.~015007, 2017.

\bibitem{Li2014}
B.-A. Li, A.~Ramos, G.~Verde, and I.~Vida{\~{n}}a, ``Topical opissue on nuclear
  symmetry energy,'' {\em The European Physical Journal A}, vol.~50, p.~9,
  2014.

\bibitem{Stein_2014}
L.~C. Stein, K.~Yagi, and N.~Yunes, ``{Three}-{hair} {relations} {for}
  {rotating} {stars}: {nonrelativistic} {limit},'' {\em The Astrophysical
  Journal}, vol.~788, p.~15, 2014.

\bibitem{Nattila16}
J.~N\"attil\"a, A.~W. Steiner, J.~J.~E. Kajava, V.~F. Suleimanov, and
  J.~Poutanen, ``Equation of state constraints for the cold dense matter inside
  neutron stars using the cooling tail method,'' {\em Astronomy and
  Astrophysics}, vol.~591, p.~A25, 2016.

\bibitem{De2018}
S.~De, D.~Finstad, J.~M. Lattimer, D.~A. Brown, E.~Berger, and C.~M. Biwer,
  ``{Tidal Deformabilities and Radii of Neutron Stars from the Observation of
  GW170817},'' {\em Physical Review Letters}, vol.~121, p.~091102, 2018.

\bibitem{Bender03}
M.~Bender, P.-H. Heenen, and P.-G. Reinhard, ``Self-consistent mean-field
  models for nuclear structure,'' {\em Reviews of Modern Physics}, vol.~75,
  p.~121, 2003.

\bibitem{Machleidt_2001}
R.~Machleidt and I.~Slaus, ``The nucleon-nucleon interaction,'' {\em Journal of
  Physics G: Nuclear and Particle Physics}, vol.~27, p.~R69, 2001.

\bibitem{Serot:1984ey}
B.~D. Serot and J.~D. Walecka, ``{The Relativistic Nuclear Many Body
  Problem},'' {\em Advances in Nuclear Physics}, vol.~16, p.~1, 1986.

\bibitem{Brockmann90}
R.~Brockmann and R.~Machleidt, ``Relativistic nuclear structure. {I. Nuclear
  matter},'' {\em Physical Review C}, vol.~42, p.~1965, 1990.

\bibitem{Dickhoff_1992}
W.~H. Dickhoff and H.~Muther, ``Nucleon properties in the nuclear medium,''
  {\em Reports on Progress in Physics}, vol.~55, p.~1947, 1992.

\bibitem{pandh81}
V.~R. Pandharipande, I.~Sick, and P.~K. A.~d. Huberts, ``Independent particle
  motion and correlations in fermion systems,'' {\em Reviews of Modern
  Physics}, vol.~69, p.~981, 1997.

\bibitem{Heiselberg00}
H.~Heiselberg and V.~Pandharipande, ``{Recent Progress in Neutron Star
  Theory},'' {\em Annual Review of Nuclear and Particle Science}, vol.~50,
  p.~481, 2000.

\bibitem{MarioArtur2}
G.~Baym and L.~P. Kadanoff, ``{Conservation Laws and Correlation Functions},''
  {\em Physical Review}, vol.~124, p.~287, 1961.

\bibitem{Artur2}
M.~Baldo, A.~Polls, A.~Rios, H.-J. Schulze, and I.~Vida\~na, ``Comparative
  study of neutron and nuclear matter with simplified {Argonne} nucleon-nucleon
  potentials,'' {\em Physical Review C}, vol.~86, p.~064001, 2012.

\bibitem{Artur1}
A.~Carbone, A.~Polls, and A.~Rios, ``Symmetric nuclear matter with chiral
  three-nucleon forces in the self-consistent {Green's} functions approach,''
  {\em Physical Review C}, vol.~88, p.~044302, 2013.

\bibitem{MarioArtur4}
A.~Carbone, A.~Cipollone, C.~Barbieri, A.~Rios, and A.~Polls,
  ``{Self-consistent Green's functions formalism with three-body
  interactions},'' {\em Physical Review C}, vol.~88, p.~054326, 2013.

\bibitem{MarioArtur3}
A.~Carbone, A.~Rios, and A.~Polls, ``{Correlated density-dependent chiral
  forces for infinite-matter calculations within the Green's function
  approach},'' {\em Physical Review C}, vol.~90, p.~054322, 2014.

\bibitem{MarioArtur1}
A.~Carbone, A.~Polls, and A.~Rios, ``Microscopic predictions of the nuclear
  matter liquid-gas phase transition,'' {\em Physical Review C}, vol.~98,
  p.~025804, 2018.

\bibitem{myers82}
W.~D. Myers and W.~J. Swiatecki, ``The macroscopic approach to nuclear masses
  and deformations,'' {\em Annual Review of Nuclear and Particle Science},
  vol.~32, p.~309, 1982.

\bibitem{brown88}
B.~A. Brown and B.~H. Wildenthal, ``Status of the nuclear shell model,'' {\em
  Annual Review of Nuclear and Particle Science}, vol.~38, p.~29, 1988.

\bibitem{rin80}
P.~Ring and P.~Schuck, {\em The nuclear many-body problem}.
\newblock Springer-Verlag (New York), 1980.

\bibitem{sly42}
E.~Chabanat, P.~Bonche, P.~Haensel, J.~Meyer, and R.~Schaeffer, ``A {Skyrme}
  parametrization from subnuclear to neutron star densities {Part} {II}.
  {Nuclei} far from stabilities,'' {\em Nuclear Physics A}, vol.~635, p.~231,
  1998.

\bibitem{Skyrme58}
T.~Skyrme, ``The effective nuclear potential,'' {\em Nuclear Physics}, vol.~9,
  p.~615, 1958.

\bibitem{BaoAnLi13}
B.-A. Li and X.~Han, ``Constraining the neutron–proton effective mass
  splitting using empirical constraints on the density dependence of nuclear
  symmetry energy around normal density,'' {\em Physics Letters B}, vol.~727,
  p.~276, 2013.

\bibitem{Vinas14}
X.~Vi{\~{n}}as, M.~Centelles, X.~Roca-Maza, and M.~Warda, ``Density dependence
  of the symmetry energy from neutron skin thickness in finite nuclei,'' {\em
  The European Physical Journal A}, vol.~50, 2014.

\bibitem{skyrmechiral}
Y.~Lim and J.~W. Holt, ``Structure of neutron star crusts from new {Skyrme}
  effective interactions constrained by chiral effective field theory,'' {\em
  Physical Review C}, vol.~95, p.~065805, 2017.

\bibitem{Msk7}
F.~Tondeur, S.~Goriely, J.~M. Pearson, and M.~Onsi, ``Towards a
  {Hartree}-{Fock} mass formula,'' {\em Physical Review C}, vol.~62, p.~024308,
  2000.

\bibitem{SIII}
M.~Beiner, H.~Flocard, N.~V. Giai, and P.~Quentin, ``Nuclear ground-state
  properties and self-consistent calculations with the {Skyrme} interaction:
  {(I)}. {Spherical} description,'' {\em Nuclear Physics A}, vol.~238, p.~29,
  1975.

\bibitem{SkP}
J.~Dobaczewski, H.~Flocard, and J.~Treiner, ``Hartree-{Fock}-{Bogolyubov}
  description of nuclei near the neutron-drip line,'' {\em Nuclear Physics A},
  vol.~422, p.~103, 1984.

\bibitem{HFB27}
S.~Goriely, N.~Chamel, and J.~M. Pearson, ``Hartree-{Fock}-{Bogoliubov} nuclear
  mass model with 0.50 {MeV} accuracy based on standard forms of {Skyrme} and
  pairing functionals,'' {\em Physical Review C}, vol.~88, p.~061302, 2013.

\bibitem{SkX}
B.~Alex~Brown, ``New {Skyrme} interaction for normal and exotic nuclei,'' {\em
  Physical Review C}, vol.~58, p.~220, 1998.

\bibitem{HFB17}
S.~Goriely, N.~Chamel, and J.~M. Pearson,
  ``Skyrme-{Hartree}-{Fock}-{Bogoliubov} {Nuclear} {Mass} {Formulas}:
  {Crossing} the 0.6 {MeV} {Accuracy} {Threshold} with {Microscopically}
  {Deduced} {Pairing},'' {\em Physical Review Letters}, vol.~102, p.~152503,
  2009.

\bibitem{sgii}
N.~V. Giai and H.~Sagawa, ``Spin-isospin and pairing properties of modified
  {Skyrme} interactions,'' {\em Physics Letters B}, vol.~106, p.~379, 1981.

\bibitem{unedf1}
M.~Kortelainen, J.~McDonnell, W.~Nazarewicz, P.-G. Reinhard, J.~Sarich,
  N.~Schunck, M.~V. Stoitsov, and S.~M. Wild, ``Nuclear energy density
  optimization: {Large} deformations,'' {\em Physical Review C}, vol.~85,
  p.~024304, 2012.

\bibitem{unedf0}
M.~Kortelainen, T.~Lesinski, J.~Mor\'e, W.~Nazarewicz, J.~Sarich, N.~Schunck,
  M.~V. Stoitsov, and S.~Wild, ``Nuclear energy density optimization,'' {\em
  Physical Review C}, vol.~82, p.~024313, 2010.

\bibitem{skms}
M.~Brack, C.~Guet, and H.-B. Håkansson, ``Selfconsistent semiclassical
  description of average nuclear properties—a link between microscopic and
  macroscopic models,'' {\em Physics Reports}, vol.~123, p.~275, 1985.

\bibitem{mska}
M.~M. Sharma, G.~Lalazissis, J.~K\"onig, and P.~Ring, ``Isospin {Dependence} of
  the {Spin-Orbit} {Force} and {Effective} {Nuclear} {Potentials},'' {\em
  Physical Review Letters}, vol.~74, p.~3744, 1995.

\bibitem{msl0}
L.-W. Chen, C.~M. Ko, B.-A. Li, and J.~Xu, ``Density slope of the nuclear
  symmetry energy from the neutron skin thickness of heavy nuclei,'' {\em
  Physical Review C}, vol.~82, p.~024321, 2010.

\bibitem{skmp}
L.~Bennour, P.-H. Heenen, P.~Bonche, J.~Dobaczewski, and H.~Flocard, ``Charge
  distributions of {$^{208}\mathrm{Pb}$}, {$^{206}\mathrm{Pb}$}, and
  {$^{205}\mathrm{Tl}$} and the mean-field approximation,'' {\em Physical
  Review C}, vol.~40, p.~2834, 1989.

\bibitem{ska}
H.~Köhler, ``Skyrme force and the mass formula,'' {\em Nuclear Physics A},
  vol.~258, p.~301, 1976.

\bibitem{rsgs}
J.~Friedrich and P.-G. Reinhard, ``Skyrme-force parametrization:
  {Least}-squares fit to nuclear ground-state properties,'' {\em Physical
  Review C}, vol.~33, p.~335, 1986.

\bibitem{ski2}
P.-G. Reinhard and H.~Flocard, ``Nuclear effective forces and isotope shifts,''
  {\em Nuclear Physics A}, vol.~584, p.~467, 1995.

\bibitem{chappert08}
F.~Chappert, M.~Girod, and S.~Hilaire, ``Towards a new {Gogny} force
  parameterization: {Impact} of the neutron matter equation of state,'' {\em
  Physics Letters B}, vol.~668, p.~420, 2008.

\bibitem{goriely09}
S.~Goriely, S.~Hilaire, M.~Girod, and S.~P\'eru, ``First
  {Gogny-Hartree-Fock-Bogoliubov} {Nuclear} {Mass} {Model},'' {\em Physical
  Review Letters}, vol.~102, p.~242501, 2009.

\bibitem{NPA591Blaizot1995}
J.~Blaizot, J.~Berger, J.~Decharg{\'e}, and M.~Girod, ``Microscopic and
  macroscopic determinations of nuclear compressibility,'' {\em Nuclear Physics
  A}, vol.~591, p.~435, 1995.

\bibitem{chappert15}
F.~Chappert, N.~Pillet, M.~Girod, and J.-F. Berger, ``Gogny force with a
  finite-range density dependence,'' {\em Physical Review C}, vol.~91,
  p.~034312, 2015.

\bibitem{Friedman81}
B.~Friedman and V.~Pandharipande, ``Hot and cold, nuclear and neutron matter,''
  {\em Nuclear Physics A}, vol.~361, p.~502, 1981.

\bibitem{xu09b}
J.~Xu, L.-W. Chen, B.-A. Li, and H.-R. Ma, ``Locating the inner edge of the
  neutron star crust using terrestrial nuclear laboratory data,'' {\em Physical
  Review C}, vol.~79, p.~035802, 2009.

\bibitem{xu10a}
J.~Xu, L.-W. Chen, C.~M. Ko, and B.-A. Li, ``Transition density and pressure in
  hot neutron stars,'' {\em Physical Review C}, vol.~81, p.~055805, 2010.

\bibitem{Krastev19}
P.~G. Krastev and B.-A. Li, ``Imprints of the nuclear symmetry energy on the
  tidal deformability of neutron stars,'' {\em Journal of Physics G: Nuclear
  and Particle Physics}, vol.~46, p.~074001, 2019.

\bibitem{chen14}
L.-W. Chen, C.~M. Ko, B.-A. Li, C.~Xu, and J.~Xu, ``Probing isospin- and
  momentum-dependent nuclear effective interactions in neutron-rich matter,''
  {\em The European Physical Journal A}, vol.~50, p.~29, 2014.

\bibitem{Chen05}
L.-W. Chen, C.~M. Ko, and B.-A. Li, ``Nuclear matter symmetry energy and the
  neutron skin thickness of heavy nuclei,'' {\em Physical Review C}, vol.~72,
  p.~064309, 2005.

\bibitem{Behera_1997}
B.~Behera, T.~R. Routray, and R.~K. Satpathy, ``Causal violation of the speed
  of sound and the equation of state of nuclear matter,'' {\em Journal of
  Physics G: Nuclear and Particle Physics}, vol.~23, p.~445, 1997.

\bibitem{Behera_2011}
B.~Behera, T.~R. Routray, and S.~K. Tripathy, ``Neutron{\textendash}proton
  effective mass splitting and thermal evolution in neutron-rich matter,'' {\em
  Journal of Physics G: Nuclear and Particle Physics}, vol.~38, p.~115104,
  2011.

\bibitem{Xu11}
C.~Xu, B.-A. Li, L.-W. Chen, and C.~M. Ko, ``Analytical relations between
  nuclear symmetry energy and single-nucleon potentials in isospin asymmetric
  nuclear matter,'' {\em Nuclear Physics A}, vol.~865, p.~1, 2011.

\bibitem{Chen12}
R.~Chen, B.-J. Cai, L.-W. Chen, B.-A. Li, X.-H. Li, and C.~Xu, ``Single-nucleon
  potential decomposition of the nuclear symmetry energy,'' {\em Physical
  Review C}, vol.~85, p.~024305, 2012.

\bibitem{Piekarewicz08}
J.~{Piekarewicz} and M.~{Centelles}, ``{Incompressibility of neutron-rich
  matter},'' {\em Physical Review C}, vol.~79, p.~054311, 2009.

\bibitem{Danielewicz13}
P.~Danielewicz and J.~Lee, ``{Symmetry Energy {II}: {Isobaric} {Analog}
  {States}},'' {\em Nuclear Physics}, vol.~A922, p.~1, 2014.

\bibitem{Tsang08}
M.~B. Tsang, Y.~Zhang, P.~Danielewicz, M.~Famiano, Z.~Li, W.~G. Lynch, and
  A.~W. Steiner, ``{Constraints on the density dependence of the symmetry
  energy},'' {\em Physical Review Letters}, vol.~102, p.~122701, 2009.

\bibitem{Zhang15}
Z.~Zhang and L.-W. Chen, ``{Electric dipole polarizability in {$^{208}$Pb} as a
  probe of the symmetry energy and neutron matter around $\rho_0/3$},'' {\em
  Physical Review C}, vol.~92, p.~031301, 2015.

\bibitem{Chen_15}
{Chen, Lie-Wen}, ``Symmetry energy systematics and its high density behavior,''
  {\em The European Physical Journal Web of Conferences}, vol.~88, p.~00017,
  2015.

\bibitem{Seif14}
W.~M. Seif and D.~N. Basu, ``Higher-order symmetry energy of nuclear matter and
  the inner edge of neutron star crusts,'' {\em Physical Review C}, vol.~89,
  p.~028801, 2014.

\bibitem{Liu18}
Z.~W. Liu, Z.~Qian, R.~Y. Xing, J.~R. Niu, and B.~Y. Sun, ``Nuclear
  fourth-order symmetry energy and its effects on neutron star properties in
  the relativistic {Hartree-Fock} theory,'' {\em Physical Review C}, vol.~97,
  p.~025801, 2018.

\bibitem{Wellenhofer2016}
C.~Wellenhofer, J.~W. Holt, and N.~Kaiser, ``Divergence of the
  isospin-asymmetry expansion of the nuclear equation of state in many-body
  perturbation theory,'' {\em Physical Review C}, vol.~93, p.~055802, 2016.

\bibitem{Tsang2012}
M.~B. Tsang, J.~R. Stone, F.~Camera, P.~Danielewicz, S.~Gandolfi, K.~Hebeler,
  C.~J. Horowitz, J.~Lee, W.~G. Lynch, Z.~Kohley, R.~Lemmon, P.~M\"oller,
  T.~Murakami, S.~Riordan, X.~Roca-Maza, F.~Sammarruca, A.~W. Steiner,
  I.~Vida\~na, and S.~J. Yennello, ``Constraints on the symmetry energy and
  neutron skins from experiments and theory,'' {\em Physical Review C},
  vol.~86, p.~015803, 2012.

\bibitem{Lattimer2013}
J.~M. Lattimer and Y.~Lim, ``{Constraining the symmetry energy of the nuclear
  interaction},'' {\em The Astrophysical Journal}, vol.~771, p.~51, 2013.

\bibitem{Roca-Maza15}
X.~{Roca-Maza}, X.~{Vi{\~n}as}, M.~{Centelles}, B.~K. {Agrawal}, G.~{Col{\`o}},
  N.~{Paar}, J.~{Piekarewicz}, and D.~{Vretenar}, ``{Neutron skin thickness
  from the measured electric dipole polarizability in {$^{68}$Ni$^{120}$Sn} and
  {$^{208}$Pb}},'' {\em Physical Review C}, vol.~92, p.~064304, 2015.

\bibitem{Lattimer2016}
J.~M. Lattimer and M.~Prakash, ``{The equation of state of hot, dense matter
  and neutron stars},'' {\em Physics Reports}, vol.~621, p.~127, 2016.

\bibitem{Birkhan16}
J.~Birkhan {\em et~al.}, ``{Electric dipole polarizability of {$^{48}$Ca} and
  implications for the neutron skin},'' {\em Physical Review Letters},
  vol.~118, p.~252501, 2017.

\bibitem{Holt2017}
J.~W. Holt and N.~Kaiser, ``Equation of state of nuclear and neutron matter at
  third-order in perturbation theory from chiral effective field theory,'' {\em
  Physical Review C}, vol.~95, p.~034326, 2017.

\bibitem{Drischler1710.08220}
C.~Drischler, K.~Hebeler, and A.~Schwenk, ``Chiral interactions up to
  next-to-next-to-next-to-leading order and nuclear saturation,'' {\em Physical
  Review Letters}, vol.~122, p.~042501, Jan 2019.

\bibitem{Loan2011}
D.~T. Loan, N.~H. Tan, D.~T. Khoa, and J.~Margueron, ``Equation of state of
  neutron star matter, and the nuclear symmetry energy,'' {\em Physical Review
  C}, vol.~83, p.~065809, 2011.

\bibitem{Baldo09}
M.~Baldo and C.~Ducoin, ``Elementary excitations in homogeneous neutron star
  matter,'' {\em Physical Review C}, vol.~79, p.~035801, 2009.

\bibitem{SellahewaPhD}
R.~Sellahewa, {\em {Isovector And Pairing Properties Of The Gogny Force In The
  Context Of Neutron Stars}}.
\newblock PhD thesis, University of Surrey, 2016.

\bibitem{Lattimer01}
J.~M. Lattimer and M.~Prakash, ``Neutron {Star} {Structure} and the {Equation}
  of {State},'' {\em The Astrophysical Journal}, vol.~550, p.~426, 2001.

\bibitem{Than2011}
H.~S. Than, E.~Khan, and N.~{Van Giai}, ``{Wigner-Seitz cells in neutron star
  crust with finite-range interactions},'' {\em Journal of Physics G: Nuclear
  and Particle Physics}, vol.~38, p.~025201, 2011.

\bibitem{Hebeler13}
K.~Hebeler, J.~M. Lattimer, C.~J. Pethick, and A.~Schwenk, ``Equation of state
  and neutron star properties constrained by nuclear physics and observation,''
  {\em The Astrophysical Journal}, vol.~773, p.~11, 2013.

\bibitem{Lattimer14}
J.~M. Lattimer and A.~W. Steiner, ``Constraints on the symmetry energy using
  the mass-radius relation of neutron stars,'' {\em European Physical Journal
  A}, vol.~50, p.~40, 2014.

\bibitem{Guillot14}
S.~Guillot and R.~E. Rutledge, ``{Rejecting proposed dense-matter equations of
  state with quiescent low-mass {X}-ray binaries},'' {\em The Astrophysical
  Journal}, vol.~796, p.~L3, 2014.

\bibitem{Heinke14}
C.~O. {Heinke}, H.~N. {Cohn}, P.~M. {Lugger}, N.~A. {Webb}, W.~C.~G. {Ho},
  J.~{Anderson}, S.~{Campana}, S.~{Bogdanov}, D.~{Haggard}, A.~M. {Cool}, and
  J.~E. {Grindlay}, ``{Improved mass and radius constraints for quiescent
  neutron stars in {$\omega$} {Cen} and {NGC} 6397},'' {\em Monthly Notices of
  the Royal Astronomical Society}, vol.~444, p.~443, 2014.

\bibitem{Ozel15}
F.~Ozel, D.~Psaltis, T.~Guver, G.~Baym, C.~Heinke, and S.~Guillot, ``{The Dense
  Matter Equation of State from Neutron Star Radius and Mass Measurements},''
  {\em The Astrophysical Journal}, vol.~820, p.~28, 2016.

\bibitem{Ozel16}
F.~Ozel and P.~Freire, ``{Masses, Radii, and the Equation of State of Neutron
  Stars},'' {\em Annual Review of Astronomy and Astrophysics}, vol.~54, p.~401,
  2016.

\bibitem{Chen15}
W.-C. Chen and J.~Piekarewicz, ``{Compactness of Neutron Stars},'' {\em
  Physical Review Letters}, vol.~115, p.~161101, 2015.

\bibitem{Jiang15}
W.-Z. Jiang, B.-A. Li, and F.~J. Fattoyev, ``{Small radii of neutron stars as
  an indication of novel in-medium effects},'' {\em The European Physical
  Journal}, vol.~A51, p.~119, 2015.

\bibitem{Tolos16}
L.~{Tolos}, M.~{Centelles}, and A.~{Ramos}, ``{Equation of State for Nucleonic
  and Hyperonic Neutron Stars with Mass and Radius Constraints},'' {\em The
  Astrophysical Journal}, vol.~834, p.~3, 2017.

\bibitem{Roca13}
X.~Roca-Maza, M.~Brenna, B.~K. Agrawal, P.~F. Bortignon, G.~Col\`o, L.-G. Cao,
  N.~Paar, and D.~Vretenar, ``{Giant quadrupole resonances in {$^{208}$Pb}, the
  nuclear symmetry energy, and the neutron skin thickness},'' {\em Physical
  Review C}, vol.~87, p.~034301, 2013.

\bibitem{Agrawal05}
B.~K. Agrawal, S.~Shlomo, and V.~K. Au, ``{Determination of the parameters of a
  Skyrme type effective interaction using the simulated annealing approach},''
  {\em Physical Review C}, vol.~72, p.~014310, 2005.

\bibitem{Piekarewicz11}
J.~Piekarewicz, ``{Pygmy resonances and neutron skins},'' {\em Physical Review
  C}, vol.~83, p.~034319, 2011.

\bibitem{Fattoyev13}
F.~J. Fattoyev and J.~Piekarewicz, ``{Has a Thick Neutron Skin in $^{208}$Pb
  Been Ruled Out?},'' {\em Physical Review Letters}, vol.~111, p.~162501, 2013.

\bibitem{Centelles09}
M.~Centelles, X.~Roca-Maza, X.~Vi{\~{n}}as, and M.~Warda, ``{Nuclear Symmetry
  Energy Probed by Neutron Skin Thickness of Nuclei},'' {\em Physical Review
  Letters}, vol.~102, p.~122502, 2009.

\bibitem{Danielewicz:2002pu}
P.~Danielewicz, R.~Lacey, and W.~G. Lynch, ``{Determination of the equation of
  state of dense matter},'' {\em Science}, vol.~298, p.~1592, 2002.

\bibitem{Hagen:2015yea}
G.~Hagen {\em et~al.}, ``{Neutron and weak-charge distributions of the
  {$^{48}$Ca} nucleus},'' {\em Nature Physics}, vol.~12, p.~186, 2015.

\bibitem{Chen:2014mza}
W.-C. Chen and J.~Piekarewicz, ``{Searching for isovector signatures in the
  neutron-rich oxygen and calcium isotopes},'' {\em Physics Letters},
  vol.~B748, p.~284, 2015.

\bibitem{Ventura94}
J.~Ventura, A.~Polls, X.~Viñas, and E.~Hernandez, ``{Spectral and
  thermodynamical properties of symmetric nuclear matter with Gogny
  interaction},'' {\em Nuclear Physics A}, vol.~578, p.~147, 1994.

\bibitem{Audi12}
G.~Audi, M.~Wang, A.~Wapstra, F.~Kondev, M.~MacCormick, X.~Xu, and B.~Pfeiffer,
  ``{The Ame2012 atomic mass evaluation},'' {\em Chinese Physics C}, vol.~36,
  p.~1287, 2012.

\bibitem{robledo02}
L.~Robledo {\em HFBaxial computer code}, 2002.

\bibitem{rob11}
L.~M. Robledo and G.~F. Bertsch, ``{Application of the gradient method to
  Hartree-Fock-Bogoliubov theory},'' {\em Physical Review C}, vol.~84,
  p.~014312, 2011.

\bibitem{Rob11b}
L.~M. Robledo and G.~F. Bertsch, ``{Global systematics of octupole excitations
  in even-even nuclei},'' {\em Physical Review C}, vol.~84, p.~054302, 2011.

\bibitem{RRG00}
R.~R. Rodr\'{\i}guez-Guzm\'an, J.~L. Egido, and L.~M. Robledo, ``{Description
  of quadrupole collectivity in $N\approx 20$ nuclei with techniques beyond the
  mean field},'' {\em Physical Review C}, vol.~62, p.~054319, 2000.

\bibitem{Hilaire.07}
S.~Hilaire and M.~Girod, ``{Large-scale mean-field calculations from proton to
  neutron drip lines using the D1S Gogny force},'' {\em The European Physical
  Journal A}, vol.~33, p.~237, 2007.

\bibitem{Baldo13a}
M.~Baldo, L.~M. Robledo, P.~Schuck, and X.~Vi{\~n}as, ``{New Kohn-Sham density
  functional based on microscopic nuclear and neutron matter equations of
  state},'' {\em Physical Review C}, vol.~87, p.~064305, 2013.

\bibitem{Rob15}
L.~M. Robledo {\em Journal of Physics G: Nuclear and Particle Physics},
  vol.~42, p.~055109, 2015.

\bibitem{Pillet17}
N.~Pillet and S.~Hilaire, ``{Towards an extended Gogny force},'' {\em The
  European Physical Journal A}, vol.~53, p.~193, 2017.

\bibitem{Brown00}
B.~A.~Brown, ``{Neutron Radii in Nuclei and the Neutron Equation of State},''
  {\em Physical Review Letters}, vol.~85, p.~5296, 2000.

\bibitem{hampel08}
M.~Hempel and J.~Schaffner-Bielich, ``Mass, radius and composition of the outer
  crust of nonaccreting cold neutron stars,'' {\em Journal of Physics G:
  Nuclear and Particle Physics}, vol.~35, p.~014043, 2007.

\bibitem{baym70}
G.~Baym, C.~Pethick, and P.~Sutherland, ``The {Ground} {State} of {Matter} at
  {High} {Densities}: {Equation} of {State} and {Stellar} {Models},'' {\em The
  Astrophysical Journal}, vol.~170, p.~299, 1971.

\bibitem{kubis06}
S.~Kubis, ``Nuclear symmetry energy and stability of matter in neutron stars,''
  {\em Physical Review C}, vol.~76, p.~025801, 2007.

\bibitem{lattimer95}
J.~M. Lattimer and F.~D. Swesty, ``A generalized equation of state for hot,
  dense matter,'' {\em Nuclear Physics A}, vol.~535, p.~331, 1991.

\bibitem{shen98a}
H.~Shen, H.~Toki, K.~Oyamatsu, and K.~Sumiyoshi, ``Relativistic equation of
  state of nuclear matter for supernova and neutron star,'' {\em Nuclear
  Physics A}, vol.~637, p.~435, 1998.

\bibitem{shen98b}
H.~Shen, H.~Toki, K.~Oyamatsu, and K.~Sumiyoshi, ``Relativistic {Equation} of
  {State} of {Nuclear} {Matter} for {Supernova} {Explosion},'' {\em Progress of
  Theoretical Physics}, vol.~100, p.~1013, 1998.

\bibitem{Carreau19}
T.~Carreau, F.~Gulminelli, and J.~Margueron, ``Bayesian analysis of the
  crust-core transition with a compressible liquid-drop model,'' {\em
  arXiv:1902.07032}, 2019.

\bibitem{kubis04}
S.~Kubis, ``Diffusive instability of a kaon condensate in neutron star
  matter,'' {\em Physical Review C}, vol.~70, p.~065804, 2004.

\bibitem{pethick95}
C.~Pethick, D.~Ravenhall, and C.~Lorenz, ``The inner boundary of a neutron-star
  crust,'' {\em Nuclear Physics A}, vol.~584, p.~675, 1995.

\bibitem{Tsaloukidis19}
L.~Tsaloukidis, C.~Margaritis, and C.~C. Moustakidis, ``Effects of the equation
  of state on the core-crust interface of slowly rotating neutron stars,'' {\em
  Physical Review C}, vol.~99, p.~015803, 2019.

\bibitem{horowitz01b}
C.~J. Horowitz and J.~Piekarewicz, ``Neutron radii of {${}^{208}\mathrm{Pb}$}
  and neutron stars,'' {\em Physical Review C}, vol.~64, p.~062802, 2001.

\bibitem{chomaz04}
P.~Chomaz, M.~Colonna, and J.~Randrup, ``Nuclear spinodal fragmentation,'' {\em
  Physics Reports}, vol.~389, p.~263, 2004.

\bibitem{providencia06}
C.~Provid\^encia, L.~Brito, S.~S. Avancini, D.~P. Menezes, and P.~Chomaz,
  ``Low-density instabilities in relativistic asymmetric matter of compact
  stars,'' {\em Physical Review C}, vol.~73, p.~025805, 2006.

\bibitem{ducoin08a}
C.~Ducoin, J.~Margueron, and P.~Chomaz, ``{Cluster formation in asymmetric
  nuclear matter: {Semi-classical} and quantal approaches},'' {\em Nuclear
  Physics A}, vol.~809, p.~30, 2008.

\bibitem{ducoin08b}
C.~Ducoin, C.~Provid\^encia, A.~M. Santos, L.~Brito, and P.~Chomaz, ``Cluster
  formation in compact stars: {Relativistic} versus {Skyrme} nuclear models,''
  {\em Physical Review C}, vol.~78, p.~055801, 2008.

\bibitem{pais10}
H.~Pais, A.~Santos, L.~Brito, and C.~Provid\^encia, ``Dynamical properties of
  nuclear and stellar matter and the symmetry energy,'' {\em Physical Review
  C}, vol.~82, p.~025801, 2010.

\bibitem{Chen09}
L.-W. Chen, B.-J. Cai, C.~M. Ko, B.-A. Li, C.~Shen, and J.~Xu, ``Higher-order
  effects on the incompressibility of isospin asymmetric nuclear matter,'' {\em
  Physical Review C}, vol.~80, p.~014322, 2009.

\bibitem{Providencia14}
C.~Provid{\^e}ncia, S.~S. Avancini, R.~Cavagnoli, S.~Chiacchiera, C.~Ducoin,
  F.~Grill, J.~Margueron, D.~P. Menezes, A.~Rabhi, and I.~Vida{\~{n}}a,
  ``Imprint of the symmetry energy on the inner crust and strangeness content
  of neutron stars,'' {\em The European Physical Journal A}, vol.~50, p.~1,
  2014.

\bibitem{Pais2016}
H.~Pais, A.~Sulaksono, B.~K. Agrawal, and C.~Provid\^encia, ``Correlation of
  the neutron star crust-core properties with the slope of the symmetry energy
  and the lead skin thickness,'' {\em Physical Review C}, vol.~93, p.~045802,
  2016.

\bibitem{negele72a}
J.~W. Negele and D.~Vautherin, ``{Density-Matrix} {Expansion} for an
  {Effective} {Nuclear} {Hamiltonian},'' {\em Physical Review C}, vol.~5,
  p.~1472, 1972.

\bibitem{negele72b}
J.~W. Negele and D.~Vautherin, ``{Density-Matrix} {Expansion} for an
  {Effective} {Nuclear} {Hamiltonian}. {II},'' {\em Physical Review C},
  vol.~11, p.~1031, 1975.

\bibitem{soubbotin00}
V.~Soubbotin and X.~Vi{\~n}as, ``Extended {Thomas–Fermi} approximation to the
  one-body density matrix,'' {\em Nuclear Physics A}, vol.~665, p.~291, 2000.

\bibitem{soubbotin03}
V.~B. Soubbotin, V.~I. Tselyaev, and X.~Vi{\~n}as, ``Quasilocal density
  functional theory and its application within the extended {Thomas-Fermi}
  approximation,'' {\em Physical Review C}, vol.~67, p.~014324, 2003.

\bibitem{krewald06}
S.~Krewald, V.~B. Soubbotin, V.~I. Tselyaev, and X.~Vi{\~n}as, ``Density matrix
  functional theory that includes pairing correlations,'' {\em Physical Review
  C}, vol.~74, p.~064310, 2006.

\bibitem{behera16}
B.~Behera, X.~Vi{\~n}as, T.~R. Routray, L.~M. Robledo, M.~Centelles, and S.~P.
  Pattnaik, ``Deformation properties with a finite-range simple effective
  interaction,'' {\em Journal of Physics G: Nuclear and Particle Physics},
  vol.~43, p.~045115, 2016.

\bibitem{Chamel2008}
N.~Chamel and P.~Haensel, ``{Physics of Neutron Star Crusts},'' {\em Living
  Reviews in Relativity}, vol.~11, 2008.

\bibitem{Andersson2012}
N.~Andersson, K.~Glampedakis, W.~C.~G. Ho, and C.~M. Espinoza, ``{Pulsar
  Glitches: The Crust is not Enough},'' {\em Physical Review Letters},
  vol.~109, p.~241103, 2012.

\bibitem{Zdunik17}
{J. L. Zdunik}, {M. Fortin}, and {P. Haensel}, ``Neutron star properties and
  the equation of state for the core,'' {\em Astronomy and Astrophysics},
  vol.~599, p.~A119, 2017.

\bibitem{HaenselPichon}
P.~{Haensel} and B.~{Pichon}, ``{Experimental nuclear masses and the ground
  state of cold dense matter},'' {\em Astronomy and Astrophysics}, vol.~283,
  p.~313, 1994.

\bibitem{Ravenhall1994}
D.~G. Ravenhall and C.~J. Pethick, ``{Neutron star moments of inertia},'' {\em
  The Astrophysical Journal}, vol.~424, p.~846, 1994.

\bibitem{Lattimer2005}
J.~M. Lattimer and B.~F. Schutz, ``{Constraining the Equation of State with
  Moment of Inertia Measurements},'' {\em The Astrophysical Journal}, vol.~629,
  p.~979, 2005.

\bibitem{Hartle1967}
J.~B. Hartle, ``{Slowly Rotating Relativistic Stars. I. Equations of
  Structure},'' {\em The Astrophysical Journal}, vol.~150, p.~1005, 1967.

\bibitem{burgay03}
M.~Burgay {\em et~al.}, ``An increased estimate of the merger rate of double
  neutron stars from observations of a highly relativistic system,'' {\em
  Nature}, vol.~426, p.~531, 2003.

\bibitem{Lyne04}
A.~G. Lyne, M.~Burgay, M.~Kramer, A.~Possenti, R.~Manchester, F.~Camilo, M.~A.
  McLaughlin, D.~R. Lorimer, N.~D{\textquoteright}Amico, B.~C. Joshi,
  J.~Reynolds, and P.~C.~C. Freire, ``{A Double-Pulsar System: A Rare
  Laboratory for Relativistic Gravity and Plasma Physics},'' {\em Science},
  vol.~303, p.~1153, 2004.

\bibitem{Yagi13a}
K.~Yagi and N.~Yunes, ``{I-Love-Q: Unexpected Universal Relations for Neutron
  Stars and Quark Stars},'' {\em Science}, vol.~341, p.~365, 2013.

\bibitem{Yagi13b}
K.~Yagi and N.~Yunes, ``{I-Love-Q relations in neutron stars and their
  applications to astrophysics, gravitational waves, and fundamental
  physics},'' {\em Physical Review D}, vol.~88, p.~023009, 2013.

\bibitem{Breu2016}
C.~Breu and L.~Rezzolla, ``{Maximum mass, moment of inertia and compactness of
  relativistic stars},'' {\em Monthly Notices of the Royal Astronomical
  Society}, vol.~459, p.~646, 2016.

\bibitem{Lattimer00}
J.~M. Lattimer and M.~Prakash, ``Nuclear matter and its role in supernovae,
  neutron stars and compact object binary mergers,'' {\em Physics Reports},
  vol.~333-334, p.~121, 2000.

\bibitem{Lattimer07}
J.~M. Lattimer and M.~Prakash, ``Neutron star observations: {Prognosis} for
  equation of state constraints,'' {\em Physics Reports}, vol.~442, p.~109,
  2007.

\bibitem{Flanagan08}
E.~E. Flanagan and T.~Hinderer, ``{Constraining neutron-star tidal Love numbers
  with gravitational-wave detectors},'' {\em Physical Review D}, vol.~77,
  p.~021502, 2008.

\bibitem{Hinderer08}
T.~Hinderer, ``{Tidal Love Numbers of Neutron Stars},'' {\em The Astrophysical
  Journal}, vol.~677, p.~1216, 2008.

\bibitem{Hinderer2010}
T.~Hinderer, B.~D. Lackey, R.~N. Lang, and J.~S. Read, ``Tidal deformability of
  neutron stars with realistic equations of state and their gravitational wave
  signatures in binary inspiral,'' {\em Physical Review D}, vol.~81, p.~123016,
  2010.

\bibitem{Piekarewicz19}
J.~Piekarewicz and F.~J. Fattoyev, ``Impact of the neutron star crust on the
  tidal polarizability,'' {\em Physical Review C}, vol.~99, p.~045802, 2019.

\bibitem{Lourenco19}
O.~Lourenço, M.~Dutra, C.~Lenzi, S.~Biswal, M.~Bhuyan, and D.~Menezes,
  ``{Consistent Skyrme parametrizations constrained by GW170817},'' {\em
  arXiv:1901.04529}, 2019.

\bibitem{Lourenco19a}
O.~Lourenço, M.~Bhuyan, C.~Lenzi, M.~Dutra, C.~Gonzalez-Boquera, M.~Centelles,
  and X.~Vi{\~n}as {\em preprint, to be published to Physics Letters B}.

\bibitem{nakada03}
H.~Nakada, ``Hartree-{Fock} approach to nuclear matter and finite nuclei with
  {M3Y}-type nucleon-nucleon interactions,'' {\em Physical Review C}, vol.~68,
  p.~014316, 2003.

\bibitem{campi78a}
X.~Campi and A.~Bouyssy, ``A simple approximation for the nuclear density
  matrix,'' {\em Physics Letters B}, vol.~73, p.~263, 1978.

\bibitem{campi78b}
A.~Bouyssy and X.~Campi, ``Nuclear densities from theory and experiment,'' {\em
  Nukleonika}, vol.~24, p.~1, 1979.

\bibitem{centelles98}
M.~Centelles, X.~Vi{\~n}as, M.~Durand, P.~Schuck, and D.~Von-Eiff,
  ``Variational {Wigner–Kirkwood} {$\hbar$} expansion,'' {\em Annals of
  Physics}, vol.~266, p.~207, 1998.

\bibitem{gridnev98}
K.~Gridnev, V.~Soubbotin, X.~Vi{\~n}as, and M.~Centelles, {\em {Quantum
  {Theory} in honour of {Vladimir} {A.} {Fock}, {Y.} {Novozhilov} and {V.}
  {Novozhilov}}}.
\newblock Publishing Group of St.Petersburg University, 1998.

\bibitem{durand93}
M.~Durand, P.~Schuck, and X.~Vi{\~n}as, ``On the nuclear curvature energy,''
  {\em Z. Physik A - Hadrons and Nuclei}, vol.~346, p.~87, 1993.

\end{thebibliography}

}

\end{document}